%% file: V-quark.tex
\documentclass[11pt,a4paper]{article}
\pdfoutput=1
\usepackage{jheppub}
\usepackage[T1]{fontenc}
\usepackage[english]{babel}
\graphicspath{{VFigures/}} 

\title{Are the CKM anomalies induced by vector-like quarks? 
Limits from flavor changing and Standard Model precision tests
 }

\author[a,b]{B. Belfatto}
\author[c,d]{and Z. Berezhiani}
\affiliation[a]{Dipartimento di Fisica “E. Fermi”, Universit\`a di Pisa, \\ Largo Bruno Pontecorvo 3, I-56127 Pisa, Italy}
\affiliation[b]{INFN, Sezione di Pisa, \\ Largo Bruno Pontecorvo 3, I-56127 Pisa, Italy}
\affiliation[c]{Dipartimento di Fisica e Chimica, Universit\`a di L'Aquila,\\ 67100 Coppito, L'Aquila, Italy}
\affiliation[d]{INFN, Laboratori Nazionali del Gran Sasso,\\ 67010 Assergi, L'Aquila, Italy}
\emailAdd{benedetta.belfatto@gssi.it}
\emailAdd{zurab.berezhiani@lngs.infn.it}

\abstract{

Recent high precision determinations of $V_{us}$ and $V_{ud}$ indicate 
towards anomalies in the first row of the CKM matrix.
Namely, determination of $V_{ud}$ from superallowed beta decays and of $V_{us}$
from kaon decays
imply a violation of first row unitarity at about $4\sigma$ level.
Moreover, there is tension between determinations of $V_{us}$ obtained from
leptonic $K\mu2$ and semileptonic $K\ell3$ kaon decays.
These discrepancies can be explained if there exist extra vector-like quarks 
at the TeV scale, which have large enough mixings with the lighter quarks. 
In particular, extra vector-like weak singlets quarks can be thought as a solution to the
CKM unitarity problem and an extra vector-like weak doublet can in principle
resolve all tensions.  
The implications of this kind of mixings are examined 
against the flavour changing phenomena and SM precision tests. 
We consider separately the effects of an extra down-type isosinglet,
up-type isosinglet and an isodoublet containing extra quarks of both up and down type, 
and determine available parameter spaces for each case.  
We find that the experimental constraints on flavor changing phenomena 
become more stringent with larger masses, so that the extra species 
should have masses no more than few TeV. 
Moreover, 
only one type of extra multiplet cannot entirely explain 
all the discrepancies, and some their combination is required, 
e.g. two species of isodoublet, or one isodoublet and one (up or down type) isosinglet. 
We show that these scenarios are testable with future experiments. 
Namely, if extra vector-like quarks are responsible for CKM anomalies,
then at least one of them should be found at scale of few TeV, 
and 
anomalous weak isospin violating $Z$-boson couplings with light quarks 
should be detected 
if the experimental precision on $Z$ hadronic decay rate is improved by a factor of $2$ or so.

}

\begin{document} 
\maketitle
\flushbottom

\section{Introduction}
The Standard Model (SM) $SU(3)\times SU(2)\times U(1)$ contains three fermion families with
left-handed (LH) components of quarks $q_{Li}=(u_L,d_L)_i$ and leptons 
$\ell_{Li}=(\nu_L,e_L)_i$, $i=1,2,3$,
forming doublets and
the right-handed (RH) components $u_{Ri}$, $d_{Ri}$, $e_{Ri}$ being
singlets of isotopic symmetry $SU(2)$ of weak interactions.
The charged current weak interactions in terms  of the quark mass eigenstates, up quarks 
$u,c,t$ and down quarks $d,s,b$, are described by the coupling
\begin{align}\label{CC}
&\frac{g}{2\sqrt{2}}\, W_\mu^{+}
\begin{array}{ccc} \overline{ (u ~~ c ~~ t ) } \end{array} \gamma^\mu 
(1-\gamma^5)\, V_\text{CKM}
\left(\begin{array}{c} d \\ s  \\ b \end{array}\right)  ~ +~ \text{h.c.}
%
\end{align}
where $V_\text{CKM}$ 
is the Cabibbo-Kobayashi-Maskawa (CKM) matrix:
\begin{align}
\label{vckmsm}
&V_\text{CKM}=\left(\begin{array}{ccc}
V_{ud} & V_{us} & V_{ub}  \\
V_{cd} & V_{cs} & V_{cb}  \\
V_{td} & V_{ts} & V_{tb} 
\end{array}\right) 
\end{align}
In the SM context  $V_\text{CKM}$ should be unitary.
Any deviation from its unitarity 
can be a signal of new physics beyond the SM (BSM).

At present,
the determinations of $|V_{us}|$ and $|V_{ud}|$ have reached high enough precision to test
with high accuracy
 the unitarity of the first row in CKM matrix (\ref{vckmsm}):
\begin{align}
\label{uns}
& |V_{ud}|^2+|V_{us}|^2+|V_{ub}|^2=1
\end{align}
Namely, precision experimental data on kaon decays, in combination with the latest
lattice QCD calculations  
of the decay constants and form-factors, provide accurate information about  $\vert V_{us}\vert $. 
On the other hand, recent calculations of short-distance radiative corrections in $\beta$-decays
substantially improved
the determination of $|V_{ud}|$. 
Since the contribution of
$\vert V_{ub} \vert \simeq 0.004$ is very small and actually negligible, 
the test of the sum rule in eq. (\ref{uns})
is practically equivalent to a Cabibbo universality check. 
 
In our previous paper ref. \cite{Belfatto:2019swo} it was pointed out that there is a significant
(about $4\sigma$) anomaly in the first row unitarity (\ref{uns}),  
after using three types of independent determinations of $|V_{us}|$ 
and $|V_{ud}|$,
which were dubbed as determinations of type A, B and C.
Specifically, determination A corresponds to the direct determination of $\vert V_{us} \vert$ from the 
kaon semileptonic ($K\ell 3$) decays, B comes from the determination of the ratio
$\vert V_{us}/V_{ud}\vert $ obtained from charged kaon leptonic ($K\mu 2$) decays by comparing them
with pion leptonic decays, and C corresponds to the direct determination
of $\vert V_{ud} \vert$ from superallowed $0^{+}$--$0^{+}$ nuclear transitions by employing 
the value of the Fermi constant obtained from the muon decay, $G_{F}=G_{\mu}$.
%


For explaining this anomaly we proposed two possible BSM scenarios. 
One is related to a new physics in the lepton sector. Namely,
we considered the horizontal gauge symmetry $SU(3)_{\ell}\times SU(3)_{e}$ 
between the lepton families, with $SU(3)_{\ell}$ acting between LH states
$\ell_{Li}=(\nu_L,e_L)_i$ and  $SU(3)_{e}$ 
acting between RH ones $e_{Ri}$.
We have shown that flavour changing gauge bosons of $SU(3)_{\ell}$ induce 
an effective operator which contributes to the muon decay in positive interference 
with the SM contribution ($W$-boson exchange). In this way,  
the muon decay constant $G_\mu$ becomes different from the Fermi constant $G_{F}$: 
$G_\mu = G_F (1+\delta_\mu)$ with $\delta_\mu =(v_w/v_F)^2$,  
where $v_w$ and $v_F$ respectively are the electroweak and horizontal symmetry breaking scales.
Since the values of $|V_{us}|$ and $|V_{ud}|$ 
are normally extracted by assuming $G_F = G_\mu $, in this scenario they
are shifted by a factor $1+\delta_\mu$ while their ratio is not affected.
CKM unitarity is recovered with
$\delta_\mu \simeq 7\times 10^{-4}$ which corresponds to a horizontal breaking scale
of about $6$ TeV.
Interesting point is that such a low mass scale for the horizontal gauge bosons is not in conflict 
with the stringent experimental limits on the lepton flavour changing processes as 
$\mu \rightarrow 3e$, $\tau\to 3\mu$ etc. \cite{Belfatto:2019swo}. 
The breaking scale of $SU(3)_{e}$ symmetry of the RH leptons 
can also be as small as few TeV without contradicting
experimental limits \cite{B1}.

Another (more straightforward) possibility discussed in ref. \cite{Belfatto:2019swo} 
is to introduce extra vector-like quarks.\footnote{After ref. \cite{Belfatto:2019swo} 
the problem of the CKM unitarity anomaly was addressed  with different approaches 
in several subsequent papers 
\cite{Grossman:2019bzp,Pagliara,Cheung:2020vqm, Endo:2020tkb, Capdevila:2020rrl,Crivellin:2020ebi,Kirk:2020wdk,Coutinho:2020xhc,Alok:2020jod,Crivellin:2021njn}. }
%
In particular, with extra isosinglet quarks of down-type $b'$ or up-type $t'$ 
one can settle the CKM unitarity problem
which results from
the determination of
$V_{ud}$ from superallowed beta decays (C) and $V_{us}$ from kaon decays (A and B),
whereas by employing the extra quarks forming the weak isodoublet $(t',b')$
all the tensions between the independent determinations A, B and C can in principle be explained.

However, large mixings with SM families induce flavour changing phenomena which can be
in potential conflict with stringent experimental limits.
In this work we give a detailed study of the effects on relevant flavour changing processes
and electroweak observables
and constrain the parameter space for each scenario 
(extra weak isosinglets of up-type or down-type or weak isodoublets).

As we will show, there still remains some available parameter space which can 
satisfy these stringent constraints but it is very limited and can be excluded with future 
experimental data. In particular, it can be excluded
if the limits on masses of extra vector-like species will increase up to $3$ TeV or so
or the limits on some relevant flavour changing phenomena or $Z$ boson physics
will further strengthen.
Therefore, all these scenarios can be falsified in close future.

The paper is organized as follows.
Since after ref. \cite{Belfatto:2019swo} some new data appeared, in section \ref{situation} we update
the analysis of the CKM first row anomalies. 
In section \ref{inter} we discuss the generalities about the role of different types of vectorlike quarks
in fixing the problem.
In sections \ref{sec-down} and \ref{sec-up} we analyze separately the scenarios with 
extra weak isosinglet quarks of down-type ($b'$) and up-type ($t'$),
by providing a detailed study of flavour changing phenomena induced in this scenarios and 
determining the available parameter space.
In section \ref{sec-doub} we perform the analysis in the case of additional
extra weak isodoublet $(t',b')$.
In section \ref{other} we discuss some combinations in case more families are introduced.
At the end, in section \ref{conclusion} we give our conclusions.

\section{Present situation in the determination of $|V_{us}|$ and $|V_{ud}|$}
\label{situation}

As already stated, the precision of
recent determinations of $|V_{us}|$ and $|V_{ud}|$ allows to test 
the first row unitarity (\ref{uns})  of CKM matrix.
Deviation from unitarity can be parameterized as
\begin{align}
\label{newundelta}
& \vert V_{ud} \vert^2 + \vert V_{us} \vert^2 +\vert V_{ub} \vert^2   = 
1-\delta_\text{CKM}
\end{align} 
Hence, the value $\delta_\text{CKM}$ shows  the measure of the unitarity deficit. 


The element $\vert V_{us} \vert $ can be directly determined from
semileptonic $K\ell3$ decays ($K_L \mu3$, $K_L e3$,  $K^{\pm}e3$, etc.) which imply
\cite{Moulson}: 
\begin{align}
\label{f+}
& f_+(0) \vert V_{us}\vert   = 0.21654\pm 0.00041 
\end{align}
where $f_+(0)$  is the 
vector form factor at zero momentum transfer which can be computed in the lattice QCD simulations. 
The average of 4-flavor computations reported by FLAG 2019  is 
$f_+(0)=0.9706(27)$ \cite{FLAG2019}. 
We combine it with the latest 4-flavor result $f_+(0) = 0.9696(19)$ \cite{Bazavov} (which was
not included in FLAG 2019 \cite{FLAG2019}) getting  $f_+(0)=0.9699(15)$.  
In this way, from eq. (\ref{f+})   we obtain 
the value of $\vert V_{us} \vert$ (determination A in  the following) as: 
\begin{align}
\label{A}
&\text{A}: \qquad \vert V_{us} \vert_\text{A} = 0.22326(55)
\end{align}

An independent information (determination B in the following) 
stems from the ratio of the kaon and pion leptonic 
decay rates $K\to \mu\nu(\gamma)$ and $\pi \to \mu\nu(\gamma)$ which implies \cite{PDG18}:
\begin{align}
\label{fK}
& \vert V_{us}/V_{ud} \vert \times (f_{K^\pm}/f_{\pi^\pm}) = 0.27599 \pm 0.00038
\end{align}
Then, by employing the 4-flavour average for 
the decay constants ratio 
 $f_{K^\pm}/f_{\pi^\pm}=1.1932(19)$  reported in FLAG 2019 \cite{FLAG2019}, 
 we obtain:
\begin{align}
\label{rapporto}
& \text{B} :  \qquad |V_{us}|/|V_{ud}|=0.23130(49)
\end{align} 

As regards the element $\vert V_{ud} \vert $, its  most precise determination is obtained from
superallowed $0^+$-- $0^+$ nuclear $\beta$-decays, 
which are pure Fermi transitions 
sensitive only to the vector coupling constant $G_V=G_F \vert V_{ud} \vert$.
The master formula reads \cite{Hardy,Hardy2}:
\begin{align}
\label{Vud-super}
& \vert V_{ud} \vert^2 = \frac{K }{ 2 G_F^2 \mathcal{F} t\, (1+ \Delta_R)}  
= \frac{0.97147(20)}{ 1+ \Delta_R } 
\end{align}
where $K = 2\pi^3 \ln 2/m_e^5 = 8120.2776(9) \times 10^{-10}$~s/GeV$^4$,
$G_F=G_\mu  = 1.1663787(6) \times 10^{-5}$~GeV$^{-2}$ is the Fermi constant determined 
from the muon decay \cite{mulan} and
$\mathcal{F} t = 3072.07(72)\, \text{s}$ is the nucleus independent value derived from 
$ft$-values of $14$ best determined superallowed $0^+$-- $0^+$ nuclear transitions  
by absorbing in the latter all transition-dependent (so called outer) corrections
 \cite{Hardy2}.
The major uncertainty is related to the transition independent short-distance 
(so called inner) radiative correction $\Delta_R$ 
which in 2006 was computed by Marciano and Sirlin  
obtaining $\Delta_R =0.02361(38)$ \cite{Marciano}.

However, a recent calculation with improved hadronic uncertainties brought to a 
drastically different value $\Delta_R = 0.02467(22)$ \cite{Seng}.
A more conservative approach of ref. \cite{Marciano2} gives a slightly lower result with
relatively larger uncertainty, $\Delta_R = 0.02426(32)$.
For our analysis we decided to use a democratic average  
of these two results, $\Delta_R = 0.02454(27)$. 
This means that we `democratically' take as uncertainty  
simply the arithmetical average of the error bars reported in original 
refs. \cite{Seng} and \cite{Marciano2}.\footnote{If we would treat the errors as statistical fluctuations and 
combine the errors in the standard way, then we would have to reduce our 
`democratic' error $(27)$  to $(18)$. 
However, since we deal with the results of  theoretical calculations, we decided 
to be conservative in treating their uncertainties.}
Then, using this average, from eq. (\ref{Vud-super}) 
we get the value of $\vert V_{ud} \vert$ (determination C in the following):
 \begin{align}
  \label{vudmedio}
 & \text{C} :  \qquad 
 \vert V_{ud} \vert = 0.97376(16)
  \end{align}

The value of $\vert V_{ud} \vert$ can be extracted  also from 
the free neutron $\beta$-decay:
\begin{align}
& \vert V_{ud} \vert^2 = \frac{K/ \ln 2 }{ G_F^2 \mathcal{F}_n \tau_n \, (1+3g_A^2)  (1+ \Delta_R)} 
=  \frac{5024.46(30)~{\rm s}}{\tau_n (1+3g_A^2)(1+ \Delta_R)} 
\end{align}
where $\mathcal{F}_n = f_n(1+\delta'_R)$ is the neutron $f$-value $f_n= 1.6887(1)$ 
corrected by the long-distance QED correction $\delta'_R = 0.01402(2)$ \cite{Hardy3}.  
However it is less precise  
due to limited accuracy in the experimental determination of the neutron lifetime $\tau_n$ and 
the axial coupling constant $g_{A}=G_{A}/G_{V}$. 
Moreover, there is an apparent tension ($4\sigma$) between the neutron lifetime measurements  
using the bottle and beam experimental methods, 
origin of which requires more profound understanding and perhaps some new physics \cite{puzzle}. 
%
%
E.g.  by combining the recent determination of axial coupling $g_A=1.27625(50)$ and 
the ``bottle" lifetime $\tau_n^{\rm bottle} = 879.4(0.6)$~s as in ref. \cite{Belfatto:2019swo},  
one gets:  
\begin{align}
& \vert V_{ud} \vert = 0.97333(33)_{\tau_n}(32)_{g_A}(11)_{\Delta_R}
\label{vudneut}
\end{align}
which is compatible with our choice (\ref{vudmedio}) but has about 3 times larger errors. 
(Interestingly, by comparing the determinations of  $\vert V_{ud} \vert$ 
from free neutron decays and superallowed $0^+$-- $0^+$ decays, the factor $1+\Delta_R$  
cancels out and one obtains an accurate determination of the neutron lifetime 
$\tau_n = 5172.0(1.1)/(1+3g_A^2) = 878.7(0.6)$ \cite{Czarnecki} which  well agrees 
with $\tau_n^{\rm bottle}$ but is in strong tension with $\tau_n^{\rm beam} = 888.0(2.0)$~s 
\cite{puzzle}.) 
The measurements of
$\pi^{+}\rightarrow\pi^{0}e^{+}\nu$ branching ratio by PIBETA experiment   \cite{Pocanic:2003pf} 
lead to the independent result $\vert V_{ud} \vert =0.9728(30)$  
which however has about 20 times larger uncertainties as compared to (\ref{vudmedio}). 
Therefore, we take in consideration only determination C 
obtained from superallowed $0^+$-- $0^+$ transitions. 

\begin{figure}[t]
\centering
\includegraphics[width=0.75\textwidth]{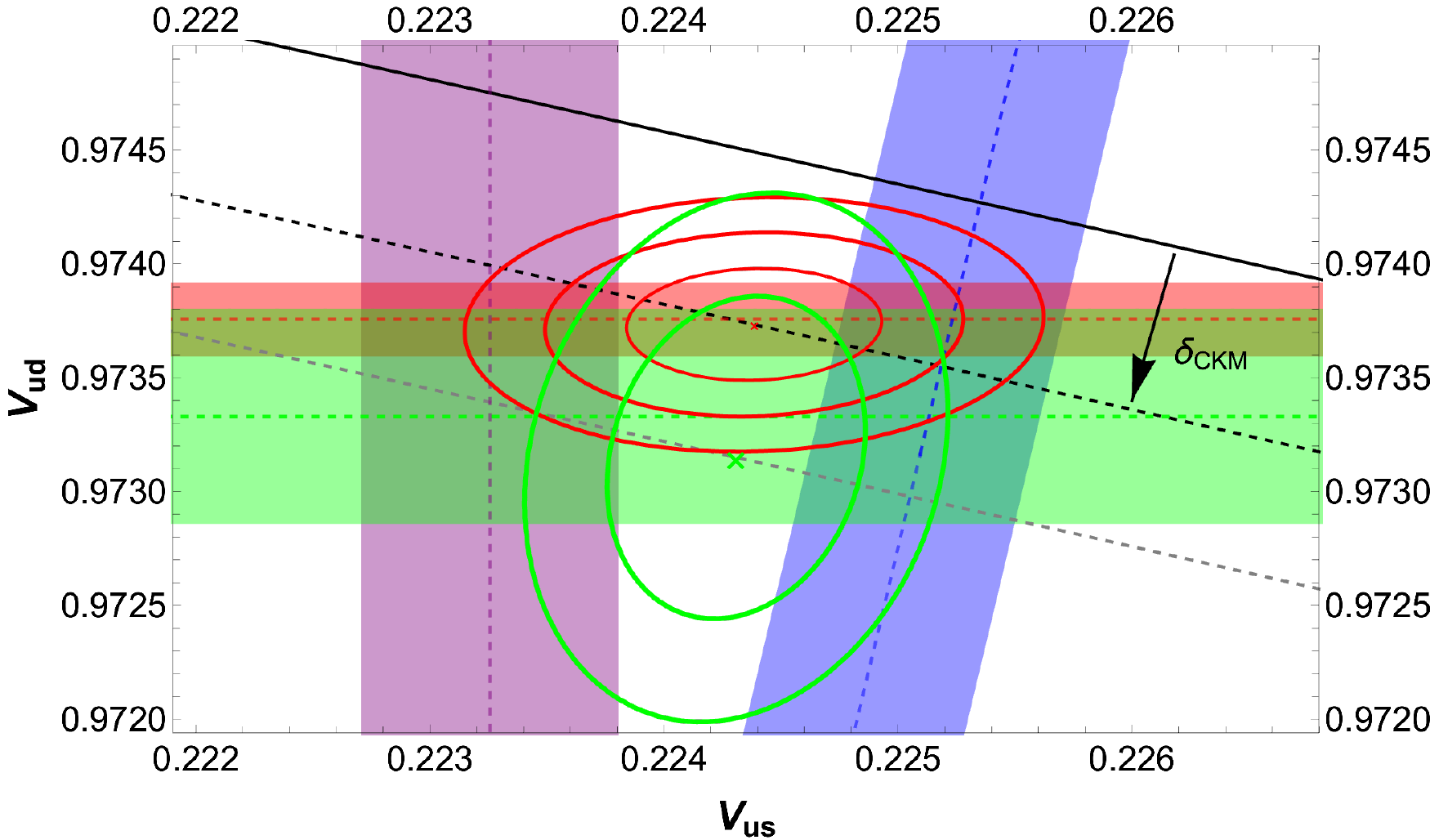}
\caption{\label{plotC3} The purple, blue 
and red bands correspond  respectively 
to the values of $\vert V_{us} \vert $ from eq. (\ref{A}), 
$\vert V_{us}/V_{ud} \vert $  from eq. (\ref{rapporto}) and
 $\vert V_{ud} \vert =0.97376(16)$ from eq. (\ref{vudmedio}). 
The best fit point and  $1$, $2$ and $3\sigma$ contours are shown (red cross and red circles)
for these values. 
The black curve corresponds to the unitarity condition (\ref{uns}).
The dashed black curve corresponds to eq. (\ref{newundelta})
with the deficit of unitarity $\delta_\text{CKM}=1.48\cdot 10^{-3}$.
The green band correspond to the value 
 $\vert V_{ud} \vert =0.97333(47)$ obtained from neutron decay (\ref{vudneut}). 
The best fit point and  $1$ and $2\sigma$ contours 
(green cross and green circles) are obtained by fitting the latter determination with the other
two determinations $\vert V_{us} \vert $ from eq. (\ref{A}), 
$\vert V_{us}/V_{ud} \vert $  from eq. (\ref{rapporto}).
The dashed gray curve corresponds to eq. (\ref{newundelta})
with the deficit of unitarity $\delta_\text{CKM}=2.6\cdot 10^{-3}$.
}
\end{figure}

The tensions between determinations A, B and C are shown in figure \ref{plotC3},
which presents the fit of 
the values (\ref{A}),  (\ref{rapporto}), (\ref{vudmedio}), 
with $V_{us}$ and $V_{ud}$ considered as independent parameters, without imposing unitarity.
The unitarity condition (\ref{uns}) is shown with the black continuous line.
 The best fit (minimum $\chi^2$) corresponds to 
\begin{align}
\label{vusvudfit}
& |V_{us}| = 0.22439(36)   \qquad  |V_{ud}| = 0.97373(16)  
\end{align}
which is $3.6\sigma$ away from the unitarity curve. 
The $\chi^2$ value is rather large, $\chi^2 =7.4$, 
due to the tension between the two determinations A and B from kaon decays.
For the deficit of CKM unitarity the best fit values (\ref{vusvudfit}) imply  
$\delta_\text{CKM}=1.48(36)\cdot 10^{-3}$.

The tensions can be manifested also in another way. 
We can take $\vert V_{us} \vert_\text{A} = 0.22326(55)$ from the direct 
determination A  while B and C can be 
also translated in $\vert V_{us}\vert $ determinations by imposing the unitarity condition  (\ref{uns}).
Namely, the value of $\vert V_{us}\vert$ obtained in this way from eq. (\ref{rapporto}) is:
\begin{align}
\label{B}
& 
\vert V_{us} \vert_\text{B} = 0.22535(45)
\end{align}
which is compatible also with a theoretical result $\vert V_{us} \vert  = 0.22567(42)$ 
from $K\mu2$ decays obtained in ref. \cite{Martinelli}. 

From determination C (\ref{vudmedio}) we get instead:
 \begin{align}
\label{c}
& 
\vert V_{us} \vert_\text{C} = 0.22756(70)
\end{align}
For completeness, in table \ref{TableC} we show the values of $|V_{ud}|$ and respective
``unitarity'' values of $|V_{us}|$ corresponding to the choices of $\Delta_R$
as reported in original refs. \cite{Marciano}, \cite{Seng}, \cite{Marciano2}, 
indicated as C$_{0}$, C$_{1}$ and C$_{2}$.

Figure \ref{situation} displays the values $|V_{us}|_\text{A}$ (\ref{A}), 
$|V_{us}|_\text{B}$ (\ref{B}) and $|V_{us}|_\text{C}$ (\ref{c}) 
with corresponding error bars (shaded areas).
Between determinations obtained from kaon decays, A and B, there is
$2.9\sigma$ tension, which maybe could disappear 
 with more accurate lattice simulations.\footnote{These determinations 
 obtained from 3-flavor lattice computations \cite{FLAG2017}  were in fact compatible 
 because of larger error-bars,  see figure \ref{situation}. } 
Therefore,  we conservatively take 
a democratic average of  $|V_{us}|_\text{A}$ and
$|V_{us}|_\text{B}$ without reducing error bars
(with the uncertainty taken as arithmetical average of two uncertainties):
\begin{align}
\label{AB}
& \overline{\text{A+B}} \, : \qquad |V_{us}|=0.22451(50)
\end{align}
We see that there is about $5\sigma$ tension between determination A and C and 
$3\sigma$ tension between B and C. 
The discrepancy of the average $\overline{\text{A+B}}$ from C
results in $3.6\sigma$.
Let us notice that if we try to fit the incompatible
determinations of $V_{us}$ $A$ (\ref{A}), $B$ (\ref{B}), $C$ (\ref{c}), we would get
the average value $|V_{us} |= 0.22511$ but with a ugly large
$\chi^2$ value, $\chi^2 =24$.

\begin{table}
\centering
\begin{tabular}{| l @{\hspace{3\tabcolsep}}  c @{\hspace{4\tabcolsep}}  c @{\hspace{4\tabcolsep}}  c |}
\hline
Determination  &  $\Delta_R $ & $|V_{ud}|$ & $|V_{us}|$ \\
\hline
C$_0$   \cite{Marciano} & $0.02361(38)$ & $0.97420(21)$ & $0.2257(9)$ \\
C$_1$ \cite{Seng} &  $0.02467(22)$  &  $0.97370(14)$ & $0.22780(60)$ \\
C$_2$ \cite{Marciano2} &  $0.02426(32)$ & $0.97389(18)$ & $0.22699(77)$ \\
C (our choice)        &$0.02454(27)$ &   $0.97376(16)$ &  $0.22756(70)$ \\ 
\hline
\end{tabular}
\caption{\label{TableC} 
Values of $\Delta_R$ reported in original references
and corresponding values of $|V_{ud}|$ obtained from eq. (\ref{Vud-super}). 
Values of $|V_{us}|$ are obtained assuming unitarity (\ref{uns}).
C represents our `democratic' average (see text).
}
\end{table}

\begin{figure}[t]
\centering
\includegraphics[width=0.8\textwidth]{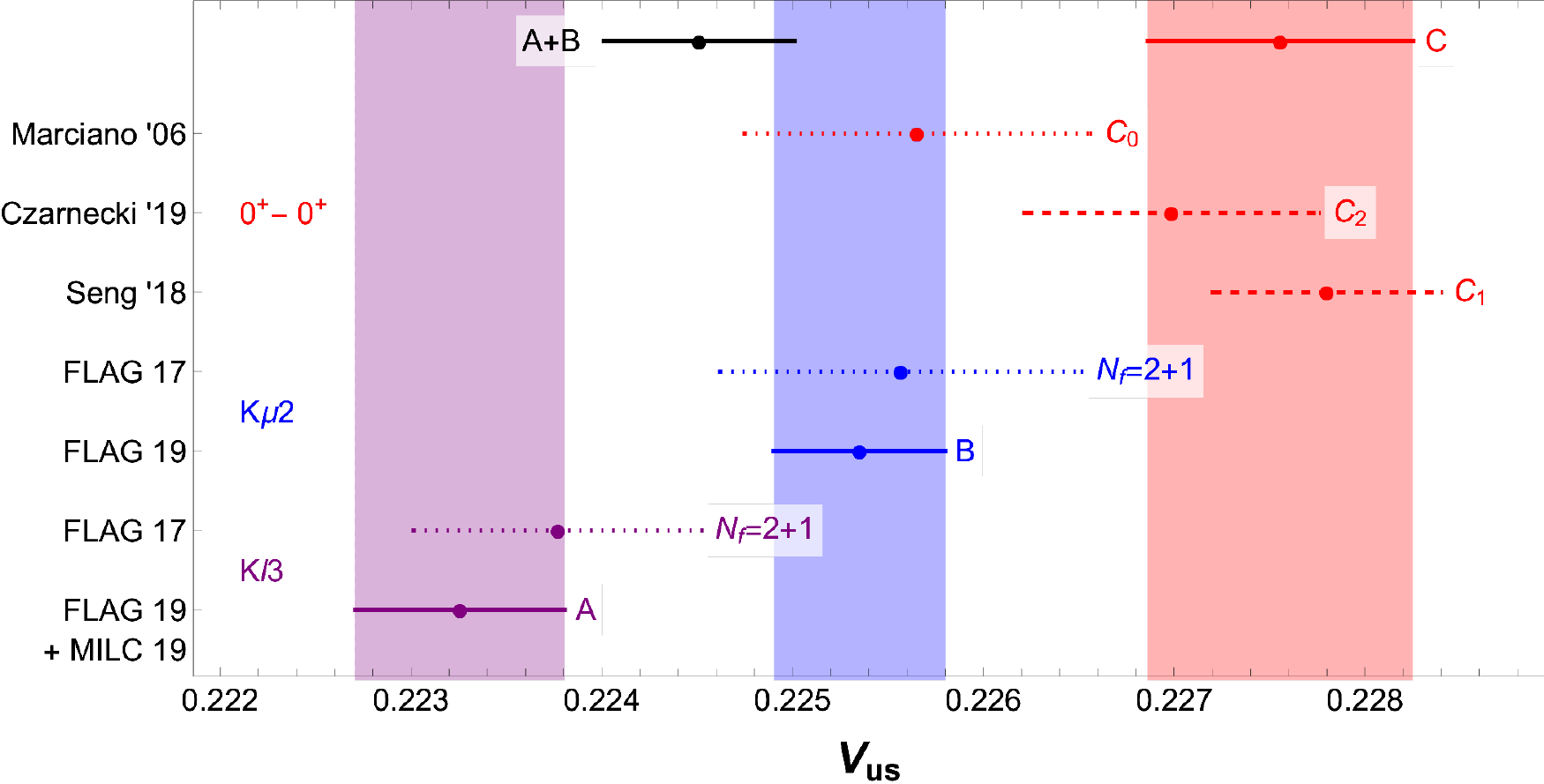}
\caption{\label{situation} 
Shaded areas 
show the values of $|V_{us}|$  obtained from determinations A (\ref{A}),
B (\ref{rapporto}) and C (\ref{vudmedio}) by assuming CKM unitarity (\ref{uns}), while
the black line corresponds to the democratic average of A and B (\ref{AB}) (see text). 
C$_{0}$, C$_{1}$, C$_{2}$ are the values of $|V_{us}|$ obtained from the values of
$\Delta_{R}$ reported by original references, as listed in table \ref{TableC}.
For comparison we also show the values of $|V_{us}|$ 
obtained using $3$ flavours lattice QCD simulations as reported in FLAG 17 \cite{FLAG2017}
and were adopted in Particle Data Group 2018. These determinations have practically no tension
with the old determination C$_{0}$ \cite{Marciano}. Hence, this picture demonstrate that
the CKM tensions in fact emerged due to improved precision of $4$-flavours computations 
\cite{FLAG2019,Bazavov} on one side, 
and due to the changes in inner radiative correction $\Delta_{R}$ \cite{Seng,Marciano2}
on the other side.
}
\end{figure}

\begin{table}
\centering
\begin{tabular}{|@{\hspace{.5\tabcolsep}} l @{\hspace{1\tabcolsep}}  c @{\hspace{1\tabcolsep}}  c   @{\hspace{1\tabcolsep}} c @{\hspace{0.3\tabcolsep}}|}
\hline
            &  A: $|V_{us}|\! =\! 0.22326(55)$ &  B: $|V_{us}|\! =\! 0.23130(49) |V_{ud}| $  & Average$^{*}$ \\
\hline
C$_1$:  $|V_{ud}| \! =\! 0.97370(14) $   
          &  $ 2.05(37)\cdot 10^{-3} $  &  $1.17(35)\cdot 10^{-3} $   & $1.55(36)\cdot 10^{-3}$  \\
C$_2$:  $|V_{ud}| \! =\! 0.97389(18)$  
          & $1.68(43)\cdot 10^{-3}$   &  $0.78(41)\cdot 10^{-3}$  &  $1.17(42)\cdot 10^{-3} $ \\
C: \hspace{2pt} $|V_{ud}| \!= \!0.97376(16) $  
          & $1.94(40)\cdot 10^{-3}$     &  $1.05(38)\cdot 10^{-3}$    &  $1.43(39)\cdot 10^{-3}$ \\
\hline
\end{tabular}
\caption{\label{Tabledelta2} 
Values of $\delta_\text{CKM}$ obtained for different choices of the values of 
$|V_{us}|$ and $|V_{ud}|$.\newline
$^{*}$ We average the values of $V_{us}$ given in the columns 
 A and B  conservatively,  taking its error bar as arithmetical average of their
 uncertainties. 
  }
\end{table}

In table \ref{Tabledelta2} we show the landscape of possible values of the unitarity deficit 
$\delta_\text{CKM}$. We see that depending on the choice of the data 
this value spans from about
$10^{-3}$ to about $2\cdot 10^{-3}$. 


In conclusion,
it is clear that the CKM unitarity condition with three families (\ref{uns})
is inconsistent with the present determinations of $|V_{us}|$ and $|V_{ud}|$.
An immediate solution would be to introduce a fourth sequential family $(t',b')$,
analogous to the three SM families,
with the LH components forming weak isodoublets and RH components being isosinglets.
Then the $3\times 3$ CKM matrix (\ref{vckmsm}) should be extended 
to a unitary $4\times 4$ matrix:
\begin{align}
& V_\text{CKM}=\left(\begin{array}{cccc}
V_{ud} & V_{us} & V_{ub} & V_{ub'} \\
V_{cd} & V_{cs} & V_{cb} & V_{cb'}  \\
V_{td} & V_{ts} & V_{tb}  & V_{tb'} \\
V_{t'd} & V_{t's} & V_{t'b}  & V_{t'b'} 
\end{array}\right).
\end{align}
and correspondingly the first row unitarity condition would be modified to:
\begin{align}
\label{newunb}
& \vert V_{ud} \vert^2 + \vert V_{us} \vert^2 +\vert V_{ub} \vert^2  + \vert V_{ub'} \vert^2 = 1
\end{align} 
Comparing this equation with eq. (\ref{newundelta}) we see 
that the parameter
$\delta_\text{CKM}$ assumes the meaning of the mixing with the fourth family, 
$\delta_\text{CKM}= |V_{ub'}|^{2}$. 
Therefore for typical values of $\delta_\text{CKM}$ given in table \ref{Tabledelta2}
we get $\vert V_{ub'}\vert \approx 0.04$,
which is comparable with $\vert V_{cb}\vert$ and 
an order of magnitude larger than $\vert V_{ub} \vert \approx 0.004$.
It looks not very natural 
that the mixing of the first family with fourth family is much stronger than
its mixing with third family,
but some models admit this possibility \cite{Bregar}.
Unfortunately,
the existence of a fourth sequential family is excluded by the limits from electroweak precision data
combined with the LHC data. 

However, vector-like quarks 
can be introduced without any contradiction with SM precision tests.
The LHC limits merely tell that their masses should be above $1$ TeV or so.

In the rest of this paper we discuss how the anomalies in the first row of CKM matrix
can be solved by introducing extra vector-like fermions.
In particular we will consider the role of  
weak isosinglets of down-type or up-type and weak isodoublets.

In fact, two approaches to the problem can be considered.
The incompatibility inside kaon physics may be attributed to some uncertainties
which can disappear maybe soon with more precise determinations,
focusing instead on the average of determinations from kaons.
Then the problem consist in solving the lack of unitarity in the first row of $V_\text{CKM}$.
The insertion of an extra vectorlike weak isosinglet, down-type or up-type, is on this line.

Otherwise the discrepancy inside kaon physics can be considered seriously 
by looking for a solution addressing the whole situation. 
However, the $V_{us}$ anomaly (discrepancy between $V_{us}$ determinations  
from the kaon semileptonic and leptonic decays) can be considered seriously,
\footnote{In fact, a recent high precision determination of $K\ell 3$ radiative corretions \cite{Seng:2021boy}
indicates that the SM electroweak effects are not large enough to account for $V_{us}$ anomaly.}  
and one has to look for a solution addressing the whole situation. 
As it will be shown, a weak isodoublet 
can in principle explain all the anomalies.

In these scenarios with extra vector-like families, 
the unitarity deficit $\delta_\text{CKM}$ will be related to the mixing
of the extra quarks with the SM families.

\section{The SM with extra vector-like fermions}
\label{inter}

The Standard Model $SU(3)\times SU(2) \times U(1)$ contains, by definition, 
three chiral families of fermions, 
the LH  quarks $q_{Li}=(u_L,d_L)_i$ and leptons $\ell_{Li}=(\nu_L,e_L)_i$ 
transforming as $SU(2)$  isodoublets and
the RH components $u_{Ri}$, $d_{Ri}$, $e_{Ri}$ being isosinglets, 
with $i=1,2,3$ being the family index.  
This set of fermions is free of gauge anomalies. 
Attractive property of the SM is that these fermions can acquire masses only 
 via the Yukawa couplings with Higgs doublet $\varphi$: 
\begin{align}
\label{SM-Yuk}
& {\cal L}_{\rm  Yuk}^{\rm SM}  
=  Y_u^{ij} \tilde\varphi\, \overline{q_{Li} } u_{Rj}  + Y_d^{ij}\varphi \, \overline{q_{Li}} d_{Rj}  + 
Y_e^{ij}\varphi \, \overline{\ell_{Li}} e_{Rj}  ~ + ~ {\rm h.c.} 
\end{align} 
where $Y_{u,d,e}^{ij}$ are the Yukawa constant matrices, and $\tilde\varphi = i\tau_2 \varphi^\ast$.  
The known species of quarks and leptons  
are eigenstates of mass matrices $M_{u,d,e} = Y_{u,d,e} v_w$ 
where $\langle \varphi^0\rangle = v_w$ is  the Higgs VEV.  
In other words, in the SM the quark and lepton masses are induced only  
after the electroweak symmetry breaking, and their values are proportional to the 
electroweak scale $v_w$. 

The quark mass matrices can be diagonalized by bi-unitary transformations 
\begin{align}\label{LLRR}
& V_{Lu}^\dagger M_u V_{Ru} = \widetilde{M}_u  = {\rm diag}(m_u,m_c,m_t), \quad 
V_{Ld}^\dagger M_d V_{Rd} = \widetilde{M}_d  = {\rm diag}(m_d,m_s,m_b) 
\end{align}  
and the weak eigenstates in terms of mass eigenstates are:
\begin{align}\label{VVVV}
&\left(\begin{array}{c} u_{L1} \\ u_{L2} \\ u_{L3}  \end{array}\right)
=V_{Lu}\left(\begin{array}{c}
u_L \\ c_L \\ t_L  \end{array}\right);   
&\left(\begin{array}{c} d_{L1} \\ d_{L2} \\ d_{L3} \end{array}\right) 
=V_{Ld}\left(\begin{array}{c} d_L \\ s_L \\ b_L  \end{array}\right) 
\end{align}
The CKM mixing matrix in $W$ boson  charged current couplings (\ref{CC}) 
emerges as a combination of  the `left' unitary transformations,  
$V_{\rm CKM} = V_{Lu}^\dagger V_{Ld}$, and thus it should be unitary. 
As for the `right' matrices $V_{Ru}$ and $V_{Rd}$, in the SM frames they 
 have no physical significance. Without loss of generality, one can choose 
 a fermion basis in which one of the Yukawa matrices $Y_u$ or $Y_d$ is diagonal  
 in which cases  we would have respectively $V_{\rm CKM} = V_{Ld}$ or 
 $V_{\rm CKM} = V_{Lu}^\dagger$. 

The SM exhibits a remarkable feature of natural suppression of flavor-changing neutral currents (FCNC) 
\cite{Glashow, Paschos}:  no flavor mixing emerges in neutral currents coupled to $Z$ boson
and Higgs boson.  In particular, this means that $Z$ boson tree level couplings 
with the fermion mass eigenstates  remain diagonal after rotations (\ref{VVVV}).   
On the other hand,  the Yukawa matrices $Y_{u,d,e}$   and mass matrices $M_{u,d,e}$ 
are proportional  and thus by transformations (\ref{LLRR}) they are diagonalized simultaneously, 
so that  the Yukawa couplings of  the Higgs boson $H$ with the fermion mass eigenstates are diagonal. 
Hence, all FCNC phenomena are suppressed at tree level and emerge exclusively 
from radiative corrections. 
At present, the majority of experimental data on flavor changing and CP violating processes 
are in good agreement with the SM predictions. 

Clearly, in the SM framework  the unitarity of the CKM matrix as well as the natural 
flavor conservation in neutral currents are direct consequences following from the fact that 
the three families are in identical representations of $SU(3)\times SU(2) \times U(1)$. 


However, in addition to three chiral families of quarks and leptons, 
there can exist extra vector-like species, 
with the LH and RH in the same representations of the SM.  
In particular, one can consider the extra fermion species in the same representations of 
$SU(3)\times SU(2) \times U(1)$ 
as standard quarks and leptons, namely in the form of weak isosinglets 
of down quark type $D_{L,R}$, up quark type $U_{L,R}$ 
and charged lepton type $E_{L,R}$, and weak isodoublets $Q_{L,R} = ({\cal U},{\cal D})_{L,R}$
and $L_{L,R} = ({\cal N},{\cal E})_{L,R}$ of quark and lepton types.\footnote{ 
Such vector-like species are predicted in some extensions 
of the Standard Model. For example, $D$ and $L$ type species emerge (per each family) 
 in the context of minimal $E_6$ \cite{Gursey,Achiman} or $SU(6)$  \cite{Dvali} 
grand unifications.    
In addition, the specifics of the latter model in which Higgs emerges as pseudo-Goldstone particle, 
requires  at least one copy of $Q$, $U$ and $E$ type species 
for inducing the fermion masses and in particular the top quark mass \cite{Barbieri,su6}. }
(Extra vector-like fermions can be introduced also in other representations as e.g.  
$SU(2)$ isotriplets which can contain quark or lepton type fragments 
but also some fragments with exotic electric charges but here we do not address these cases.) 
The mass terms of these species are not protected by the SM gauge symmetries and 
hence their masses can be (or must be) considerably larger than the electroweak scale.  

In the following we shall concentrate on the quark sector. Namely, we consider a theory which, 
besides the three chiral families of standard quarks $u_{Ri}$, $d_{Ri}$ and $q_{Li}=(u_L,d_L)_i$ 
($i=1,2,3$), includes some extra vector-like quark species 
$U_{L,R}$, $D_{L,R}$ and $Q_{L,R}=({\cal U},{\cal D})_{L,R}$ 
which in principle can be introduced in different amounts.  
%
Therefore, 
along with the standard Yukawa terms for the three chiral families:\footnote{
Hereafter indices $i=1,2,3$ of normal families as well as indices of extra species are suppressed.}
\begin{align}
\label{SM-q}
& {\cal L}_{\rm  Yuk}   
=  Y_u \tilde\varphi \, \overline{q_{L} } u_{R}  + Y_d \varphi\, \overline{q_{L}} d_{R}   ~ + ~ {\rm h.c.} 
\end{align} 
the most general Lagrangian of this system
must include the mixed Yukawa terms between the standard and extra species: 
\begin{align}
\label{BSM-Yuk}
&  {\cal L}_{\rm  Yuk}^{\rm mix} = h_U \tilde\varphi \, \overline{q_{L} } U_{R} 
+ h_D \varphi  \, \overline{q_{L} } D_{R}   
+ h_{\cal U} \tilde\varphi \, \overline{Q_{L}} u_{R}  +  h_{\cal D} \varphi \, \overline{Q_{L}} d_{R} 
 ~ + ~ {\rm h.c.} 
\end{align} 
and the mass terms 
\begin{align}
\label{BSM-mass}
& {\cal L}_{\rm mass} = 
M_U  \overline{{U}_L} {U}_R  + M_{D} \overline{{D}_L} {D}_R  
+ M_Q \overline{Q_L} Q_R + 
\mu_u  \overline{{U}_L} u_R +  \mu_d  \overline{{D}_L} d_R  
+ \mu_q  \overline{q_L} Q_R ~+~ {\rm h.c.}    
\end{align} 
where $h_{U,D}$ and $h_{u,d}$ in the Yukawa terms (\ref{BSM-Yuk}) and 
$M_{U,D,Q}$ and $\mu_{u,d,q}$ in mass terms (\ref{BSM-mass}) are 
the matrices of proper dimensions depending on the amounts of extra species. 
One could introduce also the Yukawa couplings between extra species:
\begin{align}\label{extra-Yuk}
& {\cal L}_{\rm  Yuk}^{\rm extra} = \lambda_U \tilde\varphi \, \overline{Q_{L} } U_{R}  
+ \lambda_{\cal U} \tilde\varphi \, \overline{Q_{R} } U_{L} 
+ \lambda_D \varphi  \, \overline{Q_{L} } D_{R}  +  \lambda_{\cal D} \varphi  \, \overline{Q_{R} } D_{L}  
 ~ + ~ {\rm h.c.} 
\end{align} 
However, they play no relevant role in further discussions and for simplicity we neglect them. 

The vector-like quarks are key players in models with horizontal inter-family symmetries 
\cite{PLB83,Dimopoulos,PLB85,Bagger,Khlopov1,Khlopov2,King,so10}, 
in some models of the axion \cite{Kim,Shifman,Khlopov,Zphys,Sakharov} 
as well as in axionless (Nelson-Barr type) models for solving the strong CP  problem
via spontaneous CP or P violation  \cite{Nelson,Barr,Babu,MPL,Senjanovic}. 
In these models the  values $M_{U,D,Q}$ and $\mu_{u,d,q}$ in mass terms (\ref{BSM-mass})
are related to the breaking  scales of  respective symmetries, and  
in some models e.g. \cite{Babu,Senjanovic,Berezhiani:1990} 
they can be as low as few TeV. 

Interestingly, some of these symmetries (e.g. flavor symmetry or Peccei-Quinn symmetry) 
may forbid the direct Yukawa terms (\ref{SM-q}) 
but allow the mixed ones (\ref{BSM-Yuk})   
while  mass terms $\mu$ and $M$ in (\ref{BSM-Yuk})  
can be originated from some physical scales.\footnote{ E.g. in the `seesaw' model of ref. \cite{PLB83}  
the values of $\mu_{u,d}$ and $M_{U,D}$  are  respectively determined 
by the breaking scales of left-right symmetry and family symmetry,  
i.e. by the VEV of the `right'  Higgs doublet 
and VEVs of flavon scalars which break  the horizontal symmetry.}
Nevertheless,  despite that the original constants $Y_{u,d}$ in (\ref{SM-q}) are vanishing, 
the SM Yukawa terms (\ref{SM-Yuk}) for normal fermions  
will be induced after integrating out the heavy states. In particular,    
provided that mixing mass terms $\mu_{u,d,q}$  are smaller than $M_{U,D,Q}$,  we obtain  
\begin{align}
\label{seesaw}
& Y_u \simeq h_U M_U^{-1} \mu_u + \mu_q M_Q^{-1} h_{\cal U}, \quad  
Y_d \simeq h_D M_D^{-1} \mu_d + \mu_q M_Q^{-1} h_{\cal D}  
\end{align}
In other words, the non-zero quark masses are induced via the mixings 
with the extra vector-like species. 
Such a scenario known as `universal' seesaw mechanism \cite{PLB83,Rajpoot,Davidson,Rattazzi}  
is commonly used in predictive model building for fermion masses and mixings as e.g. 
\cite{PLB83,Rattazzi,Rattazzi2,tanbeta,Anderson:1993fe,Berezhiani:1996bv,Koide1,Koide2,Nesti}.  
In the context of supersymmetric models with flavor symmetry this mechanism can 
give a natural realization of  the minimal flavor violation scenario via the alignment 
of soft supersymmetry breaking terms with the Yukawa terms \cite{MFV,MFV1,MFV2}. 

In the following we are not  interested in the model details and in possible dynamical 
effects of the underlying symmetries broken at higher scales, but only in the effects of the mixing 
between the three normal (chiral) and extra (vector-like) quarks. 
Therefore, we can conveniently redefine the fermion basis.  
Namely, the species $u_R$ and $U_R$, $d_R$ and $D_R$, and $q_L$ and $Q_L$, 
 are in the identical representations of  $SU(3)\times SU(2) \times U(1)$. 
 Thus, by redefining these species, one can eliminate mixed mass terms $\mu_u$, 
 $\mu_d$ and $\mu_q$ in (\ref{BSM-mass}) by `absorbing' them respectively 
 in the mass terms $M_U$, $M_D$ and $M_Q$
 (this means that e.g. from $3+n$ RH species with quantum numbers of $d_R$ we can 
 always select $n$ their combinations which `marry'  $n$ species of LH fermions ${\cal D}_L$
 via mass terms while the remaining $3$ combinations have no mass terms). 
 In addition,  without losing generality, 
 the `heavy' mass matrices $M_U$, $M_D$ and $M_Q$ can be taken to be diagonal and real. 
 
 In this basis the total mass matrices of up type $(u,U,{\cal U})_{L,R}$ and down type 
$(d,D,{\cal D})_{L,R}$ quarks, after substituting the Higgs VEV $\langle \phi \rangle = v_w$,  read: 
 \begin{align}\label{matrices}
& {\cal M}_{\rm up} = \left(\begin{array}{ccc}
Y_u v_w& h_Uv_w  & 0 \\
0 & M_U & 0   \\
h_{\cal U} v_w & 0 & M_Q  
\end{array}\right),   \quad \quad 
{\cal M}_{\rm down}  = \left(\begin{array}{ccc}
Y_d v_w & h_Dv_w  & 0 \\
0 & M_D & 0   \\
h_{\cal D} v_w & 0 & M_Q  
\end{array}\right) 
\end{align} 
where the  blocks $Y_{u,d}$ are matrices of dimensions $3\times 3$.  
Assuming that the numbers of extra species $U$, $D$ and $Q$ are respectively 
$p$, $n$ and $m$, then blocks $M_U$, $M_D$ and $M_Q$ should be 
correspondingly of dimensions $p\times p$, $n\times n$ and $m\times m$. 
Thus, $ {\cal M}_{\rm up}$ and $ {\cal M}_{\rm down}$ respectively are 
$(3+p+m)\times (3+p+m)$ and $(3+n+m)\times (3+n+m)$ matrices. 


The mass matrices (\ref{matrices}) can be brought to the diagonal forms via bi-unitary transformations 
$({\cal V}^{L}_{\rm up})^\dagger {\cal M}_{\rm up} {\cal V}^R_{\rm up}=\widetilde{\cal M}_{\rm up}$ and  
$({\cal V}^{L}_{\rm down})^\dagger {\cal M}_{\rm down} {\cal V}^R_{\rm down}=\widetilde{\cal M}_{\rm down}$. 
In this way, the  initial states 
of e.g. down-type quarks are related to their physical states (mass eigenstates) as 
  \begin{align}\label{unitary-D}
& 
\left(\begin{array}{c} d \\ D  \\ {\cal D}   \end{array}\right)_{\!\!L,R} = 
{\cal V}_{\rm down}^{L,R} 
\left(\begin{array}{c} d'  \\ D'  \\ {\cal D}'  \end{array}\right)_{\!\!L,R}=
\left(\begin{array}{ccc}
V_{dd'}^{L,R} & V_{dD'}^{L,R} & V_{d{\cal D}' }^{L,R} \\
V_{Dd'}^{L,R} & V_{DD'}^{L,R} & V_{D{\cal D}' }^{L,R} \\
V_{{\cal D}d'}^{L,R} & V_{{\cal D}D'} ^{L,R} &  V_{{\cal D}{\cal D}'}^{L,R}
\end{array}\right) 
\left(\begin{array}{c} d'   \\ D'   \\ {\cal D}'   \end{array}\right)_{\!\!L,R}    
\end{align} 
Here $d  = (d_1,d_2,d_3)^T$ are initial states and  $d' = (d,s,b)^T$ 
are the mass eigenstates, and similarly  for heavy species $D$ and ${\cal D}$. 
Analogously, unitary matrices ${\cal V}^{L,R}_{\rm up}$ connect the initial up-quark type states 
$u, U, {\cal U}$ with their mass eigenstates $u', U', {\cal U}'$, where $u= (u_1,u_2,u_3)^T$ 
and $u' = (u,c,t)^T$. 

Since we have 
$\widetilde{\cal M}_{\rm down}^2 = 
({\cal V}^{L}_{\rm down})^\dagger \, {\cal M}_{\rm down}{\cal M}_{\rm down}^\dagger \, {\cal V}^L_{\rm down}=
({\cal V}^{R}_{\rm down})^\dagger\, {\cal M}_{\rm down}^\dagger {\cal M}_{\rm down}\, {\cal V}^R_{\rm down}  $, 
unitary matrices $ {\cal V}^{L}_{\rm down}$ and $ {\cal V}^{R}_{\rm down}$ can be determined 
by considering the hermitean squares of ${\cal M}_{\rm down}$: 
\begin{align}\label{matrice-squares}
&& {\cal M}_{\rm down} {\cal M}_{\rm down}^\dagger = \left(\begin{array}{ccc}
v_w^2 Y_dY_d^\dagger  + v_w^2 h_D h_D^\dagger & v_w  h_D  M_D & v_w^2  Y_d h_{\cal D}^\dagger \\
v_w M_D h_D^\dagger  & M_D^2 & 0   \\
 v_w^2 h_{\cal D} Y_d^\dagger & 0 & M_Q^2 + v_w^2 h_{\cal D} h_{\cal D}^\dagger  \end{array}\right),  
  \nonumber \\
&& {\cal M}_{\rm down}^\dagger {\cal M}_{\rm down}  = \left(\begin{array}{ccc}
v_w^2 Y_d^\dagger Y_d +v_w^2 h_{\cal D}^\dagger h_{\cal D}   & v^2_w Y_d^\dagger h_D  & 
v_w h_{\cal D}^\dagger M_Q \\
v_w^2 h_D^\dagger Y_d  & M_D^2 + v_w^2 h_D^\dagger h_D  & 0   \\
 v_w  M_Q h_{\cal D} & 0 & M_Q^2  
\end{array}\right) \, . 
\end{align} 
The off-diagonal entries of these matrices  are fixed by the value $v_w$,    
so that the elements of the off-diagonal blocks $V_{dD'}$,  $V_{d{\cal D}'}$ etc. in (\ref{unitary-D}) 
are determined by the ratio of the electroweak scale to the masses of extra quark species. 
In the limit when the latter are very heavy they decouple and 
their mixings with light quarks become negligibly small. 
Thus, in this limit 
 $3\times 3$ block $V_{dd'}$  becomes unitary. 
 The same is true for analogous $V_{uu'}$ block in up quark mixing.  
 However, if the extra quarks are not that heavy and off-diagonal blocks 
 are not negligible, then $V_{dd'}$ and $V_{uu'}$ blocks are no more unitary.  


The present experimental limits on the extra quark masses 
are  $M_{U,D,Q}>(1\div 1.5)$~TeV, depending on their type and decay modes \cite{PDG18}.
Therefore, the ratios  $\epsilon_u = v_w/M_U$, $\epsilon_d= v_w/M_D$ and 
$\epsilon_q= v_w/M_Q$ can be considered as small parameters, $\epsilon_{u,d,q} \leq 0.1$ or so. 
By inspecting the matrices (\ref{matrice-squares}), one can estimate
the elements of the off-diagonal blocks  in 
${\cal V}_{L,R}^{\rm down}$ and ${\cal V}_{L,R}^{\rm up}$ as  
  \begin{align}\label{orders} 
& 
\vert V^L_{dD'} \vert \sim \epsilon_d, \quad   \vert V^L_{uU'} \vert \sim \epsilon_u  , &&
\qquad \vert V^L_{d{\cal D}'} \vert , \vert V^L_{u{\cal U}'} \vert \sim \epsilon_q^2 \qquad
\nonumber \\
&\qquad \vert V^R_{d{\cal D}'} \vert,  \vert V^R_{u{\cal U}'} \vert \sim \epsilon_q, 
&& \vert V^R_{dD'} \vert \sim \epsilon_d^2, \quad \vert V^R_{uU'} \vert  \sim \epsilon_u^2  \qquad
\end{align}
modulo the Yukawa constants which are assumed to be $\leq 1$ for perturbativity. 
Therefore,  the deviation from unitarity of the ``left" matrices $V_{dd'}^L$ and $V_{uu'}^L$ blocks are 
$\sim \epsilon^2_{u,d} \leq 10^{-2}$.   
E.g. the first row unitarity of the matrix ${\cal V}^{L}_{\rm down}$  implies 
$\vert V_{1d} \vert^2 + \vert V_{1s} \vert^2 +\vert V_{1b} \vert^2  = 1 - \vert V_{1D'} \vert^2  
- \vert V_{1{\cal D}'} \vert^2 = 1 - \delta_d$. 
Taking into account the above estimations, we see that the deviation can be as large as 
 $\delta_d \sim \epsilon_d^2 \sim 10^{-2}$.   
Let us recall that the CKM  unitarity deficit $\delta_{\rm CKM}$ estimated in previous section is about   
$(1 \div 2) \times 10^{-3}$, 
see table \ref{Tabledelta2}. 
Thus, for accounting for the above values of $\delta_{\rm CKM}$, one would need $\epsilon_d = 0.03-0.05$. 
As for  $\epsilon^4 \sim 10^{-4}$ contributions,  they are irrelevant and 
so order $\epsilon^2$ mixings as  $V_{1{\cal D}'}$ etc. can be safely neglected.

Let us assume, for simplicity, that each of $U$, $D$ and $Q$ type species is present in one copy,  
i.e.  $p=m=n=1$ 
(our discussion can be extended in a straightforward way for arbitrary number of extra species). 
In this case the off-diagonal blocks proportional to $v_w$ 
in $5\times 5$ matrices (\ref{matrices})   become columns 
as e.g. $h_D = (h_{1D}, h_{2D},h_{3D})^T$  or rows as e.g. 
$h_{\cal D} = (h_{{\cal D}1}, h_{{\cal D}2},h_{{\cal D}3})$.  
The Yukawa couplings $Y_{u,d}$ can be presented in the form  
$Y_u= V_{Lu} \widetilde{Y}_u V_{Ru}^\dagger$  and $Y_d = V_{Ld} \widetilde{Y}_d V_{Rd}^\dagger$ 
where $\widetilde{Y}_{u,d}$ are diagonal $3\times 3$ matrices, 
$\widetilde{Y}_u = {\rm diag}(y_u,y_c,y_t)$ and $\widetilde{Y}_d = {\rm diag}(y_d,y_s,y_b)$. 
Let us denote also 
$h_U = V_{Lu} \widetilde{h}_U$,  $h_D = V_{Ld} \widetilde{h}_D$, 
$h_{\cal U}= \widetilde{h}_{\cal U} V_{Ru}^\dagger$ and $h_{\cal D} = \widetilde{h}_{\cal D} V_{Rd}^\dagger$. 

Then for $5\times 5$ unitary matrix of `left' rotations we obtain, with the precision up to $\epsilon^2$ terms:
\begin{align}\label{V-d}
& {\cal V}_{\rm down}^L  = \left(\begin{array}{ccc}
V_{Ld} & 0 & 0  \\
0  & 1 & 0   \\
0 & 0 & 1  \end{array}\right)  \!
\left(\begin{array}{ccc}
\!\! \! \sqrt{1- S_D S_D^\dagger }   &  S_D  & 0\\
-S_D^\dagger  & \!\! \sqrt{1-   S_D^\dagger S_D } & 0   \\
0 & 0 & 1 
\end{array}\right)  \!
\left(\begin{array}{ccc}
\!\! \! \sqrt{1- S_{\cal D} S_{\cal D}^\dagger }   &  0  & S_{\cal D} \\
0  & 1 & 0   \\
-S_{\cal D}^\dagger  & 0 & \sqrt{1-  S_{\cal D}^\dagger S_{\cal D} } 
\end{array}\right)    
  \nonumber \\
& \quad \quad \quad 
= \left(\begin{array}{ccc}
V_{Ld} \big[ 1-\frac12 S_D S_D^\dagger  - \frac12 S_{\cal D} S_{\cal D}^\dagger + \dots \big] &  
V_{Ld}  S_D  & V_{Ld} S_{\cal D} \\
-S_D^\dagger  & 1- \frac12 S_D^\dagger S_D  
+ \dots   & O(\epsilon_d\epsilon_q^2)   \\
-S_{\cal D}^\dagger  & 0 & 1 - \frac12  S_{\cal D}^\dagger  S_{\cal D} + \dots  
\end{array}\right) 
%
%
\end{align} 
where the column $S_D = \epsilon_d \widetilde{h}_D$ describes the light quark ($d,s,b$) mixings 
with the extra isosinglet species $D$. As for their  mixings with ${\cal D}\subset Q$ from extra isodoublet, 
$S_{\cal D} = \epsilon_q^2 \widetilde{Y}_d \widetilde{h}_{\cal D}^\dagger$,  it can be neglected  
since, apart of $\epsilon^2$ suppression, these are proportional to 
the small Yukawa constants $y_d,y_s,y_b$ in $\widetilde{Y}_d$. 
Clearly, the unitarity conditions for rows and columns of this matrix is fulfilled with the 
precision up to $\sim \epsilon_d^4$ terms.  The matrix ${\cal V}_{\rm up}^L$ can be 
presented in an analogous form. 


Let us discuss now charged current interactions. 
Considering that $q_L$ and $Q_L$ are $SU(2)$ doublets while $U_L$ and $D_L$  
are singlets, the LH charged current  interacting with $W$ boson 
in terms of initial states  and mass eigenstates reads:  
  \begin{align}\label{CC-L} 
& \frac{g}{\sqrt2}\, W^+_\mu \, \overline{(u_L ~ U_L  ~ {\cal U}_L ) } \, \gamma^\mu \, J_{L} 
\left(\begin{array}{c} d_L  \\ D_L  \\ {\cal D}_L  \end{array}\right) = 
 \frac{g}{2\sqrt2}\, W^+_\mu \, \overline{(u'  ~~ U'   ~~ {\cal U}' ) } \, \gamma^\mu (1-\gamma^5) 
 \, {\cal V}_{L}^{\rm mix}
\left(\begin{array}{c} d'  \\ D'  \\ {\cal D}'   \end{array}\right)
%
\end{align}  
where $J_L = {\rm diag}(1,1,1,0,1)$ and ${\cal V}_{L}^{\rm mix} = 
({\cal V}_{L}^{\rm up})^\dagger J_L {\cal V}_{L}^{\rm down}$, or explicitly  
  \begin{align}\label{mix-L}
& {\cal V}^{\rm mix}_{L} =
\left(\begin{array}{ccc}
\!\!(1-\frac12 \Delta_U) \widetilde{V} (1-\frac12 \Delta_D)  
& (1-\frac12 \Delta_U) \widetilde{V} S_D &  O( \epsilon^2)  \\
S_U^\dagger \widetilde{V} (1-\frac12 \Delta_D) & S_U^\dagger \widetilde{V} S_D & O( \epsilon^3) \\ 
O(\epsilon^2) & O( \epsilon^3) & 1 
\end{array}\right)\! , \quad  \widetilde{V}  = V_{Lu}^\dagger V_{Ld}
%
\end{align} 
where $\Delta_U = S_U S_U^\dagger = \epsilon_u^2  \widetilde{h}_U  \widetilde{h}_U^\dagger$ 
and $\Delta_D = S_D S_D^\dagger = \epsilon_d^2  \widetilde{h}_D  \widetilde{h}_D^\dagger$ 
as far as  $O(\epsilon_q^4)$ contributions $S_{\cal U} S_{\cal U}^\dagger$ and  
$S_{\cal D} S_{\cal D}^\dagger$ can be neglected. 

 We are interested in its $3\times 3$ block which describes the transitions between 
 the quark mass eigenstates $u'=(u,c,t)$ and $d'=(d,s,b)$ in charged current: 
\begin{align}\label{CKM-new}
&V_\text{CKM}= \big(1-\frac12 \Delta_U\big)  
\, \widetilde{V} \, \big(1-\frac12 \Delta_D \big)  
=\left(\begin{array}{ccc}
V_{ud} & V_{us} & V_{ub}  \\
V_{cd} & V_{cs} & V_{cb}  \\
V_{td} & V_{ts} & V_{tb} 
\end{array}\right) 
\end{align} 
While $\widetilde{V} = V_{Lu}^\dagger V_{Ld}$ is unitary $3\times 3$ matrix, 
the `corrected' matrix  $V_{\rm CKM} $ is not. In particular, deviation from the unitarity 
for its rows or columns read, up to order $\epsilon^2$ terms,  respectively as\footnote{Obviously, 
the `large' mixing matrix ${\cal V}^{\rm mix}_{L}$ is not unitary in itself because of the non-unitary factor $J_L$ 
`sandwiched' between the unitary matrices ${\cal V}^{\rm up}_{L}$ and  ${\cal V}^{\rm down}_{L}$.  }
\begin{align}\label{rows}
& V_{\rm CKM} V_{\rm CKM}^\dagger= 1-\epsilon_u^2 \widetilde{h}_U  \widetilde{h}_U^\dagger 
 - \epsilon_d^2 \widetilde{V} \widetilde{h}_D  \widetilde{h}_D^\dagger \widetilde{V}^\dagger , \nonumber \\
& V_{\rm CKM}^\dagger  V_{\rm CKM}= 1-\epsilon_d^2 \widetilde{h}_D  \widetilde{h}_D^\dagger 
 - \epsilon_u^2 \widetilde{V} \widetilde{h}_U  \widetilde{h}_U^\dagger \widetilde{V}^\dagger 
\end{align} 
In particular, the unitarity deficit for first row (\ref{newundelta}) we obtain 
$\delta_{CKM} = 1-  \vert V_{ud} \vert^2 + \vert V_{us} \vert^2 +\vert V_{ub} \vert^2 \approx 
\epsilon_u^2 \vert \widetilde{h}_{U1} \vert^2 + \epsilon_d^2 \vert \widetilde{h}_{D1} \vert^2$, 
which can fall in the range of $(1\div 2) \times 10^{-3}$ provided that $\epsilon_u$ and/or 
$\epsilon_d$ are $\sim 0.1$ and the Yukawa constants $\widetilde{h}_{U1}$ and $\widetilde{h}_{D1}$  
are large enough.


Let us discuss the RH sector. 
Considering that $Q_R$ is an $SU(2)$ doublet while $u_{R}$, $d_{R}$, $U_R$ and $D_R$  
are singlets, the RH charged current interacting with $W$ boson 
in terms of initial states  and mass eigenstates reads:  
  \begin{align}\label{CC-R} 
& \frac{g}{\sqrt2}\, W^+_\mu \, \overline{(u_R ~ U_R  ~ {\cal U}_R ) } \, \gamma^\mu \, J_{R} 
\left(\begin{array}{c} d_R  \\ D_R  \\ {\cal D}_R  \end{array}\right) = 
 \frac{g}{2\sqrt2}\, W^+_\mu \, \overline{(u'  ~~ U'   ~~ {\cal U}' ) } \, \gamma^\mu (1+\gamma^5) 
 \, {\cal V}_{R}^{\rm mix}
\left(\begin{array}{c} d'  \\ D'  \\ {\cal D}'   \end{array}\right)
%
\end{align}  
where $J_R = {\rm diag}(0,0,0,0,1)$ and ${\cal V}_{R}^{\rm mix} = 
({\cal V}_{R}^{\rm up})^\dagger J_R {\cal V}_{R}^{\rm down}$. 
Thus, presenting matrices ${\cal V}_{R}^{\rm down}$  which diagonalizes 
${\cal M}_{\rm down}^\dagger {\cal M}_{\rm down}$ (\ref{matrice-squares}) 
in the form similar to ${\cal V}_{L}^{\rm down}$ (\ref{V-d}), and analogously doing for 
and ${\cal V}_{R}^{\rm up}$, we get up to $O(\epsilon^2)$ terms: 
  \begin{align}\label{mix-R}
& {\cal V}^{\rm mix}_{R} =
\left(\begin{array}{ccc}
S_u S_d^\dagger& 0 &  -S_u (1- \frac12  S_d^\dagger S_d)  \\
0 & 0 & 0 \\ 
-(1- \frac12 S_u^\dagger S_u ) S_d^\dagger \; & 0 &\; 1 -  \frac12 S_d^\dagger S_d -  \frac12S_u^\dagger  S_u 
\end{array}\right) 
\end{align} 
where $S_d = \epsilon_q \widetilde{h}_{\cal D}^\dagger$ and $S_u = \epsilon_q \widetilde{h}_{\cal U}^\dagger$. 
Thus, we see that the mixing with the weak isodoublet $Q$-type species induces RH charged 
current interactions 
between the quark mass eigenstates $u'=(u,c,t)$ and $d'=(d,s,b)$, 
given by the (non-unitary) $3\times 3$ matrix:
  \begin{align}\label{CKM-R}
\Delta = 
\left(\begin{array}{ccc}
\Delta_{ud} & \Delta_{us} & \Delta_{ub}  \\
\Delta_{cd} & \Delta_{cs} & \Delta_{cb}  \\
\Delta_{td} & \Delta_{ts} & \Delta_{tb} 
\end{array}\right) = 
S_u S_d^\dagger = \epsilon_q^2 \widetilde{h}_{u}^\dagger  \widetilde{h}_{d} = \epsilon_q^2
\left(\begin{array}{ccc}
h_{u}^\ast {h}_{d} & h_{u}^\ast {h}_{s}  &  h_{u}^\ast {h}_{b} \\
h_{c}^\ast {h}_{d}  & h_{c}^\ast {h}_{s}   & h_{c}^\ast {h}_{b}  \\
h_{t}^\ast {h}_{d} & h_{t}^\ast {h}_{s} &  h_{t}^\ast {h}_{b}
\end{array}\right)
%
\end{align} 
where we denote the elements of row vectors $\widetilde{h}_{\cal U,D}$ in the light quark mass basis 
as $\widetilde{h}_{\cal U}= (h_u, h_c, h_t)$
and $\widetilde{h}_{\cal D}= (h_d, h_s, h_b)$. 
Therefore, mixing with $Q$-type fermions violates pure $V-A$ character of the quark interactions 
with $W$ boson, and vector and axial couplings for each transition are not equal anymore but 
have a difference $O(\epsilon_q^2)$.   

In fact, instead of purely $V-A$ couplings (\ref{CC}), now we have 
  \begin{align}\label{CC-VA} 
&  \frac{g}{2\sqrt2}\, W^+_\mu \, \overline{(u  ~~ c   ~~ t ) } \,  
 \big[ \gamma^\mu (V_{\rm CKM} + \Delta) - \gamma^\mu \gamma^5 (V_{\rm CKM} - \Delta ) \big]
 \, \left(\begin{array}{c} d  \\ s  \\ b   \end{array}\right)
%
\end{align}

The presence of RH couplings has a direct implications for our problem. 
In particular,  determination C from purely Fermi $0^+-0^+$ transitions now fixes the 
vector coupling constant $G_V = G_F \vert V_{ud} + \Delta_{ud} \vert$, instead  of $G_F \vert V_{ud} \vert$. Analogously, determination A from semileptonic decays $K\ell 3$, transforms in the determination 
of the vector coupling $\vert V_{us} + \Delta_{us} \vert$, instead  of $\vert V_{us} \vert$. 
On the other hand, since the leptonic decays $K\mu2$ and $\pi\mu2$ are contributed only by axial current, 
determination B instead of the ratio $\vert V_{us}/V_{ud}\vert $ fixes now the combination 
$\vert V_{us} - \Delta_{us}\vert / \vert V_{ud} - \Delta_{ud} \vert$. 

Therefore, instead of (\ref{vudmedio}), (\ref{A}) and (\ref{rapporto}), now we have:
\begin{align}
& \text{C} \, : \qquad 
\vert V_{ud} + \Delta_{ud} \vert = \vert V_{ud}\vert  \, \vert 1+ \delta_{ud} \vert  = 0.97376(16) 
\label{C-new} \\
&\text{A}: \qquad \vert V_{us} + \Delta_{us} \vert =  \vert V_{us} \vert \, \vert 1 + \delta_{us} \vert  = 0.22326(55) 
\label{A-new} \\
& \text{B} :  \qquad 
\left \vert \frac{V_{us} - \Delta_{us} } {V_{ud}-\Delta_{ud} } \right\vert = 
\left \vert \frac{V_{us} } {V_{ud} } \right\vert \, 
\left \vert \frac{1 - \delta_{us} } {1 -\delta_{ud} } \right\vert 
=0.23130(49)
\end{align} 
where $\delta_{ud} = \Delta_{ud}/V_{ud} = \epsilon_q^2 \, h_u^\ast h_d/V_{ud}$ and 
$\delta_{us} = \Delta_{us}/V_{us} = \epsilon_q^2 \, h_u^\ast h_s/V_{us}$
are in general complex numbers.   
Hence, $Q$-type extra fermion can have interesting implications and potentially 
it can  resolve all tensions between A, B and C determinations.\footnote{In principle, one can obtain
the RH weak currents also from left-right 
symmetric models. However, strong limits on $W_{R}$ mass 
imply that the mixing $W_{L}$-$W_{R}$ is too small to give the RH contributions needed to explain the anomalies.}

\begin{figure}[t]
\centering
\includegraphics[width=0.48\textwidth]{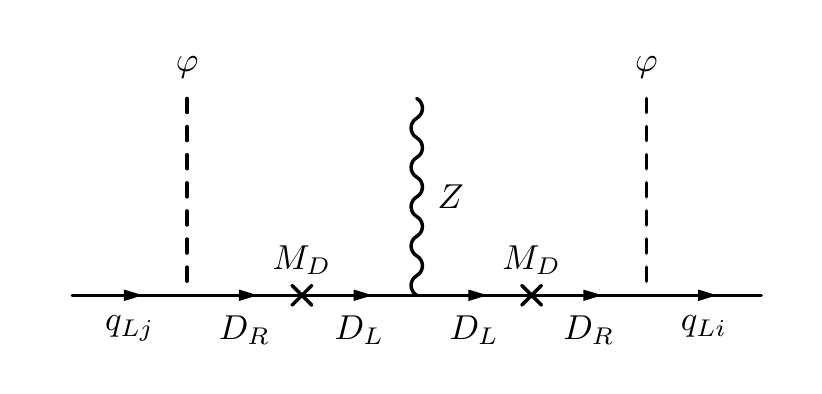}
\includegraphics[width=0.48\textwidth]{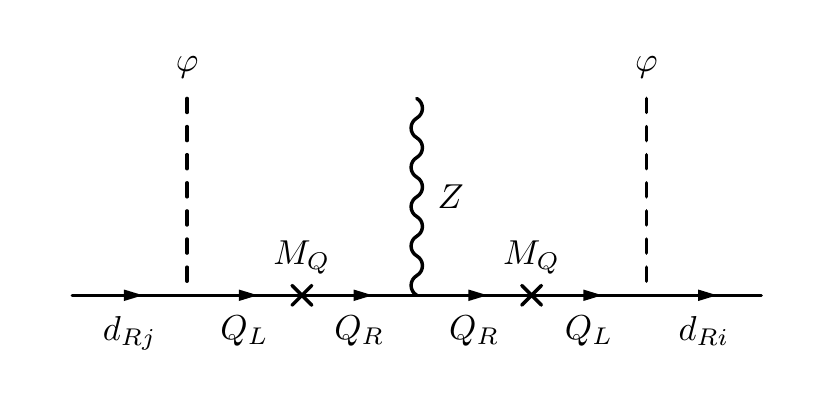}
\caption{\label{ddZ} Anomolous flavour non-diagonal couplings of $Z$-boson with SM families, 
contributing to flavour changing processes at tree level.  }
\end{figure}

\begin{figure}[t]
\centering
\includegraphics[width=0.33\textwidth]{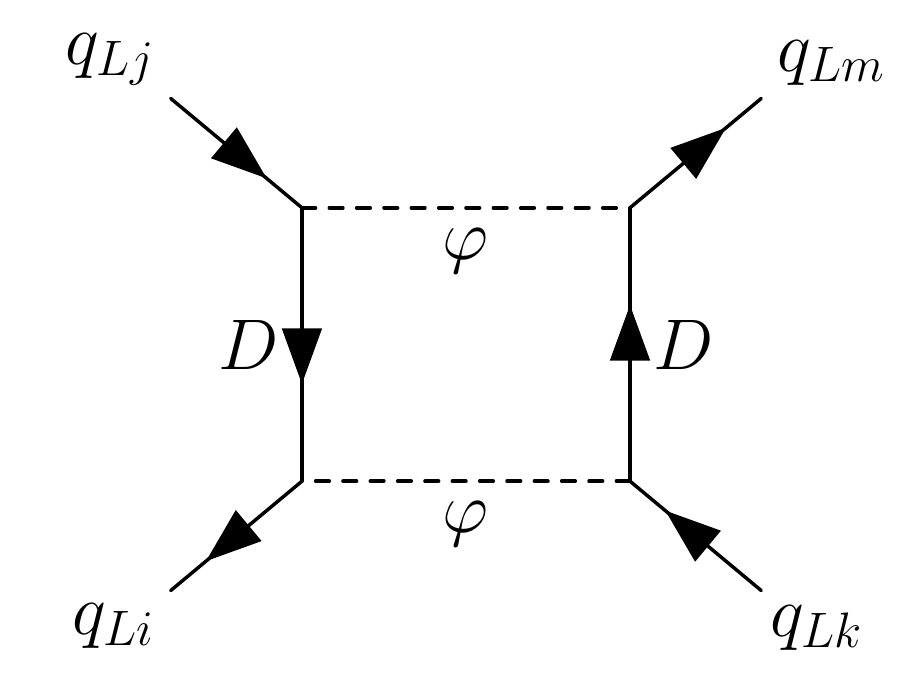}
\includegraphics[width=0.33\textwidth]{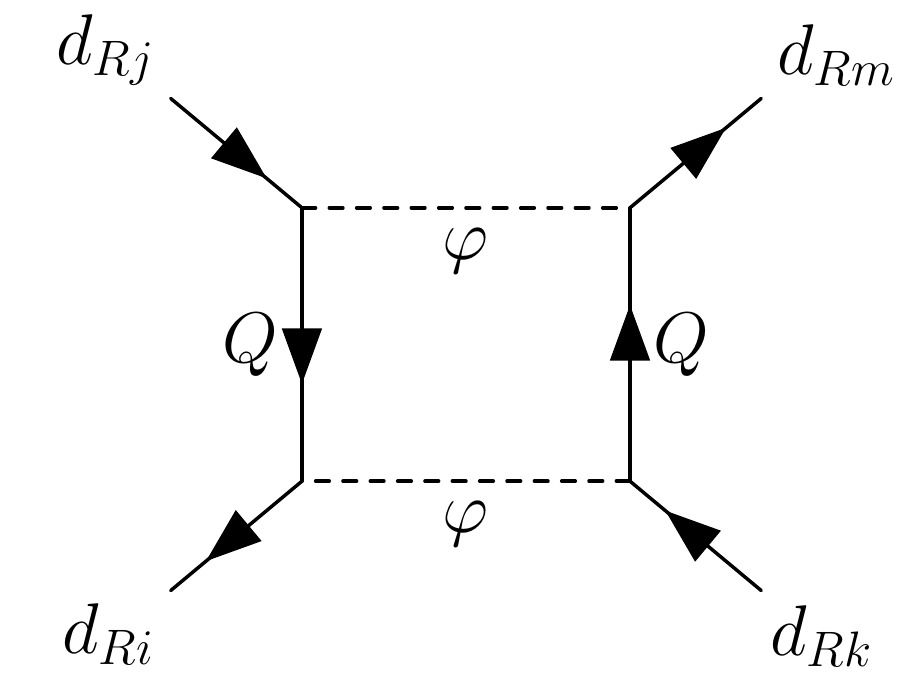}
\includegraphics[width=0.33\textwidth]{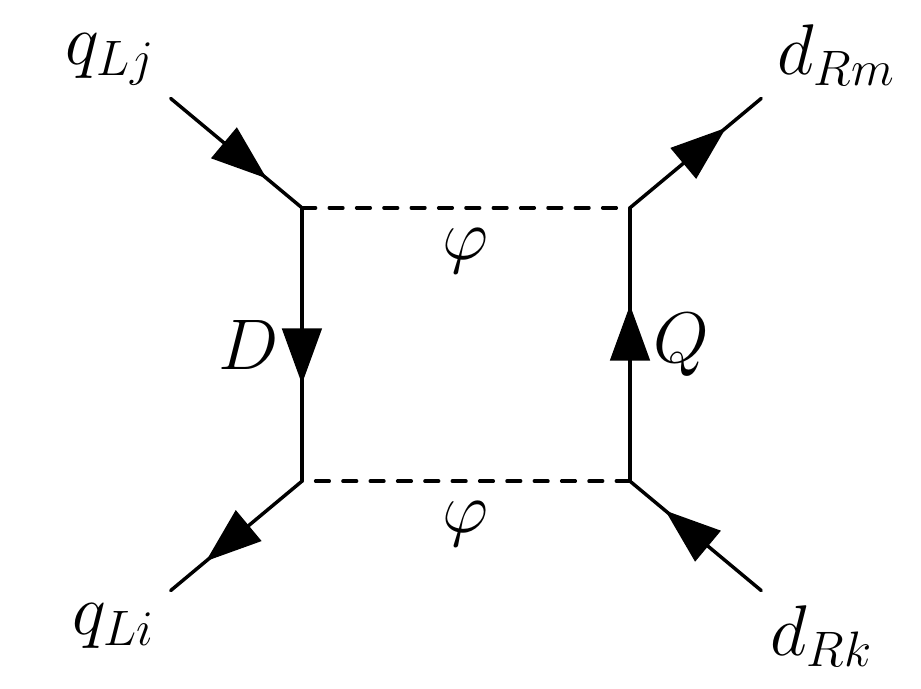}
\caption{\label{dddd} Box diagrams induced by the presence of vector-like quarks.  }
\end{figure}


However, mixing with extra vector-like species (with non-standard isosipin content) 
affects the natural flavor conservation of the SM. 
Namely, by integrating out the heavy isosinglet states $D$ and $U$  
one induces the following effective operators for the quark couplings with $Z$-boson
\begin{align}\label{Z-new-UD}
& \frac{g}{\cos\theta_W}\overline{q_{Li} }\gamma^\mu
 \left[ (T_{3}  -\text{Q}\sin\theta_{W}^{2})\delta_{ij}-  
T_{3}\frac{h_{Di}h_{Dj}^\dagger  \varphi^\dagger \varphi}{M_D^2} 
\right] 
 q_{Lj} Z_\mu \, , \nonumber \\
& \frac{g}{\cos\theta_W} \overline{q_{Li} }\gamma^\mu 
\left[ (T_{3}  -\text{Q}\sin\theta_{W}^{2})\delta_{ij}-  
T_{3}\frac{h_{Ui} h_{Uj}^\dagger  \tilde\varphi^\dagger \tilde\varphi}{M_U^2}
\right]  q_{Lj} Z_\mu ,
\end{align}
as it is shown in figure \ref{ddZ}, where $T_{3}$ is the weak isospin and Q the electric charge,
while mixing with the extra doublet $Q$ induces anomalous 
isospin violating couplings with the RH states 
\begin{align}\label{Z-new-Q}
&  \frac{g}{\cos\theta_W}
\left[ 
-\frac{1}{2}\frac{h_{di}^\dagger h_{dj} \varphi^\dagger \varphi}{M_Q^2} +\frac{1}{3}\sin\theta_{W}^{2}\delta_{ij}\right]
\overline{d_{Ri} }\gamma^\mu d_{Rj} Z_\mu ,  \nonumber\\ &
 \frac{g}{\cos\theta_W}\left[ 
\frac{1}{2} \frac{h_{ui}^\dagger h_{uj} \tilde\varphi^\dagger \tilde\varphi}{M_Q^2} -\frac{2}{3}\sin\theta_{W}^{2}\delta_{ij}\right]
 \overline{u_{Ri} }\gamma^\mu u_{Rj} Z_\mu ,
\end{align}
Thus, such anomalous couplings, flavor-non-diagonal between the mass eigenstates, 
after substitution of the VEV $\langle \phi \rangle = v_w$, 
contribute at tree level in the flavor changing phenomena as $K^0-\overline{K}^0$ mixing etc. 
inducing four-fermion effective operators 
\begin{align}\label{new-UD}
& -\frac{h_{Di} h_{Dj}^\dagger h_{Dk} h_{Dm}^\dagger v_w^2   }{4\, M^4_D} 
(\overline{d_{Li} }\gamma^\mu d_{Lj})( \overline{d_{Lk} }\gamma_\mu d_{Lm} ) , 
\quad \quad 
-\frac{h_{Ui} h_{Uj}^\dagger h_{Uk} h_{Um}^\dagger v_w^2 }{4\, M^4_U} 
\overline{u_{Li} }\gamma^\mu u_{Lj} \overline{u_{Lk} }\gamma^\mu u_{Lm} 
\end{align}
and analogously for RH states.
These operators parametrically are order $G_F (v_w/M)^4 \sim v_w^2/M^4$. 
On the other hand, box diagrams shown in the upper-left part of 
figure \ref{dddd} (and analogously for up-type quarks)
induce operators which 
parametrically are order $1/M^2$: 
\begin{align}\label{new-UD}
&  \frac{h_{Di} h_{Dj}^\dagger h_{Dk} h_{Dm}^\dagger }{128 \pi^2 \, M^2_D} 
(\overline{d_{Li} }\gamma^\mu d_{Lj})( \overline{d_{Lk} }\gamma_\mu d_{Lm} ) , 
\quad \quad 
\frac{h_{Ui} h_{Uj}^\dagger h_{Uk} h_{Um}^\dagger  }{128 \pi^2 \, M^4_U} 
\overline{u_{Li} }\gamma^\mu u_{Lj} \overline{u_{Lk} }\gamma^\mu u_{Lm} 
\end{align}
but are suppressed by a loop factor. 
Regarding the RH sector, as in the upper-right part of figure \ref{dddd} we have:
\begin{align}
&  \frac{h_{di}^\dagger h_{dj} h_{dk}^\dagger h_{dm} }{64 \pi^2 \, M^2_D} 
(\overline{d_{Ri} }\gamma^\mu d_{Rj})( \overline{d_{Rk} }\gamma_\mu d_{Rm} ) , 
\quad \quad 
\frac{h_{ui}^\dagger  h_{uj} h_{uk}^\dagger h_{um}  }{64 \pi^2 \, M^4_U} 
\overline{u_{Ri} }\gamma^\mu u_{Rj} \overline{u_{Rk} }\gamma^\mu u_{Rm} 
\end{align}
In the presence of all type of fermions, 
also left-right operators can be induced, as shown in figure \ref{dddd} (bottom).

In addition, extra vector-like leptons induce analogous contributions in lepton 
sector. However, in the following, for brevity, we shall concentrate only on quark sector, 
and study one by one implications of $D$-type, $U$-type and $Q$-type 
extra fermions.  

Namely, in each of these cases we put limits that emerge from flavour changing phenomena
($K^{0}$-$\bar{K}^{0}$, $D^{0}$-$\bar{D}^{0}$, $B^{0}$-$\bar{B}^{0}$, 
flavour changing leptonic and semileptonic meson decays)
and flavour conserving observables ($Z$-boson physics, low energy 
electroweak observables).
For down-type weak singlets these limits were first discussed in ref. \cite{Lavoura}.
However in that work contributions of box diagrams involving heavy states were not discussed,
whereas, as we will show, these diagrams
become dominant if heavy vector-like quarks have masses larger than $\sim 3$ TeV.

\section{Extra down-type isosinglet}
\label{sec-down}

Let us examine in details the implications of the addition of a down-type vector-like weak isosinglet 
(D-type) couple of quarks $D_{L}=d_{L4}$ and $D_{R}=d_{R4}$ 
involved in the mixing with the SM three families 
$q_{Li}=(u_{Li},d_{Li})^{T}$, $u_{Ri}$ and $d_{Ri}$, $i=1,2,3$.
New Yukawa terms and Dirac mass terms should be added to the Lagrangian density besides
the standard Yukawa terms:
$
h_{d j}' \varphi\overline{q_{Lj}}d_{Ri}'+
 m_{i}\overline{d_{L4}}d_{Ri}'+\text{h.c.} $.
Since the four species of right-handed singlets $d_{Ri}'$ have identical quantum numbers,
a unitary transformation can be applied on the four components $d_{Ri}'$ so that
 $m_j=0$ for $j=1,2,3$ and $d_{R4}$ is identified with the combination making the Dirac mass term
 with the left handed singlet $d_{L4}$.
Thus the 
Yukawa Lagrangian of this system can be written as:
\begin{align}
\label{yd}
&y^{u}_{ij}\tilde{\varphi}\overline{q_{Li}}u_{Rj}+y^{d}_{ij}\varphi\overline{q_{Li}}d_{Rj}+
h_{dj}\varphi\overline{q_{Lj}}d_{R4}+M_{4}\overline{d_{L4}}d_{R4}+\text{h.c.}
\end{align}
where, without losing generality, the mass term $M_{4}$ can be taken real and positive. 
Then the down-type quarks mass matrix looks like:
\begin{align}
\label{md}
&\overline{d_{Li}}\mathbf{m}^{(d)}_{ij}d_{Rj}+\text{h.c.}= \nonumber \\
=&(\overline{d_{L1}}, \overline{d_{L2}}, \overline{d_{L3}}, \overline{d_{L4}})
\left(\begin{array}{ccc|c}
 & & & h_{d1}v_{w} \\   & \mathbf{y}^{(d)}_{3\times 3}v_{w} & & h_{d2}v_{w} \\  & & & h_{d3}v_{w} \\
 \hline
 0 & 0 & 0 & M_{4}
\end{array}\right)\left(\begin{array}{c}
d_{R1} \\ d_{R2} \\ d_{R3} \\ d_{R4}
\end{array}\right)   +\text{h.c.}
\end{align}
where $v_w=174$ GeV is the SM Higgs vacuum expectation value (VEV) 
(for a convenience,  we use this normalization of the Higgs  VEV 
instead of the ``standard" normalization $\langle \phi\rangle = v/\sqrt2$, i.e. $v=\sqrt2 v_{\rm w}$)
and $\mathbf{y}^{(d)}_{3\times 3}$ is the $3\times 3 $ 
matrix of Yukawa couplings.
The mass matrix $\mathbf{m}^{(d)}$ can be diagonalized with positive eigenvalues 
by a biunitary transformation:
\begin{align}
V_L^{(d)\dagger}\mathbf{m}^{(d)}V_R^{(d)}=\mathbf{m}^{(d)}_\text{diag}=
\text{diag}(y_dv_w,y_sv_w,y_bv_w,M_{b'})
\end{align}
where $V^{(d)}_{L,R}$ are two unitary $4\times 4 $ matrices.
$\mathbf{m}^{(d)}_\text{diag}$
is the diagonal matrix of mass eigenvalues
$m_{d,s,b}=y_{d,s,b}v_w$ and $M_{b'}\approx M_4$.
Weak eigenstates in terms of mass eigenstates are:
\begin{align}
\label{vld}
&\left(\begin{array}{c}
d_{L1} \\ d_{L2} \\ d_{L3} \\ d_{L4}
\end{array}\right)
=V_L^{(d)}\left(\begin{array}{c}
d_{L} \\ s_{L} \\ b_{L} \\ b'_{L} \end{array}\right) \, , \qquad V^{(d)}_{L}=\left(\begin{array}{cccc}
V_{L1d} & V_{L1s} & V_{L1b} & V_{L1b'} \\
V_{L2d} & V_{L2s} & V_{L2b} & V_{L2b'} \\
V_{L3d} & V_{L3s} & V_{L3b} & V_{L3b'} \\
V_{L4d} & V_{L4s} & V_{L4b} & V_{L4b'} 
\end{array}\right)
\end{align}
As for up-type quarks, the up-quark Yukawa matrix can be taken diagonal, 
$\mathbf{y}^{u}=\text{diag}(y_{u},y_{c},y_{t})$, so that also the mass matrix 
$\mathbf{m}_{u}=\text{diag}(m_{u},m_{c},m_{t})=\text{diag}(y_{u},y_{c},y_{t})\times v_{w}$
is diagonal, and $u_{1}, u_{2}, u_{3}$ correspond to the mass eigenstates $u,c,t$.
Since only the three down-type quarks $d_{L1},d_{L2},d_{L3}$ 
couple with $W$-bosons, the Lagrangian for the charged weak interactions 
expressed in terms of mass eigenstates become:
\begin{align}
\mathcal{L}_\text{cc}&=
\frac{g}{\sqrt{2}}\left(\begin{array}{ccc}
\overline{u_{L1}} & \overline{u_{L2}} & \overline{u_{L3}}
\end{array}\right)\gamma^\mu
\left(\begin{array}{c}
d_{L1} \\ d_{L2} \\ d_{L3} 
\end{array}\right) W_\mu^+ +\text{h.c.} = \nonumber \\
&=\frac{g}{\sqrt{2}}\left(\begin{array}{ccc}
\overline{u_L} & \overline{c_L} & \overline{t_L}
\end{array}\right)\gamma^\mu
\tilde{V}_\text{CKM}
\left(\begin{array}{c}
d_{L} \\ s_{L} \\ b_{L} \\ b'_{L}
\end{array}\right) W_\mu^+ +\text{h.c.} 
\end{align}
where 
\begin{align}
\tilde{V}_\text{CKM}=&
\tilde{V}_L^{(d)}=
\left(\begin{array}{cccc}
V_{L1d} & V_{L1s} & V_{L1b} & V_{L1b'} \\
V_{L2d} & V_{L2s} & V_{L2b} & V_{L2b'} \\
V_{L3d} & V_{L3s} & V_{L3b} & V_{L3b'} 
\end{array}\right)= 
\left(\begin{array}{ccc|c}
 & & & V_{ub'} \\ & V_\text{CKM} & & V_{cb'} \\ & & & V_{tb'}
\end{array}\right)=
\left(\begin{array}{cccc}
V_{ud} & V_{us} & V_{ub} & V_{ub'} \\ V_{cd} & V_{cs} & V_{cb} & V_{cb'} \\ 
V_{td} & V_{ts} & V_{tb} & V_{tb'}
\end{array}\right)
\label{vckm}
\end{align}
is the $3\times 4 $ submatrix of $V_L^{(d)}$ in eq. (\ref{vld}), obtained by cutting the last row. 
Obviously, $\tilde{V}_\text{CKM}$ is not unitary anymore. More precisely, 
although the unitarity condition is violated for columns, it still holds for rows:
$\tilde{V}^{(d)}_L\tilde{V}^{(d)\dag}_L=1_{3\times 3}$, with 
$1_{3\times 3}$ being the $3\times 3$ identity matrix. 
In particular, 
the first row unitarity condition of CKM matrix is modified to 
\begin{align}
\label{newun}
& \vert V_{ud} \vert^2 + \vert V_{us} \vert^2 +\vert V_{ub} \vert^2  + \vert V_{ub'} \vert^2 = 1
\end{align} 
which is the extended unitarity condition for the first row.
The elements of the fourth column of $\tilde{V}_\text{CKM}$ determine the strength of 
the violation of SM CKM unitarity. 
%
%

\begin{figure}[t]
\centering
\includegraphics[width=0.6\textwidth]{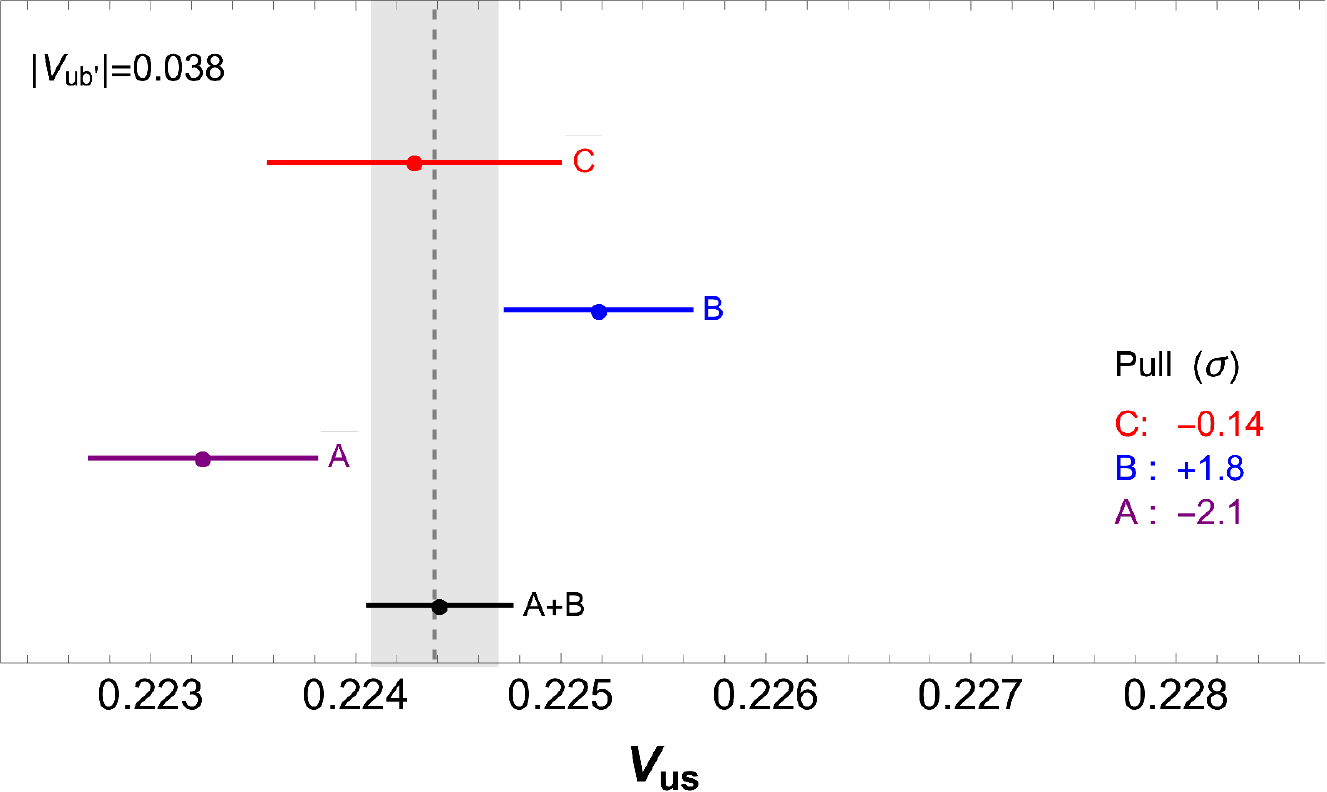}
\caption{\label{vusudm43plot} Determinations of $\vert V_{us} \vert$  
obtained  
using eq. (\ref{newun}) with $\vert V_{ub'} \vert = 0.038$,
with the dataset in eqs. (\ref{A}), (\ref{rapporto}), 
(\ref{c}), to be compared with figure \ref{situation}.   }
\end{figure}
 \begin{table}
\centering
\begin{tabular}{| lcc |}
\hline
Determination  &  $|V_{us}|$ value & $|V_{ub'}|=0.038$\\
\hline
A &  $0.22326(55)$  & $0.22326(55)$ \\
B   &$0.22535(45)$ &   $0.22518(45)$  \\ 
C  &   $0.22756(70)$ & $0.22429(71)$  \\
A+B   & $0.22451(35)$ &  $0.22441(35)$ \\
A+B+C  & $0.22511(31)$ & $0.22439(31)$ \\
\hline
\end{tabular}
\caption{\label{Table23} 
Values of $V_{us}$ obtained from the dataset in eqs. (\ref{A}), (\ref{rapporto}), (\ref{c}).
In the first column the SM unitarity of CKM matrix is used,
while in the second column the extended unitarity (\ref{newun}) is used
with $\vert V_{ub'} \vert = 0.038$.
  }
\end{table}
The dataset A, B, C from eqs. (\ref{A}), (\ref{rapporto}), (\ref{vudmedio}) 
can be fitted in this scenario
by using the extended unitarity (\ref{newun}). 
The best fit point ($\chi^2_\text{dof}=7.4$) is obtained in 
$|V_{us}| = 0.22439(36)$, with: 
\begin{align}
\label{newunsol}
& 
|V_{ub'}|^{2}=1.48(36)\times 10^{-3} \; , \qquad \quad
|V_{ub'}| =0.038(5)
\end{align}
At $95$\% C.L. the needed additional mixing is $|V_{ub'}|=0.038^{+0.008}_{-0.011}$.
In figure \ref{vusudm43plot} it is shown how the present situation 
would look like 
by choosing $|V_{ub'}|=0.038$. 
The data are listed in table \ref{Table23}.
Determinations B and C are shifted with respect to the ones in figure \ref{situation},
while A remains unchanged.
Determination C, which is 
obtained from $V_{ud}$ 
from superallowed beta decays,
becomes perfectly aligned with the average 
of the determinations of $V_{us}$ obtained from kaon decays, because of the 
extended unitarity relation (\ref{newun}).
The $\chi^{2}$ is rather large $\chi^2_\text{dof}=7.4$, 
because of the remaining
tension between the determinations of $V_{us}$ extracted from kaon physics (A and B).
%
%
%
%
%
%
%

However,
the mixing $V^{(d)}_L$ induces non-standard couplings of $Z$-boson with the LH down quarks, since the normal families $d_{Li}$, $i=1,2,3$ reside in doublets while $d_{L4}$ is a weak singlet
(the mixing $V^{(d)}_R$ of the RH quarks does not give the same effect since all RH states $d_{Ri}$
are in identical representations of the SM). In fact, $Z$-boson couples to a fermion species $f$ (LH or RH) as $Z_{\mu}\overline{f}(T_{3}+Q\sin^{2}\theta_{W})f$, where $T_{3}$ is the weak isospin projection and $Q$ the electric charge. Therefore, $Q$-dependent couplings remain diagonal between the mass eigenstates $d,s,b,b'$, while the $T_{3}$ dependent part gets non-diagonal couplings.
The weak neutral current Lagrangian for down quarks reads:
\begin{align}
\label{nc}
\mathcal{L}_\text{nc} =&
\frac{g}{\cos\theta_W}\left[-\frac{1}{2}
\left(\begin{array}{ccc}
\overline{d_{L1}} & \overline{d_{L2}} & \overline{d_{L3}}
\end{array}\right)\gamma^\mu
\left(\begin{array}{c}
d_{L1} \\ d_{L2} \\ d_{L3} 
\end{array}\right)
+\frac{1}{3}\sin^2\theta_W\left(\overline{\mathbf{d}_L}\gamma^\mu \mathbf{d}_L+
\overline{\mathbf{d}_R}\gamma^\mu \mathbf{d}_R\right)\right]Z_\mu= \nonumber \\
 =& \frac{g}{\cos\theta_W}\left[-\frac{1}{2}
\left(\begin{array}{cccc}
\overline{d_L} & \overline{s_L} & \overline{b_L} & \overline{b'_L}
\end{array}\right)\gamma^\mu
V^{(d)}_\text{nc}
\left(\begin{array}{c}
d_{L} \\ s_{L} \\ b_{L} \\ b'_{L}
\end{array}\right)
+\frac{1}{3}\sin^2\theta_W\left(\overline{\mathbf{d}_L}\gamma^\mu \mathbf{d}_L+
\overline{\mathbf{d}_R}\gamma^\mu \mathbf{d}_R\right)\right]Z_\mu \nonumber \\
 V^{(d)}_\text{nc}= & V_L^{(d)\dag}\text{diag}(1,1,1,0)V_L^{(d)}=\tilde{V}_L^{(d)\dag}\tilde{V}_L^{(d)}
\end{align}
where $\mathbf{d}$ is the column vector of the four down-type quarks $d,s,b,b'$.
As comes out from eq. (\ref{nc}), 
the non-unitarity of $\tilde{V}^{(d)}_L$ is at the origin of 
non-diagonal couplings with $Z$ boson, which means
flavor changing neutral currents (FCNC)
at tree level. 
Explicitly, the weak isospin dependent part of the $Z$ coupling is given by:
\begin{align}
\label{vnc}
&V_\text{nc}^{(d)}=V_L^{(d)\dag}\text{diag}(1,1,1,0)V_L^{(d)}= \nonumber\\
&=\! \left(\begin{array}{c@{\hspace{1\tabcolsep}}c@{\hspace{1\tabcolsep}}c@{\hspace{1\tabcolsep}}c}
1&0&0&0\\0&1&0&0\\0&0&1&0\\0&0&0&0
\end{array}\right)\!+\!
\left(\begin{array}{c@{\hspace{1\tabcolsep}}c@{\hspace{1\tabcolsep}}c@{\hspace{1.5\tabcolsep}}c}
-|V_{L4d}|^2 & -V_{L4d}^*V_{L4s} & -V_{L4d}^*V_{L4b} & -V_{L4d}^*V_{L4b'} \\
 -V_{L4s}^*V_{L4d} & -|V_{L4s}|^2 & -V_{L4s}^*V_{L4b} & -V_{L4s}^*V_{L4b'} \\
  -V_{L4b}^*V_{L4d} & -V_{L4b}^*V_{L4s} & -|V_{L4b}|^2 &  -V_{L4b}^*V_{L4b'} \\
  -V_{L4b'}^*V_{L4d} & -V_{L4b'}^*V_{L4s} &  -V_{L4b'}^*V_{L4b} & 
  |V_{L1b'}|^2\!+\!|V_{L2b'}|^2\!+\!|V_{L3b'}|^2 
\end{array}\right)
\end{align}
%
%

$V^{(d)}_L$ can be parameterized by $6$ angles and $10$ phases.
However, four phases can be eliminated by phase transformations of $d,s,b,b'$ states, 
so that $V^{(d)}_L$ can be presented as:
\begin{align}
&V^{(d)}_L=\left(\begin{array}{cccc}
V_{L1d} & V_{L1s} & V_{L1b} & V_{L1b'} \\
V_{L2d} & V_{L2s} & V_{L2b} & V_{L2b'} \\
V_{L3d} & V_{L3s} & V_{L3b} & V_{L3b'} \\
V_{L4d} & V_{L4s} & V_{L4b} & V_{L4b'} 
\end{array}\right)  \simeq  
V_{3L}^{(d)}L^{(d)}=  \nonumber \\
&= 
\left(\begin{array}{c@{\hspace{0.5\tabcolsep}}c@{\hspace{0.5\tabcolsep}}c@{\hspace{1\tabcolsep}}c}
&  &  & 0 \\
 & V_{3\times 3}^{(d)} &  & 0 \\
 & &  & 0 \\
0 & 0 & 0 & 1
\end{array}\right)\!
\left(\begin{array}{cccc}
1  & 0 & 0    & 0 \\
 0 & 1 & 0    & 0 \\
 0 & 0 & c^d_{L3} & -\tilde{s}^d_{L3} \\
0  & 0 & \tilde{s}^{d*}_{L3} & c^d_{L3}
\end{array}\right) \!
\left(\begin{array}{cccc}
1 & 0     & 0 & 0\\
 0 & c^d_{L2} & 0 & -\tilde{s}^d_{L2} \\
 0 & 0    & 1 & 0 \\
0 & \tilde{s}^{d*}_{L2}  & 0 & c^d_{L2}
\end{array}\right) \!
\left(\begin{array}{cccc}
c^d_{L1} & 0 & 0 & -\tilde{s}^d_{L1} \\
 0   & 1 & 0 & 0 \\
 0   & 0 & 1 & 0 \\
\tilde{s}^{d*}_{L1} & 0 & 0 & c^d_{L1}
\end{array}\right)\! \approx \nonumber \\
&\approx 
\left(\begin{array}{c@{\hspace{0.5\tabcolsep}}c@{\hspace{0.5\tabcolsep}}c@{\hspace{1\tabcolsep}}c}
&  &  & 0 \\
 & V_{3\times 3}^{(d)} &  & 0 \\
 & &  & 0 \\
0 & 0 & 0 & 1
\end{array}\right)
\left(\begin{array}{cccc}
c^d_{L1}  & 0 & 0    & -\tilde{s}^d_{L1} \\
- \tilde{s}^d_{L2}\tilde{s}^{d*}_{L1} & c^d_{L2} & 0    & -\tilde{s}^d_{L2} \\
- \tilde{s}^d_{L3}\tilde{s}^{d*}_{L1} & -\tilde{s}^{d*}_{L2}\tilde{s}^d_{L3} &  c^d_{L3} & -\tilde{s}^d_{L3} \\
\tilde{s}^{d*}_{L1}  & \tilde{s}^{d*}_{L2} & \tilde{s}^{d*}_{L3} & c^d_{L1}c^d_{L2}c^d_{L3}
\end{array}\right) 
\label{conangoli}
\end{align}
$c^d_{Li}$ are cosines and
$\tilde{s}^d_{Li}$ are complex sines of angles in the $1\,4$, $2\,4$, $3\,4$ family 
planes parameterizing the mixing of the first three families
with the vector-like quark:
\begin{align}
\label{tildes}
&\tilde{s}^d_{Li}=\sin\theta^d_{L i4} e^{i \delta^d_{Li}}=s^d_{Li} e^{i \delta^d_{Li}} \; ,
\qquad c^d_{Li}=\cos\theta^{d}_{Li4}
\nonumber\\
& \delta^{d}_{Lij}=\delta^{d}_{Li}-\delta^{d}_{Lj} 
\end{align}
and corresponding to the elements of the last row:
\begin{align}
& V_{L4d}\approx \tilde{s}_{L1}^{d*} \, , &&  V_{L4s}\approx \tilde{s}_{L2}^{d*}\, ,  &&  V_{L4b}= \tilde{s}_{L3}^{d*}
\label{4row}
\end{align}
Since 
it is the relative phase of the elements which will come into play, 
we also defined the relative phase of the elements in eq. (\ref{tildes}).
%
$V_{3L}^{(d)}$ contains $3$ angles and $3$ phases; it diagonalizes the $3\times 3$ 
Yukawa matrix $\mathbf{y}^{(d)}_{3\times 3}$ in eq. (\ref{md}).
After applying this $3\times 3$ diagonalization, 
since the elements in $V_{3L}^{(d)}$ are small, 
the mixing of the three SM
families with the vector-like species is still described by the couplings:
$h_{di}=V^{(d)}_{Lij}\tilde{h}_{dj}\approx \tilde{h}_{di}$.
Equation (\ref{conangoli}) is true at order 
$O(|h_{di}|(y_{i})^2\frac{v_w^3}{M_{b'}^3}+|h_{dj}|^{2}|h_{di}|\frac{v_w^3}{M_{b'}^3})$,
$i,j=1,2,3$, $y_{i}=y_{d,s,b}$,
with:
\begin{align}
&\frac{1}{2}\tan(2\theta^{d}_{Li4})
\approx  \frac{|h_{di}|v_{w}}{M_{b'}} \; , \qquad \quad 
 \delta^d_{Li}=\arg ( h_{di})-\pi 
\end{align}
that is:
\begin{align}
\label{si}
& \tilde{s}^d_{Li}\approx -\frac{h_{di}v_{w}}{M_{b'}}
\end{align}
Because of small mixing angles, 
$V_{3\times 3}^{(d)}$ is practically equal to the $3\times 3$ submatrix of $V^{(d)}_L$ in (\ref{vld}).
In fact, for example in the chosen parameterization (\ref{conangoli}),
the main corrections regard the elements:
$V_{L3 d}\simeq V_{3\times 3\, 3d}-\tilde{s}^d_{L1}\tilde{s}^d_{L3}$, 
$ V_{L2 d}\simeq V_{3\times 3\, 2d}-\tilde{s}^d_{L1}\tilde{s}^d_{L2} $, 
$V_{L3 s}\simeq V_{3\times 3\, 3s}-\tilde{s}^d_{L2}\tilde{s}^d_{L3} $, 
$V_{L1d}\simeq V_{3\times 3\, 1d} \, c_{L1}^{d} $.
%
However,
it will be shown that, in order to have $s^d_{L1}\approx |V_{L4d}| \approx |V_{ub'}|\approx 0.03$,
it should be that
at most 
$s^d_{L3}=|V_{L4b}|<7.4 \times 10^{-3}$ 
and $s^d_{L2}\approx |V_{L4s}|<5.8\times 10^{-4}$.
Then:
\begin{align}
& [V_L^{(d)}]_{i\beta}\simeq [V^{(d)} _{3\times 3}]_{ i\beta}
\end{align}
%
%

As regards charged currents, 
$\tilde{V}_\text{CKM}$ in (\ref{vckm}) can be described by $6$ moduli and $9$ phases, 
$6$ of which can be absorbed into the quark fields. 
For the submatrix $V_\text{CKM}$ in (\ref{vckm}), it holds that:
\begin{align}
& [V_\text{CKM}]_{\alpha\beta}=\sum_{i=1}^3V^{(u)*}_{L\, i\alpha}V_{L\, i\beta}^{(d)}
\simeq \sum_{i=1}^3V^{(u)*}_{L\, i\alpha}V_{3\times 3\, i\beta}^{(d)}
\label{elvckm}
\end{align}
Then,
after rephasing the quark fields, $V_\text{CKM}$ can be in the
usual parameterization with
$3$ angles and one phase. Also another phase can be absorbed, 
so we can always choose $\delta^{d}_{L1}=0$ without loss of generality.
%
%
%
%
From (\ref{elvckm}),
for the elements of the fourth column of $\tilde{V}_\text{CKM}$ in (\ref{vckm}) 
it holds that:
\begin{align}
\label{4col}
&V_{ub'}
\approx
-V_{L4d}^{*}V_{ud}-V_{L4s}^{*}V_{us}-V_{L4b}^{*}V_{ub}
\approx -V_{L4d}^{*} 
\\ \label{Vcbp}
 &V_{cb'}\approx
-V_{L4d}^{*}V_{cd} -V_{L4s}^{*}V_{cs} -V_{L4b}^{*}V_{cb}
\\ 
 &V_{tb'}
 \approx
   -V_{L4d}^{*}V_{td} -V_{L4s}^{*}V_{ts} -V_{L4b}^{*}V_{tb}
 \simeq -V_{L4b}^{*} \label{4col3}
\end{align}
where the last approximations come from the mentioned constraints on the mixings.

It should be noted that also the couplings of quarks with the real Higgs are not diagonal
if the vector-like species are added.
In fact, the matrix of Yukawa couplings and the mass matrix (\ref{md}) are not proportional anymore,
and then they are not diagonalized by the same transformation.
In particular, 
left-handed light quarks are coupled with $b'_{R}$ with coupling constants which can be in principle 
of order $O(1)$.
In fact, the Higgs couplings with quarks are:
\begin{align}
\label{Higgs}
& 
(\overline{d_{L1}}, \overline{d_{L2}}, \overline{d_{L3}}, \overline{d_{L4}})
\left(\begin{array}{ccc|c}
 & & & \tilde{h}_{d1} \\   & \mathbf{y}^{(d)}_{3\times 3} & & \tilde{h}_{d2} \\  & & & \tilde{h}_{d3} \\
 \hline
 0 & 0 & 0 & 0
\end{array}\right)\left(\begin{array}{c}
d_{R1} \\ d_{R2} \\ d_{R3} \\ d_{R4}
\end{array}\right) \frac{H^{0}}{\sqrt{2}} +\text{h.c.}
\approx    \nonumber
\\
\approx &(\overline{d_{L}}, \overline{s_{L}}, \overline{b_{L}}, \overline{b'_{L}}) L^{(d)\dag}
\left(\begin{array}{ccc c}
y_{d} & 0 & 0  & h_{d1} \\  0 & y_{s} &0 & h_{d2} \\ 0 & 0 & y_{b} & h_{d3} \\
 0 & 0 & 0 & 0
\end{array}\right) R^{(d)} \left(\begin{array}{c}
d_{R} \\ s_{R} \\ b_{R} \\ b'_{R}
\end{array}\right) \frac{H^{0}}{\sqrt{2}} +\text{h.c.} \approx  \nonumber\\
\approx &(\overline{d_{L}}, \overline{s_{L}}, \overline{b_{L}}, \overline{b'_{L}}) 
\left(\begin{array}{ccc c}
c_{L1}^{d}y_{d} & \tilde{s}_{R2}^{d*}h_{d1} & \tilde{s}_{R3}^{d*}h_{d1}  & h_{d1} \\  
\tilde{s}_{R1}^{d*}h_{d2} & y_{s} &\tilde{s}_{R3}^{d*}h_{d2} & h_{d2} \\ 
\tilde{s}_{R1}^{d*}h_{d3} & \tilde{s}_{R2}^{d*}h_{d3} & y_{b} & h_{d3} \\
 -\tilde{s}_{L1}^{d*}y_{d} &  -\tilde{s}_{L2}^{d*}y_{s} &  -\tilde{s}_{L3}^{d*}y_{b} & 
  \sum_{i=1}^{3}(-\tilde{s}_{Li}^{d*}h_{di})
\end{array}\right) 
\left(\begin{array}{c}
d_{R} \\ s_{R} \\ b_{R} \\ b'_{R}
\end{array}\right) \frac{H^{0}}{\sqrt{2}} +\text{h.c.} 
\end{align}
where $L^{(d)}$ is defined in (\ref{conangoli}) and $R^{(d)}$ is analogously defined.
Then in principle FC couplings between light quarks emerge at tree level.
However the mixing angles of the SM right-handed quarks with $b'_{R}$
are much smaller than angles in the left-handed sector, 
$s_{Ri}^{d}\approx \frac{y_{i}|h_{di}|v_{w}^{2}}{M^{2}_{b'}}$, where $s_{Ri}^{d}$ is defined 
in the same way as $s_{Li}^{d}$ in eqs. (\ref{conangoli}), (\ref{tildes}).

It should also be noticed that,
because of the large mixing with the first family, the extra quark $b'$ would mainly decay
into $u$ or $d$ quark via the couplings with $W$, $Z$, $H$.
The 
CMS experiment put lower limits on the mass of vector-like quarks coupling to light quarks,
which in our scenario imply
$M_{b'} \gtrsim 700$~GeV \cite{CMS}. 
It should be noticed that, with this constraint,
$\vert V_{ub'}\vert  \simeq  0.03$ 
can be obtained if $|h_{d 1}| \gtrsim 0.1$, 
much larger than the Yukawa constant of the bottom quark $y_{b}$. 
In turn, by taking $\vert V_{ub'} \vert  > 0.03$ in  
$M_{b'} =|h_{d 1}|v_{w}/\vert V_{ub'}\vert$, 
and assuming (for the perturbativity) $|h_{d 1} |< 1$, 
there is an upper limit on the extra quark mass, $M_{b'} < 5.8 $ TeV. 

In the following sections experimental limits from FCNC and electroweak observables are examined.
The results are summarized in section \ref{downend}, 
in table \ref{limiti} and figures \ref{fcplotd}, \ref{vcbv4s}.

Table \ref{values} displays the values used in computations.
Since the effects of new physics on effective operators generated at the tree-level within the SM
can in general be neglected, the experimental determinations of the CKM matrix quantities
derived by tree level processes can be used in our BSM scenarios.

\begin{table}[t]
\centering
\begin{tabular}{| lr  @{\hspace{4\tabcolsep}}  l r |}
\hline
Quantity & Value & Quantity & Value \\
\hline
$|V_{cd}|$ & $0.221(4)$  &  $M_t$[GeV] & $172.76(30)$ \\
$|V_{cb}|$ & $0.0410(14)$  & $m_c(m_c)$[GeV] & $1.27(2)$ \\
$|V_{ts}|$ & $0.0388(11)$ &  $m_s(2 \text{GeV})$[MeV] & $93.12(69)$ \cite{FLAG2019} \\
$|V_{td}|$ & $0.0080(3)$   &  $m_b(m_b)$[GeV] & $4.18(3)$ \\
$|V_{ub}|$ & $0.00382(24)$ &   $m_{K^0}$[MeV] & $497.611(13)$  \\
$|V_{us}|$ & $0.2245(8)$   &   $\tau_{K_L}$[s] & $5.116(21) \cdot 10^{-8}$  \\
$|V_{cs}|$ & $0.987(11)$  &    $\tau_{K_S}$[s] & $8.954(4) \cdot 10^{-11}$  \\
$\lambda$ & $0.22653(48)$  &  $M_{K^+}$[MeV] & $493.677(16)$  \\
$\bar{\rho}$ & $0.123(32)$ & $\tau_{K^+}$[s] & $1.2380(20) \cdot 10^{-8}$  \\
$A$ & $0.799(28)$  & $m_\mu$[MeV] & $105.6583745(24)$  \\
$\bar{\eta}$ & $0.382(29)$  &  $\tau_{D^+}$[s] & $1.040(7)\cdot 10^{-12}$  \\
$\alpha_s(M_Z)$ & $0.1185(16)$  & $M_{D^+}$[MeV] & $1869.65(5)$ \\
$\alpha(M_Z)^{-1}$ & $127.952(9)$  &  $\tau_{D^0}$[s] & $4.101(15)\cdot 10^{-13}$  \\
$\sin^2\theta_W(M_Z)$ & $0.23121(4)$ &   $m_{D^0}$[MeV] & $1864.83(5)$  \\
$G_F[\text{GeV}^{-2}]$ & $1.1663787(6)\cdot 10^{-5} $ &  $\tau_{B^0_{d}}$[s] & $1.519(4)\cdot 10^{-12}$  \\
$m_W$[GeV] & $80.379(12)$ & $M_{B^0_{d}}$[MeV] & $5279.65(12)$  \\
$m_Z$[GeV] & $91.1876(21)$ &  $\tau_{B^0_{sH}}$[s] & $1.620(7)\cdot 10^{-12}$  \\
$m_H$[GeV] & $125.10(14)$ 
&  $M_{B^0_{s}}$[MeV] & $5366.88(14)$ \vspace{1pt}  \\
\hline
\end{tabular}
\caption{\label{values}  The central values are employed in the computations. All values are taken from ref. \cite{PDG18}, except 
$m_s(2 \text{GeV})=93.12(69)$ MeV from ref. \cite{FLAG2019}.
The adopted values of $\lambda$, $A$, $\bar{\rho}$ and $\bar{\eta}$ are those received using
only tree-level inputs in the global fit \cite{PDG18}.
}
\end{table}

\subsection{Limits from rare kaon decays}

\subsubsection{$K^+\rightarrow \pi^+\nu\bar{\nu}$}

The decay $K^+\rightarrow \pi^+\nu\bar{\nu}$ is one of the golden modes for testing the SM, since
long-distance contributions are negligibly small.
The effective Lagrangian comes from a combination of $Z$-penguin and box-diagram and it
is given by \cite{Buchalla:1995vs}:
\begin{align}
\label{kpnnsm}
\mathcal{L}_\text{SM}&=-
\frac{4G_F}{\sqrt{2}}\frac{\alpha(M_Z)}{2\pi\sin^2\theta_\text{W}}
\sum_{\ell=e,\mu,\tau}
[V_{cs}^*V_{cd}X^\ell(x_c)+V_{ts}^*V_{td}X^{\ell}(x_t)]
(\overline{s_{L}}\gamma^\mu d_{L})
(\overline{\nu_{\ell L}}\gamma_\mu\nu_{\ell L}) +\text{h.c.}
\end{align}
 $X(x_a)$ are the Inami-Lim function including QCD and electroweak corrections,
 with $x_a=m^2_a/M^2_W$, $a=c,t$.
The index $\ell$ denotes the lepton flavor. 
The dependence on the charged lepton mass 
is negligible for the top contribution due to $m_{t}\gg m_{\ell}$,
$X^{\ell}(x_{t})=X(x_{t})$,
whereas, being $m_{\tau}$ comparable with $m_{c}$,
for the c-quark contribution
the box with $\ell=\tau$ gives somewhat different contribution
from $\ell=e,\mu$ \cite{Buchalla:1995vs}, 
$X^e(x_c)=X^e(x_c)\neq X^\tau(x_c)$. Therefore, it is used an averaged value
$\bar{X}=\frac{1}{3}\left(X^e(x_c)+X^\mu(x_c)+X^\tau(x_c)\right)$.
Then the effective Lagrangian (\ref{kpnnsm}) can also be written as:
\begin{align}
\label{kpnnsm2}
\mathcal{L}_\text{SM} 
&=-\frac{4G_F}{\sqrt{2}}\mathcal{F}_{K}(\overline{s_{L}}\gamma^\mu d_{L})\sum_{e,\mu,\tau}
(\overline{\nu_{\ell L}}\gamma_\mu\nu_{\ell L}) \nonumber \\
 \mathcal{F}_{K} &= \frac{\alpha(M_Z)}{2\pi\sin^2\theta_\text{W}}\Big(V_{ts}^*V_{td}X(x_t)+
(V_{cs}^*V_{cd}) \bar{X}(x_{c})\Big)= \nonumber \\
&=-\frac{\alpha(M_Z)}{2\pi\sin^2\theta_\text{W}}\lambda^{5}\Big(
A^2 (1-\rho-i\eta)X(x_t)+
(1-\frac{\lambda^2}{2}) P_{c}(X)\Big)
\end{align}
where $P_c(X)=\frac{1}{\lambda^4}\bar{X}$, and
we used the Wolfenstein parameterization:
\begin{align}
& \text{Re}(V_{cs}^*V_{cd})=-\lambda(1-\frac{\lambda^2}{2}) \, , &&
 \text{Re}(V_{ts}^*V_{td})=-(1-\frac{\lambda^{2}}{2})A^2\lambda^5(1-\rho) \, , &&
 \text{Im}(V_{ts}^*V_{td})=A^2\lambda^5\eta \, .
\label{vtsvtdIm}
\end{align}
The top contribution gives $X(x_{t})=1.481$ \cite{Buras15} (central value).
For the charm contribution,
using the values in table \ref{values}, from the formula in ref. \cite{Brod1}
we obtain $P_c(X)\approx 0.351 $. 
By using $\rho=0.126$, $\eta=0.392$ and
the central values in table \ref{values} we have
$\mathcal{F}_{K}\approx (-3.7+i\, 1.2)\times 10^{-6}$,
$|\mathcal{F}_{K}|\approx 3.9\times 10^{-6}$.
%
%
The predicted SM contribution for the branching ratio can be written as:
\begin{align}
&\text{Br}(K^+\rightarrow \pi^+\nu\bar{\nu})_\text{SM}\approx 
k_{K^{+}} |\mathcal{F}_{K}|^{2} \: ,
&& k_{K^{+}}= \frac{G_{F}^{2}\tau_{K^+}M^5_{K^+}f^{2}_+(0)\mathcal{I}_\nu^+}{64\pi^3} 
\approx 5.43
\end{align}
where $f_+(0)=0.9699(15)$ is the form factor,
$\mathcal{I}_\nu^+=0.15269$ is the phase space integral \cite{Mescia:2007kn}
and we have used the values in table \ref{values} for the kaon mass and mean life,.
In this way, we obtain our benchmark value 
$\text{Br}(K^+\rightarrow \pi^+\nu\bar{\nu})_\text{SM}\approx 0.82\cdot 10^{-10}$, 
which is within the range of 
the estimate 
reported by PDG
Br$(K^+\rightarrow \pi^+\nu\bar{\nu})_\text{SM}=(0.85\pm 0.05)\cdot 10^{-10}$
\cite{PDG18}.
The SM expectation is compatible with
the experimental branching ratio \cite{PDG18}:
\begin{align}
\label{kpexp}
\text{Br}(K^+\rightarrow \pi^+\nu\bar{\nu})_\text{exp}=(1.7\pm 1.1)\cdot 10^{-10}
\end{align}
With future experimental precision,
any deviation from the SM prediction of this golden mode branching ratio 
would indicate towards new physics.

\begin{figure}[t]
\centering
\includegraphics[width=0.33\textwidth]{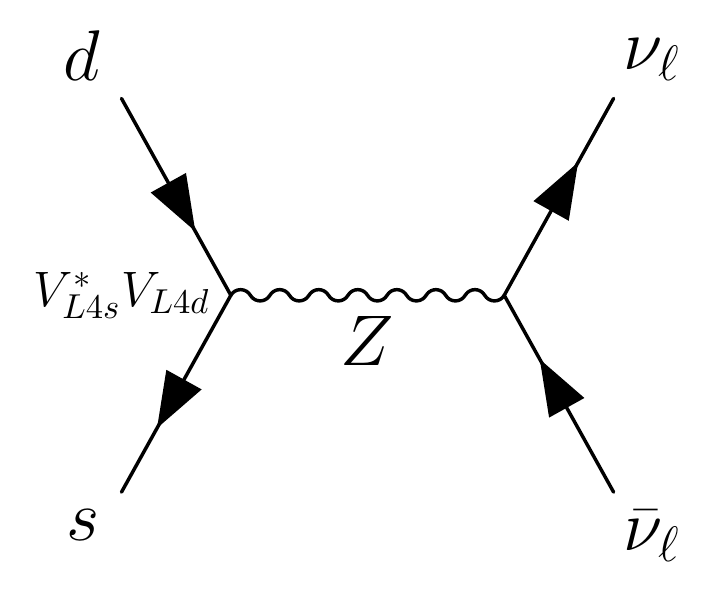}
\caption{\label{knn} Tree level contribution to the rare kaon decays 
$K^+\rightarrow \pi^+\nu\bar{\nu}$, $K^{0}\rightarrow \pi^0\nu\bar{\nu}$ arising from the mixing of the SM families with the extra downtype vectorlike quark.   }
\end{figure}

In our BSM scenario with extra vectorlike $b'$ quark,
the non-diagonal couplings of $Z$-boson with light quarks
in eqs. (\ref{nc})
induce at tree level the 
operator with the same structure as the SM one (\ref{kpnnsm}),
as shown in the diagram in figure \ref{knn}:
\begin{align}
\label{Lnew}
\mathcal{L}_\text{new}=
&-\frac{g^2}{M^2_Z\cos^2\theta_W}
\left(\frac{1}{2} V_{L4s}^*V_{L4d}\right) (\overline{s_L}\gamma^\mu d_L)
\frac{1}{2}\sum_{e,\mu,\tau}(\overline{\nu_{\ell\text{L}}}\gamma_\mu\nu_{\ell\text{L}})= 
\nonumber \\
=&-\frac{4G_F}{\sqrt{2}}\frac{1}{2}V_{L4s}^*V_{L4d}
(\overline{s_L}\gamma^\mu d_L)
\sum_{e,\mu,\tau}(\overline{\nu_{\ell\text{L}}}\gamma_\mu\nu_{\ell\text{L}} )
\end{align}
Thus, 
the new operator (\ref{Lnew}) contributes to the decay 
$K^+\rightarrow \pi^+\nu\bar{\nu}$, in interference with the SM (\ref{kpnnsm}). 
The total branching ratio becomes:
\begin{align}
 &\text{Br}(K^+\rightarrow \pi^+\nu\bar{\nu})_{\text{tot}}  \approx 
\, k_{K^{+}}  \left|\frac{1}{2}V_{L4s}^{*}V_{L4d}+\mathcal{F}_{K}\right|^{2}=    
 \nonumber \\ &= \text{Br}(K^+\rightarrow \pi^+\nu\bar{\nu})_\text{SM} 
 \left|\frac{\frac{1}{2}V_{L4s}^{*}V_{L4d}}{\mathcal{F}_{K}}+1\right|^{2} 
\end{align}
The experimental result (\ref{kpexp}) implies, at $95\%$ C.L., the upper limit:
\begin{align}
\label{Brmax}
& \text{Br}(K^+\rightarrow \pi^+\nu\bar{\nu})_{\text{tot}}  <   3.9 \cdot 10^{-10}
\end{align}
Then 
 we have:
\begin{align}
\label{kpnncon0}
&  \left|\frac{\frac{1}{2}V_{L4s}^{*}V_{L4d}}{\mathcal{F}_{K}}+1\right|< 2.2
\end{align}
figure \ref{fcplotd} shows the constraint (\ref{kpnncon0}) in terms of the modulus and 
phase of the elements $V_{L4d}$, $V_{L4s}$,
using $V_{L4s}^{*}V_{L4d}=|V_{L4s}^{*}V_{L4d}|e^{i\delta^{d}_{L21}}$, 
$\delta^{d}_{L21}=\delta^{d}_{L2}-\delta^{d}_{L1}$, from eqs. (\ref{tildes}) and (\ref{4row}).
%
%
Depending on the unknown relative phase, the limit results in the constraint on the modulus:
\begin{align}
|V_{L4s}^*V_{L4d}|<(0.9 \div 2.5 )\cdot 10^{-5}
\end{align}
%
As can be seen,
in the most conservative case of destructive interference, the new amplitude can be up to
three times the SM amplitude:
\begin{align}
\frac{|V_{L4s}^*V_{L4d}|}{ 2 | \mathcal{F}_{K} | } & \lesssim \Delta_{K^{+}} =3.2
\end{align}
In this case we can express the experimental limit 
in terms of the ratio $\Delta_{K^{+}}$:
 \begin{align}
|V_{L4s}^*V_{L4d}| & \lesssim 2.5 \cdot 10^{-5} 
\left[\frac{\Delta_{K^{+}}}{3.2}  \right]
 \left[\frac{|\mathcal{F}_{K}|}{ 3.9 \cdot 10^{-6} }  \right] 
 \end{align}

\subsubsection{$K_\text{L}\rightarrow \pi^0\nu\bar{\nu}$}
The second golden mode is the
decay $K_\text{L}\rightarrow \pi^0\nu\bar{\nu}$. 
In the Standard Model it is described by the same Lagrangian as in 
(\ref{kpnnsm2}). However this decay 
proceeds almost entirely through direct CP violation,
hence it is
completely dominated by short-distance loop diagrams with top quark exchanges and the charm contribution can be neglected \cite{Buras:1998raa}.
In fact,
with the phase convention
$C|K^0\rangle =|\bar{K}^0\rangle$ (or $CP|K^0\rangle =-|\bar{K}^0\rangle$)
we have:
\begin{align}
&\langle\pi^0|(\bar{d}s)_\text{V$-$A}|\bar{K}^0\rangle=
-\langle\pi^0|(\bar{s}d)_\text{V$-$A}|K^0\rangle  \\
& K_\text{L}=\frac{1}{\sqrt{2}}[(1+\epsilon)K^0+(1-\epsilon)\bar{K}^0] 
\end{align}
However, the terms of indirect ($\Delta S=2$) CP violation (proportional to $\epsilon$)
can be neglected in this decay and the dominant contribution comes from the imaginary part
of the operator (\ref{kpnnsm2}):
\begin{align}
\langle \pi^0\nu\bar{\nu} | \mathcal{L}_\text{SM}  |\frac{K^0+\bar{K}^0}{\sqrt{2}}\rangle&= 
-\frac{4G_F}{\sqrt{2}}\frac{i\,2\text{Im}\mathcal{F}_{K}}{\sqrt{2}} 
\langle\pi^0| (\overline{s_L}\gamma^\mu d_L) |K^0\rangle 
\sum_{\ell=e,\mu,\tau}\langle\overline{\nu_{\ell\text{L}}}\gamma_\mu\nu_{\ell\text{L}}\rangle
\end{align}
The SM predicted branching ratio can be written as: 
\begin{align}
&\text{Br}(K_{L}\rightarrow \pi^0\nu\bar{\nu})_\text{SM}\approx 
k_{K_{L}} |\mathcal{F}_{K}|^{2} \: ,
&& k_{K_{L}}= \frac{G_{F}^{2}\tau_{K_{L}}M^5_{K^0}f^{2}_+(0)\mathcal{I}_\nu^L}{64\pi^3} 
\end{align}
where $\mathcal{I}_\nu^+=0.16043(31)$ is the phase space integral  \cite{Mescia:2007kn}.
Putting the values of kaon mass and lifetime (see table \ref{values})
we obtain $\text{Br}(K^+\rightarrow \pi^+\nu\bar{\nu})_\text{SM}\approx 3.5\times 10^{-11}$,
which practically agrees with the estimate reported by PDG 
$\text{Br}(K_\text{L}\rightarrow \pi^0\nu\bar{\nu})_\text{SM}=(3.0\pm 0.2)\times 10^{-11}$
\cite{PDG18}. 

On the other hand, the experimental limit on this decay 
is \cite{PDG18}:  
\begin{align}
\label{kLexp}
\text{Br}(K_\text{L}\rightarrow \pi^0\nu\bar{\nu})_\text{exp}<3.0\times 10^{-9} \quad (90\%\, \text{C.L.})
\end{align}
which is two orders of magnitude larger than 
the SM expectation. Therefore, there is still much room for new physics.

In our BSM scenario with extra vectorlike $b'$ quark,
the new contribution comes from the imaginary part of
the Lagrangian in eq. (\ref{Lnew}) 
Since the interference term with the SM contribution can be neglected,
the experimental limit can be directly applied to the new contribution:
\begin{align}
\label{Brmax0}
& 
\text{Br}(K_\text{L}\rightarrow \pi^0\nu\bar{\nu})_\text{new}\approx
k_{K_{L}}\frac{1}{4} \big(\text{Im}(V_{L4s}^*V_{L4d})\big)^2 <\text{Br}_{\text{exp}}
\end{align}
So in the given parameterization (\ref{tildes}), (\ref{4row}) we have:
\begin{align}
\label{kLpnn}
&|\text{Im}(V_{L4s}^*V_{L4d})|= |V_{L4s}^*V_{L4d}| |\sin(\delta^d_{L21})|<2.2\times 10^{-5}  \left[\frac{\text{Br}_{\text{exp}}}{3.0 \times 10^{-9}}  \right]^{\frac{1}{2}}
\end{align}
This condition is shown in figure \ref{fcplotd}.
Then, the present experimental limit allows
the new contribution to be one order of magnitude larger than the SM contribution:
\begin{align}
\frac{|\text{Im}(V_{L4s}^*V_{L4d})|}{2|\text{Im}\mathcal{F}_{K}|}<9.3
\left[\frac{\text{Br}_{\text{exp}}}{3.0 \cdot 10^{-9}}  \right]^{\frac{1}{2}}
\end{align}
So the discovery of the decay $K_{L}\rightarrow\pi^{0}\nu\bar{\nu}$ 
with branching ratio larger than the SM expectation can be a signal for BSM physics.

\subsubsection{$K_\text{L}\rightarrow \pi^0e^+e^-$}
The decay $K_\text{L}\rightarrow \pi^0e^+e^-$ contains a direct CP violating contribution, 
indirect CP violating contribution, interference between them, 
and also a small CP conserving contribution \cite{PDG18}.
The CP conserving contribution to the amplitude is dominated by a two photon exchange
$K_L \rightarrow \pi^{0}\gamma\gamma\rightarrow \pi^{0}e^{+}e^{-}$,
with both an absorptive and a dispersive part. 
Using the the decay $K_L \rightarrow \pi^{0}\gamma\gamma$,
it is estimated that the CP-conserving branching ratio is of order $\sim O(10^{-13})$ \cite{PDG18}.
The indirectly CP violating amplitude also derives from the coupling of leptons to photons, 
it is given by the long-distance dominated $K_S \rightarrow \pi^{0}e^{+}e^{-}$
amplitude times the CP parameter $\epsilon_{K}$  \cite{Buchalla:1995vs}.
The complete CP-violating contribution to the rate assuming a positive sign for the interference term  
is estimated to be 
$\text{Br}(K_\text{L}\rightarrow \pi^0e^+e^-)_\text{CPV}\approx  (3.1 \pm 0.9) \times 10^{-11}$, where 
the three contributions from indirect, interference and direct CP violation are
$(1.76,\, 0.9, \, 0.45) \times 10^{-11}$ respectively \cite{PDG18}.

Only the direct CP-violating amplitude can be calculated in detail within the SM, since 
it is short distance dominated. 
The relevant operator contains the vector part of the hadronic current and both 
axial and vector component of the leptonic current, and
it is given by:
\begin{align}
\mathcal{L}_\text{SM,SD}&=-\frac{G_F\alpha(M_Z)}{\sqrt{2}\, 2\pi}V_{ts}^*V_{td}
(\overline{s}\gamma_\mu d)
\left(\tilde{y}_{V}(\overline{e}\gamma^\mu e)+\tilde{y}_A
(\overline{e}\gamma^\mu\gamma^5 e)\right)+\text{h.c.}
\end{align}
where $\tilde{y}_{V,A}$ are a combination of 
Inami-Lim functions of box and $Z$ and $\gamma$ penguin diagrams, as
defined in ref. \cite{Buras:1997fb}.
The direct CP-violating branching ratio in the SM can be written as: 
\begin{align}
&\text{Br}(K_\text{L}\rightarrow \pi^0e^+e^-)_\text{SM,SD}=
k_{Ke}
\left(\frac{G_F\alpha(M_Z)}{\sqrt{2}\, 2\pi}\right)^2\left(\frac{2\text{Im}V_{ts}^*V_{td}}{\sqrt{2}}\right)^2
(\tilde{y}_{A}^2+\tilde{y}_{V}^2) \\
&k_{Ke}= \frac{1}{2}\frac{M^5_{K^0}\tau_{K_L}f^{2}_+(0)\mathcal{I}_e}{192\pi^3}
\end{align}
where $\mathcal{I}_e=0.16043(31)$ \cite{Mescia:2007kn} is the phase space integral. 

\begin{figure}[t]
\centering
\includegraphics[width=0.3\textwidth]{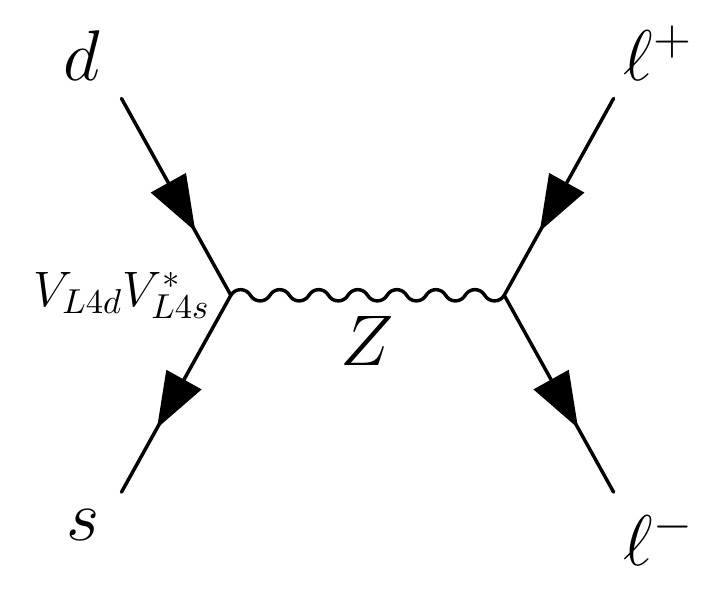}
\caption{\label{kll} Tree level contribution to the rare kaon decays
$K^{0}\rightarrow \pi^0 \ell^+ \ell^-$, $K^{0}\rightarrow \mu^+\mu^-$ ($\ell=e,\mu$) arising from the mixing of the SM families with the extra downtype vector-like quark.   }
\end{figure}
The new effective Lagrangian contributing to this decay (as shown in figure \ref{kll}) is:
\begin{align}
\label{kpeenew}
\mathcal{L}_\text{new}=&-\frac{G_F}{\sqrt{2}}V_{L4s}^{*}V_{L4d}
(\overline{s}\gamma^\mu d) 
\left[g_{V}\overline{e}\gamma_\mu e+
g_{A}\overline{e}\gamma_\mu\gamma_5 e\right]+\text{h.c.}
%
\end{align}
where $g_{A}= \frac{1}{2}$, $g_{V}=-\frac12+2\sin^2\theta_\text{W}\approx -0.038$.
The experimental limit on this decay is \cite{PDG18}:
\begin{align}
& \text{Br}(K_\text{L}\rightarrow \pi^0e^+e^-)_\text{exp}<2.8\times 10^{-10} \quad (90\% \, \text{C.L.})
\end{align}
which is one order of magnitude bigger than the SM expectation.
Then, we can consider the SM contribution as negligible and
let the new contribution to be the dominant one,
imposing a limit directly on the new contribution:
\begin{align}
&\text{Br}(K_\text{L}\rightarrow \pi^0e^+e^-)_\text{new}= 
k_{Ke} G_F^2\,[\text{Im}(V_{L4s}V_{L4d}^*)]^2
\left[g_{V}^2+g_{A}^{2}\right]
< \text{Br}_{\text{exp}}
\end{align}
In the given parameterization (\ref{tildes}), (\ref{4row}) then:
\begin{align}
\label{kLpee}
&|\text{Im}(V_{L4s}V_{L4d}^*)|=|V_{L4s}V_{L4d}^*||\sin(\delta^d_{L12})|<1.7\times 10^{-5} 
\left[\frac{\text{Br}_{\text{exp}}}{2.8\cdot 10^{-10}}  \right]^{\frac{1}{2}}
\end{align}
where the central values of all the quantities have been used.
Also the decay $K_{L}\rightarrow\pi^{0}\mu^{+}\mu^{-}$ gives a comparable constraint,
being the experimental limit
$\text{Br}(K_\text{L}\rightarrow \pi^0\mu^+\mu^-)_\text{exp}<3.8\times 10^{-10} \quad (90\% \, \text{C.L.})$,
to be compared with the SM expectation 
$\text{Br}(K_\text{L}\rightarrow \pi^0\mu^+\mu^-)=(1.5\pm 0.3) \times 10^{-11}$ (assuming positive interference between the direct- and indirect-CP violating components) \cite{PDG18}.



\subsubsection{$K_\text{L}\rightarrow \mu^+\mu^-$}
The rare decay $K_\text{L}\rightarrow \mu^+\mu^-$ is a CP conserving decay.
Its short-distance part is given by $Z$-penguins and box diagrams.
However, this decay 
is dominated by a long-distance contribution from 
a two-photon intermediate state. 
In fact, the full branching ratio can be written as \cite{Buchalla:1995vs}:
\begin{align}
& \text{Br}(K_\text{L}\rightarrow \mu^+\mu^-) = |\text{Re}A|^2 + |\text{Im}A|^2 \, ,
&& \text{Re}A = A_\text{SD} + A_\text{LD}
\end{align}
with Re$A$ and Im$A$ denoting the dispersive and absorptive contributions, respectively. 
The absorptive (imaginary) part of the long-distance component is determined by the measured rate for $K_L \rightarrow \gamma\gamma$ to be 
$\text{Br}_\text{abs}(K_L \rightarrow \mu^+\mu^-)=(6.64 \pm 0.07) \cdot 10^{-9}$ and it almost 
completely saturates the observed rate 
Br$(K_\text{L}\rightarrow \mu^+\mu^-)_\text{exp}=(6.84\pm 0.11)\cdot 10^{-9}$ \cite{PDG18}.
The real part of the long-distance amplitude cannot be derived directly from experiment.
However in ref. \cite{Isidori} it is estimated an upper bound 
on the short distance contribution:
\begin{align}
\label{SDtot}
&\text{Br}(K_\text{L}\rightarrow \mu^+\mu^-)_\text{SD}<2.5\cdot 10^{-9}
\end{align}
As shown below,
the SM prediction for the short distance contribution results:
$\text{Br}(K_\text{L}\rightarrow \mu^+\mu^-)_\text{SD,SM}\approx 0.9 \cdot 10^{-9}$  
\cite{Buras:1997fb}. Then the condition (\ref{SDtot}) provides 
a constraint
on new physics scenarios.

Since only the axial component of both hadronic and lepton current contributes to the decay,
the effective Lagrangian describing the short distance contribution to the decay in the SM
can be written as \cite{Buchalla:1995vs}:
\begin{align}
\mathcal{L}_\text{SM,SD}&=\frac{G_F}{\sqrt{2}}\frac{\alpha(M_Z)}{2\pi\sin^2\theta_\text{W}}
\big(V_{cs}^*V_{cd}Y(x_c)+V_{ts}^*V_{td}Y(x_t)\big)
(\overline{s}\gamma^\mu\gamma_{5} d)(\overline{\mu}\gamma_\mu\gamma_{5}\mu)+\text{h.c.}
= \nonumber \\
&=\frac{G_F}{\sqrt{2}}\mathcal{F}_{L2}(\overline{s}\gamma^\mu\gamma_{5} d)
(\overline{\mu}\gamma_\mu\gamma_{5}\mu) + \text{h.c.}
\label{kmumusm}
\end{align}
$Y(x_a)$, $x_a=m^2_a/M^2_W$,
are the Inami-Lim functions including QCD and electroweak corrections,  
whose leading order term 
is a linear combination of the axial 
components of $Z$-penguins and box-diagrams.
Numerically 
$Y(x_c)\approx 3.3\times 10^{-4}$, $Y(x_t)\approx 0.97$
\cite{Buchalla:1995vs}.
In eq. (\ref{kmumusm}) we defined the constant 
$\mathcal{F}_{L2}\approx(-2.1+i\, 0.78)\times 10^{-6}$.

In the scenario with extra isosinglet quark,
the non-diagonal couplings of $Z$-boson with SM families due to the mixing of 
light quarks with the $b'$-quark
leads to the Lagrangian analogous to (\ref{kpeenew}), contributing to the decay
$K_\text{L}\rightarrow \mu^+\mu^-$ at tree level (figure \ref{kll}):
\begin{align}
\label{kmumunew}
\mathcal{L}_\text{new}&=
\frac{G_F}{\sqrt{2}}V_{L4s}^{*}V_{L4d}(\overline{s}\gamma_\mu\gamma_{5} d)
\frac{1}{2}(\overline{\mu}\gamma^\mu\gamma_{5}\mu)+\text{h.c.} \\
\end{align}
%
By using again the phase convention $C|K^0\rangle=+|\bar{K}^0\rangle$
we have:
\begin{align}
&\langle 0|\bar{d}\gamma_{\mu}\gamma_{5} s |\bar{K}^0\rangle=
+\langle 0|\bar{s} \gamma_{\mu}\gamma_{5}  d |K^0\rangle
\end{align}
and neglecting indirect CP violation:
\begin{align}
& K_L\approx K_2 =\frac{1}{\sqrt{2}}(K^0+\bar{K}^0) 
\end{align}
Then we get:
\begin{align}
& \langle \mu^{+}\mu^{-}|\mathcal{L}_\text{SM,SD}|K_{L}\rangle 
= \frac{G_F}{\sqrt{2}}
\frac{2\text{Re}(\mathcal{F}_{L2})}{\sqrt{2}}
\langle 0|\bar{s} \gamma_{\mu}\gamma_{5}  d |K^0\rangle
[\bar{u}(p^{(\mu)})\gamma^\mu\gamma_{5}v(p^{(\bar{\mu})})] \\
& \langle \mu^{+}\mu^{-}|\mathcal{L}_\text{new}|K_{L}\rangle 
= \frac{G_F}{\sqrt{2}}
\frac{\text{Re}(V_{L4s}^{*}V_{L4d})}{\sqrt{2}}
\langle 0|\bar{s} \gamma_{\mu}\gamma_{5}  d |K^0\rangle
[\bar{u}(p^{(\mu)})\gamma^\mu\gamma_{5}v(p^{(\bar{\mu})})]
\end{align}
Then we can define the branching ratio given by the amplitude of the short distance contribution:
\begin{align}
\label{Brkmm}
& \text{Br}(K_\text{L}\rightarrow \mu^+\mu^-)_\text{SD}\! = 
k \left[\text{Re}(2\mathcal{F}_{L2}+V_{L4s}^{*}V_{L4d})\right]^2 
 , &&
 k\! = \! \frac{ \text{Br}(K^+\rightarrow  \mu^+\nu_\mu)\tau_{K_L}M_{K^0}\sqrt{1\! -\! 4\frac{m^2_\mu}{M^2_{K^0}}}}
{\tau(K^+)M_{K^+}\left(1\! -\! \frac{m^2_\mu}{M^2_{K^+}}\right)^2|V_{us}|^2} 
\end{align}
where we have used $\langle 0|\bar{s} \gamma_{\mu}\gamma_{5}  d |K^0\rangle= ip_{\mu}f_{K}$,
$\text{Br}(K^+\rightarrow \mu^+\nu_\mu)=0.6356(11)$ and $|V_{us}|=0.2252$. 
In absence of new physics, the short distance contribution coincides with the SM expectation
$\text{Br}(K_\text{L}\rightarrow \mu^+\mu^-)_\text{SD,SM}$.

By using the upper bound in eq. (\ref{SDtot}) on the branching ratio (\ref{Brkmm})
we have:
\begin{align}
\label{kmumud}
 \left|\text{Re}(2\mathcal{F}_{L2}+V_{L4s}^{*}V_{L4d})\right|& <6.9\times 10^{-6}
 \end{align}
This constraint is shown in figure \ref{fcplotd} in terms of the modulus $|V_{L4s}^{*}V_{L4d}|$
and relative phase $\delta_{L21}^{d}$, as parameterized in (\ref{tildes}), (\ref{4row}).
In terms of the ratio between the new amplitude (\ref{kmumunew}) and 
the SM amplitude (\ref{kmumusm}), 
eq. (\ref{kmumud}) implies:
\begin{align}
&  \left|1+\frac{\text{Re}(V_{L4s}^{*}V_{L4d})}{2\text{Re}(\mathcal{F}_{L2})}\right| < 1.7 \, ,
\qquad 0.7< \frac{|\text{Re}(V_{L4s}^{*}V_{L4d})|}{2|\text{Re}\mathcal{F}_{L2}|}< 2.7 
\end{align}
or, by using 
$ \text{Re}(V_{L4s}^{*}V_{L4d})= |V_{L4s}^{*}V_{L4d}| \cos\delta^d_{L21}$ we have:
\begin{align}
& 
-0.28 \times 10^{-5}< |V_{L4s}^{*}V_{L4d}| \cos\delta^d_{L21}<1.11\times 10^{-5} \, .
\end{align}
where we used the Wolfenstein parameterization (\ref{vtsvtdIm}).

\subsubsection{$K_\text{S}\rightarrow \mu^+\mu^-$}
The effective Lagrangian (\ref{kmumusm}) gives the short distance contribution to
the decay $K_\text{S}\rightarrow \mu^+\mu^-$:
\begin{align}
& \text{Br}(K_\text{S}\rightarrow \mu^+\mu^-)_\text{SD,SM} = 
4 k_{K_{S}} \left[\text{Im}(\mathcal{F}_{L2})\right]^2 
\end{align}
where $k_{K_{S}}$ is the same as $k$ in eq. (\ref{Brkmm}) with the change 
$\tau_{K_{L}}\rightarrow\tau_{K_{S}}$.
It is obtained $\text{Br}(K_\text{S}\rightarrow \mu^+\mu^-)_\text{SD} \approx 0.2  \cdot 10^{-12}$.
There are also long distance contributions arising from the two photon intermediate state 
which result in the rate 
$\text{Br}(K_\text{S}\rightarrow \mu^+\mu^-)_\text{LD} \approx 5.1  \cdot 10^{-12}$ \cite{Cirigliano}.
The experimental upper limit for this decay was found by
the LHCb collaboration 
\cite{LHCbmumupub}: 
\begin{align}
& \text{Br}(K_\text{S}\rightarrow \mu^+\mu^-)_\text{exp} < 1.0 \cdot 10^{-9} \quad (95\%\, \text{C.L.})
\end{align}
Then,  
an extra contribution
is allowed to be higher than both
the short and long-distance contributions to the decay, so we can impose an upper limit on the
new contribution arising in the scenario with the extra isosinglet.
The new decay channel is described by the effective Lagrangian
(\ref{kmumunew}) and gives the rate: 
\begin{align}
\label{kSmumu}
&\text{Br}(K_\text{S}\rightarrow \mu^+\mu^-)_\text{new}=
k_{K_{S}}
 \left[\text{Im}(V_{L4s}^{*}V_{L4d})\right]^2 < \text{Br}_\text{exp}
\end{align}
We have:
\begin{align}
\label{kSmm}
&\text{Br}(K_\text{S}\rightarrow \mu^+\mu^-)_\text{new}<1.0\cdot 10^{-9} \\
&|\text{Im}(V_{L4s}^*V_{L4d})|=  |V_{L4s}^{*}V_{L4d}||\sin\delta^d_{L21}|<
1.0\times 10^{-4}  \left[\frac{\text{Br}_{\text{exp}}}{1.0 \cdot 10^{-9}}  \right]^{1/2}
\end{align}
where 
we used the parameterization in 
eqs. (\ref{tildes}), (\ref{4row}).

\subsection{Limits from neutral mesons systems}

\subsubsection{$K^0$-$\bar{K}^0$ mixing}
\label{kksec}

\begin{figure}[t]
\centering
\includegraphics[width=0.33\textwidth]{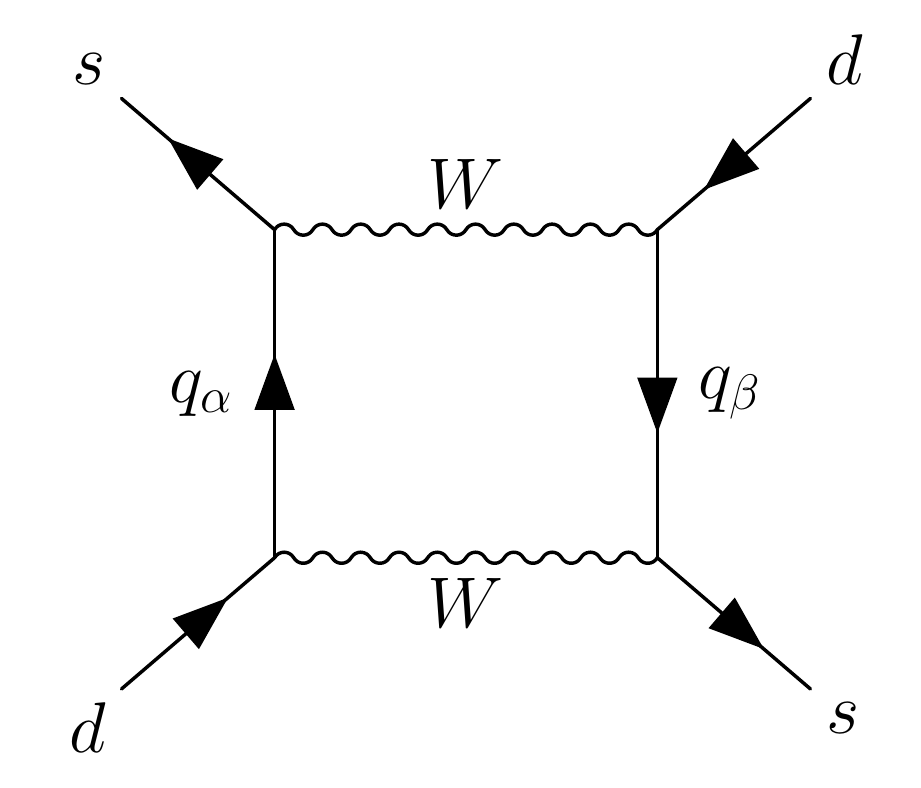}
\caption{\label{kksm} Box diagram for the transition $K^{0}\rightarrow \bar{K}^{0}$ in the SM,
$q_{\alpha},q_{\beta}=u,c,t$.   }
\end{figure}

In the SM the short-distance contribution to the transition 
$K^0(d\bar{s})\leftrightarrow \bar{K}^0(\bar{d}s)$
arises from weak box diagrams (figure \ref{kksm}).
The effective Lagrangian describing this contribution is given by:
\begin{align}
\label{smkk}
\mathcal{L}^\text{SM}_{\Delta S=2}&=-\frac{G^2_F m^2_W}{4\pi^2}\left[\lambda^{2}_c S_0(x_c)+\lambda^{2}_t S_0(x_t)+2\lambda_c\lambda_tS_0(x_c,x_t)\right](\overline{s_L}\gamma^\mu d_L)^{2}
+\text{h.c.} 
\end{align}
where $\lambda_a=V_{as}^*V_{ad}$, $x_a=\frac{m_a^2}{m_W^2}$ and $S_0(x_i)$ are the Inami-Lim
functions \cite{Inami:1980fz}:
\begin{align}
 S_0(x)=x & \left( \frac{4-11x+x^2}{4(1-x)^2}-\frac{3x^2\ln x}{2(1-x)^3} \right)
\label{S0x} \\
 S_0(x_j,x_k)=x_jx_k &\left[\left(\frac{1}{4}-\frac{3}{2(x_j-1)}-\frac{3}{4(x_j-1)^2}\right)
\frac{\log x_j}{x_j-x_k}+\right. \nonumber \\
&\left.\left(\frac{1}{4}-\frac{3}{2(x_k-1)}-\frac{3}{4(x_k-1)^2}\right)\frac{\log x_k}{x_k-x_j}\right.
\nonumber \\
&\left. -\frac{3}{4(x_j-1)(x_k-1)}\right] 
\label{S0xy}
\end{align}
which, being $x_c= m_c^2/m_W^2 \ll 1$, for the c-quark contribution reduce to:
\begin{align}
& S_{0}(x_{c})\approx x_{c} \\
&S_0(x_c,x_t)\approx x_c\left[\ln\frac{x_t}{x_c}-\frac{3x_t}{4(1-x_t)}-\frac{3x_t^2\ln x_t}{4(1-x_t)^2}\right] 
\label{S0xx}
\end{align}
keeping only linear terms in $x_c$ \cite{Buchalla:1995vs}.

The weak short-distance contribution to the mass splitting $\Delta m_K=m_{K_L}-m_{K_S}$
and the CP-violating parameter $\epsilon_K$ 
are described by
the off-diagonal term $M_{12}$ of the mass matrix of neutral kaons, which is given by:
\begin{align}
& M_{12}=-\frac{1}{2m_{K^{0}}} \langle K^0|\mathcal{L}_{\Delta S=2}|\bar{K}^0\rangle
\label{M12L}
\end{align}
The hadronic matrix element in vacuum insertion approximation (VIA) is written as:
\begin{align}
&\langle K^0 |(\overline{d_L}\gamma^\mu s_L)^2| \bar{K}^0\rangle =\frac{2}{3}
f^2_Km^{2}_{K^{0}}
\end{align}
with the phase choice $CP|K^{0}\rangle =-|\bar{K}^{0}\rangle$ \cite{Branco}. 
%
$f_K$ is the kaon decay constant, which can be estimated in lattice QCD to be
$f_K=155.7(0.7)$ MeV 
\cite{FLAG2019}, 
and $m_{K^{0}}$ is the neutral kaons mass. Then, the SM Lagrangian gives:
\begin{align}
M_{12}^\text{SM}&=
\frac{G^2_F m^2_W}{12\pi^2}(\eta_1\lambda^{*2}_c S_0(x_c)+\eta_2\lambda^{*2}_t S_0(x_t)+2\eta_3\lambda^*_c\lambda^*_tS_0(x_c,x_t))f^2_Km_{K^{0}}B_K
\label{M12sm}
\end{align}
where we also added the factors
$\eta_1 = 1.38\pm 0.20 $, $ \eta_2 = 0.57 \pm 0.01$, $\eta_3 = 0.47\pm 0.04 $ 
which
describe short-distance QCD effects \cite{Buchalla:1995vs}. 
The factor $B_{K}$ is the correction to the VIA which is estimated from
lattice QCD calculations, giving
$B_K=0.7625(97)$ 
\cite{FLAG2019}.

The modulus $|M_{12}|$ and the imaginary part $\text{Im}M_{12}$
respectively describe short-distance contributions in
the mass splitting and CP-violation in $\bar{K}^0\rightarrow K^0$ transition.
In the SM,
in the standard parameterization of $V_\text{CKM}$
we have:
\begin{align}
\label{M12mod}
|M_{12}^\text{SM}| \, \approx \, |\text{Re}M_{12}^\text{SM}| \, \approx \frac{G^2_Fm^2_W}{12\pi^2}\eta_1 & |\lambda^{2}_c|S_0(x_c) \approx \frac{G^2_F}{12\pi^2}\eta_1|\lambda^{2}_c| m^2_c f^2_Km_{K^{0}}B_K \\
\text{Im}M_{12}^\text{SM}=-\frac{G_F^2m_W^2}{6\pi^2}f^2_Km_{K^{0}}B_K
&  \left[
\eta_{1}\text{Re}(\lambda_{c})\text{Im}(\lambda_{c})S_0(x_c)+ \right. \nonumber \\
& \left.  +\big( \eta_2\text{Re}(\lambda_{t})S_0(x_t)+
\eta_3 \text{Re}(\lambda_{c}) S_0(x_c,x_t)\big)  \text{Im}(\lambda_{t}) \right]
\end{align}
The mass difference between mass eigenstates is given by
\cite{Buchalla:1995vs}:
\begin{align}
&\Delta m_K \approx 2|M_{12}|
+\Delta m_\text{LD}
\end{align}
where $\Delta m_\text{LD}$ is the long-distance contribution, which is difficult to evaluate.
Nevertheless, the short distance contribution 
$\Delta m_K \approx 2|M_{12}^\text{SM}| \simeq 2.2 \cdot 10^{-15} \, \text{GeV} $ 
gives a dominant contribution to
the experimentally measured value \cite{PDG18}:
\begin{align}
&\Delta M_K=(3.484\pm 0.006) \times 10^{-15} \,\text{GeV} 
\end{align}
%
The CP violation is parameterized by $\epsilon_K$, which, 
with the phase choice $CP|K^{0}\rangle =-|\bar{K}^{0}\rangle$,
is almost determined by short distance physics
\cite{Branco}:
\begin{align}
&|\epsilon_K| \approx \frac{|\text{Im}[ M_{12}]|}{\sqrt{2}\Delta m_K}
\end{align}
in the standard parameterization of $V_\text{CKM}$.
From experimental data it is obtained
\cite{PDG18}:
\begin{align}
& |\epsilon_K|=(2.228 \pm 0.011) \times 10^{-3}
\label{epsexp}
\end{align}
\begin{figure}[t]
\centering
\includegraphics[width=0.33\textwidth]{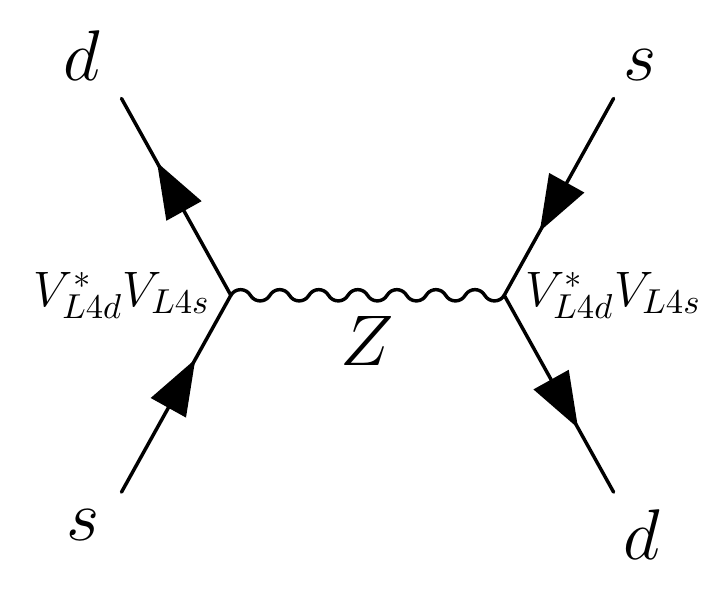}
\hspace{1cm}
\includegraphics[width=0.35\textwidth]{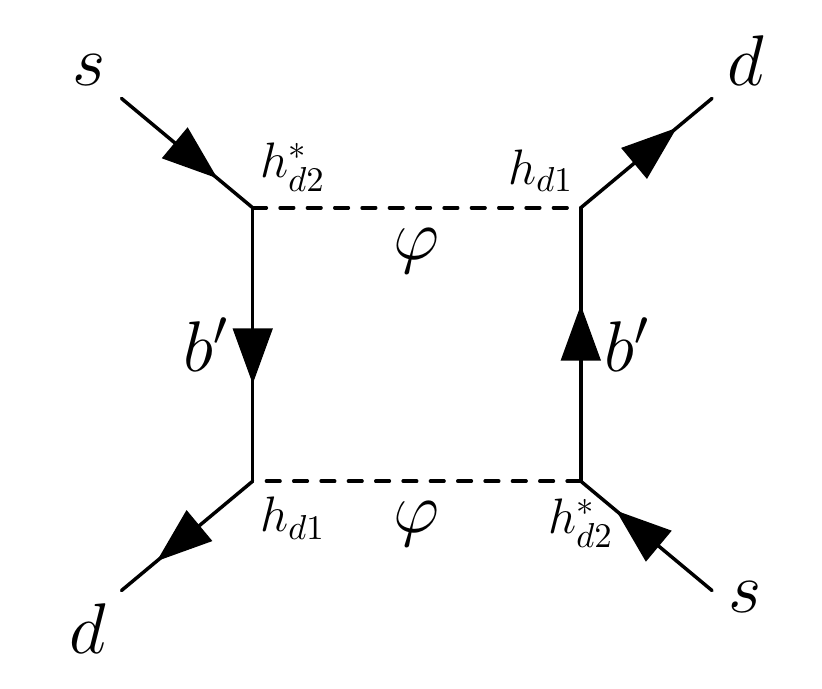}
\caption{\label{K0K0} Contributions to $\bar{K}^{0}$-$K^{0}$ transition due to the insertion of 
the additional vectorlike down-type weak isosinglet couple $b'_{L},b'_{R}$ in quark mixing.  }
\end{figure}

In our BSM scenario with extra vectorlike $b'$ quark,
the non-diagonal couplings of $Z$ and Higgs bosons with light quarks
in eqs. (\ref{nc}), (\ref{Higgs})
induce the same operator as in (\ref{smkk}), both at tree level and loop level, 
interfering with the SM in
the transition $\bar{K}^{0} \leftrightarrow K^{0}$, as shown in figure \ref{K0K0}. 

The new tree level contribution translates into the effective Lagrangian:
\begin{align}
\label{newtkk}
\mathcal{L}^\text{tree}_{\Delta S=2}=&-\frac{G_F}{\sqrt{2}}(V_{L4d}^*V_{L4s})^2(\overline{d_L}\gamma^\mu s_L)^2+\text{h.c.} 
\end{align}
However if $M_{b'}$ is of order few TeV, the loop contribution to $\bar{K}^0$-$K^{0}$ mixing
becomes important. In particular, for $M_{b'}=3.1$ TeV the 
effective operator for the 
box diagram with Higgs and $Z$ bosons exchange in figure \ref{K0K0}
gives the same contribution as the effective operator describing the tree level diagram. 
Moreover the box diagram contribution 
grows as $\propto M^{2}_{b'}$. 
In fact, the effective Lagrangian describing the loop contribution in figure \ref{K0K0} is:
\begin{align}
\label{looph}
\mathcal{L}^{\text{box}}_{\Delta S=2}
\approx & - \frac{(h_{d1}h_{d2}^{*})^{2}}{128 \pi^{2}M^{2}_{b'}} 
(\overline{d_L}\gamma^\mu s_L)^2+\text{h.c.} 
\end{align}
where $h_{di}$ are the Yukawa couplings defined in eq. (\ref{yd}).
However, we are interested in fixing a value for the elements of the mixing matrix rather than
setting the Yukawas $h_{di}$. Then, it is convenient to express the Lagrangian explicitly
in terms of the mixing elements.
From eqs. (\ref{4row}), (\ref{si}), we have:
\begin{align}
\label{hd}
& h_{d1}\approx -V_{L4d}^{*}\frac{M_{b'}}{v_{w}}\, ,
\qquad h_{d2}\approx -V_{L4s}^{*}\frac{M_{b'}}{v_{w}}\, .
\end{align}
Then, 
using $G_{F}/\sqrt{2}=1/4v_{w}^{2}$,
the Lagrangian can be written as:
\begin{align}
\label{Lbox}
&\mathcal{L}^{\text{box}}_{\Delta S=2}\approx - \frac{G^{2}_{F}\,M^{2}_{b'}}{16 \pi^{2}} 
(V_{L4d}^*V_{L4s})^{2}
(\overline{d_L}\gamma^\mu s_L)^2+\text{h.c.} 
\end{align}
Hence, the complete new contribution is:
\begin{align}
\label{newkk}
  \mathcal{L}^\text{new}_{\Delta S=2}&= 
\mathcal{L}^\text{tree}_{\Delta S=2}+ \mathcal{L}^\text{box}_{\Delta S=2}\approx
 \nonumber \\ &
\approx 
-\frac{G_F}{\sqrt{2}}(V_{L4d}^*V_{L4s})^2\left(1+\frac{G_{F}M^{2}_{b'}}{8\sqrt{2}\pi^{2}}  \right)
(\overline{d_L}\gamma^\mu s_L)^2+\text{h.c.} =
 \nonumber \\ &
=-\frac{G_F}{\sqrt{2}}(V_{L4d}^*V_{L4s})^2
f(M_{b'})
(\overline{d_L}\gamma^\mu s_L)^2+\text{h.c.}  \\ 
 f(M_{b'})&=1+\left(\frac{M_{b'} }{3.1 \, \text{TeV}}\right)^{2}
\label{fMb}
\end{align}
where we defined the function $f(M_{b'})$ of the mass of the extra quark $b'$.
From the Lagrangian (\ref{newkk}),
there is an additive contribution $M_{12}^\text{new}$ to 
the mixing mass $M_{12}$ (see eq. (\ref{M12L})), participating
both in the mass splitting and CP-violating effects. 
Then, the new contribution is constrained by both CP-conserving and CP-violating observables.

Regarding the CP-conserving part, which originates the mass splitting $\Delta m_{K}$,
we can confront the moduli of the two components of the mixing mass and
impose that the new contribution 
$|M_{12}^\text{new}|$ 
is less than a fraction $\Delta_{K}$ of the short-distance SM contribution
$ |M_{12}^\text{SM}|$: 
\begin{align}
\label{mkcon1}
&|M_{12}^\text{new}|<|M_{12}^\text{SM}|\,\Delta_{K} 
\end{align}
where we defined real and positive $\Delta_{K}$.
This is analogous to comparing 
the modulus of the coefficients of the effective operators in 
eqs. (\ref{smkk}) and (\ref{newkk}). 
We evaluate the constraint at leading order of both SM and new physics contributions, 
neglecting QCD corrections.
Hence, eq. (\ref{mkcon1}) corresponds to:
\begin{align}
\label{mkcon2}
&\frac{G_F}{\sqrt{2}} f(M_{b'})
\left|V_{L4d}^*V_{L4s}\right|^{2}<
\, \frac{G^2_F m^2_c}{4\pi^2}|\lambda^2_c| \, \Delta_{K} 
\end{align}
eq. (\ref{mkcon1}) can also be translated into 
a constraint on the scale of the new contribution:
\begin{align}
&\frac{G_F}{\sqrt{2}} f(M_{b'})
\left|(V_{L4d}^*V_{L4s})^2\right|< \frac{1}{\: \Lambda_{sd}^{2}}  \, , \qquad
 \Delta_{K} = \left(\frac{1.9\cdot 10^{3}\, \text{TeV}}{\Lambda_{ds}}\right)^{2}
\end{align}
where we defined the minimum allowed equivalent scale $\Lambda_{sd}^{2}$ of the new contribution.
We estimate the constraint with $\Delta_{K}=1$, that is by imposing the condition 
$|M_{12}^\text{new}|<|M_{12}^\text{SM}|$, meaning that
the additional processes contribute to $\Delta m_{K}$ at most as the SM 
short-distance contribution.
Anyway, we leave 
$\Delta_{K}$ as free parameter in next equations 
so to allow to reevaluate the results with a different constraint.
Then, using
eqs. (\ref{tildes}), (\ref{4row}),
from the condition (\ref{mkcon2}) 
we have:
\begin{align}
\label{vsdmk}
& |V_{L4d}^*V_{L4s}| 
< 1.7 \times 10^{-4} 
\left[ \frac{f(1 \,  \text{TeV})}{f(M_{b'})} \right]^{1/2}
\left[ \Delta_{K} \right]^{1/2}
\end{align}
where we evaluated the constraint for the benchmark value $M_{b'}=1$~TeV.

Regarding the contribution to the CP-violating parameter $\epsilon_K$,
the imaginary part of the new contribution to the mixing mass $\text{Im}M_{12}^\text{new}$ 
can be constrained to be a fraction $\Delta_{\epsilon_{K}}$ of the SM contribution:
\begin{align}
\label{epskcon1}
|\text{Im}M_{12}^\text{new}|<|\text{Im}M_{12}^\text{SM}| \, \Delta_{\epsilon_{K}} \, ,
\end{align}
where we defined real and positive $\Delta_{\epsilon_{K}}$.
At leading order of 
the new physics contribution, 
eq. (\ref{epskcon1}) is equivalent to 
comparing the magnitude of the imaginary part of operators 
(\ref{smkk}) and (\ref{newkk}):
\begin{align}
\frac{G_F}{\sqrt{2}}
f(M_{b'})
|\text{Im}[(V_{L4d}^*V_{L4s})^{2}]|
<&\, \frac{G_F^2m^2_W}{4\pi^2}  2  \left|
\eta_{1}\text{Re}(\lambda_{c})\text{Im}(\lambda_{c})S_0(x_c)+ \right. \nonumber \\
& \left.  +\big( \eta_{2}\text{Re}(\lambda_{t})S_0(x_t)+
 \eta_{3}\text{Re}(\lambda_{c}) S_0(x_c,x_t)\big)  \text{Im}(\lambda_{t})
\right| \, \Delta_{\epsilon_{K}} 
\label{epskcon}
\end{align}
eq. (\ref{epskcon}) can also be translated into 
a constraint on the scale of the new contribution:
\begin{align}
\label{Lambdaeps}
&\frac{G_F}{\sqrt{2}} f(M_{b'})
\left|\text{Im}[(V_{L4d}^*V_{L4s})^2]\right|< \frac{1}{\: \Lambda_{sd,\text{Im}}^{2}}  \, , \qquad
 \Delta_{\epsilon_{K}} = \left(\frac{1.6\cdot 10^{4}\, \text{TeV}}{\Lambda_{ds,\text{Im}}}\right)^{2}
\end{align}
where we defined the minimum allowed equivalent scale $\Lambda_{sd,\text{Im}}^{2}$ of the new contribution.
The numerical value is obtained by using the Wolfenstein parameterization as in eq. 
(\ref{vtsvtdIm}), with $\text{Im}(V_{cs}^{*}V_{cd})=-A^{2}\lambda^{5}\eta$.
We make an estimation choosing $\Delta_{\epsilon_{K}}=0.4$,
corresponding to $\Lambda_{ds,\text{Im}}=2.5\cdot 10^{4}\, \text{TeV}$.
%
Using eqs. (\ref{tildes}) and (\ref{4row}), from the condition (\ref{epskcon}) we have:
\begin{align}
\label{eps041}
& |V_{L4d}^*V_{L4s}|\sqrt{|\sin[2(\delta^d_{L12})]|}< 1.3 \cdot 10^{-5} 
\left[ \frac{f(1 \,  \text{TeV})}{f(M_{b'})} \right]^{1/2}
\left[ \frac{\Delta_{\epsilon_{K}}}{0.4} \right]^{1/2} 
\end{align}
where we evaluated the constraint for the benchmark value $M_{b'}=1$~TeV.
The constraints are also 
shown in figure \ref{fcplotd} for $M_{b'}=1$~TeV, in terms of the modulus and relative phase of the elements $V_{L4d}^*V_{L4s}$.

Let us elaborate a little more about 
the two contributions to $ \mathcal{L}^\text{new}_{\Delta S=2}$
from the tree and box diagrams.
The tree level operator (\ref{newkk}) 
can be written in terms of the Yukawa constants $h_{di}$,
in order to be compared with
the operator (\ref{looph}).
Using $G_{F}/\sqrt{2}=1/4v_{w}^{2}$ and eq. (\ref{hd}), we have:
\begin{align}
& \mathcal{L}^\text{tree}_{\Delta S=2}=
-\frac{(h_{d1}h_{d2}^{*})^{2}}{4}\frac{v_{w}^{2}}{M_{b'}^{4}}
(\overline{d_L}\gamma^\mu s_L)^2+\text{h.c.} 
\end{align}
which parametrically is a factor $v_{w}^{2}/M_{b'}^{2}$ less than the box operator (\ref{looph}).
That is why, after compensating 
the loop factor, for $M_{b'}>3.1$ TeV the box contribution becomes
comparable or 
more important than the tree level contribution.

\subsubsection{$B^0_{d,s}$-$\bar{B}_{d,s}^0$ mixing}
\label{bbsec}

\begin{figure}[t]
\centering
\includegraphics[width=0.33\textwidth]{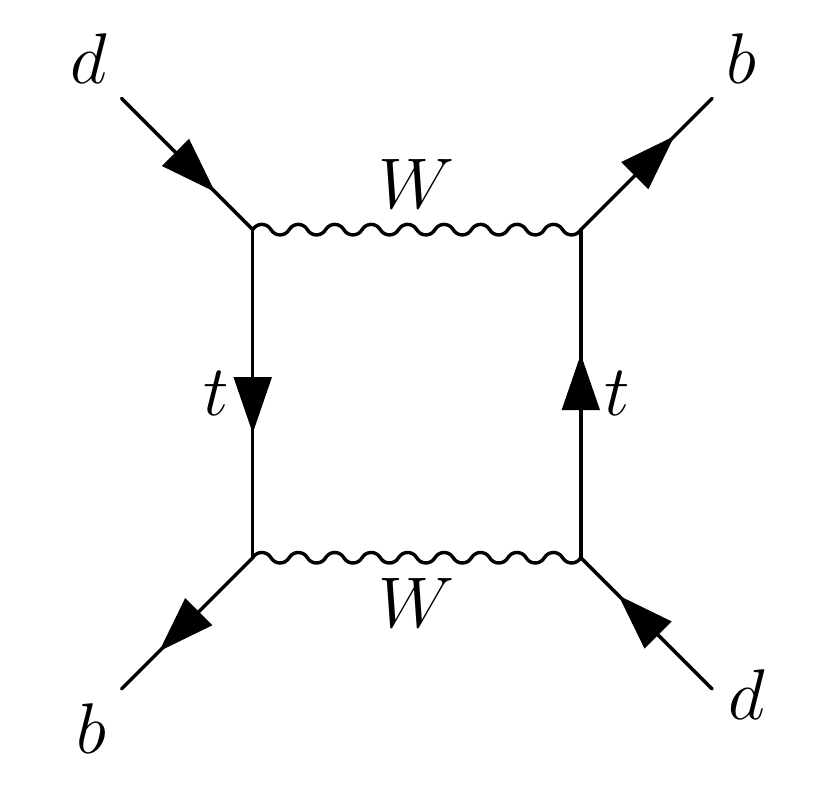}
\caption{\label{bbsm} Box diagram for the transition $B^{0}\rightarrow \bar{B}^{0}$ in the SM.   }
\end{figure}

In the SM the dominant contribution to 
$B^0_{d}(\bar{b}d)$-$\bar{B}_{d}^0(\bar{d}b)$ and $B^0_{s}(\bar{b}s)$-$\bar{B}_{s}^0(\bar{s}b)$ 
mixings comes from 
box-diagrams with internal top-quark, as shown in figure \ref{bbsm}, while
the charm quark and the mixed top-charm contributions are entirely negligible.
The effective Lagrangians for $B^0_{d,s}$-$\bar{B}_{d,s}^0$ mixings are
\cite{Buchalla:1995vs}:
\begin{align}
\label{BBsm}
&\mathcal{L}^\text{SM}_{\Delta B(d)=2}=-\frac{G_F^2}{4\pi^2}m^2_W S_{0}(x_t)(V_{tb}^*V_{td})^2
(\overline{b_L}\gamma_\mu d_L)^2+\text{h.c.} \\
&\mathcal{L}^\text{SM}_{\Delta B(s)=2}=-\frac{G_F^2}{4\pi^2}m^2_W S_{0}(x_t)(V_{tb}^*V_{ts})^2
(\overline{b_L}\gamma_\mu s_L)^2+\text{h.c.}
\label{BBsSM}
\end{align}
where $S_{0}(x)$ is the function in eq. (\ref{S0x}).
Then, analogously to the kaons system,
in VIA:
\begin{align}
\label{Bbraket}
&\langle \bar{B}_{d,s}^0 |(\overline{b_L}\gamma^\mu 
q_L)^2|B^0_{d,s}\rangle =\frac{2}{3}f^2_{B_{d,s}}M^2_{B^{0}_{d,s}}
\end{align}
and
\begin{align}
\label{m12bb}
&2M_{B^{0}_{d,s}}M_{12}^{(d,s)*}=\langle \bar{B}_{d,s}^0 |-\mathcal{L}_{\Delta B(d,s)=2}|B^0_{d,s}\rangle \\
&\Delta M_{d,s}=2|M_{12}^{(d,s)}| 
\label{dmm12bb}
\end{align}
Putting together (\ref{Bbraket}), (\ref{m12bb}), (\ref{dmm12bb}), in the SM:
\begin{align}
\label{dmBB}
&\Delta M_{d,s}^\text{SM}=\frac{G_F^2m_W^2}{6\pi^2}f^2_{B_{d,s}}m_{B_{d,s}}B_{B_{d,s}}|(V_{tb}V_{td/s}^*)^2|\eta_BS_{0}(x_t)
\end{align}
where $\eta_B$ is the QCD factor, $\eta_B=0.551$ \cite{Buchalla:1995vs}, and the factors
$B_{B_{d,s}}$ are correcting the VIA. From lattice QCD calculations:
$f_{B_d}\sqrt{B_{B_d}}=225(9)$ MeV, $f_{B_s}\sqrt{B_{B_s}}=274(8)$ MeV \cite{FLAG2019}.
In contrast to $\Delta m_{K}$, long distance contributions are estimated to be very small 
in neutral B meson systems.
and $\Delta M_{d,s}$ is very well approximated by the relevant box diagrams \cite{Buchalla:1995vs}.
The experimental results are \cite{PDG18}:
\begin{align}
&\Delta M_{d\,\text{exp}}=(3.334\pm 0.013) \cdot 10^{-13} \, \text{GeV}  \\
&\Delta M_{s\,\text{exp}}=(1.1683\pm 0.0013) \cdot 10^{-11} \, \text{GeV}  
\end{align}


\begin{figure}[t]
\centering
\includegraphics[width=0.33\textwidth]{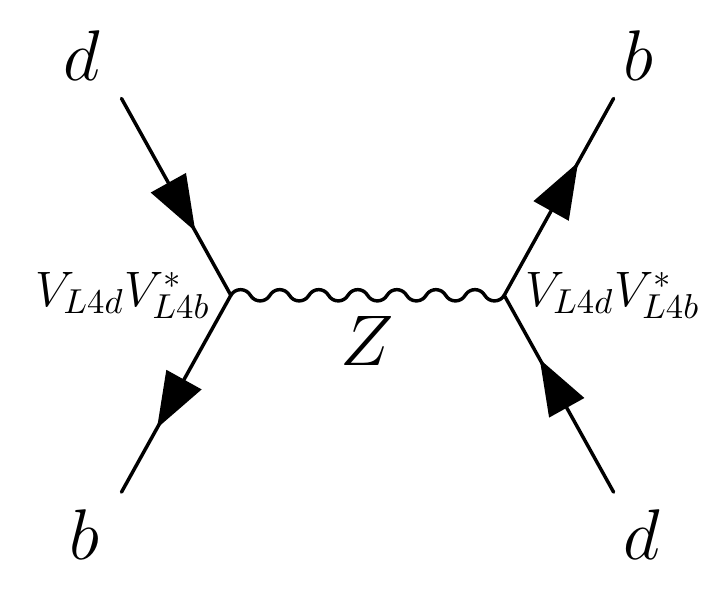}
\hspace{1cm}
\includegraphics[width=0.35\textwidth]{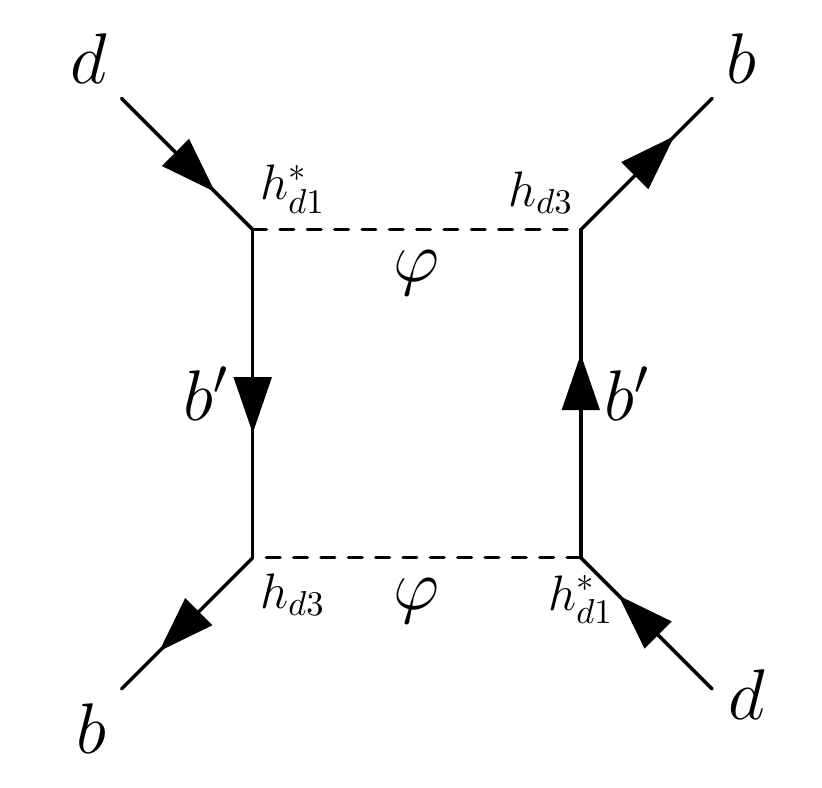}
\caption{\label{BB} Contributions to $B^{0}$-$\bar{B}^{0}$ transition due to the insertion of 
the additional vectorlike down-type weak isosinglet couple $b'_{L},b'_{R}$ in quark mixing.  }
\end{figure}

In our BSM scenario with extra vectorlike $b'$ quark,
the non-diagonal couplings of $Z$-boson and Higgs boson with light quarks
in eqs. (\ref{nc}), (\ref{Higgs})
induce the 
operator with the same structure as in (\ref{BBsm}), 
interfering with the SM in
the transition $\bar{B}^{0}_{d,s} \leftrightarrow B^{0}_{d,s}$. 
The new contributions at tree level and loop level are analogous to the operators giving
$K^{0}$ mixing 
(\ref{newtkk}), (\ref{Lbox}),
as shown in diagrams in figure \ref{BB}. 
Then, analogously to the neutral kaons system,
the new contribution to neutral B-mesons mixing is:
\begin{align}
\label{newBB}
 & \mathcal{L}^\text{new}_{\Delta B(d)=2}
\approx -\frac{G_F}{\sqrt{2}}(V_{L4d}^*V_{L4b})^2
f(M_{b'})
(\overline{d_L}\gamma^\mu b_L)^2+\text{h.c.}  \\ 
\label{newBBs}
&  \mathcal{L}^\text{new}_{\Delta B(s)=2}
\approx -\frac{G_F}{\sqrt{2}}(V_{L4s}^*V_{L4b})^2
f(M_{b'})
(\overline{s_L}\gamma^\mu b_L)^2+\text{h.c.}  \\ 
 & f(M_{b'})=1+\left(\frac{M_{b'} }{3.1 \, \text{TeV}}\right)^{2}
\end{align}
%
From the Lagrangians (\ref{newBB}), (\ref{newBBs}),
there is an additive contribution $M_{12}^\text{new}$ to 
the mixing mass $M_{12}^{(d,s)}$ (\ref{m12bb}), and thus to the mass difference
$\Delta M_{d,s}$ (\ref{dmm12bb}). 
We can constrain 
the new contribution to be 
less than a fraction $\Delta_{B_{(d,s)}}$ of the SM contribution
$\Delta M_{d,s}^\text{SM}= 2|M_{12}^{(d,s)\text{SM}}|$ given in eq. (\ref{dmBB}).
This is analogous to comparing the coefficients of the effective operators in 
eqs. (\ref{newBB}), (\ref{newBBs}) with the SM ones (\ref{BBsm}), (\ref{BBsSM}):
\begin{align}
&\frac{G_F}{\sqrt{2}}f(M_{b'})|V_{L4d}^*V_{L4b}|^2<
\frac{G_F^2}{4\pi^2}m^2_W|V_{td}|^2S(x_t)\cdot \Delta_{B_d}= 
\frac{1}{\Lambda_{bd}^{2}} \\
&\frac{G_F}{\sqrt{2}}f(M_{b'})|V_{L4s}^*V_{L4b}|^2<\frac{G_F^2}{4\pi^2}m^2_W|V_{ts}|^2S(x_t)\cdot \Delta_{B_s}= 
\frac{1}{\Lambda_{bs}^{2}}
\end{align}
where we evaluate the constraint at leading order of both SM and new physics contributions, 
neglecting QCD corrections.
We use $\Delta_{B_{(d,s)}}=0.3$ as a benchmark value,
corresponding to the scales 
$\Lambda_{bd}=1.0\cdot 10^{3}\,\text{TeV}$, $\Lambda_{bs}=2.1\cdot 10^{2}\,\text{TeV}$. 
Then we obtain
(with $m_t(m_t)=163$ GeV):
\begin{align}
&|V_{L4d}^*V_{L4b}| 
<3.3 \times 10^{-4} 
\left[ \frac{f(1 \,  \text{TeV})}{f(M_{b'})} \right]^{1/2}
\left[\frac{\Delta_{B_d}}{0.3}\right]^{1/2}
\\ 
&|V_{L4s}^*V_{L4b}|  
< 1.6\times 10^{-3}
\left[ \frac{f(1 \,  \text{TeV})}{f(M_{b'})} \right]^{1/2}
\left[\frac{\Delta_{B_s}}{0.3}\right]^{1/2}
\end{align}

\subsubsection{$D^0$-$\bar{D}^0$ mixing}
\label{DDdown}
In the SM $D^0(c\bar{u})$-$\bar{D}^0(\bar{c}u)$ mixing receives contributions from box diagrams, 
dipenguin diagrams and from long-distance effects \cite{Branco}.
In box diagrams only the internal strange and down quarks contribute effectively.
Differently from the case of kaons and B-mesons, the masses of $s$ and $d$ are small compared to the mass of the external $c$ quarks, which then cannot be neglected and 
an extra operator appears in the effective Lagrangian. In VIA approximation, 
the box contribution results 
$|M_{12}| \sim 10^{-17}\div 10^{-16}$ GeV \cite{Branco}.
Also 
dipenguin diagrams can give a contribution 
not much smaller than 
box diagrams \cite{Petrov}.
Long-distance effects 
are expected to be large \cite{Branco}. 
However, since 
their contribution is non-perturbative and cannot be computed reliably,
there is room for new physics, which in principle can be the dominant contribution to 
the mass difference $\Delta m_{D}$ in $D^{0}$ system.
%
In fact, the experimental value of the mass difference in the $D^0$ mesons system is \cite{PDG18}:
\begin{align}
\label{dmdexp}
&\Delta m_{D\,\text{exp}}=(6.25^{+2.70}_{-2.90})\times 10^{-15} \text{GeV}
\end{align}
which at $95\%$ C.L. 
allows values 
two orders of magnitude bigger
than the SM short-distance expectation.

\begin{figure}[t]
\centering
\includegraphics[width=0.33\textwidth]{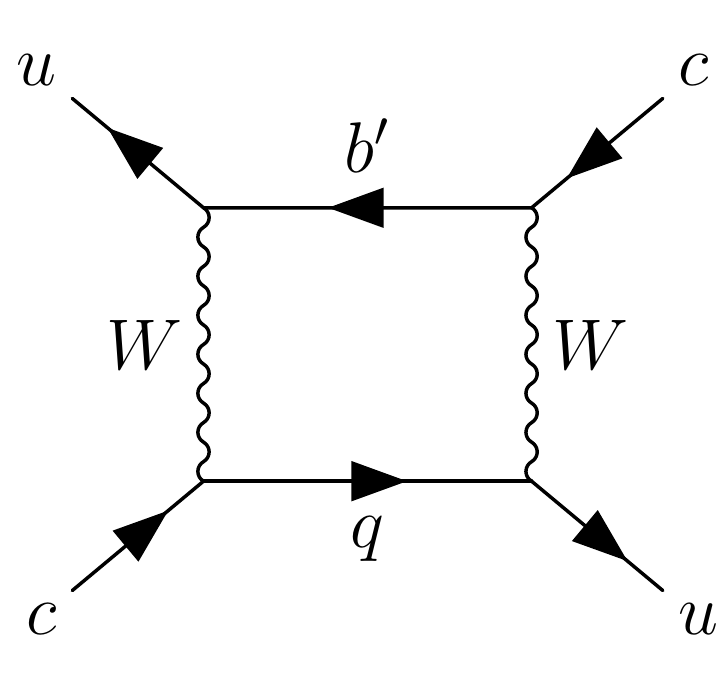}
\caption{New contribution to $D^0-\bar{D^0}$ mixing, $q=d,s,b,b'$.}
\label{DD}
\end{figure}
In the scenario with extra down-type quark,
the additional box diagrams with internal $b'$ quarks 
contribute to 
$D$ mesons mixing, as shown in 
figure (\ref{DD}). The corresponding effective Lagrangian is:
\begin{align}
\label{DDLloop}
\mathcal{L}^\text{new}_{\Delta C=2} \approx &-\frac{G^2_F m^2_W}{4\pi^2}
(V_{ub'}V_{cb'}^*)^2 S_0(x_{b'}) (\overline{u_L}\gamma^\mu c_L)^{2}
+\text{h.c.}  
\end{align}
with $x_{b'}=\frac{M_{b'}^2}{m_W^2}$ (taking $M_{b'}\simeq 1$ TeV, the two mixed contributions 
would be competitive 
only if $|V_{ub'}V_{cb'}^*|<2\cdot 10^{-7}$, but 
we are considering scenarios with larger mixings).
Since the new effective operator (\ref{DDLloop}) originates from 
charged currents, this process involves directly
the elements $V_{ub'}$, $V_{cb'}$ of the enlarged CKM matrix $\tilde{V}_\text{CKM}$ (\ref{vckm}).
This operator gives an additional contribution $M^{D}_{12\,\text{new}}$ 
to the mixing mass
and thus contributes to the mass difference $\Delta m_{D}$:
\begin{align}
\label{dmdnewloop}
&\Delta m_{D \text{new}}=2|M^{D}_{12\,\text{new}}| \simeq 
\frac{G^2_Fm^2_W}{6\pi^2}(V_{ub'}V_{cb'}^*)^2 S_0(x_{b'})f^2_Dm_{D^{0}}
\end{align}
where $f_D=212.0\pm 0.7$ MeV \cite{FLAG2019} is the decay constant.
Since the new contribution can be regarded as
the dominant one, we can think that the new operator 
can account up to the entire 
mass difference in D-mesons system. Then 
we can set a constraint directly on the new contribution.
We can constrain 
$\Delta m_{D\, \text{new}}$ by using
use the $95\%$ C.L. experimental limit obtained from eq. (\ref{dmdexp}):
\begin{align}
& 
\Delta m_{D \text{new}} < \Delta m_{D\,\text{exp}} =
1.2 \times 10^{-14} \, \text{GeV} \\
& |V_{ub'}V_{cb'}^*|< 4.6 \times 10^{-4} \left[ \frac{\Delta m_{D\,\text{exp}}}{1.2\cdot 10^{-14}}  \right]^{1/2} \left[ \frac{1\,\text{TeV}}{M_{b'}}  \right]  \label{vcbpvubpdd} 
\end{align}
The last scaling in eq. (\ref{vcbpvubpdd}) 
holds 
since $S_0(x_{b'})\sim \frac{1}{4}x_{b'}$ for $M_{b'}\gtrsim 2$ TeV.

\subsection{Limits from rare $B$ mesons decays}

\subsubsection{Rare semileptonic B decays}

Results on rare B-decays can constrain mixings of the extra vector-like quark with SM families
because of the new contributions to
FCNC processes involving $b\rightarrow s(d)$ 
originated at tree level by non-diagonal couplings of $Z$-boson
with SM quarks.

The new effective Lagrangian contributing to $b\rightarrow s(d)\ell^+\ell^-$ decays are:
\begin{align}
\label{Bnew1}
 \mathcal{L}_\text{new}=&-\frac{4G_F}{\sqrt{2}}V_{L4d}V_{L4b}^*
(\overline{b_L}\gamma^\mu d_L)\left[(-\frac{1}{2}+\sin^2\theta_W)(\overline{\ell_L}\gamma_\mu\ell_L)
+\sin^2\theta_W(\overline{\ell_R}\gamma_\mu\ell_R)\right] 
\\
\mathcal{L}_\text{new}=&-\frac{4G_F}{\sqrt{2}}V_{L4b}^*V_{L4s}
(\overline{b_L}\gamma^\mu s_L)\left[(-\frac{1}{2}+\sin^2\theta_W)(\overline{\ell_L}\gamma_\mu\ell_L)
+\sin^2\theta_W(\overline{\ell_R}\gamma_\mu\ell_R)\right] 
\label{Bnew2}
\end{align}
giving 
both exclusive decays, such as $B\rightarrow \pi\ell^+\ell^-$ and $B\rightarrow K\ell^+\ell^-$, and
inclusive decays $B\rightarrow X_{s(d)}\ell^+\ell^-$,
where $X_{s(d)}$ stands for anything with a 
$s(d)$ quark.

As regards exclusive decays,
experimental branching fractions are below SM predictions \cite{PDG18}.
The decays $B^0\rightarrow K^{0*}\ell^+\ell^-$ and $B^+\rightarrow K^+\ell^+\ell^-$
are the best studied and 
tensions with SM were found related to the quantity $P'_5$ and to lepton universality test
\cite{PDG18} (and references therein).

Inclusive decays are also studied in order to avoid hadronic form factors.
As regards $B\rightarrow X_s\ell^+\ell^-$, 
branching fractions can be analyzed both from a theoretical and experimental point of view in 
low and high dilepton invariant mass regions, that is
$1\,\text{GeV}^2<q^2<6\,\text{GeV}^2$ and
$q^2>14.4\text{GeV}^2$ 
regions (the intermediate region is dominated by charmonia resonances).
Experimental measurements of the inclusive decay were made by both 
Belle \cite{Belle} and BaBar \cite{BaBar}. Their 
data are consistent with SM expectations at $95\%$ C.L. \cite{Huber:2015sra}.
In order to have a rough estimate on the constraint for the mixings 
$|V_{L4d}V_{L4b}^*|$, $|V_{L4s}V_{L4b}^*|$, we can consider the 
total branching fraction averaged between electrons and muons, which experimentally
results \cite{PDG18}: 
\begin{align}
\label{BXexp}
&\text{Br}(B\rightarrow X_s\ell^+\ell^-)_\text{exp} = (5.8 \pm 1.3) \times 10^{-6}
\end{align}
SM calculations predict 
$\text{Br}(B\rightarrow X_s\ell^+\ell^-)_\text{SM} = (4.6 \pm 0.8) \times 10^{-6}$
\cite{Ghinculov:2003qd}. 
Let us define
the branching ratio which would 
be given only by the new contribution:
%
%
%
\begin{align}
\label{maxB}
&|A_\text{new}|^{2}=\text{Br}(B\rightarrow X_s\ell^+\ell^-)_\text{new} \simeq \nonumber\\
&\simeq \text{Br}(B\rightarrow X_c \ell^+\nu_\ell)
\frac{|V_{L4b}^*V_{L4s}|^2[(-\frac{1}{2}+\sin^2\theta_W)^2+\sin^4\theta_W]}{|V_{cb}|^2} 
<\text{Br}(B\rightarrow X_s\ell^+\ell^-)_{\text{max}} 
\end{align}
$\text{Br}(B\rightarrow X_s\ell^+\ell^-)_{\text{max}}$ is defined here as the maximal allowed value
for the branching ratio generated by the amplitude of the new contribution alone and 
we use the experimental branching ratio
$\text{Br}(B\rightarrow X_c\ell^+\nu_\ell)=0.1065(16)$ \cite{PDG18}.
Similarly $|A_\text{SM}|^{2}=\text{Br}(B\rightarrow X_s\ell^+\ell^-)_\text{SM}$. Then, at $95\%$ C.L.
of the experimental result (\ref{BXexp}), we should have:
\begin{align}
& 3.25\times 10^{-6}< |A_\text{new}+A_\text{SM}|^{2} <8.35\times 10^{-6}
\end{align}
Then, in order to set the least stringent constraint, we can consider the case of destructive interference:
\begin{align}
& |A_\text{new}|<\sqrt{8.35\times 10^{-6}}\left( 1+\frac{|A_\text{SM}|}{\sqrt{8.35\times 10^{-6}}} \right)
\end{align}
By using the $95\%$ C.L. interval also for the SM branching ratio, $|A_\text{SM}|^{2}=6.2\times 10^{-6}$,
the resulting constraint is:
\begin{align}
& 
|V_{L4b}^*V_{L4s}|< 1.9\times 10^{-3}
\left[\frac{\text{Br}(B\rightarrow X_s\ell^+\ell^-)_{\text{max}}}{2.9\times 10^{-5}}  \right]^{\frac{1}{2}}
\end{align}
and, for this case of destructive interference, $|V_{L4b}^*V_{L4s}|>1.5 \cdot 10^{-3}$.
As regards $b\rightarrow d\ell^+\ell^-$ transitions, 
the LHCb collaboration measured the branching ratio for the decay 
$B^+\rightarrow \pi^+\mu^+\mu^-$ \cite{PDG18}, \cite{LHCbpi0}:
\begin{align}
\label{expBpmm}
& \text{Br}(B^+\rightarrow \pi^+\mu^+\mu^-)_\text{exp}=(1.75\pm0.22)\cdot10^{-8}
\end{align}
which agrees with SM predictions, for example 
$\text{Br}(B^+\rightarrow \pi^+\mu^+\mu^-)_\text{SM}=1.88^{+0.32}_{-0.21}\cdot10^{-8}$ \cite{Ali:2013zfa}
$\text{Br}(B^+\rightarrow \pi^+\mu^+\mu^-)_\text{SM}=(2.04\pm0.21)\cdot10^{-8}$ \cite{Bailey:2015nbd}.
Other than that, upper limits on exclusive decays include 
Belle constraint $\text{Br}(B^+\rightarrow \pi^+\ell^+\ell^-)<4.9\cdot 10^{-8}$ ($90\%$ C.L.)
\cite{Belle3} and BaBar result \cite{BaBar3}:
\begin{align}
\label{expBp0ll}
& \text{Br}(B^0\rightarrow \pi^0\ell^+\ell^-)_\text{exp}<5.3\cdot 10^{-8} \qquad 90\%\, \text{C.L.}
\end{align}
$\ell=e$ or $\mu$.
The SM expectation for the $B^{0}$ decay can be obtained from the predictions of $B^{+}$ 
branching ratios by multiplying by the factor $(\tau_{B^{0}}/\tau_{B^{+}})/2$, where
$\tau_{B^{0}}/\tau_{B^{+}}=1.076(4)$ \cite{PDG18}.
Let us define as before the decay rate which would arise only from the new contribution: 
\begin{align}\label{expB1}
&|A^\text{new}_{B^\pm\rightarrow \pi^\pm\ell^+\ell^-}|^{2}= 
\text{Br}(B^\pm\rightarrow \pi^\pm\ell^+\ell^-)_\text{new}\simeq 
k_{B^{\pm}}|V_{L4d}V_{L4b}^*|^2  
\\
&|A^\text{new}_{B^0\rightarrow \pi^0\ell^+\ell^-}|^{2}=
 \text{Br}(B^0\rightarrow \pi^0\ell^+\ell^-)_\text{new}\simeq 
k_{B^{0}}|V_{L4d}V_{L4b}^*|^2  
\label{expB2}
\end{align}
with
\begin{align}\label{expB1k}
& k_{B^{\pm}}=2\frac{\text{Br}(B^\pm\rightarrow \pi^0 \ell^+\nu_\ell)[(-\frac{1}{2}+\sin^2\theta_W)^2+\sin^4\theta_W]}{|V_{ub}|^2} \\
& k_{B^{0}}= \frac{1}{2}\frac{\text{Br}(B^0\rightarrow \pi^- \ell^+\nu_\ell)[(-\frac{1}{2}+\sin^2\theta_W)^2+\sin^4\theta_W]}{|V_{ub}|^2} 
\label{expB2k}
\end{align}
%
The factors of $2$ take into account the isospin symmetry relation.
We also take advantage of the 
results in \cite{PDG18}
$\text{Br}(B^0\rightarrow \pi^- \ell^+\nu_\ell)=(1.50\pm 0.06)\cdot 10^{-4}$,
$\text{Br}(B^\pm\rightarrow \pi^0 \ell^+\nu_\ell)=(7.80\pm 0.27)\cdot 10^{-5}$, $\ell=e$ or $\mu$.
Then we should have:
\begin{align}
 1.3\times 10^{-8}<\, &|A_{B^\pm\rightarrow \pi^\pm\ell^+\ell^-}^\text{new}+A_{B^\pm\rightarrow \pi^\pm\ell^+\ell^-}^\text{SM}|^{2}  <  2.2\times 10^{-8}  \\
&|A_{B^0\rightarrow \pi^0\ell^+\ell^-}^\text{new}+A_{B^0\rightarrow \pi^0\ell^+\ell^-}^\text{SM}|^{2} <5.3\cdot 10^{-8}
\end{align}
where we defined $|A_\text{SM}|^{2}=\text{Br}_\text{SM}$. We used the $95\%$ C.L. of the 
experimental result (\ref{expBpmm}) and the upper limit (\ref{expBp0ll}).
In order to set the constraint, we can consider the case of destructive interference between the new
contribution and the SM one:
\begin{align}\label{expB1}
&|V_{L4d}V_{L4b}^*|<\frac{|A_{B^\pm\rightarrow \pi^\pm\ell^+\ell^-}^\text{SM}| }{\sqrt{k_{B^{\pm}}}}\left(1+\frac{\sqrt{2.2\times 10^{-8}} }{|A_{B^\pm\rightarrow \pi^\pm\ell^+\ell^-}^\text{SM}|}\right) 
\\
&|V_{L4d}V_{L4b}^*|<\sqrt{\frac{5.3\times 10^{-8}}{k_{B^{0}}}}\left(1+\frac{ |A_{B^0\rightarrow \pi^0\ell^+\ell^-}^\text{SM}|}{\sqrt{5.3\times 10^{-8}}}\right)
\end{align}
In order to have the least stringent constraint, we can use the $95\%$ C.L. interval of the SM expectations,
$|A_{B^\pm\rightarrow \pi^\pm\ell^+\ell^-}^\text{SM}|^{2}=2.5\times 10^{-8}$ and
$|A_{B^0\rightarrow \pi^0\ell^+\ell^-}^\text{SM}|^{2}=1.3\times 10^{-8}$. Then, we obtain:
\begin{align}
&|V_{L4d}V_{L4b}^*|< 
2.6\times 10^{-4}\left[\frac{\text{Br}(B^\pm\rightarrow \pi^\pm \ell^+\ell^{-})_{\text{max}}}{9.3\cdot 10^{-8}}  \right]^{\frac{1}{2}} \\
&|V_{L4d}V_{L4b}^*| 
< 4.3 \cdot 10^{-4} 
\left[\frac{\text{Br}(B^0\rightarrow \pi^0\ell^+\ell^-)_{\text{max}}}{1.2\cdot 10^{-7}}  \right]^{\frac{1}{2}}
\end{align}
where $\text{Br}_{\text{max}}$ is defined as in eq. (\ref{maxB}) as the maximal allowed value
for the branching ratio generated by the amplitude of the new contribution alone.
Regarding the measured decay $B^\pm\rightarrow \pi^\pm\ell^+\ell^-$,
in this case of destructive interference, we should also have $|V_{L4b}^*V_{L4d}|>2.3 \cdot 10^{-4}$.

%

\subsubsection{$B^0_{d,s}\rightarrow\mu^+\mu^-$}
The decays $B^0_{d,s}\rightarrow\ell^+\ell^-$, $\ell=e,\mu,\tau$,
are dominated by the $Z$-penguin and box diagrams involving top quark exchanges.
The charm contributions are fully negligible here and the 
effective Lagrangian 
in the SM is 
 \cite{Buras:1997fb}:
\begin{align}
\label{Bmumusm}
&\mathcal{L}_\text{SM}=
\frac{G^2_Fm^2_W}{2\pi^2}V_{tb}^*V_{tq}Y(x_{t})
(\overline{b}\gamma^\mu\gamma^5 q)(\bar{\ell}\gamma_\mu\gamma_5 \ell)
\end{align}
with $q=d,s$.
$Y(x_{t})$ is the Inami-Lim function, including QCD and electroweak corrections,
which at leading order is a linear combination of Z-penguin and box diagrams,
the same as in eq. (\ref{kmumusm}). 
 Here we use $Y(x_{t})=0.935$  
 from ref. \cite{Bobeth:2013uxa}, inserting the values in table \ref{values}.

The non-diagonal couplings of $Z$-boson with SM families lead to the same extra tree level contributions as in eqs. (\ref{Bnew1}), (\ref{Bnew2}). In this case only the axial part of both quark and leptons is involved:
\begin{align}
\label{Bmumunew}
&\mathcal{L}_\text{new}=\frac{G_F}{\sqrt{2}} \frac{1}{2}  V_{L4b}^*V_{L 4 q}
(\overline{b}\gamma^\mu\gamma^5 q)(\bar{\ell}\gamma_\mu\gamma_5 \ell)
\end{align}
for $q=d,s$.
Considering the case $\ell=\mu$, the branching ratios of the decays 
$B^0_{d,s}\rightarrow\mu^+\mu^-$ would be:
\begin{align}
\label{brBmumu}
& \text{Br}(B^0_{d}\rightarrow\mu^+\mu^-)_\text{tot}= k_{d} \frac{G_F^{2}}{2}
\left|\frac{\alpha}{2\pi\sin^{2}\theta_{W}}Y(x_{t})V_{tb}^{*}V_{td}+\frac{1}{2}V_{L4b}^{*}V_{L4d}\right|^2
 \\
& \text{Br}(B^0_{s}\rightarrow\mu^+\mu^-)_\text{tot}= k_{s}\frac{G_F^{2}}{2}
\left|\frac{\alpha}{2\pi\sin^{2}\theta_{W}}Y(x_{t})V_{tb}^{*}V_{ts}+\frac{1}{2}V_{L4b}^{*}V_{L4s}\right|^2
\label{brBmumus}
\end{align}
where $B_{q}$ is the flavour eigenstate $(\bar{b}q)$,
$f_{B_s}=230.3(1.3)$ MeV, $f_{B_d}=190.0(1.3)$ MeV \cite{FLAG2019} are the decay
constants defined by
$\langle 0|\bar{b} \gamma_{\mu}\gamma_{5}  q |B^0_{q}(p)\rangle= ip_{\mu}f_{B_{q}}$, and:
\begin{align}
& k_{d}=\tau_{B^{0}_{d}}
\frac{1}{2\pi}  f^2_{B_{d}} m^2_\mu M_{B^0_{d}}\sqrt{1-4m_\mu^2/M^2_{B^0_{d}}} \: ,
&& k_{s}=\tau_{B^{0}_{sH}}
\frac{1}{2\pi}  f^2_{B_{s}} m^2_\mu M_{B^0_{s}}\sqrt{1-4m_\mu^2/M^2_{B^0_{s}}}
\end{align}
In absence of new physics, the branching ratios (\ref{brBmumus}), (\ref{brBmumu}) 
give the SM expectations
which are respectively:
$\text{Br}(B^0_{d}\rightarrow\mu^+\mu^-)_\text{SM}\approx 8.6 \times 10^{-11}$, 
$\text{Br}(B^0_{s}\rightarrow\mu^+\mu^-)_\text{SM}\approx 3.2  \times 10^{-9}$, 
obtained using the central values of the quantities in table \ref{values}.

The experimental branching ratio of the decay $B^0_{s}\rightarrow\mu^+\mu^-$ is \cite{PDG18}:
\begin{align}
& \text{Br}(B^0_{s}\rightarrow\mu^+\mu^-)_\text{exp}=(3.0\pm 0.4) \times 10^{-9}
\end{align}
which is in agreement with the SM expectation.
Regerding the decay $B^0_{d}\rightarrow\mu^+\mu^-$, the experimental limit is \cite{Atlas}:
\begin{align}
& \text{Br}(B^0_d\rightarrow\mu^+\mu^-)_\text{exp}<2.1\times 10^{-10} \qquad 95\%\, \text{C.L.}
\end{align}

Then, we should set a limit on the decay rates (\ref{brBmumu}), (\ref{brmumus}),
in order to not contradict experimental results. At $95\%$ C.L. we can take: 
\begin{align}
\label{brdmm}
& \text{Br}(B^0_{d}\rightarrow\mu^+\mu^-)_\text{tot}
< 2.1\times 10^{-10}
\\
2.2 \times 10^{-9} < \; & \text{Br}(B^0_{s}\rightarrow\mu^+\mu^-)_\text{tot}
< 3.8 \times 10^{-9}
\label{brsmm}
\end{align}
Depending on the relative unknown phase of the mixing elements $V_{L4b}$, $V_{L4d/s}$, 
the above upper limits correspond to:
\begin{align}
\label{bdmm}
& |V_{L4b}^{*}V_{L4d}|<(0.4 \div 2.2 )\times 10^{-4} \\
& |V_{L4b}^{*}V_{L4s}|<(0.3 \div 8.1 )\times 10^{-4} 
\label{bsmm}
\end{align}
For the decay $B^0_{s}\rightarrow\mu^+\mu^-$, depending on the relative phase, 
there is also a lower limit for $-1<\cos(\delta^{d}_{L32})<-0.56$, where
$\delta^{d}_{L32}$ is the relative phase of the mixing elements $V_{L4b}$, $V_{L4s}$
in terms of the parameterization (\ref{tildes}), (\ref{4row}).
in case of negative interference 
We can express the limits (\ref{brdmm}), (\ref{brsmm}) in terms of comparison between
the new amplitude and the SM amplitude.
By defining:
\begin{align}
&\mathcal{F}_{B^{0}_{q}}=\frac{\alpha}{2\pi\sin^{2}\theta_{W}}Y(x_{t})V_{tb}^{*}V_{tq}
\end{align}
we can constrain the ratio between the magnitude of the two effective operators 
(\ref{Bmumusm}) and (\ref{Bmumunew}):
\begin{align}
& \frac{|V_{L4b}^*V_{L4q}|}{ 2|\mathcal{F}_{B^{0}_{q}}|} < \Delta_{B^{0}_{q}}   
\end{align}
In the case of destructive interference for both the decays, 
the limits (\ref{bdmm}), (\ref{bsmm}) correspond to:
\begin{align}
 &|V_{L4b}^*V_{L4d}| < 2.2 \times 10^{-4} \left[\frac{\Delta_{B^{0}_{d}}}{2.7}\right]
\left[\frac{\mathcal{F}_{B^{0}_{d}}}{4.1\times 10^{-5}}\right] \\ 
\left[\frac{\Delta_{B^{0}_{s}}}{1.8}\right] 7.1 \times 10^{-4} <\: & |V_{L4b}^{*}V_{L4s}|< 8.1 \times 10^{-4} 
\left[\frac{\Delta_{B^{0}_{s}}}{2.1}\right]
\left[\frac{\mathcal{F}_{B^{0}_{s}}}{2.0\times 10^{-4}}\right]
\end{align}
for $\delta^{d}_{L31}=2.72$, $\delta^{d}_{L32}=\pi$ respectively.

\subsection{Limits from $Z$-boson physics}
\label{Zdown}

\begin{table}[h]
\centering
\begin{tabular}{| lll |}
\hline
Quantity & Experimental value & SM prediction \\
\hline
$\Gamma_Z$ & $2.4952\pm 0.0023$ GeV & $2.4942\pm 0.0009$ GeV \\
$\Gamma(\text{had})$ & $1.7444\pm 0.0020$ GeV & $1.7411\pm 0.0008$ GeV \\
$R_b$ & $0.21629\pm0.00066$ & $0.21581\pm0.00002$ \\
$R_c$ & $0.1721\pm 0.0030$ & $0.17221\pm 0.00003$ \\
$A_{FB}^{(0,b)}$ & $0.0992\pm 0.0016$ & $0.1030\pm 0.0002$ \\
$A_{FB}^{(0,c)}$ & $0.0707\pm 0.0035$ & $0.0736\pm 0.0002$ \\
$A_{FB}^{(0,s)}$ & $0.0976\pm 0.0114$ & $0.1031\pm 0.0002$ \\
$A_{b}$ & $0.923\pm 0.020$ & $0.9347$ \\
$A_{c}$ & $0.670\pm 0.027$ & $0.6677\pm 0.0001$ \\
$A_{s}$ & $0.895\pm 0.091$ & $0.9356$ 
 \rule[-1.ex]{0pt}{0pt}\\
 $g^{eu}_{AV}$ & & $-0.1888$ \\
 $g^{ed}_{AV}$ & & $0.3419$ \\
 $Q_W(Cs)$ & $-72.82\pm 0.42$ & $-73.23\pm 0.01$ \\
$Q_W(T\ell)$ & $-116.4\pm 3.6$ & $-116.87\pm 0.02$  \\
 $g^{ep}_{AV}$ &$ -0.0356 \pm 0.0023$ & $-0.0357$ \\
 $g^{en}_{AV}$ &$ 0.4927\pm 0.0031$ & $0.4950$ \\
$2 g^{eu}_{AV}-g^{ed}_{AV}$ &$- 0.7165\pm 0.0068$ & $-0.7195$ \\
\hline
\end{tabular}
\caption{\label{Zvalues} Values of interest from Particle Data Group \cite{PDG18}.}
\end{table}
The presence of additional vector-like quarks also affects 
the diagonal couplings of Z-boson with standard quarks,
changing the prediction of many observables related to the $Z$-boson physics e.g.
the $Z$ total width $\Gamma_Z$,
the partial decay width into hadrons $\Gamma(Z\rightarrow\text{had})$,
the partial decay widths 
$R_c=\Gamma(c\bar{c})/\Gamma(Z\rightarrow\text{had})$, $R_b=\Gamma(b\bar{b})/\Gamma(Z\rightarrow\text{had})$, 
$\Gamma(Z\rightarrow q\bar{q})$, $q=u,d,s,c,b$. 
Constraints obtained from these quantities are analyzed in the following sections.
%
%
Experimental values and SM predictions 
are taken from
Particle Data Group \cite{PDG18}, as reported in table \ref{Zvalues}. 


%
%
%
The predicted partial decay width of Z-boson decaying into $b\bar{b}$ is 
$\Gamma(Z\rightarrow b\bar{b})_\text{SM}\approx 375.75\mp 0.18$ MeV, 
which should be compared to the experimental value 
$\Gamma(Z\rightarrow b\bar{b})_\text{exp}\simeq 377.3\mp 1.2$ MeV,
where data from PDG \cite{PDG18} are used, as reported in table \ref{Zvalues}.
The SM prediction of the partial decay rate of $Z\rightarrow b\bar{b}$
at tree level is given by:
\begin{align}
\label{ZbSM}
&\Gamma(Z\rightarrow b\bar{b})_{\text{SM}}=\frac{G_FM^3_Z}{\sqrt{2}\pi}\left[\left(-\frac{1}{2}+\frac{1}{3}\sin^2\theta_W\right)^2+\left(\frac{1}{3}\sin^2\theta_W\right)^2\right]
\end{align}
In order to compare with the experimental result, we should include QCD corrections, which
are given by a multiplicative factor $\approx 1.021$ \cite{PDG18}.
By inserting the vector-like isosinglet down-quark, the decay rate at tree level
changes in:
\begin{align}
\label{Zbnew}
&\Gamma(Z\rightarrow b\bar{b})=\frac{G_FM^3_Z}{\sqrt{2}\pi}\left[\left(-\frac{1}{2}(1-|V_{L4b}|^2)+\frac{1}{3}\sin^2\theta_W\right)^2+\left(\frac{1}{3}\sin^2\theta_W\right)^2\right]
\end{align}
which means:
\begin{align}
& \Gamma(Z\rightarrow b\bar{b})-\Gamma(Z\rightarrow b\bar{b})_{\text{SM}}\approx
\frac{G_FM^3_Z}{\sqrt{2}\pi}\left(-\frac{1}{2}+\frac{1}{3}\sin^2\theta_W\right) |V_{L4b}|^2<0
\end{align}
So the prediction for the decay rate is lowered, not going towards the direction of
a better agreement with the experimental value. 
Then, the extra contribution to the rate should be constrained:
\begin{align}
& \left| \Gamma(Z\rightarrow b\bar{b})-\Gamma(Z\rightarrow b\bar{b})_{\text{SM}}\right|
<\Delta \Gamma_{Zbb}
\end{align}
We may choose $\Delta \Gamma_{Zbb}$
so that $\Gamma(Z\rightarrow b\bar{b})_{\text{tot}}$ 
lays in the $95\%$ C.L. interval of the experimental value, 
which implies:
\begin{align}
&\left| \Gamma(Z\rightarrow b\bar{b})-\Gamma(Z\rightarrow b\bar{b})_\text{SM}\right|
< 8.6 \times 10^{-4} \, \text{GeV}
\\
& |V_{L4b}|< 3.2\cdot 10^{-2}  
\left[ \frac{\Delta \Gamma_{Zbb}}{8.6 \times 10^{-4} \, \text{GeV}}\right]^{1/2}
\label{zbb}
\end{align}


The SM predictions for the Z decay rate and partial decay rate into hadrons are \cite{PDG18}:
\begin{align}
&\Gamma(Z)_\text{SM}=2.4942\pm0.0009 \,  \text{GeV} \, , \qquad
\Gamma(Z\rightarrow\text{hadr})_\text{SM}=1.7411\pm0.0008  \,  \text{GeV}
\label{GZsm}
\end{align}
to be compared with the experimental results \cite{PDG18}:
\begin{align}
& \Gamma(Z)_\text{exp}=2.4952\pm0.0023 \,  \text{GeV} \, , \qquad
\Gamma(Z\rightarrow\text{hadr})_\text{exp}=1.7444\pm0.0020  \,  \text{GeV}
\label{GZexp}
\end{align}
In this BSM scenario,
the deviation 
from the SM expectation of the new predicted $Z $
partial decay rate into hadrons $\Gamma(Z\rightarrow\text{had})$
(which also corresponds to the deviation of the total $Z$ decay rate $\Gamma(Z)$, 
since there are not additional leptons)
 is:
\begin{align}
& \Gamma(Z\rightarrow\text{had})-\Gamma(Z\rightarrow\text{had})_\text{SM}=
\Gamma(Z)-\Gamma(Z)_\text{SM}= 
 \nonumber \\
&=\frac{G_FM^3_Z}{\sqrt{2}\pi}\left[ \sum_{i,j=d,s,b}
\left| -\frac{1}{2}\sum_{k=1}^3 V_{Lki}^{*}V_{Lkj}+\frac{1}{3}\sin^2\theta_W\delta_{ij} \right|^2
-3\left( -\frac{1}{2}+\frac{1}{3}\sin^2\theta_W \right)^2  \right]
\approx \nonumber \\
& 
\approx \frac{G_FM^3_Z}{\sqrt{2}\pi}
 \left(-\frac{1}{2}+\frac{1}{3}\sin^2\theta_W\right)\left(|V_{L4d}|^2+|V_{L4s}|^2+|V_{L4b}|^2\right)<0
\label{Zd}
\end{align}
QCD corrections should also be included, which amount to a multiplicative factor $\approx 1.041$
for $d,s$-quarks and $\approx 1.021$ for $b$-quark \cite{PDG18}.
As shown in eq. (\ref{Zd}),
the prediction for the decay rate is lowered with respect to the SM expectation
$\Gamma(Z\rightarrow\text{hadr})_\text{SM}$.
Then, since the SM expectation (\ref{GZsm}) is below
the experimental result (\ref{GZexp}), the contribution of the extra quarks is
not leading towards 
a better agreement. 
Therefore, 
in order to set a constraint on the new expected decay rate, 
we can 
impose that the new expectation
$\Gamma(Z\rightarrow\text{had})$ should be
in the $95\%$ C.L. interval of the experimental value $\Gamma(Z\rightarrow\text{had})_\text{exp}$,
assuming in eq. (\ref{Zd}) the limit value for the SM prediction
$\Gamma(Z\rightarrow\text{had})_\text{SM}= 1.7419$ GeV.
That gives: 
\begin{align}
&
\Gamma(Z\rightarrow\text{had})_\text{SM}-\Gamma(Z\rightarrow\text{had})<
\Delta\Gamma_{Z} \nonumber \\
& |V_{L4d}|^2+|V_{L4s}|^2+|V_{L4b}|^2
< 1.7 \times 10^{-3} 
 \left[ \frac{\Delta\Gamma_{Z}}{1.4\times 10^{-3} \,\text{GeV}} \right]
\end{align}
With $|V_{L4d}|=0.03$, the constraint means $|V_{L4s}|^2+|V_{L4b}|^2<0.0008$,
which is satisfied for example if both $|V_{L4s}|,|V_{L4b}|<0.02$. 
If $V_{L4s}=V_{L4b}=0$, this constraint implies: 
\begin{align}
& |V_{L4d}|^{2} < 1.7 \times 10^{-3} \; , \qquad \quad
|V_{L4d}|< 0.041 
\label{v4dZ}
\end{align}
which is extremely close to the value needed to solve the CKM unitarity problem
(for example, at $95\%$ C.L. $|V_{L4d}|=0.038^{+0.008}_{-0.011}$ (\ref{newunsol}) 
using our conservative averages for $V_{ud}$ and $V_{us}$ values (\ref{vudmedio}), (\ref{c})). 
This means that an extra weak singlet 
could not completely explain the CKM unitarity 
anomaly with more extreme values of the determinations of CKM elements, 
as can be seen for example by comparing eq. (\ref{v4dZ})
to the needed values $|V_{L4d}|^{2} =\delta_{\text{CKM}}$ 
displayed in table \ref{Tabledelta2}.


Constraints are expected also from 
Z-pole asymmetry analyses of $e^+e^-\rightarrow ff$
processes. 
In particular, 
left-right asymmetries $A_{LR}$, forward-backward asymmetries $A_{FB}$ 
and left-right forward-backward asymmetries $A_{LRFB}$ \cite{ALEPH:2005ab} 
were measured at LEP.
Cross sections for $Z$-boson exchange are usually written 
in terms of the asymmetry parameters $A_f$, 
$f=e,\mu,\tau,b,c,s,q$, which contain final-state couplings. 
For example, they are related as 
$A_{FB}^{(0,f)}=\frac{3}{4}A_eA_f$, 
$A_{LRFB}^{(0,f)}=\frac{3}{4}A_f$ 
(where the superscript $0$ indicates the quantity corrected for radiative effects).
The presence of an additional isosinglet changes the couplings of quarks with the $Z$ boson 
as in eq. (\ref{nc}). Consequently the predictions for the asymmetries are also changed:
\begin{align}
\label{Aqnewsing}
& A_q=\frac{(1-|V_{L4q}|^2)^2-4|Q_q|\bar{s}^2_q(1-|V_{L4q}|^2)}
{(1-|V_{L4q}|^2)^2-4|Q_q|\bar{s}^2_q(1-|V_{L4q}|^2)+8Q_q^2\bar{s}^4_q} 
\end{align}
where $\bar{s}_f^2$ are the effective weak angles which take into account EW radiative corrections.
Then, the mixing $|V_{L4q}|$ with the isosinglet quark makes the prediction for $A_{q}$ lower than
the SM one. 
The most precise result for quarks regards the asymmetry for b-quark final state.
%
%
Taking the data from Particle Data Group \cite{PDG18} (also listed in table \ref{Zvalues}),
in principle the mixing $|V_{L4b}|$ 
modifies the couplings in the ``right'' direction 
with respect to the experimental determinations of both $A_b$ and $A_{FB}^{(0,b)}$.
However, considering the other constraints from $Z$ decays,
a mixing $|V_{L4b}|\lesssim 0.03$ (\ref{zbb}) would only give a relative change to 
$A_{b}$ at most of $\sim 0.016\%$,
two orders of magnitude less than the relative experimental error.

\subsection{Summary of experimental limits}
\label{downend}

\input{tab/tabelladown2}

\begin{figure}[t]
\centering
\includegraphics[width=0.8\textwidth]{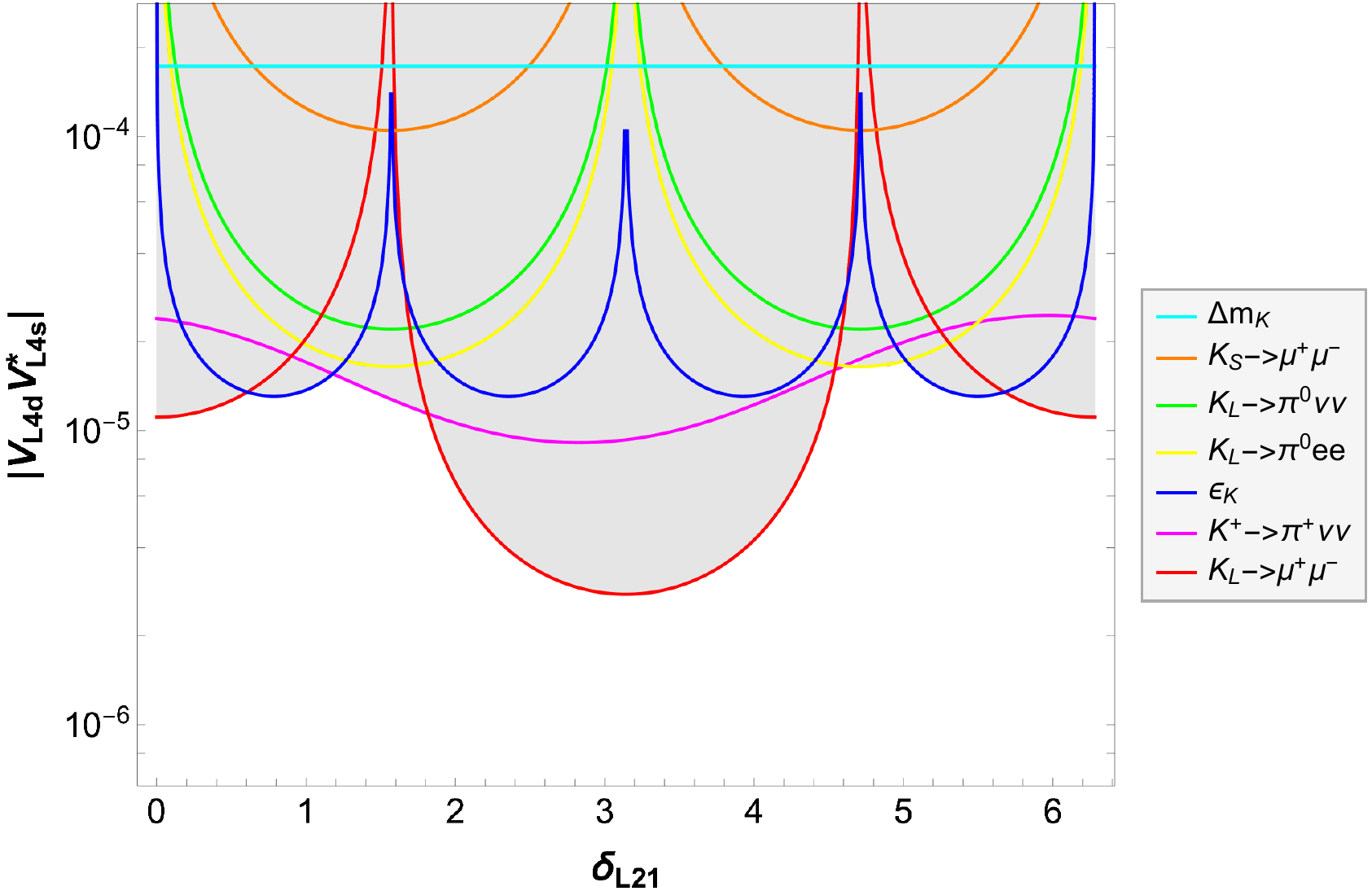}
\caption{\label{fcplotd} 
Upper limits obtained from kaon decays and neutral kaon mixing
on the product $|V_{L4s}^*V_{L4d}|$ of the
elements 
of the mixing matrix $V_{L}^{(d)}$ (\ref{vld}),
as a function of their relative phase $\delta^d_{L21}$.
}
\end{figure}

\begin{figure}[t]
\centering
\includegraphics[width=0.49\textwidth]{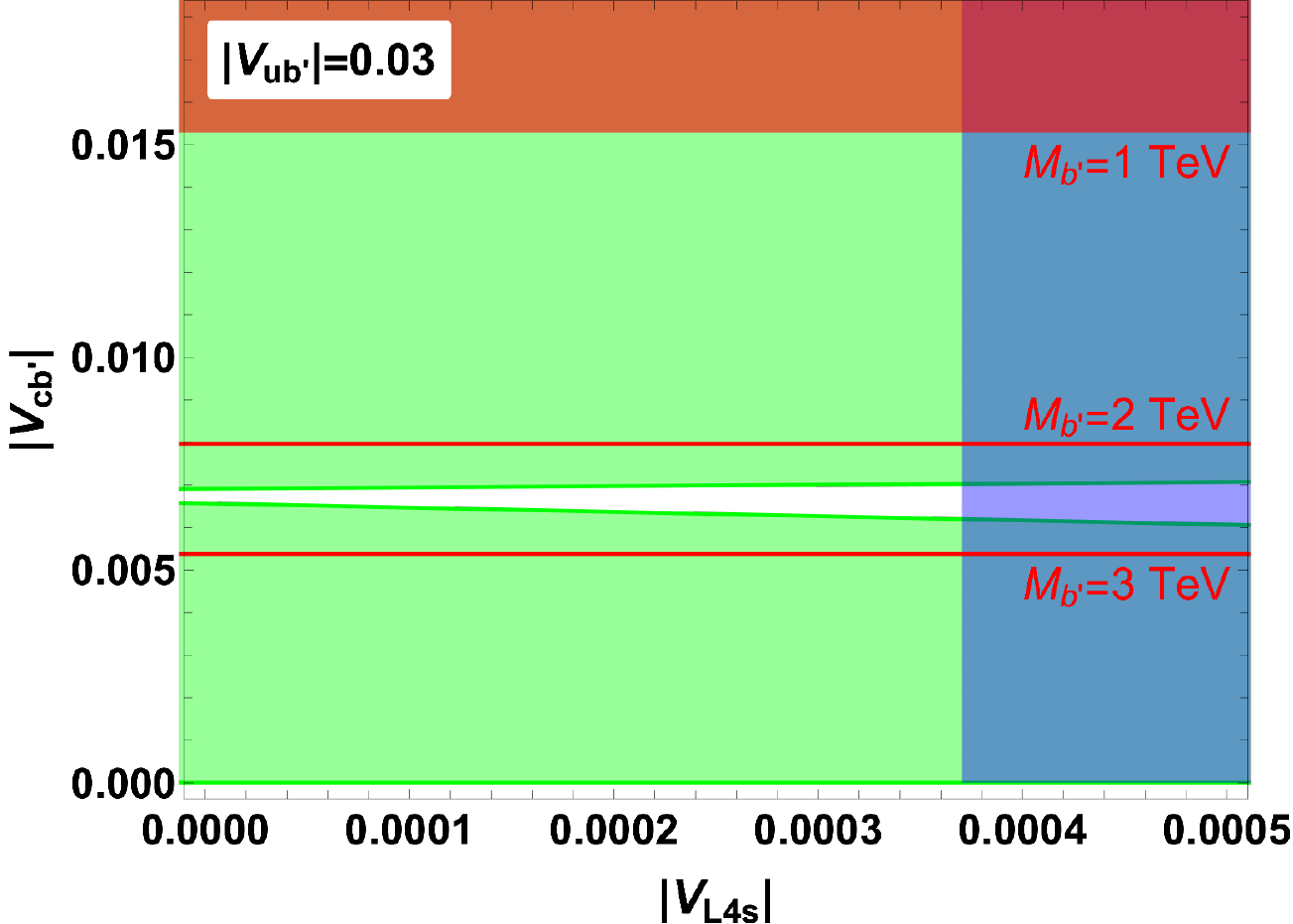}
\includegraphics[width=0.49\textwidth]{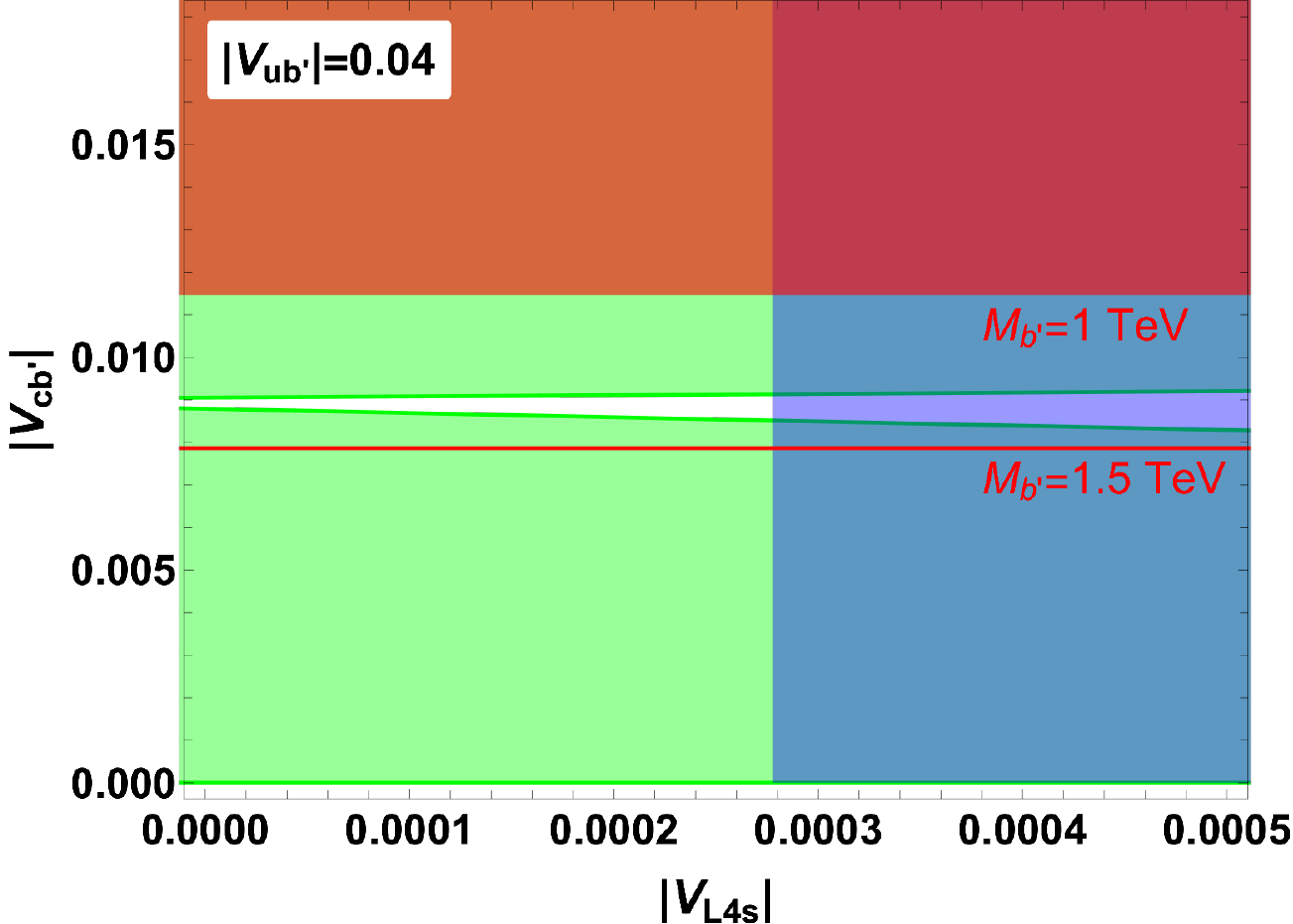}
\caption{\label{vcbv4s} Constraints on $|V_{cb'}|$ and $|V_{L4s}|$.
On the left,
$|V_{ub'}|\approx |V_{L4d}|=0.03$ is set as a benchmark value.
The red area is excluded by the constraints on
$|V_{cb'}|$ from $D^0$ systems, with $M_{b'}=1$ TeV, as reported in table \ref{limiti}.
The red lines show where the boundary shifts
if the mass of the extra quark is taken as $M_{b'}=2$ and $3$ TeV.
The blue area is excluded by flavour changing processes involving K-mesons,
using the result $|V_{L4s}| < 3.7 \cdot 10^{-4}$, from figure \ref{fcplotd}, eq. (\ref{vl4stot}),
which is valid for $0\lesssim\delta^{d}_{L21}\lesssim 1.82$, 
$4.46\lesssim\delta^{d}_{L21}\lesssim 2\pi$.
The green region is excluded by 
the relation in eq. (\ref{vcbp2}), for the considered values of the relative phase $\delta^{d}_{L21}$.
On the right, the same constraints are shown setting $|V_{ub'}|\approx |V_{L4d}|=0.04$.
The constraint on $|V_{cb'}|$ from D-mesons mixing (red area) is lowered to the red line
taking the mass of the extra quark as $M_{b'}=1.5$ TeV.
}
\end{figure}


As illustrated in section \ref{situation}, 
the analysis 
of the latest determinations of $V_{us}$ obtained from kaon decays 
and the ones of $V_{ud}$ from beta decays 
results in a deviation from unitarity in the first row of the CKM matrix.
The SM unitarity relation (\ref{uns}) can be modified into
the relation in eq. (\ref{newun}) if an extra vector-like down-type quark $d_{L4},d_{R4}$ participates in the mixing 
of the SM families. In section \ref{sec-down} it was shown that a quite large mixing with the first 
family, $|V_{ub'}|\approx |V_{L4d}| \approx 0.038(5)$ (\ref{newunsol}), 
is needed in order to explain the data 
(see also 
table \ref{Tabledelta2},
where in this case $\delta_\text{CKM}= |V_{ub'}|^{2}$).
Then, we need to verify if such large mixing is compatible with experimental constraints from flavour changing decays and electroweak observables.

In table \ref{limiti} the relevant constraints extracted in this section are listed.

Limits from flavour changing kaon decays are summarized in figure \ref{fcplotd}, 
where $V_{L4d}$ and $V_{L4s}$ are elements of the mixing matrix $V_{L}^{(d)}$ (\ref{vld}).
It results that, depending on the relative phase, 
the product $|V_{L4d}V_{L4s}^*|$ 
cannot exceed the limit:
\begin{align}
& |V_{L4d}V_{L4s}^*|\lesssim (0.3 \div 1.7) \times 10^{-5}
\label{v4dv4send}
\end{align}
taking
$M_{b'}=1$ TeV. Using the parameterization in eqs. (\ref{tildes}), (\ref{4row}),
the maximum and minimum values are obtained for 
$\delta^d_{L21} \approx 5.0$ and 
$\delta^d_{L21} = \pi$ respectively.
Taking into account the needed values to recover the unitarity of the CKM matrix
in eq. (\ref{newunsol}),
we use $|V_{ub'}|\approx |V_{L4d}|=0.03$ as a conservative benchmark value.
Then, from eq. (\ref{v4dv4send}) we have: 
\begin{align}
\label{vl4stot}
& |V_{L4s}| < (0.9\div 5.8) \times 10^{-4} \left[\frac{0.03}{|V_{ub'}|}   \right]
\end{align}
Constraints on the value of $|V_{tb'}|\approx  |V_{L4b}|$ are obtained from 
flavour changing B-mesons decays, in particular from the decay $B^{0}\rightarrow\mu^{+}\mu^{-}$
(\ref{bdmm}):
\begin{align}
&  |V_{L4b}^{*}V_{L4d}|<(0.4 \div 2.2 )\times 10^{-4} 
\, , \qquad \quad |V_{L4b}| < (1.4 \div 7.4) \times 10^{-3} \left[\frac{0.03}{|V_{ub'}|}   \right]
\label{v4btot}
\end{align}
depending on the relative phase $\delta^{d}_{L31}$, where the maximum and 
minimum value are obtained respectively for $\delta^{d}_{L31}=2.72$ and 
$\delta^{d}_{L31}=-0.42$. 
Then, 
taking into account that 
(\ref{4row}), (\ref{si}):
\begin{align}
\label{v4h}
& V_{L4d}\approx -\frac{h_{d1}^{*}v_{w}}{M_{b'}} \; , \qquad   
 V_{L4s}\approx -\frac{h_{d2}^{*}v_{w}}{M_{b'}} \; , \qquad   
V_{L4b}\approx -\frac{h_{d3}^{*}v_{w}}{M_{b'}} 
\end{align}
where $h_{di}$ are the Yukawa couplings defined in eq. (\ref{yd}),
from eqs. (\ref{vl4stot}), (\ref{v4btot}) it follows that
the Yukawa couplings $h_{d2}$ and $h_{d3}$ should be respectively 
at least $50$ times and $4$ times smaller than the coupling of the first family $h_{d1}$.

As regards the new column of the enlarged CKM matrix (\ref{vckm}),
from eq. (\ref{Vcbp}) we have:
\begin{align}
\label{vcbp2}
 | V_{cb'} |\approx &\, | -V_{L4d}^{*}V_{cd} -V_{L4s}^{*}V_{cs} -V_{L4b}^{*}V_{cb}| = \nonumber \\
= &\, \Big| |V_{cd}| |V_{ub'}| -|V_{L4s}|e^{i\delta^{d}_{L21}} - |V_{cb}| |V_{L4b}| e^{i\delta^{d}_{L31}}  \Big|
\end{align}
Then, taking into account the constraints (\ref{vl4stot}), (\ref{v4btot}), it is obtained:
\begin{align}
& | V_{cb'} |\approx |V_{cd}| |V_{ub'}| 
\label{vcbp0}
\end{align}
This means that, considering the constraint from $D^{0}$ mesons mixing (\ref{vcbpvubpdd}):
\begin{align}
& |V_{ub'}V_{cb'}^*|< 4.6 \times 10^{-4}
\left[ \frac{\Delta m_{D\,\text{exp}}}{1.2\cdot 10^{-14}}  \right]^{1/2} \left[ \frac{1\,\text{TeV}}{M_{b'}}  \right]
\label{vcbpvubpdd2}
\end{align}
it should be that:
\begin{align}
&  |V_{ub'} |< 0.046  \left[ \frac{\Delta m_{D\,\text{exp}}}{1.2\cdot 10^{-14}}  \right]^{1/4} \left[ \frac{1\,\text{TeV}}{M_{b'}}  \right]^{1/2} 
\end{align}
In fact, 
there is no much room in the parameter space 
to accommodate the relation (\ref{vcbp2}) without contradicting experimental constraints
from $D^0$ mixing and flavour changing kaon decays.
For $|V_{ub'}|\approx 0.03$ and $M_{b'}\approx 1$ TeV, we have:
\begin{align}
& |V_{cb'}|\approx (6.6\pm 0.4) \times 10^{-3} \, , \qquad |V_{tb'}|< (1.4 \div 7.4) \times 10^{-3}
\label{vcbptbp}
\end{align}
The first condition is shown in figure \ref{vcbv4s}. 
There, the blue area is excluded by flavour changing processes involving K-mesons 
(\ref{vl4stot}).
We use the upper limit obtained for $\delta^{d}_{L21}=\text{Arg}(V_{L4s}^{*}V_{L4d})=0$,
$|V_{L4s}^{*}V_{L4d}|<1.1\times 10^{-5}$,  
which is valid for 
$0<\delta^{d}_{L21}<1.82$, $4.46<\delta^{d}_{L21}<2\pi$
(for other values of the phase $\delta^{d}_{L21}$ the constraint should be more stringent,
see figure \ref{fcplotd}).
The green region is excluded by the relation in eq. (\ref{vcbp2}), 
for any value of the relative phase $\delta^{d}_{L31}$
and for the considered values of the relative phase $\delta^{d}_{L21}$
(other values of $\delta^{d}_{L21}$ only allow a slightly larger opening angle of the allowed 
``cone'' on the upper side, which does not widen the allowed range of values of $V_{cb'}$,
because 
the constraint from kaon physics is more stringent for those other values of the phase $\delta^{d}_{L21}$).
The red area is excluded by the constraints on
$|V_{cb'}|$ from $D^0$ systems (\ref{vcbpvubpdd}) with 
$M_{b'}=1$ TeV: 
\begin{align}
& |V_{cb'}|< 1.5 \cdot 10^{-2} \left[\frac{0.03}{|V_{ub'}|}   \right]
\left[ \frac{1\,\text{TeV}}{M_{b'}}  \right]
\label{vcbpdd2}
\end{align}
As discussed in section \ref{DDdown}, 
the constraint is obtained by allowing the new contribution to the mass difference in 
$D^{0}$ mesons system
to account for the experimental value $\Delta m_{D\,\text{exp}}$, at $95\%$ C.L..
This means that the extra singlet quark can be thought as a way to
explain the value of the mass difference in neutral D-mesons system.

As can be seen, 
a narrow allowed region (including $V_{L4s}=0$) can be found.
However, 
the mass of the extra quark cannot exceed few TeV.
In fact, with larger values of $M_{b'}$
the constraint from $D^{0}$ mixing becomes more stringent.
For example,
as shown in figure \ref{vcbv4s}, the allowed region vanishes if $M_{b'}=3$ TeV 
with the choice $|V_{ub'}|=0.03$, 
while $M_{b'}=1.5$ TeV is large enough to make the 
allowed area disappear if $|V_{ub'}|=0.04$ is needed.
Also the constraint from flavour changing processes of kaons 
shifts towards lower values with larger values of the extra quark mass.
Moreover, with $|V_{ub'}|=0.03$, $M_{b'} < 5.8 \, \text{TeV}$ is needed
for the perturbativity ($|h_{d\, 1} |< 1$). 
Anyway, also by taking $M_{b'}=1$ TeV, the allowed parameter space vanishes with 
$|V_{ub'}|>0.046$.

In addition to that, results on $Z$-boson decay rate into hadrons imply that
(\ref{v4dZ}):
\begin{align}
& |V_{L4d}|^{2} < 1.7 \times 10^{-3} \; , \qquad \quad 
|V_{ub'}|\approx |V_{L4d}|< 0.041
\end{align}
which is in the range of values needed to solve the CKM unitarity problem
(for example, at $95\%$ C.L. $|V_{L4d}|=0.038^{+0.008}_{-0.011}$ (\ref{newunsol}) 
using our conservative average for $V_{ud}$ (\ref{vudmedio})). 
This means that an extra weak singlet would not be 
a good solution of the CKM 
unitarity problem with the more extreme values of the determinations of CKM elements, 
as can also be seen by comparing eq. (\ref{v4dZ})
to the needed values $|V_{L4d}|^{2} =\delta_{\text{CKM}}$ 
displayed in table \ref{Tabledelta2}.

From eq. (\ref{vcbptbp}) we can 
notice that the mixing of 
$b'$ with c-quark and t-quark should be at least four times smaller 
than the mixing with the u-quark. 
Moreover, $\vert V_{ub'} \vert \sim 0.03 $ is comparable to $\vert V_{cb} \vert$ 
and ten times larger than $\vert V_{ub} \vert$.
However, although it may seem unnatural to expect
a larger mixing of the 4th state with the lightest family than with the heavier ones,
it cannot be excluded.

\section{Extra up-type isosinglet}
\label{sec-up}

The case of the addition of a up-type vector-like isosinglet couple of quarks
$(u_{4L},u_{4R})$ is examined in this section.
Analogously to the case of the extra down-type quark, 
there is an additional piece in the Yukawa Lagrangian:
\begin{align}
\label{yu}
& &y^{u}_{ij}\tilde{\varphi}\overline{q_{Li}}u_{Rj}+y^{d}_{ij}\varphi\overline{q_{Li}}d_{Rj}+
h_{ui}\tilde{\varphi}\overline{q_{Li}}u_{R4}+M_{4u}\overline{u_{L4}}u_{R4}+\text{h.c.}
\end{align}
with $i,j=1,2,3$ and
$\tilde{\varphi}=i\tau_2\varphi^*$. Then the up-type quarks mass matrix looks like:
\begin{align}
\label{mu}
&\overline{u_{Li}}\mathbf{m}^{(u)}_{ij}u_{Rj}+\text{h.c.}= \nonumber \\
=&(\overline{u_{L1}}, \overline{u_{L2}}, \overline{u_{L3}}, \overline{u_{L4}})
\left(\begin{array}{ccc|c}
 & & & h_{u1}v_{w} \\   & \mathbf{y}^{(u)}_{3\times 3}v_{w} & & h_{u2}v_{w} \\  & & & h_{u3} v_{w}\\
 \hline
 0 & 0 & 0 & M_{4u}
\end{array}\right)\left(\begin{array}{c}
u_{R1} \\ u_{R2} \\ u_{R3} \\ u_{R4}
\end{array}\right)   +\text{h.c.}
\end{align}
where $v_w=174$ GeV is the SM Higgs vacuum expectation value (VEV) 
and $\mathbf{y}^{(u)}_{3\times 3}$ is the $3\times 3 $ 
matrix of Yukawa couplings.
The mass matrix $\mathbf{m}^{(u)}$ can be diagonalized with positive eigenvalues by a biunitary transformation:
\begin{align}
& V_L^{(u)\dagger}\mathbf{m}^{(u)}V_R^{(u)}=\mathbf{m}^{(u)}_\text{diag}=
\text{diag}(y_u v_w, y_c v_w, y_t v_w,M_{t'})
\end{align}
where $V^{(u)}_{L,R}$ are two unitary $4\times 4$ matrices. $\mathbf{m}^{(u)}_\text{diag}$
is the diagonal matrix of mass eigenvalues 
$m_{u,c,t}=y_{u,c,t}v_w$
and $M_{t'}\approx M_{4u}$.
The mass eigenstates are:
\begin{align}
&\left(\begin{array}{c}
u_{L} \\ c_{L} \\ t_{L} \\ t'_{L}
\end{array}\right)=V_L^{(u)\dag}\left(\begin{array}{c}
u_{L1} \\ u_{L2} \\ u_{L3} \\ u_{L4}
\end{array}\right) \, , \qquad
V^{(u)}_L=\left(\begin{array}{cccc}
V_{L1u} & V_{L1c} & V_{L1t} & V_{L1t'} \\
V_{L2u} & V_{L2c} & V_{L2t} & V_{L2t'} \\
V_{L3u} & V_{L3c} & V_{L3t} & V_{L3t'} \\
V_{L4u} & V_{L4c} & V_{L4t} & V_{L4t'} 
\end{array}\right)
\label{vlu}
\end{align}
while the three up-type quarks involved in charged weak interactions expressed in terms of mass
eigenstates are:
\begin{align}
&\left(\begin{array}{c}
u_{L1} \\ u_{L2} \\ u_{L3}
\end{array}\right)=\tilde{V}^{(u)}_L\left(\begin{array}{c}
u_{L} \\ c_{L} \\ t_{L} \\ t'_{L} 
\end{array}\right)_L=\left(\begin{array}{cccc}
V_{L1u} & V_{L1c} & V_{L1t} & V_{L1t'} \\
V_{L2u} & V_{L2c} & V_{L2t} & V_{L2t'} \\
V_{L3u} & V_{L3c} & V_{L3t} & V_{L3t'} 
\end{array}\right)\left(\begin{array}{c}
u_{L} \\ c_{L} \\ t_{L} \\ t'_{L} 
\end{array}\right)
\end{align}
where $\tilde{V}^{(u)}_L$ is the $3\times 4 $ submatrix of $V^{(u)}_L$ without the last row.
Although it still holds that $\tilde{V}^{(u)}_L\tilde{V}^{(u)\dag}_L=1_{3\times 3}$ ($1_{3\times 3}$ being the $3\times 3$ identity matrix), 
$\tilde{V}^{(u)}_L$ is not unitary $\tilde{V}^{(u)\dag}_L\tilde{V}^{(u)}_L\neq 1_{3\times 3}$.

The Lagrangian for the charged current interaction is:
\begin{align}
\label{ccu}
\mathcal{L}_\text{cc}&=
\frac{g}{\sqrt{2}}\left(\begin{array}{ccc}
\overline{u_{L1}} & \overline{u_{L2}} & \overline{u_{L3}}
\end{array}\right)\gamma^\mu
\left(\begin{array}{c}
d_{L1} \\ d_{L2} \\ d_{L3} 
\end{array}\right) W_\mu^+ +\text{h.c.} = \nonumber \\
&=\frac{g}{\sqrt{2}}\left(\begin{array}{cccc}
\overline{u_L} & \overline{c_L} & \overline{t_L} & \overline{t'_L}
\end{array}\right)\gamma^\mu
\tilde{V}_\text{CKM}
\left(\begin{array}{c}
d_{L} \\ s_{L} \\ b_{L} 
\end{array}\right) W_\mu^+ +\text{h.c.} 
\end{align}
where 
\begin{align}
&\tilde{V}_\text{CKM}=\tilde{V}^{(u)\dag}_L V_L^{(d)}=
\left(\begin{array}{ccc}
V_{ud} & V_{us} & V_{ub}  \\ V_{cd} & V_{cs} & V_{cb} \\ 
V_{td} & V_{ts} & V_{tb}  \\ V_{t'd} & V_{t's} & V_{t'b} 
\end{array}\right)
\label{vckmu}
\end{align}
is a $4\times 3$ matrix, $V^{(d)}_L$ being the unitary $3\times 3 $ matrix diagonalizing the 
down-type quark mass matrix from the left.
Again $\tilde{V}_\text{CKM}$ is not unitary:
\begin{align}
&\tilde{V}_\text{CKM}\tilde{V}_\text{CKM}^\dag= \tilde{V}_L^{(u)\dag}\tilde{V}_L^{(u)}\neq\mathbf{1}
\end{align}
In particular, for the first row it holds that
\begin{align}
\label{newunup}
& |V_{ud}|^2+|V_{us}|^2+|V_{ub}|^2=
[\tilde{V}_\text{CKM}\tilde{V}_\text{CKM}^\dag]_{11}
=[\tilde{V}_L^{(u)\dag}\tilde{V}_L^{(u)}]_{11}=1-|V_{L4u}|^2
\end{align}
which is the same extended unitarity condition as in eqs. (\ref{newundelta}) and (\ref{newun}),
where $\delta_\text{CKM}=|V_{L4u}|^2$ and $|V_{L4u}|$ has
the same effect on the unitarity of the first row as $|V_{ub'}|$.
Then, the analysis 
of the determinations of $V_{us}$ obtained from 
leptonic and semileptonic kaon decays
and of $V_{ud}$ from beta decays 
gives the same result
as in section \ref{sec-down}, leading to the best fit point in eq. (\ref{newunsol}):
\begin{align}
\label{newunsolup}
& 
|V_{L4u}|^{2}=1.48(36)\times 10^{-3} \; , \qquad \quad
|V_{L4u}| =0.038(5)
\end{align}
which shifts the values of $V_{us}$ obtained from the three determinations
as shown in figure \ref{vusudm43plot}.

The mixing matrix $V^{(u)}_L$ induces non-standard couplings of $Z$-boson with the LH up quarks, 
since $u_{L4}$ is a weak singlet while the first three (SM) families act as weak doublets.
In fact, $Z$-boson couples 
as $Z_{\mu}\overline{f}(T_{3}+Q\sin^{2}\theta_{W})f$, where $T_{3}$ is the weak isospin projection and $Q$ the electric charge. Therefore, $Q$-dependent couplings remain diagonal between the mass eigenstates $u,,t,t'$, while the $T_{3}$ dependent part gets non-diagonal couplings.
The weak neutral current Lagrangian for up quarks reads:
\begin{align}
\label{ncu}
&\mathcal{L}_\text{nc}=
\frac{g}{\cos\theta_W}\left[\frac{1}{2}
\left(\begin{array}{ccc}
\overline{u_{L1}} & \overline{u_{L2}} & \overline{u_{L3}} 
\end{array}\right)\gamma^\mu
\left(\begin{array}{c}
u_{L1} \\ u_{L2} \\ u_{L3} 
\end{array}\right)
-\frac{2}{3}\sin^2\theta_W\left(\overline{\mathbf{u}_L}\gamma^\mu \mathbf{u}_L+
\overline{\mathbf{u}_R}\gamma^\mu \mathbf{u}_R\right)\right]Z_\mu =\nonumber \\
&=\frac{g}{\cos\theta_W}\! \left[\frac{1}{2}
\left(\begin{array}{cccc}
\overline{u_L} & \overline{c_L} & \overline{t_L} & \overline{t'_L}
\end{array}\right)\gamma^\mu\tilde{V}_L^{(u)\dag}\tilde{V}_L^{(u)}
\left(\begin{array}{c}
u_{L} \\ c_{L} \\ t_{L} \\ t'_{L} 
\end{array}\right)
-\frac{2}{3}\sin^2\theta_W\left(\overline{\mathbf{u}_L}\gamma^\mu \mathbf{u}_L+
\overline{\mathbf{u}_R}\gamma^\mu \mathbf{u}_R\right)\right]Z_\mu 
\end{align}
where $\mathbf{u}$ is the column vector of the four up-type quarks $u,c,t,t'$.
Then, as shown in eq. (\ref{ncu}), the non-unitarity of $\tilde{V}^{(u)}_L$ is at the origin of 
non-diagonal couplings with $Z$ boson, which means
flavor changing neutral currents (FCNC)
at tree level. 
Explicitly, the weak isospin dependent part of the $Z$ coupling is given by:
\begin{align}
\label{vunc}
&V_\text{nc}^{(u)}=\tilde{V}_L^{(u)\dag}\tilde{V}_L^{(u)}= 
V^{(u)\dag}_{L}\text{diag}(1,1,1,0)V^{(u)}_{L}= \nonumber \\
&=\! \left(\begin{array}{c@{\hspace{1\tabcolsep}}c@{\hspace{1\tabcolsep}}c@{\hspace{1\tabcolsep}}c}
1&0&0&0\\0&1&0&0\\0&0&1&0\\0&0&0&0
\end{array}\right)\!+\!
\left(\begin{array}{c@{\hspace{1\tabcolsep}}c@{\hspace{1\tabcolsep}}c@{\hspace{1.5\tabcolsep}}c}
-|V_{L4u}|^2 & -V_{L4u}^*V_{L4c} & -V_{L4u}^*V_{L4t} & -V_{L4u}^*V_{L4t'} \\
 -V_{L4c}^*V_{L4u} & -|V_{L4c}|^2 & -V_{L4c}^*V_{L4t} & -V_{L4c}^*V_{L4t'} \\
  -V_{L4t}^*V_{L4u} & -V_{L4t}^*V_{L4c} & -|V_{L4t}|^2 &  -V_{L4t}^*V_{L4t'} \\
  -V_{L4t'}^*V_{L4u} & -V_{L4t'}^*V_{L4c} &  -V_{L4t'}^*V_{L4t} & 
  |V_{L1t'}|^2\! +\! |V_{L2t'}|^2\! +\! |V_{L3t'}|^2  
\end{array}\right)
\end{align}
$V^{(u)}_L$ can be parameterized analogously to eqs. (\ref{conangoli}), (\ref{tildes}):
\begin{align}
\label{conangoliu}
V_{L}^{(u)}&\simeq
\left(\begin{array}{cccc} & & & 0 \\ & V_{3\times 3}^{(u)} & & 0 \\ & & & 0 \\ 0& 0& 0& 1 
\end{array}\right)
\left(\begin{array}{cccc}
c_{L1}^u  & -\tilde{s}_{L2}^{u}\tilde{s}_{L1}^{u*}  & -\tilde{s}_{L1}^{u*}\tilde{s}_{L3}^{u} & \tilde{s}_{L1}^{u*}      \\
0  & c_{L2}^u &  -\tilde{s}_{L2}^{u*}\tilde{s}_{L3}^{u}   & \tilde{s}_{L2}^{u*}     \\
 0   &0 &  c_{L3}^u & \tilde{s}_{L3}^{u*}  \\
- \tilde{s}_{L1}^{u}   & - \tilde{s}_{L2}^{u}  &  - \tilde{s}_{L3}^{u}   &c_{L1}^u c_{L2}^u c_{L3}^u    
\end{array}\right) 
\end{align}
Then the elements of the last row correspond to:
\begin{align}
& V_{L4u}= -\tilde{s}_{L1}^{u} \, , &&  V_{L4c}\approx -\tilde{s}_{L2}^{u}\, ,  
&&  V_{L4t}\approx -\tilde{s}_{L3}^{u}
\label{4rowu}
\end{align}
$c^u_{Li}$ are cosines and $\tilde{s}^u_{Li}$ are complex sines 
(one of which can be made real) of angles in the $1\,4$, $2\,4$, $3\,4$ family 
planes parameterizing the mixing of the first three families with the vector-like species:
\begin{align}
\label{tildesu}
&\tilde{s}^u_{Li}=\sin\theta^u_{Li4} e^{i \delta^u_{Li}}=s^u_{Li} e^{i \delta^u_{Li}}  \; , \qquad \quad
c^u_{Li}=\cos\theta^u_{Li4}
\nonumber\\
& \delta^u_{Lij}=\delta^u_{Li}-\delta^u_{Lj}
\end{align}
and 
\begin{align}
&\tilde{s}_{Li}^u\approx \frac{h^{*}_{ui}v_w}{M_{t'}}
\label{siu}
\end{align}
where we recalled $h_{ui}$ the couplings after the diagonalization 
$V^{(u)}_{Lij} h_{uj}\approx h_{ui}$.
$V^{(d)}_L$ can be naturally chosen with small mixings, so that, in the parameterization
of eq. (\ref{conangoliu}),
$V_{3\times 3}^{(u)}$ is basically equal to the $3\times 3$ submatrix of $\tilde{V}_{L}^{(u)}$ 
and 
diagonalizes the $3\times 3$ submatrix $\mathbf{y}^{(u)}_{3\times 3}$ in eq. (\ref{mu}).
Moreover,
as it will be shown, in order to have $V_{L4u}\approx 0.03$ it should be
$|V_{L4c}|<5\times 10^{-3} $,
$|V_{t'b}|< 8.5\times 10^{-2}$.
Then, $\tilde{V}_\text{CKM}$ can be parameterized as:
\begin{align}
\tilde{V}_\text{CKM}&=
\left(\begin{array}{ccc}
V_{ud} & V_{us} & V_{ub}  \\
V_{cd} & V_{cs} & V_{cb} \\
V_{td} & V_{ts} & V_{tb} \\
 V_{t'd} &V_{t's} &V_{t'b}
\end{array}\right)  
\simeq\left(\begin{array}{ccc}
c_{L1}^u           & 0           & 0            \\
 -\tilde{s}_{L2}^{u*}\tilde{s}_{L1}^u   & c_{L2}^u       & 0     \\
 -\tilde{s}_{L1}^u\tilde{s}_{L3}^{u*}   & -\tilde{s}_{L2}^u\tilde{s}_{L3}^{u*} &  c_{L3}^u  \\
\tilde{s}_{L1}^u       & \tilde{s}_{L2}^u      & \tilde{s}_{L3}^u 
\end{array}\right) \cdot
\left(\begin{array}{ccc}
V_{ud} & V_{us} & V_{ub}  \\
V_{cd} & V_{cs} & V_{cb} \\
V'_{td} 
& V_{ts} & V_{tb} 
\end{array}\right)\simeq \nonumber \\
&\simeq
\left(\begin{array}{ccc}
V_{ud} & V_{us} & V_{ub}  \\
V_{cd} & V_{cs} & V_{cb} \\
V_{td} & V_{ts} & V_{tb} \\
\tilde{s}_{L1}^u       
&\; \tilde{s}^u_{L1}V_{us} +\tilde{s}_{L2}^u V_{cs}+\tilde{s}_{L3}^u V_{ts}\;  & \tilde{s}_{L3}^u 
\end{array}\right)
\label{vup}
\end{align}
where in this parameterization the $3\times 3$ submatrix (made of the first three rows)
contains
$3$ angles and one phase. 
Also another phase can be absorbed, 
so we can always choose $\delta^{u}_{L1}=0$ without loss of generality.
%
Hence, for
the elements of the fourth row of $\tilde{V}_\text{CKM}$ in (\ref{vup}) 
it holds that:
\begin{align}
&V_{t'd}  \approx   -V_{L4u}  \nonumber \\  
&V_{t's}\approx -V_{L4u}V_{us} -V_{L4c} V_{cs}-V_{L4t} V_{ts} \label{4colu}\\ 
 &V_{t'b} \approx -V_{L4t} 
 \nonumber
\end{align}

Also the couplings of quarks with the real Higgs are not diagonal
in presence of the vector-like up-quark.
In fact, the matrix of Yukawa couplings and the mass matrix (\ref{mu}) are not proportional anymore,
and then they are not diagonalized by the same transformation.
In particular, similarly to eq. (\ref{Higgs}), 
left-handed SM quarks are coupled with $t'_{R}$ with coupling constants which can be in principle 
of order $O(1)$:
\begin{align}
\label{Higgsu}
& 
(\overline{u_{1L}}, \overline{u_{2L}}, \overline{u_{3L}}, \overline{u_{4L}})
\left(\begin{array}{ccc|c}
 & & & h_{u1} \\   & \mathbf{y}^{(u)}_{3\times 3} & & h_{u2} \\  & & & h_{u3} \\
 \hline
 0 & 0 & 0 & 0
\end{array}\right)\left(\begin{array}{c}
u_1 \\ u_2 \\ u_3 \\ u_4
\end{array}\right)_R \frac{H^{0}}{\sqrt{2}} +\text{h.c.}
\approx \\
&(\overline{u_{L}}, \overline{c_{L}}, \overline{t_{L}}, \overline{t'_{L}}) 
\left(\begin{array}{ccc c}
y_{1}^{u} & 0 & 0  & h_{u1} \\  0 & y^{u}_{2} &0 & h_{u2} \\ 0 & 0 & y^{u}_{3} & h_{u3} \\
0 & 0 & 0 & 
  \sum_{i=1}^{3}(\tilde{s}_{Li}^{u} h_{u i})
\end{array}\right)  
 \left(\begin{array}{c}
u_{R} \\ c_{R} \\ t_{R} \\ t'_{R}
\end{array}\right) 
\frac{H^{0}}{\sqrt{2}} +\text{h.c.} \nonumber
\end{align}

Let us also notice that, 
because of the large mixing with the first family, the extra quark $t'$ would mainly decay
into $u$ or $d$ quark via the couplings with $W$, $Z$, $H$.
The 
CMS experiment put a lower limit 
$M_{t'} \gtrsim 685$~GeV \cite{CMS},  which
implies that 
$\vert V_{t'd}\vert  \simeq  0.03$ 
can be obtained if $|h_{u1}|  \gtrsim 0.1$, much larger than the Yukawa constant of the bottom quark.
In turn, by taking $\vert V_{L4u} \vert  > 0.03$ in  
$M_{t'} =|h_{u 1}| v_{w}/|V_{t'd}|$, 
 and assuming (for the perturbativity) $h_{u1} \lesssim 1$, 
there is an upper limit on the extra quark mass, $M_{t'}< 5.8 $ TeV.

In the following sections experimental limits from FCNC and electroweak observables are examined.
The results are summarized in section \ref{upend}, 
in table \ref{limiti2} and figures \ref{fcplotuT}, \ref{vtsv4cplot}.

\subsection{Limits from rare $D$ mesons decays}
Mixings of the standard quarks
with the additional vector-like singlet up-quark $t'$
should be constrained by D-mesons rare decays.
In fact, the non-diagonal couplings of $Z$-boson with light quarks
in eqs. (\ref{ncu}) induce at tree level flavour changing leptonic and semileptonic decays
of 
D-mesons:
\begin{align}\label{D}
& \mathcal{L}_\text{new}=\frac{4G_F}{\sqrt{2}}V_{L4u}^*V_{L4c}
(\overline{u_L}\gamma^\mu c_L)\left[(-\frac{1}{2}+\sin^2\theta_W)(\overline{\ell_L}\gamma_\mu\ell_L)
+\sin^2\theta_W(\overline{\ell_R}\gamma_\mu\ell_R)\right] + \text{h.c.}
\end{align}

The most stringent constraint from 
semileptonic decays comes from the experimental limit \cite{PDG18}:
\begin{align}
& \text{Br}(D^+\rightarrow \pi^+ \mu^+ \mu^-)_\text{exp}<7.3 \cdot 10^{-8} \qquad 90\% \, \text{C.L.}
\end{align}
Neglecting the SM contribution, we can impose the experimental limit on the new contribution:
\begin{align}
\text{Br}(D^+\rightarrow \pi^+\mu^+\mu^-)_\text{new} &\simeq \text{Br}(D^+\rightarrow \pi^0 \mu^+\nu_\mu)
\frac{2|V_{L4u}^*V_{L4c}|^2[(-\frac{1}{2}+\sin^2\theta_W)^2+\sin^4\theta_W]}{|V_{cd}|^2} \nonumber\\
&<7.3\cdot 10^{-8}
\end{align}
from which 
\begin{align}
&|V_{L4u}^*V_{L4c}|< 2.0 \cdot 10^{-3}
\left[\frac{\text{Br}_{\text{max}}}{7.3\cdot 10^{-8}}  \right]^{\frac{1}{2}}
\end{align}
where 
we take advantage of the experimental branching ratio
$\text{Br}(D^+\rightarrow \pi^0 \mu^+\nu_\mu)=(3.50\pm 0.15)\cdot 10^{-3}$ \cite{PDG18}.
$\text{Br}_{\text{max}}$ is defined as in eq. (\ref{Brmax}) as the chosen maximum value 
allowed for the rate which would be generated by the beyond SM amplitude alone.
The experimental constraint on the leptonic decay $D^0\rightarrow \mu^+\mu^-$
is \cite{PDG18}:
\begin{align}
& \text{Br}(D^0\rightarrow \mu^+\mu^-)_\text{exp}<6.2\cdot 10^{-9} \qquad 90\% \, \text{C.L.}
\end{align}
Neglecting the SM short- and long- distance contribution, we can limit 
the new contribution obtained from the operator (\ref{D}):
\begin{align}\label{kmumu}
&\text{Br}(D^0\rightarrow \mu^+\mu^-)_\text{new}=
\text{Br}(D^+\rightarrow  \mu^+\nu_\mu)\frac{1}{2}
\frac{\tau(D^0) M_{D^0}\sqrt{1-4\frac{m^2_\mu}{M^2_{D^0}}}}
{\tau(D^+)M_{D^+}\left(1-\frac{m^2_\mu}{M^2_{D^+}}\right)^2}
\frac{|V_{L4u}^*V_{L4c}|^2}{|V_{cd}|^2} <6.2\cdot 10^{-9}
\end{align}
It is obtained that:
\begin{align}
& |V_{L4u}^*V_{L4c}|  < 2.0 \cdot 10^{-3}
\left[\frac{\text{Br}_{\text{max}}}{6.2\cdot 10^{-9}}  \right]^{\frac{1}{2}}
\end{align}
using the values in table \ref{values}, with 
$\text{Br}(D^+\rightarrow  \mu^+\nu_\mu)=(3.74\pm 0.17)\cdot 10^{-4}$.
$\text{Br}_{\text{max}}$ is 
the maximum value 
allowed for the rate which would be generated by the beyond SM amplitude alone.

%

\subsection{Limits from neutral mesons systems}
\subsubsection{$D^0$-$\bar{D}^0$ mixing}

As described in Section \ref{DDdown}, the experimental result (\ref{dmdexp}) for the mass difference 
in the $D^{0}$ system
allows values 
which can be
two orders of magnitude higher than the SM short-distance expectation, 
while long-distance effects cannot be computed reliably. 
Then,
new physics scenarios 
can in principle account for the experimental mass difference 
$\Delta m_{D\,\text{exp}}$ in eq. (\ref{dmdexp}).

The non-diagonal couplings of Higgs and $Z$ bosons with quarks
in eqs. (\ref{ncu}), (\ref{Higgsu})
bring additional contributions
to the transition $\bar{D}^{0} \leftrightarrow D^{0}$, both at tree level and loop level.
Analogously to eq. (\ref{newkk}), the corresponding effective Lagrangian 
reads:
\begin{align}
\label{newddtree}
\mathcal{L}^\text{new}_{\Delta C=2} & \approx 
-\frac{G_F}{\sqrt{2}}(V_{L4u}^*V_{L4c})^2\left(1+\frac{G_{F}M^{2}_{t'}}{8\sqrt{2}\pi^{2}}  \right)
(\overline{u_L}\gamma^\mu c_L)^2+\text{h.c.} = \nonumber \\
&= -\frac{G_F}{\sqrt{2}}(V_{L4u}^*V_{L4c})^2 f(M_{t'})
(\overline{u_L}\gamma^\mu c_L)^2+\text{h.c.} \\
 f(M_{t'})&=1+\left(\frac{M_{t'} }{3.1 \, \text{TeV}}\right)^{2}
 \label{fmtprimo}
\end{align}
Hence, analogously to the kaons case, the additional contribution to the mixing mass is:
\begin{align}
& 2m_{D^{0}}M_{12\,\text{new}}^{D*}=-\langle \bar{D}^0|\mathcal{L}^\text{new}_{\Delta C=2}|D^0\rangle 
\end{align}
where in VIA approximation:
\begin{align}
&\langle \bar{D}^0 |(\overline{u_L}\gamma^\mu c_L)^2|D^0\rangle =\frac{2}{3}f^2_Dm^2_{D^{0}} \, .
\end{align}
Then, 
the new operator (\ref{newddtree})
can generate the mass difference:
\begin{align}
\label{dmdnewtree}
&\Delta m_{D\,\text{new}} \approx 2|M_{12\,\text{new}}^{D}|\simeq
\frac{2}{3}\frac{G_F}{\sqrt{2}}f(M_{t'})|V_{L4u}^*V_{L4c}|^2f^2_Dm_{D^{0}} 
\end{align}
We can assume that $\Delta m_{D\,\text{new}}$ is the dominant contribution to the mass difference.
Then, 
by using the $95\%$ C.L. limit obtained from eq. (\ref{dmdexp}) we have:
\begin{align}
&\Delta m_{D\,\text{new}} < 1.2\times 10^{-14}
\end{align}
leading to:
\begin{align}
\label{DDupcon}
&|V_{L4u}^*V_{L4c}|< 1.5 \times 10^{-4} \left[\frac{f(1\,\text{TeV})}{f(M_{t'})} \right]^{1/2}
\left[ \frac{\Delta m_{D\text{max}}}{1.2\cdot 10^{-14}} \right] 
\end{align}

\subsubsection{$K^0$-$\bar{K}^0$ mixing}
\label{KKsecup}

\begin{figure}[t]
\centering
\includegraphics[width=0.3\textwidth]{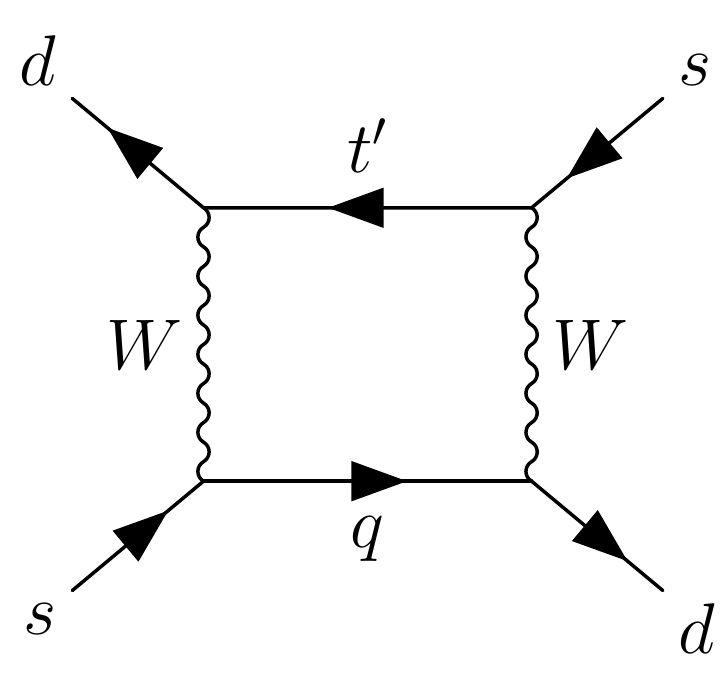}
\caption{New contribution to $\bar{K}^{0}\rightarrow K^0$ mixing, $q=u,c,t,t'$.}
\label{KK}
\end{figure}

In the SM, the short-distance contribution to the transition 
$K^0(d\bar{s})\leftrightarrow \bar{K}^0(\bar{d}s)$
arises from weak box diagrams (figure \ref{kksm}), which corresponds to the 
effective Lagrangian in eq. (\ref{smkk}).

\begin{figure}[t]
\centering
\includegraphics[width=0.7\textwidth]{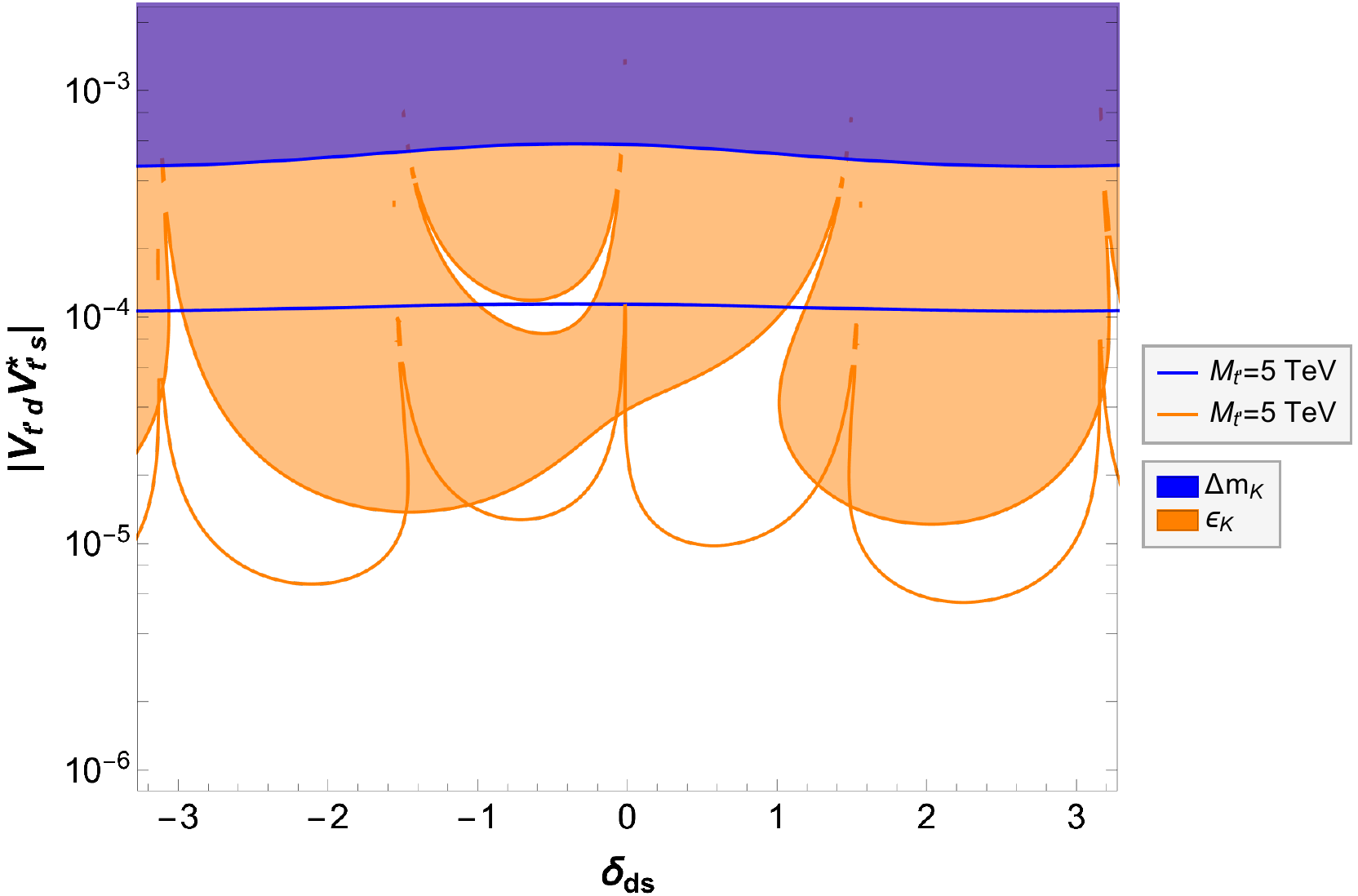} 
\caption{\label{fcplotuT} 
Upper limits on the product $|V_{t's}^*V_{t'd}|$ from neutral kaon mixing,
obtained from the constraint 
in eqs. eqs. (\ref{dmkup1}), (\ref{epskcon1up}) 
with $\Delta_{K}=1$, $\Delta_{\epsilon_{K}}=0.4$,
as a function of the relative phase $\delta_{ds}=\arg(V_{t's}^*V_{t'd})$.
The blue and orange curves indicate where 
the upper limit shifts 
for $M_{t'}=5$ TeV. 
 }
\end{figure}

In the scenario with extra up-type quark, the mixing of the SM quarks with the $t'$-quark 
in the charged current in eq. (\ref{ccu}) gives additional contributions to the $K^{0}$ mixing,
in interference with the SM.
In fact, the same operator as in (\ref{smkk}) is originated by box diagrams with internal
$t'$-quark running in the loop, as shown in figure \ref{KK}:
\begin{align}
\mathcal{L}^\text{new}_{\Delta S=2} = -\frac{G^2_F m^2_W}{4\pi^2}
& \Big((V_{t's}^*V_{t'd})^2 S(x_{t'})+
2(V_{t's}^*V_{t'd})(V_{cs}^*V_{cd})S(x_c,x_{t'})+  \nonumber \\
&\quad +2(V_{t's}^*V_{t'd})(V_{ts}^*V_{td})S(x_t,x_{t'})\Big)
 (\overline{s_L}\gamma^\mu d_L)(\overline{s_L}\gamma^\mu d_L) 
+ \text{h.c.}
\label{kkloopnew}
\end{align}
where $x_a=\frac{m_a^2}{m_W^2}$ and 
$S_0(x,y)$ are the Inami-Lim functions \cite{Inami:1980fz} in eqs. (\ref{S0x}), (\ref{S0xy}).

As described in section \ref{kksec}, 
the mass splitting $\Delta m_K=m_{K_L}-m_{K_S}$
and the CP-violating parameter $\epsilon_K$ 
are dominated by the short distance contribution, which is given by
the off-diagonal term $M_{12}$ of the mass matrix of neutral kaons: 
\begin{align}
& M_{12}=-\frac{1}{2m_{K^{0}}} \langle K^0|\mathcal{L}_{\Delta S=2}|\bar{K}^0\rangle \nonumber\\
&\Delta m_K \approx 2|M_{12}| 
+\Delta m_\text{LD}
\nonumber \\
&|\epsilon_K| \approx \frac{|\text{Im}[ M_{12}]|}{\sqrt{2}\Delta m_K}  
\end{align}
where $\Delta m_\text{LD}$ is the unknown long distance contribution.
Then, in order to not contradict experimental results, 
the new contribution (\ref{kkloopnew}), and in particular the mixing elements
$V_{t's}^*V_{t'd}$, should be constrained.

The mixing mass in the SM $M_{12}^\text{SM}$ is given in eq. (\ref{M12sm}).
Regarding the CP-conserving part, 
as in section \ref{kksec}, 
in order to estimate the constraint on the new mixing elements
we can impose:
\begin{align}
\label{dmkup1}
&\left|M_{12}^\text{new}\right|<\left|M_{12}^\text{SM}\right|\,\Delta_{K}
\end{align}
where $\Delta_{K}$ is real and positive and we will evaluate for $\Delta_{K}=1$.
This is analogous to comparing 
the coefficients of the effective operators in 
eqs. (\ref{smkk}) and (\ref{kkloopnew}):
\begin{align}
\label{dmkup}
& \Big\vert  (V_{t's}^*V_{t'd})^2 S_{0}(x_{t'})+
2(V_{t's}^*V_{t'd})(V_{cs}^*V_{cd})S_{0}(x_c,x_{t'})
+ \nonumber \\ &  \hspace{4cm}
+ 2(V_{t's}^*V_{t'd})(V_{ts}^*V_{td})S_{0}(x_t,x_{t'})
\Big\vert < 
S_0(x_c)|V_{cs}^*V_{cd}|^{2} \, \Delta_{K}
\end{align}
In this case the upper bound
is basically determined by 
constraining the contribution of
the box diagram with $t'$ running in the loop: 
\begin{align}
& 
|V_{t's}^*V_{t'd}|^2   
S_{0}(x_{t'})  < S_0(x_c)(V_{cs}^*V_{cd})^{2}\, \Delta_{K}    \nonumber \\
\label{vtsvtdkk}
& |V_{t's}^*V_{t'd}|  
< 5.2 \times 10^{-4} 
\left[ \frac{1 \,\text{TeV}}{M_{t'}}  \right] 
\left[\Delta_{K}  \right]^{1/2} 
\end{align}
The scaling with $t'$ mass in eq. (\ref{vtsvtdkk}) 
holds since $S_0(x_{t'})\sim \frac{1}{4}x_{t'}$ for $M_{t'}\gtrsim 2$ TeV.

Regarding the contribution to the CP-violating parameter $\epsilon_K$,
we can estimate the upper bound on the new operator by constraining
the imaginary part of the new contribution to the mixing mass $|\text{Im}M_{12}^\text{new}|$ 
to be a fraction $\Delta_{\epsilon_{K}}$ of the SM contribution,
as in eq. (\ref{epskcon1}):
\begin{align}
|\text{Im}M_{12}^\text{new}|<
|\text{Im}M_{12}^\text{SM}| \, \Delta_{\epsilon_{K}}
\label{epskcon1up}
\end{align}
At leading order of 
the new physics contribution, 
eq. (\ref{epskcon1up}) is equivalent to 
comparing the magnitude of the imaginary part of operators 
(\ref{smkk}) and (\ref{kkloopnew}), in the standard parameterization of CKM matrix:
\begin{align}
&  \left| \text{\large{Im}}\Big( (V_{t's}^*V_{t'd})^2 S(x_{t'})+ 
2(V_{t's}^*V_{t'd})(V_{cs}^*V_{cd})S_{0}(x_c,x_{t'})
+\right. \nonumber \\ &\left. \hspace{5cm}
+2(V_{t's}^*V_{t'd})(V_{ts}^*V_{td})S_{0}(x_t,x_{t'}) 
\Big)\right|<  
\left|\text{\large{Im}}(\mathcal{C}_\text{SM})\right|  \, 
\Delta_{\epsilon_{K}} 
\label{epskconup}
\end{align}
with
\begin{align}
&\mathcal{C}_\text{SM}= \eta_{1} (V_{cs}^*V_{cd})^{2}S_0(x_c)+  
 \eta_{2}(V_{ts}^{*}V_{td})^{2}S_0(x_t)+
2 \eta_{3}(V_{cs}^*V_{cd})(V_{ts}^{*}V_{td}) S_0(x_c,x_t)
\end{align}
This constraint also depends on the mass of the extra quark,
since the first term of eq. (\ref{kkloopnew}) scales as
$S_0(x_{t'})\sim \frac{1}{4}M_{t'}^{2}/m_{W}^{2}$ for $M_{t'}\gtrsim 2$ TeV.
while the second and third components are logarithmically growing with $t'$ mass.
We make an estimation choosing $\Delta_{\epsilon_{K}}=0.4$,
corresponding to an equivalent scale of the new operator 
$\Lambda_{ds,\text{Im}}=2.5\cdot 10^{4}\, \text{TeV}$, as defined in eq. (\ref{Lambdaeps}).
The 
upper limit on the product $V_{t's}^*V_{t'd}$
obtained from the conditions in eqs. (\ref{dmkup}), (\ref{epskconup}) 
with $\Delta_{K}=1$, $\Delta_{\epsilon_{K}}=0.4$
is shown in figure \ref{fcplotuT}, 
where we used the Wolfenstein parameterization as in eq. 
(\ref{vtsvtdIm}) and $\delta_{ds}=\arg(V_{t's}^*V_{t'd})$.
The excluded region is shown for $M_{t'}=1$ TeV, while the lines show the lowered upper limit
for $M_{t'}=6$ TeV.

\subsubsection{$B_{s,d}^0$-$\bar{B}_{s,d}^0$ mixing}
\label{BBupsec}

As illustrated in section \ref{bbsec},
in the SM the dominant contribution to 
$B^0_{d}(\bar{b}d)$-$\bar{B}_{d}^0(\bar{d}b)$ and $B^0_{s}(\bar{b}s)$-$\bar{B}_{s}^0(\bar{s}b)$ 
mixings comes from  
box-diagrams with internal top-quark, as shown in figure \ref{bbsm}, with effective Lagrangian in eqs.
(\ref{BBsm}), (\ref{BBsSM}),
and, analogously to the kaons system:
\begin{align}
&2M_{B^{0}_{d,s}}M_{12}^{(d,s)*}=\langle \bar{B}_{d,s}^0 |-\mathcal{L}_{\Delta B(d,s)=2}|B^0_{d,s}\rangle \nonumber \\
&\Delta M_{d,s}=2|M_{12}^{(d,s)}| 
\end{align}

In the scenario with extra up-type quark, the mixing of the SM quarks with the $t'$-quark 
in the charged current in eq. (\ref{ccu}) 
contribute to neutral B-mesons mixing, in interference with the SM.
In fact, the same operator as in (\ref{BBsm}) is originated by box diagrams with internal
$t'$ and $t,t'$-quarks:
\begin{align}
\label{bb0t}
\mathcal{L}^\text{new}_{\Delta B(d)=2}=-\frac{G^2_F m^2_W}{4\pi^2}&
\left((V_{t'b}^*V_{t'd})^2 S_{0}(x_{t'})+
2(V_{t'b}^*V_{t'd})(V_{tb}^*V_{td})S_{0}(x_t,x_{t'})\right)
(\overline{b_L}\gamma^\mu d_L)^{2}+\text{h.c. } \\
\mathcal{L}^\text{new}_{\Delta B(s)=2}=-\frac{G^2_F m^2_W}{4\pi^2}&
\left((V_{t'b}^*V_{t's})^2 S_{0}(x_{t'})+
2(V_{t'b}^*V_{t's})(V_{tb}^*V_{ts})S_{0}(x_t,x_{t'})\right)
(\overline{b_L}\gamma^\mu s_L)^{2}+\text{h.c. }
\label{bb0ts}
\end{align}
where $x_a=\frac{m_a^2}{m_W^2}$ and $S(x_a)$ are the Inami-Lim functions \cite{Inami:1980fz} given in eqs. (\ref{S0x}), (\ref{S0xy}). 
The c-quark contribution is negligible.

\begin{figure}[t]
\centering
\includegraphics[width=0.47\textwidth]{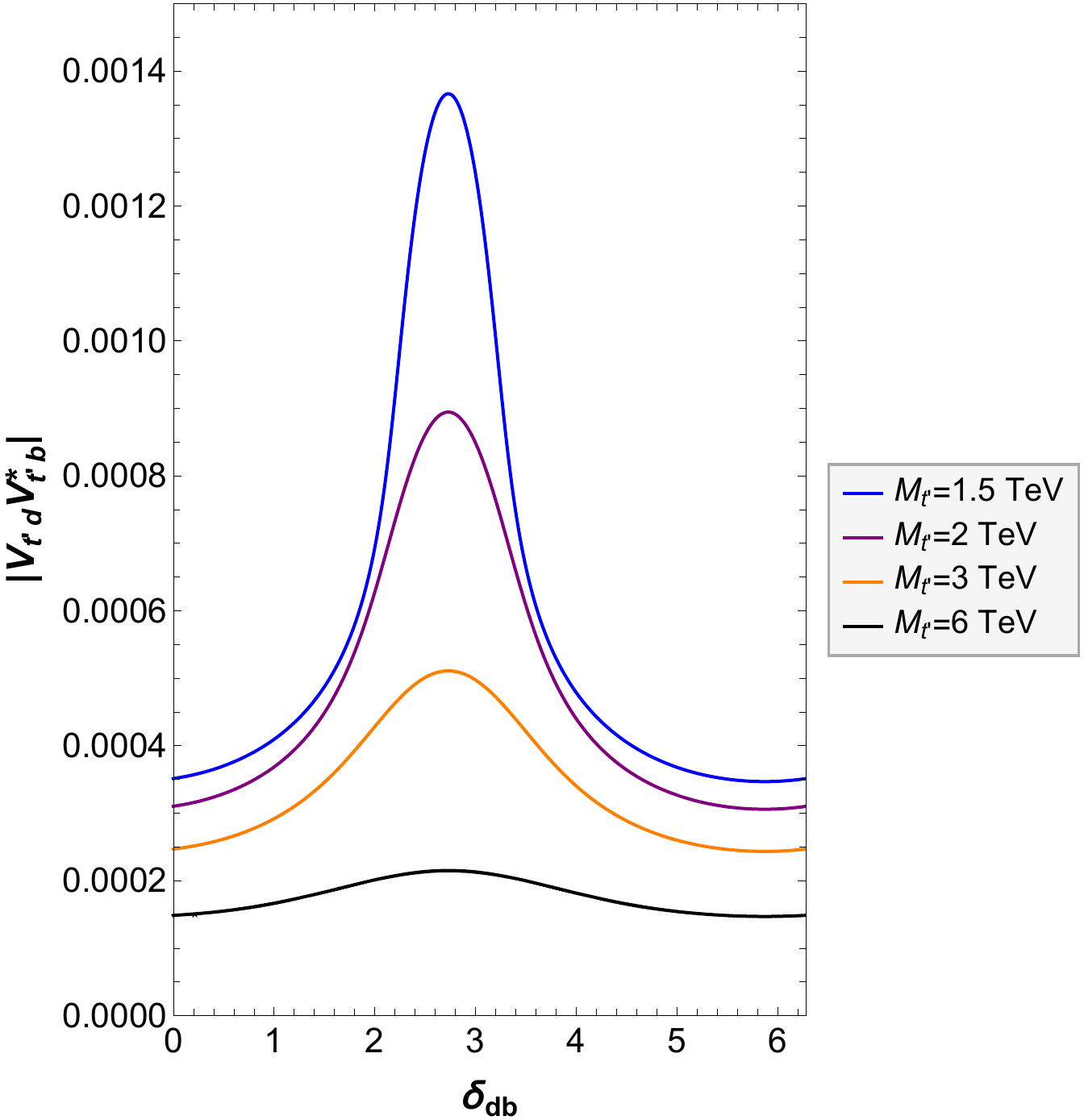}
\includegraphics[width=0.34\textwidth]{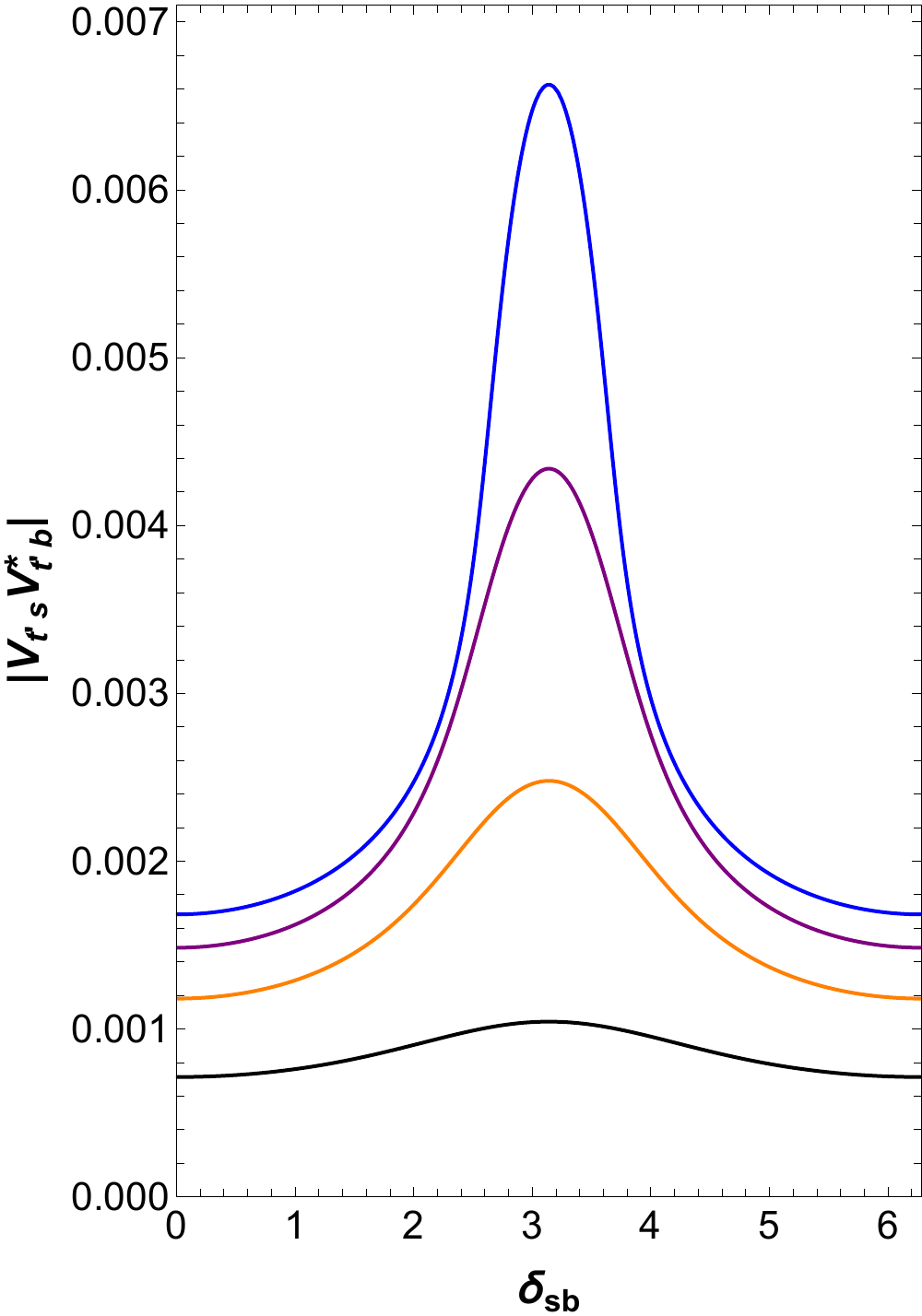}
\caption{\label{vtdsb} Upper limits on the products of the mixing elements 
$|V_{t's}^*V_{t'd}|$, $|V_{t's}^*V_{t'd}|$ 
from neutral B-mesons mixing,
obtained from the constraint 
in eqs. eqs. (\ref{BBupcon})
with $ \Delta_{B_{d,s}}=0.3$,
as a function of the relative phases 
$\delta_{db}=\arg(V_{t'b}^*V_{t'd})$, $\delta_{sb}=\arg(V_{t'b}^*V_{t's})$,
evaluated for the values of extra quark masses $M_{t'}=1.5$, $2$, $3$, $6$ TeV
(blue, purple, orange and black curves respectively).
}
\end{figure}

We can constrain 
the new contribution to be 
less than a fraction $\Delta_{B_{(d,s)}}$ of the SM contribution
$\Delta M_{d,s}^\text{SM}= 2|M_{12}^{(d,s)\text{SM}}|$ given in eq. (\ref{dmBB}), as:
\begin{align}
& | M_{12}^{(d,s)\text{new}} |<
|M_{12}^{(d,s)\text{SM}} |\, \Delta_{B_{d,s}}
\label{BBupcon1}
\end{align}
This is analogous to comparing the coefficients of the effective operators in 
eqs. (\ref{bb0t}), (\ref{bb0ts}) with the SM ones (\ref{BBsm}), (\ref{BBsSM}):
\begin{align}
&\Big|(V_{t'b}^*V_{t'd/s})^2 S_{0}(x_{t'})+2(V_{t'b}^*V_{t'd/s})(V_{tb}^*V_{td/s})S_{0}(x_t,x_{t'})
\Big|
< S(x_t)|V_{tb}^*V_{td/s}|^2 \, 
\Delta_{B_{d,s}}
\label{BBupcon}
\end{align}
where we evaluate the constraint at leading order of both SM and new physics contributions.
As in section \ref{bbsec}, we use $\Delta_{B_{(d,s)}}=0.3$ as a benchmark value.
Then, by taking $M_{t'}=1$ TeV, we obtain
(with $m_t(m_t)=163$ GeV):
\begin{align}
\label{vtbvtdBB}
&|V_{t'b}^*V_{t'd}| < 0.4\div 2.6 \times 10^{-3} 
\\ 
&|V_{t'b}^*V_{t's}| <0.2 \div 1.2 \times 10^{-2} 
\label{vtbvtdBBs}
\end{align}
depending on the relative phases
$\delta_{db}=\arg(V_{t'b}^*V_{t'd})$, $\delta_{sb}=\arg(V_{t'b}^*V_{t's})$.
The constraints become more stringent for higher values of $t'$ mass. 
The contribution from the mixed box diagrams
with $t,t'$ grows logarithmically with $M_{t'}$, but the
contribution from the box-diagram with $t'$ increases linearly with the mass for $M_{t'}\gtrsim 2$ TeV. 
Then, for $M_{t'}\gtrsim 3$ TeV the upper limits start to decrease 
as $M_{t'}^{-1}$.
The upper bound on the mixing elements $|V_{t'b}^*V_{t'd}|$, $|V_{t'b}^*V_{t'd}|$ as a function of the 
relative phase of the elements is shown in
figure \ref{vtdsb} for increasing values of the extra quark $t'$ mass.


\subsection{$\Gamma(Z\rightarrow\text{hadr})$, $\Gamma(Z)$}
\label{Zup}

The presence of 
vector-like quarks affect both
off-diagonal couplings and diagonal couplings 
of Z-boson with quarks, changing the prediction of the observables 
related to the $Z$ boson physics. 

The SM predictions for the Z decay rate and partial decay rate into hadrons are \cite{PDG18}:
\begin{align}
&\Gamma(Z)_\text{SM}=2.4942\pm0.0009 \,  \text{GeV} \, , \qquad
\Gamma(Z\rightarrow\text{hadr})_\text{SM}=1.7411\pm0.0008  \,  \text{GeV}
\label{GZsm2}
\end{align}
to be compared with the experimental results \cite{PDG18}:
\begin{align}
& \Gamma(Z)_\text{exp}=2.4952\pm0.0023 \,  \text{GeV} \, , \qquad
\Gamma(Z\rightarrow\text{hadr})_\text{exp}=1.7444\pm0.0020  \,  \text{GeV}
\label{GZexp2}
\end{align}
In this BSM scenario, the
partial decay rate into hadrons $\Gamma(Z\rightarrow\text{had})$
(which also corresponds to the deviation of the total $Z$ decay rate $\Gamma(Z)$, 
since there are not additional leptons)
would be changed with respect to the SM expectation by:
\begin{align}
&\Gamma(\text{had})-\Gamma(\text{had})_\text{SM}=
 \Gamma(Z)-\Gamma(Z)_\text{SM}= 
 \nonumber \\
=&\frac{G_FM^3_Z}{\sqrt{2}\pi}\left[ \sum_{i,j=u,c}
\left| \frac{1}{2}\sum_{k=1}^3 V_{Lki}^{*}V_{Lkj}-\frac{2}{3}\sin^2\theta_W\delta_{ij} \right|^2
-2\left( \frac{1}{2}-\frac{2}{3}\sin^2\theta_W \right)^2  \right]= \nonumber \\
=&\frac{G_FM^3_Z}{\sqrt{2}\pi}
\left[\sum_{q=u,c}\left(\frac{1}{2}(1-|V_{L4q}|^2)-\frac{2}{3}\sin^2\theta_W\right)^2  
+\frac{1}{2}|V_{L4u}^* V_{L4c}|^2 -2\left( \frac{1}{2}-\frac{2}{3}\sin^2\theta_W \right)^2  \right] 
\approx \nonumber \\
\approx &\frac{G_FM^3_Z}{\sqrt{2}\pi} 
\left(-\frac{1}{2}+\frac{2}{3}\sin^2\theta_W\right)\left(|V_{L4u}|^2+|V_{L4c}|^2\right)  <0
\label{Zupg}
\end{align}
where the QCD corrections factor $\approx 1.050$ should be included.
The extra contribution to the decay rate
should be constrained: 
\begin{align}
&
|\Gamma(Z\rightarrow\text{had})-\Gamma(Z\rightarrow\text{had})_\text{SM}|<
\Delta\Gamma_{Z}
\end{align}
As shown in eq. (\ref{Zupg}), 
the prediction for the decay rate is lowered with respect to the SM one.
Since the SM expectation (\ref{GZsm2}) is below the experimental result (\ref{GZexp2}),
we can choose to use the limit value of the SM prediction
$\Gamma(Z\rightarrow\text{had})_\text{SM}= 1.7419$ GeV in eq. (\ref{Zupg}). 
Then we can impose that
$\Gamma(Z\rightarrow\text{had})$ should stay
in the $95\%$ C.L. interval of the experimental value $\Gamma(Z\rightarrow\text{had})_\text{exp}$, 
which means:
\begin{align}
&|V_{L4u}|^2+|V_{L4c}|^2<  2.0\cdot 10^{-3} 
\left[ \frac{\Delta\Gamma_{Z}}{1.4\times 10^{-3} \,\text{GeV}} \right]
\end{align}
The last constraint is satisfied if both $|V_{L4u,c}|\lesssim 0.032$. 
With $|V_{L4u}|=0.03$, it means $|V_{L4c}|< 0.034$. 
With $V_{L4c}=0$, the constraint implies: 
\begin{align}
& |V_{L4u}|^2 <  2.0\cdot 10^{-3} \; , \qquad \quad
|V_{L4u}|< 0.044 
\label{v4uZ}
\end{align}
which is extremely close to the value needed to solve the CKM unitarity problem
(for example, at $95\%$ C.L. $|V_{L4u}|=0.038^{+0.008}_{-0.011}$ (\ref{newunsol}) 
using our conservative average for $V_{ud}$ (\ref{vudmedio})). 

\subsection{Summary of experimental limits}
\label{upend}

\input{tab/tabellaup2}


\begin{figure}[t]
\centering
\includegraphics[width=0.8\textwidth]{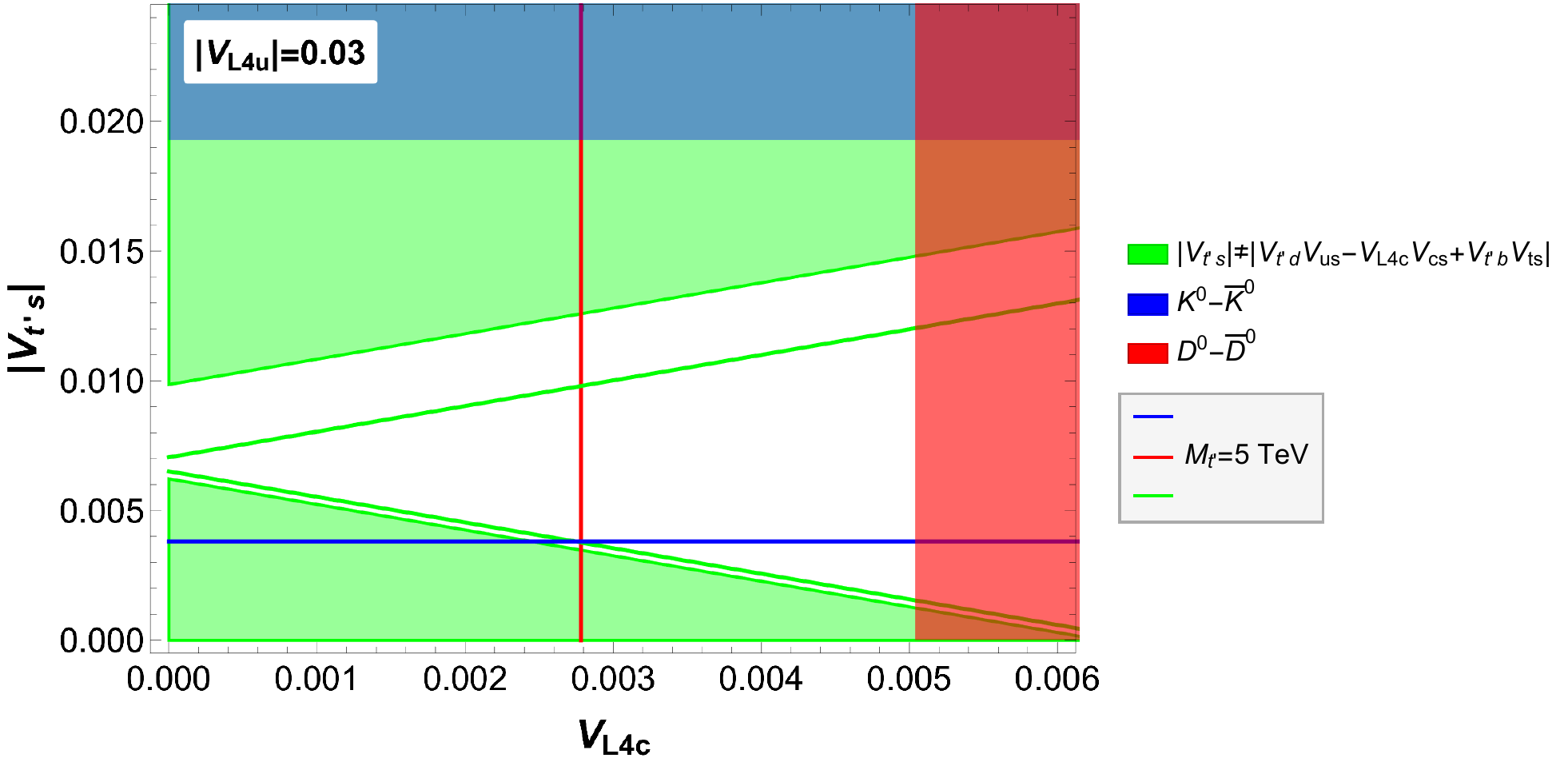}
\caption{\label{vtsv4cplot} Allowed area for $|V_{t's}|$ and $|V_{L4c}|$.
$|V_{t'd}|=0.03$ 
is set as a conservative benchmark value.
The red area is excluded by the constraints on
$|V_{L4c}|$ from $D^0$ systems with $M_{t'}=1$ TeV, as reported in 
eq. (\ref{DDupcon}). 
The red line shows where the boundary is shifted
if the mass of the extra quark is taken as $M_{t'}=5$ TeV. 
The blue area is excluded by neutral kaons mixing,
using the result $|V_{t's}| < 1.9 \cdot 10^{-2}$ from eq. (\ref{vtskk}).
The blue line indicates where the forbidden region spread 
if the mass of the extra quark is taken as $M_{t'}=5$ TeV. 
The green region is excluded by 
the relation in eq. (\ref{vtpsf}), for any value of the relative phases,
using the maximum value allowed for $V_{t'b}$ by the limit from the B-mesons system.
}
\end{figure}
\begin{figure}[t]
\centering
\includegraphics[width=0.8\textwidth]{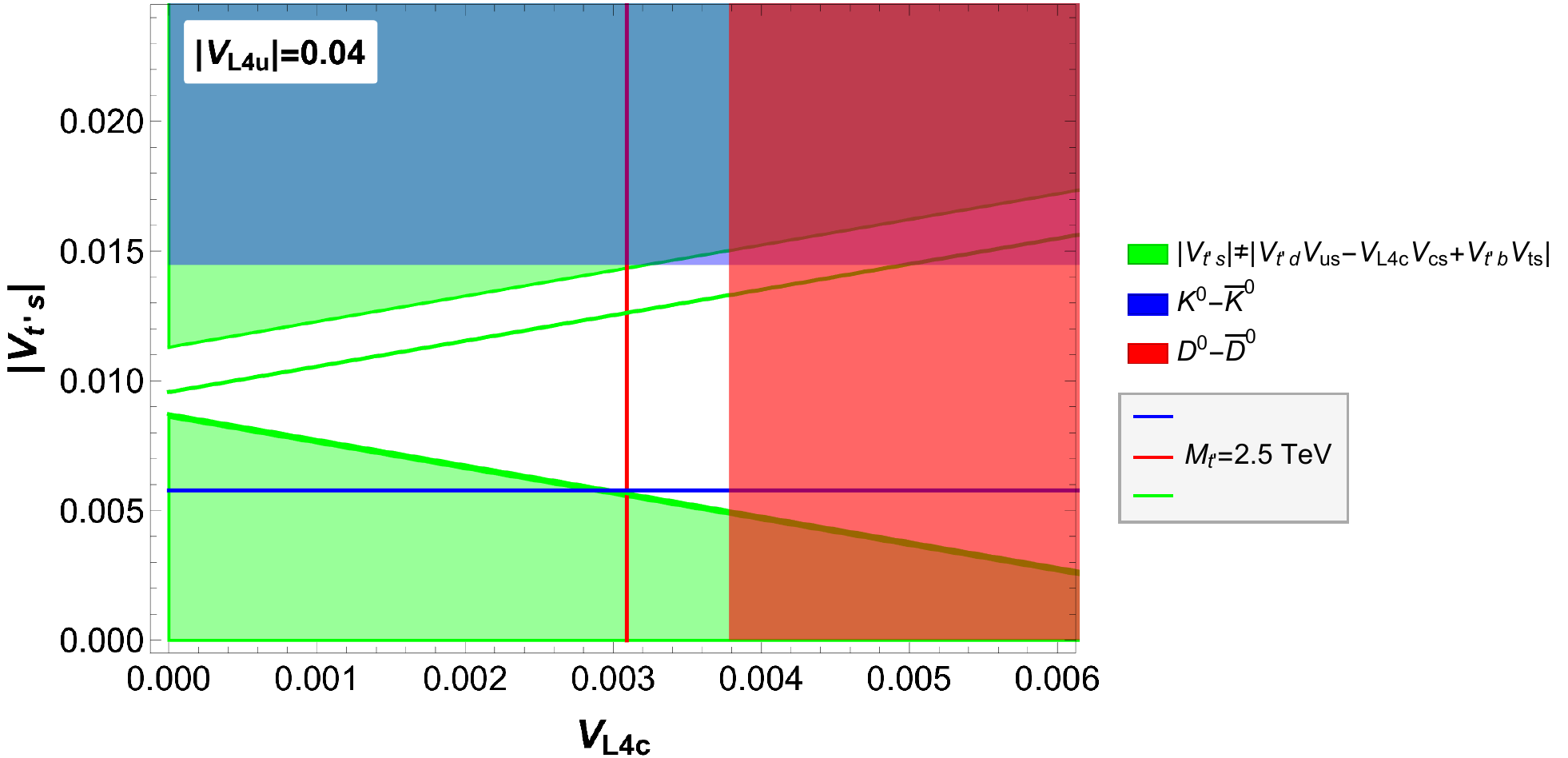}
\caption{\label{vtsv4cplot4} Allowed area for $|V_{t's}|$ and $|V_{L4c}|$,
with $|V_{t'd}|=0.04$. 
The red area is excluded by the constraints on
$|V_{L4c}|$ from $D^0$ systems with $M_{t'}=1$ TeV. as reported in 
eq. (\ref{DDupcon}). 
The red line shows where the boundary is shifted
if the mass of the extra quark is taken as $M_{t'}=2.5$ TeV.
The blue area is excluded by neutral kaons mixing.
using the result $|V_{t's}| < 1.4 \cdot 10^{-2}$ from eq. (\ref{vtskk}).
The blue line indicates where the forbidden region expands
if the mass of the extra quark is taken as $M_{t'}=2.5$ TeV.
The green region is excluded by 
the relation in eq. (\ref{vtpsf}), for any value of the relative phases,
using the 
values for $V_{t'b}$ allowed by the limit from the B-mesons system 
in eq. (\ref{vtbvtdBB}), 
figure \ref{vtdsb}. 
}
\end{figure}

As illustrated in section \ref{situation}, 
the analysis 
of the latest determinations of $V_{us}$ obtained from kaon decays
and of $V_{ud}$ from beta decays results in
a deviation from unitarity of the first row of the CKM matrix.
The SM unitarity relation (\ref{uns}) can be modified into
the relation in eq. (\ref{newunup}) 
if an extra vector-like up-type quark $u_{L4},u_{R4}$ participates in the mixing 
of the SM families. 
A quite large mixing with the first 
family, $|V_{t'd}|\approx |V_{L4u}|\approx  0.038(5)$ (\ref{newunsolup}), 
is needed in order to explain the data, 
see eq. (\ref{newunsolup}) 
(see also 
table \ref{Tabledelta2},
where in this scenario $\delta_\text{CKM}= |V_{L4u}|^{2}$).
In this way, the values of $V_{us}$ obtained from the three determinations can be shifted
as shown in figure \ref{vusudm43plot},
where $|V_{t'd}|\approx |V_{L4u}|$ plays the same role as $|V_{ub'}|$.
Then, we need to verify if such large mixing is compatible with experimental constraints from flavour changing decays and electroweak observables.

In table \ref{limiti2} 
constraints obtained in this section on the mixing of the new up-type vector-like quark
with the SM families are summarized
($|V_{t'd}|\approx |V_{L4u}|=0.03$ is used as a conservative benchmark value).

In figures \ref{vtsv4cplot} and \ref{vtsv4cplot4} 
the allowed area in the parameter space of $|V_{t's}|$ and $|V_{L4c}|$
is shown, using $|V_{t'd}|=0.03$ and $|V_{t'd}|=0.04$ respectively. 

The blue area is excluded by constraints on neutral K-mesons system, as
obtained in section \ref{KKsecup} and 
summarized in figure \ref{fcplotuT}. In figures \ref{vtsv4cplot}, \ref{vtsv4cplot4} we use
the result: 
\begin{align}
& |V_{t's}|  
< 1.9 \times 10^{-2} \left[ \frac{0.03}{|V_{L4u}|}\right]
\label{vtskk}
\end{align}
obtained for $M_{t'}=1$ TeV, 
with the most convenient choice of the phase $\delta_{ds}=\arg(V_{t's}^*V_{t'd})=-0.04$.
The blue line indicates where the forbidden region expand 
if the mass of the extra quark is taken as $M_{t'}=5$ TeV and $M_{t'}=2.5$ TeV
in the two figures respectively.
In fact,
for specific values of the phase $\delta_{ds}=\arg(V_{t's}^*V_{t'd})$,
the upper limit is basically due to the restriction (\ref{vtsvtdkk}): 
\begin{align}
& |V_{t's}^{*}V_{t'd}|^{2} S_{0}(x_{t'}) < |V_{cs}^{*}V_{cd}|^{2}S_{0}(x_{c}) 
\end{align}
which gives:
\begin{align}
&|V_{t's}^{*}V_{t'd}| < 5.2 \times 10^{-4} \left[ \frac{1 \,\text{TeV}}{M_{t'}}  \right]  \; , \qquad \quad
|V_{t's}| <1.7 \cdot 10^{-2} 
\left[ \frac{0.03}{|V_{L4u}|}\right]
\left[ \frac{1 \,\text{TeV}}{M_{t'}}  \right] 
\label{vtptot}
\end{align}
%
%
%

The red area is excluded by the constraints on
$|V_{L4c}|$ from $D^0$ systems.
The new contribution to the mass difference is allowed to account for
the experimental limit on $\Delta m_{D\, \text{exp}}$ at $95\%$ C.L., which means that
the extra singlet quark can be thought as a way to explain the experimental result.
Then, as reported in eq. (\ref{DDupcon}), the constraint is:
\begin{align}
&|V_{L4u}^*V_{L4c}|< 1.5 \times 10^{-4} \left[\frac{f(1\,\text{TeV})}{f(M_{t'})} \right]^{1/2}
\; , \qquad 
 |V_{L4c}|< 5.0 \cdot 10^{-3} \left[ \frac{0.03}{|V_{L4u}|} \right]
\label{DD4c}
\end{align}
where $f(M_{t'})$ is defined in eq. (\ref{fmtprimo}).
The red line shows where the boundary is shifted
if the mass of the extra quark is taken as $M_{t'}=5$ TeV and $M_{t'}=2.5$ TeV
in the two cases respectively.
We can also notice that,
taking into account the relations 
(\ref{4rowu}), (\ref{siu}):
\begin{align}
\label{v4hu}
&V_{L4u}\approx -\frac{h_{u1}^{*}v_{w}}{M_{t'}} \; , \qquad \quad 
V_{L4c}\approx -\frac{h_{u2}^{*}v_{w}}{M_{t'}}
\end{align}
where $h_{di}$ are the Yukawa couplings defined in eq.  (\ref{yu}),
from eq. (\ref{DD4c}) follows that 
the coupling of the 4th species with the 2nd family $h_{d2}$ 
should be at least six times smaller  
than the coupling with the first one $h_{d1}$. 

From eq. (\ref{4colu}) we have:
\begin{align}
\label{vtpsf}
 |V_{t's}| & \approx  |V_{t'd}V_{us} -V_{L4c}V_{cs} +V_{t'b}V_{ts}| \approx \nonumber \\
& \approx 
\left| |V_{t'd}| V_{us} + e^{i\delta^{u}_{21}} |V_{L4c}| V_{cs} - e^{-i \delta_{db}} |V_{t'b}||V_{ts}| \right|
\end{align}
where we use the parameterization in eqs. (\ref{4rowu}) and (\ref{tildesu}), and
$\delta_{d}=\arg(V_{t'd})\approx \delta^{u}_{L1}$, 
$-\delta_{db}=\arg(V_{t'b}V_{t'd}^{*}) \approx \delta^{u}_{L31}=\delta^{u}_{L3}-\delta^{u}_{L1} $, 
and we can choose $\delta^{u}_{L1}=0$ without loss of generality.
The green region is not allowed by 
the relation (\ref{vtpsf}) for any value of the relative phases,
using the 
values for $V_{t'b}$ allowed by the limit from the B-mesons system 
in eq. (\ref{vtbvtdBB}), 
figure \ref{vtdsb}: 
\begin{align}
\label{vtpb}
& |V_{t'b}^*V_{t'd}| < 0.4\div 2.6 \times 10^{-3} \; , \qquad 
|V_{t'b}| < (1.3\div 8.5) \times 10^{-2}   \left[ \frac{0.03}{|V_{t'd}|}\right]
\end{align}

Then, as regards the fourth row of the extended CKM matrix in eq. (\ref{ccu}),
with $|V_{t'd}|=0.03$ and $M_{t'} =1 \, \text{TeV}$, we get:
\begin{align}
& |V_{t's}| < 1.5 \cdot 10^{-2} &&  |V_{t'b}| < 8.5 \cdot 10^{-2} 
\label{tpstpb}
\end{align}
These conditions also satisfy the constraint on the product $V_{t'b}^{*}V_{t's}$
(\ref{vtbvtdBBs}) for the relative phase 
needed in order to have the upper bounds (\ref{tpstpb}),
$\delta_{sb}=\text{Arg}(V_{t'b}^{*}V_{t's})=\delta_{db}-\delta_{ds}\approx 2.77$.

However,
there is no much room 
to accommodate the relation in eq. (\ref{vtpsf})
without contradicting experimental constraints from neutral mesons mixing.
In particular, as shown in figure \ref{vtsv4cplot}, for $|V_{L4u}|=0.03$ the mass of the extra quark 
cannot exceed $\sim 5$ TeV in order to leave allowed values in the parameter space.
If $M_{t'}\approx 5$ TeV, 
it should be that:
\begin{align}
& |V_{t's}| \sim 3.8 \cdot 10^{-3} &&  |V_{t'b}| \sim 5.7 \cdot 10^{-3} 
\end{align}
which also satisfies the constraint (\ref{vtbvtdBBs}).
This also means that in this case the mixing of 
$t'$ with the second and third families is respectively eight times and five times smaller  
than the mixing with the d-quark. 
Moreover $\vert V_{t'd} \vert \simeq 0.03 $ is comparable to $\vert V_{cb} \vert$ 
and ten times larger than $\vert V_{ub} \vert$.
This situation
may seem unnatural, but it cannot be excluded.

If $\vert V_{L4u} \vert \simeq 0.04 $ is needed in the unitarity relation (\ref{newunup}), 
then the constraints are more stringent,
and the available parameter space is remarkably reduced.
As can be inferred from 
figure \ref{vtsv4cplot4}, in this case the mass of the extra quark 
should not exceed $\sim 2.5$ TeV.
Moreover, 
the constraint from K-mesons mixing in eq. (\ref{vtptot}) requires a specific 
value for the phase $\delta_{ds}$ in order to avoid constraints from CP violation.
In any case, 
it should be
$M_{t'} < 5.8 \, \text{TeV}$ with $|V_{t'd}|=0.03$ and $M_{t'} < 4.4 \, \text{TeV}$ with
$|V_{t'd}|=0.04$ 
for the perturbativity ($|h_{u\, 1} |< 1$). 

In addition to that, the result for $Z$-boson decay rate into hadrons implies that
(\ref{v4uZ}):
\begin{align}
& |V_{L4u}|^2 <  2.0\cdot 10^{-3} \; , \qquad \quad
|V_{L4u}|< 0.044 
\end{align}
which can be in the range of values needed to solve the CKM unitarity problem
(for example, at $95\%$ C.L. $|V_{L4u}|=0.038^{+0.008}_{-0.011}$ (\ref{newunsol}) 
using our conservative average for $V_{ud}$ (\ref{vudmedio})).


\section{Extra weak isodoublet}
\label{sec-doub}

Weak isosinglets can be investigated in order to solve the problem of lack of unitarity 
in $V_\text{CKM}$
when $V_{ud}$ is extracted from superallowed beta decays and $V_{us}$ from kaon decays.
However, also between the results from kaon physics there is inconsistency.
In particular, it can be noticed that the lack of compatibility lies between vector and 
and axial-vector couplings in weak charged currents.
In fact, determination $B$ (\ref{B}) 
is extracted from the axial-vector coupling of leptonic kaon decays 
(as also determination $\vert V_{us} \vert  = 0.22567(42)$ 
from $K\mu2$ decays \cite{Martinelli},
 which agrees with determination $B$),
while determination $A$ (\ref{A}) results from the vector coupling of semileptonic kaon decays.
Besides, determination $C$ (\ref{c}) is obtained from the vector coupling of beta decays. 
As displayed in Fig \ref{situation},
determination $A$ and $C$ are deviated from the axial-vector determination in opposite directions.
In fact, there is $2.7\sigma$ discrepancy between determination $B$ (\ref{B}) and $C$ (\ref{c})
and $3\sigma$ discrepancy between determinations $A$ (\ref{A}) and $B$.
As it will be shown, the insertion of an extra vector-like isodoublet generates weak right 
currents which modify vector and axial-vector couplings thus offering the possibility to 
explain both the gaps.
However, also in this case FCNC emerge at tree level and 
predictions for electroweak processes are modified, so experimental limits
must be checked.

Let us introduce the additional vector-like $SU(2)$-doublet family:
\begin{align}
& q_{L4}=\left(\begin{array}{c}
u_{L4} \\ d_{L4} 
\end{array}\right) \; , \qquad q_{R4}=\left(\begin{array}{c}
u_{R4} \\ d_{R4} 
\end{array}\right)
\end{align}
New Yukawa couplings and mass terms should appear in the Lagrangian:
$y^{u \prime}_{ij}\tilde{\varphi}\overline{q_{Li}}'u_{Rj}+
y^{d\prime}_{ij}\varphi\overline{q_{Li}}'d_{Rj}+
m_i\overline{q_{Li}}'q_{R4}
+\text{h.c.}$
with $i=1,2,3,4$, $j=1,2,3$.
As regards the mass terms
$m_{i}\overline{q'_{Li}}q_{R4}+\text{h.c.}$,
since the four species of right-handed singlets $q_{Li}'$ have identical quantum numbers,
a unitary transformation can be applied on the four components $q_{Li}'$ so that
 $m_i=0$ for $i=1,2,3$.
Then the Yukawa couplings and the extra mass term are:
\begin{align}
&\sum_{i=1}^4\sum_{j=1}^3\left[ y_{ij}^u\tilde{\varphi}\overline{q_{Li}}u_{Rj}+
y_{ij}^d\varphi\overline{q_{Li}}d_{Rj}\right]+
M_{4q}\overline{q_{L4}}q_{R4}+\text{h.c.}
\end{align}
The down quark mass matrix looks like:
\begin{align}
&\overline{d_{Li}}\mathbf{m}^{(d)}_{ij}d_{Rj}+\text{h.c.}= \nonumber \\
&=\left(\begin{array}{cccc} \overline{d_{L1}} & \overline{d_{L2}} & \overline{d_{L3}} & \overline{d_{L4}}
\end{array}\right)\left(\begin{array}{cccc}
& & & 0 \\ & \mathbf{y}^{(d)}_{3\times 3}v_w & & 0 \\& & & 0 \\
 y^d_{41}v_w & y^d_{42}v_w & y^d_{43}v_w & M_{4q}
\end{array}\right)\left(\begin{array}{c} d_{R1} \\ d_{R2} \\ d_{R3} \\ d_{R4} \end{array}\right)
+\text{h.c.}
\label{massdoub}
\end{align}
where $v_w=174$ GeV and 
$\mathbf{y}^{(d)}_{3\times 3}$ is the $3\times 3 $ 
matrix of Yukawa couplings, and similarly for the up-type quarks.
The mass matrices can be diagonalized with positive eigenvalues by biunitary transformations:
\begin{align}
&V_L^{(d)\dagger}\mathbf{m}^{(d)}V_R^{(d)}=\mathbf{m}^{(d)}_\text{diag} = 
\text{diag}(y_{d}v_w\, , y_{s}v_w\, , y_{b}v_w\, , M_q)   \nonumber \\
& V_L^{(u)\dagger}\mathbf{m}^{(u)}V_R^{(u)}=\mathbf{m}^{(u)}_\text{diag}=
\text{diag}(y_{u}v_w\, , y_{c}v_w\, , y_{t}v_w\, ,  M_q)
\label{massdiag}
\end{align}
$\mathbf{m}^{(d,u)}_\text{diag}$ are the diagonal matrices of mass eigenvalues
$m_{d,s,b}=y_{d,s,b}v_w$ and $m_{u,c,t}=y_{u,c,t}v_w$ and $M_{q}\approx M_{4q}$.
$V^{(d,u)}_{L,R}$ are unitary $4\times 4 $ matrices:
\begin{align}
\label{vrd2}
&V^{(d)}_{R}=\left(\begin{array}{cccc}
V_{R\, 1d} & V_{R\, 1s} & V_{R\, 1b} & V_{R\, 1b'} \\
V_{R\, 2d} & V_{R\, 2s} & V_{R\, 2b} & V_{R\, 2b'} \\
V_{R\, 3d} & V_{R\, 3s} & V_{R\, 3b} & V_{R\, 3b'} \\
V_{R\, 4d} & V_{R\, 4s} & V_{R\, 4b} & V_{R\, 4b'} 
\end{array}\right) \, ,
&& V^{(u)}_{R}=\left(\begin{array}{cccc}
V_{R\, 1u} & V_{R\, 1c} & V_{R\, 1t} & V_{R\, 1t'} \\
V_{R\, 2u} & V_{R\, 2c} & V_{R\, 2t} & V_{R\, 2t'} \\
V_{R\, 3u} & V_{R\, 3c} & V_{R\, 3t} & V_{R\, 3t'} \\
V_{R\, 4u} & V_{R\, 4c} & V_{R\, 4t} & V_{R\, 4t'} 
\end{array}\right)
\end{align}
and analogously for the matrices diagonalizing from the left.
Weak eigenstates in terms of mass eigenstates are:
\begin{align}
& \left(\begin{array}{c} d_1 \\ d_2 \\ d_3 \\ d_4 \end{array}\right)_{L,R}
= V^{(d)}_{L,R}\left(\begin{array}{c}
d \\ s \\ b \\ b' \end{array}\right)_{L,R} \, , && 
\left(\begin{array}{c} u_1 \\ u_2 \\ u_3 \\ u_4 \end{array}\right)_{L,R}
= V^{(u)}_{L,R}\left(\begin{array}{c}
u \\ c \\ t \\ t' \end{array}\right)_{L,R}
\end{align}
The charged-current Lagrangian is changed in:
\begin{align}
\mathcal{L}_{cc}=& \frac{g}{\sqrt{2}}\sum_{i=1}^4(\overline{u_{Li}}\gamma^\mu d_{Li})W_\mu^+
+\frac{g}{\sqrt{2}}\overline{u_{4R}}\gamma^\mu d_{4R}W_\mu +\text{h.c.}  = \nonumber \\
=& \frac{g}{\sqrt{2}}\left(\begin{array}{cccc} 
\overline{u_L} & \overline{c_L} & \overline{t_L} & \overline{t_L'}\end{array}\right)
\gamma^\mu V_\text{CKM,L}
\left(\begin{array}{c} d_L \\ s_L \\ b_L \\ b_L' \end{array}\right)W_\mu^++ \nonumber\\ 
&+\frac{g}{\sqrt{2}}\left(\begin{array}{cccc} 
\overline{u_R} & \overline{c_R} & \overline{t_R} & \overline{t_R'}\end{array}\right)
\gamma^\mu  
V_\text{CKM,R}\left(\begin{array}{c} d_R \\ s_R \\ b_R \\ b_R' \end{array}\right)W_\mu^+ 
+\text{h.c.} 
\label{doubletcc}
\end{align}
where $V_\text{CKM,L}=V_L^{(u)\dag}V_L^{(d)}$ is a $4\times 4$ unitary matrix:
\begin{align}
\label{unleft}
& V_\text{CKM,L}^\dag V_\text{CKM,L}= V_\text{CKM,L} V_\text{CKM,L}^\dag = \mathbf{1} \\
&V_\text{CKM,L}=V_L^{(u)\dag}V_L^{(d)}=\left(\begin{array}{cccc}
V_{L\, ud} & V_{L\, us} & V_{L\, ub} & V_{L\, ub'}  \\
V_{L\, cd} & V_{L\, cs} & V_{L\, cb} & V_{L\, cb'}  \\
V_{L\, td} & V_{L\, ts} & V_{L\, tb} & V_{L\, tb'} \\
V_{L\, t'd} & V_{L\, t's} & V_{L\, t'b} & V_{L\, t'b'}
\end{array}\right)
\label{vckmL}
\end{align}
In addition to that, 
in this case, as shown in eq. (\ref{doubletcc}), 
the charged current Lagrangian $\mathcal{L}_{cc}$ also involves non-diagonal 
right weak charged currents 
originated by the mixing of the vector-like family with the SM families,
with 
mixing matrix $V_\text{CKM,R}$: 
\begin{align}
V_\text{CKM,R} &= V_R^{(u)\dag}\text{diag}(0,0,0,1)V_R^{(d)}= 	\nonumber \\
&=\left(\begin{array}{cccc}
V_{R\, 4u}^*V_{R\, 4d} & V_{R\, 4u}^*V_{R\, 4s} & V_{R\, 4u}^*V_{R\, 4b} & V_{R\, 4u}^*V_{R\, 4b'} \\
V_{R\, 4c}^*V_{R\, 4d} & V_{R\, 4c}^*V_{R\, 4s} & V_{R\, 4c}^*V_{R\, 4b} & V_{R\, 4c}^*V_{R\, 4b'} \\
V_{R\, 4t}^*V_{R\, 4d} & V_{R\, 4t}^*V_{R\, 4s} & V_{R\, 4t}^*V_{R\, 4b} &  V_{R\, 4t}^*V_{R\, 4b'} \\
V_{R\, 4t'}^*V_{R\, 4d} & V_{R\, 4t'}^*V_{R\, 4s} &  V_{R\, 4t'}^*V_{R\, 4b} & V_{R\, 4t'}^*V_{R\, 4b'} 
\end{array}\right)= \nonumber\\
&= \left(\begin{array}{cccc}
V_{R\, ud} & V_{R\, us} & V_{R\, ub} & V_{R\, ub'}  \\
V_{R\, cd} & V_{R\, cs} & V_{R\, cb} & V_{R\, cb'}  \\
V_{R\, td} & V_{R\, ts} & V_{R\, tb} & V_{R\, tb'} \\
V_{R\, t'd} & V_{R\, t's} & V_{R \,t'b} & V_{R\, t'b'}
\end{array}\right)
\label{vckmR}
\end{align}
where 
$V_{R4\alpha}$ are the elements of the fourth row of the 
matrices $V_{R}^{(d)}$, $V_{R}^{(u)}$ in eq. (\ref{vrd2}),
and we defined the elements $V_{R\alpha\beta}$. 
Clearly, $V_\text{CKM,R}$ is not unitary.
%

Regarding the LH particles, 
the matrices $V^{(d)}_{L}$, $V^{(u)}_{L}$ 
can be parameterized with the same parameterization as in eq. (\ref{conangoli}).
However,
in this scenario mixings in the left-handed sector are much smaller and indeed negligible:
\begin{align}
\label{siL}
& \tilde{s}^{u,d}_{Li}\approx -\frac{y^{u,d}_i y^{u,d*}_{4i} v^2_w}{M^2_q}
\end{align}
where $y^{d}_i=y_{d,s,b}$, $y^{u}_i=y_{u,c,t}$, and
$V^{(u,d)}_{Lij}y^{u,d}_{4j}\approx y^{u,d}_{4i} $.
%
As regards the right-handed sector, without loss of generality, the basis can be chosen in which
the mixing between the SM three families has been diagonalized:
\begin{align}
&V^{(d)}_R=\left(\begin{array}{cccc}
V_{R\, 1d} & V_{R\, 1s} & V_{R\, 1b} & V_{R\, 1b'} \\
V_{R\, 2d} & V_{R\, 2s} & V_{R\, 2b} & V_{R\, 2b'} \\
V_{R\, 3d} & V_{R\, 3s} & V_{R\, 3b} & V_{R\, 3b'} \\
V_{R\, 4d} & V_{R\, 4s} & V_{R\, 4b} & V_{R\, 4b'} 
\end{array}\right) 
\approx \left(\begin{array}{cccc}
c_{R 1}^d  & 0 & 0    & -\tilde{s}^d_{R 1} \\
- \tilde{s}^d_{R 2}\tilde{s}^{d*}_{R 1} & c^d_{R 2} & 0    & -\tilde{s}^d_{R 2} \\
- \tilde{s}^d_{R 3}\tilde{s}^{d*}_{R 1} & -\tilde{s}^{d*}_{R 2}\tilde{s}^d_{R 3} &  c^d_{R 3} & 
-\tilde{s}^d_{R 3} \\
\tilde{s}^{d*}_{R 1}  & \tilde{s}^{d*}_{R 2} & \tilde{s}^{*d}_{R 3} & 
c_{R 1}^dc^d_{R 2}c^d_{R 3}
\end{array}\right) 
\label{vrd}
\end{align}
where 
$c^d_{Ri}$ are cosines and
$\tilde{s}^d_{Ri}$ are complex sines of angles in the 
$1\,4$, $2\,4$, $3\,4$ family 
planes parameterizing the mixing of the first three families
with the vector-like doublet: 
\begin{align}
\label{tildesdo}
&\tilde{s}^d_{Ri}=\sin\theta^d_{Ri4} e^{i \delta^d_{Ri}}=s^d_{Ri} e^{i \delta^d_{Ri}} 
\; , \qquad \quad c^d_{Ri}=\cos\theta^d_{Ri4} 
\nonumber \\
& \delta^d_{Rij}=\delta^d_{Ri}-\delta^d_{Rj}
\end{align}
and similarly for $V_{R}^{(u)}$.
As it can be seen by comparing the mass matrix in eq. (\ref{massdoub}) with the matrices in eqs.
(\ref{md}) and (\ref{mu}), 
in this case the 
angles parameterizing the mixing of the first three families with the fourth family
in the right-handed sector are analogous in magnitude to 
the ones 
of the left-handed sector in the previous sections:
\begin{align}
\label{sidou}
&\tilde{s}^{u,d}_{Ri} \approx -\frac{y^{u,d*}_{4i} v_w}{M_q} 
\end{align}
at order $O(|y^{u,d}_{4i}|(y^d_{i})^2\frac{v_w^3}{M_{b'}^3}+|y^{u,d}_{4j}|^{2}|y^{u,d}_{4i}|\frac{v_w^3}{M_{b'}^3})$, $i,j=1,2,3$, $y^{d}_i=y_{d,s,b}$, $y^{u}_i=y_{u,c,t}$.
The elements of the last row (and also of the last column) in $V^{(d)}_R$ and $V^{(u)}_R$
correspond to the mixing angles of the SM families with the vector-like one:
\begin{align}
& V_{R4d}\approx \tilde{s}_{R1}^{d*} \, , &&  V_{R4s}\approx \tilde{s}_{R2}^{d*}\, ,  &&  V_{R4b}= \tilde{s}_{R3}^{d*} \nonumber \\
& V_{R4u}\approx \tilde{s}_{R1}^{u*} \, , &&  V_{R4c}\approx \tilde{s}_{R2}^{u*}\, ,  &&  V_{R4t}= \tilde{s}_{R3}^{u*}
\label{4rowdo}
\end{align}

\begin{figure}[t]
\centering
\includegraphics[width=0.6\textwidth]{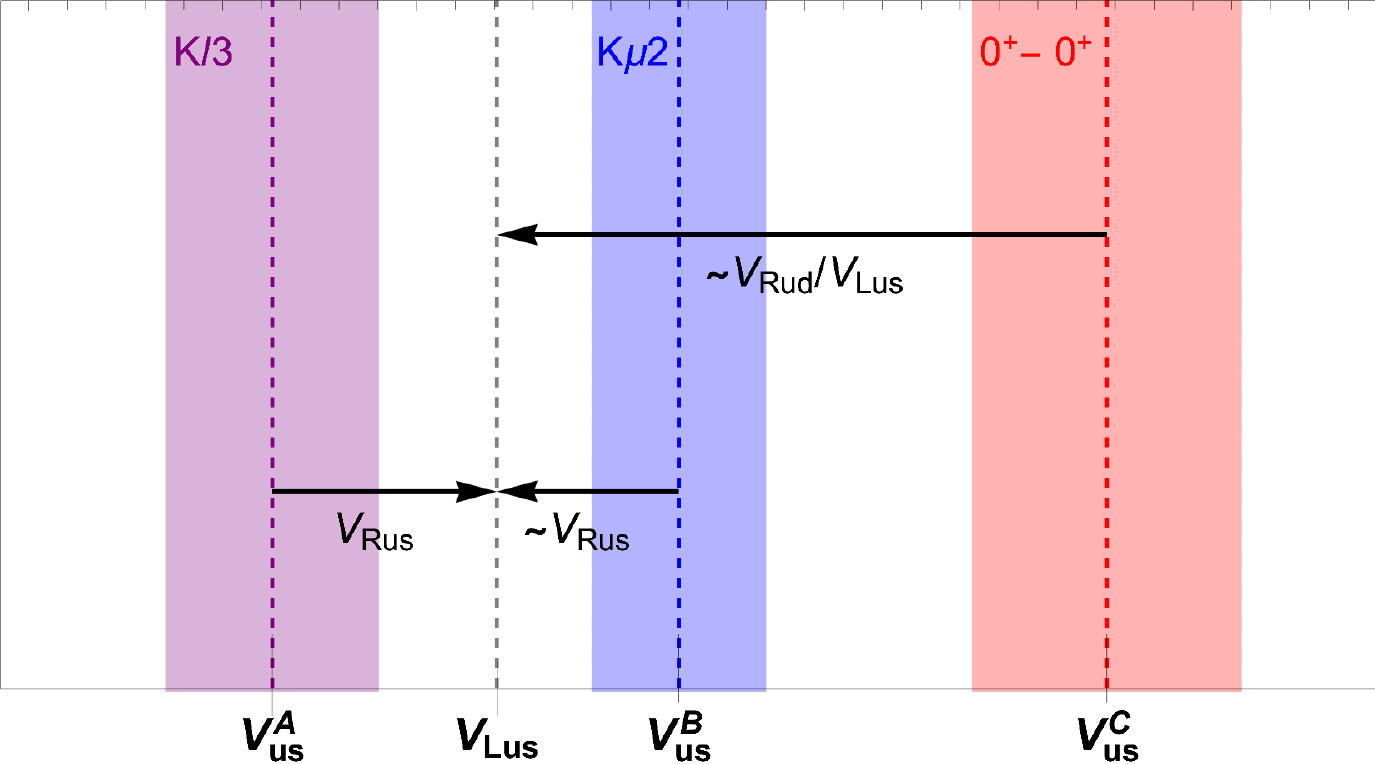}
\caption{\label{VLRusplot} 
Explanation of the anomalies of $V_{us}$ determinations in the SM extension 
with an extra vector-like isodoublet $(b',t')$. 
The three different determinations 
$V_{us}^{A}$, $V_{us}^{B}$ and $V_{us}^{C}$ in eqs. (\ref{A}), (\ref{B}), (\ref{c})
are obtained from the experimental results in eqs.
(\ref{A}), (\ref{rapporto}), (\ref{vudmedio}) using CKM unitarity in the context of the SM.
However, in presence of an extra isodoublet mixing with SM families, the three observables
would correspond to the couplings in eqs. (\ref{AD}), (\ref{BD}), (\ref{CD}).
}
\end{figure}

Let us focus on
the piece of the charged current Lagrangian $\mathcal{L}_{cc}$ in eq. (\ref{doubletcc}) 
determining the couplings of $u$-quark with down type quarks 
(which in the SM would correspond with the determination of the first row of CKM matrix):
\begin{align}
&\frac{g}{\sqrt{2}}(\overline{u_L}\gamma^\mu V_{L\,ud}d_L+
\overline{u_R}V_{R\, ud}\gamma^\mu d_R)\, W_\mu+  \nonumber \\
&+\frac{g}{\sqrt{2}}(\overline{u_L}\gamma^\mu V_{L\,us}s_L+
\overline{u_R}V_{R\, us}\gamma^\mu s_R)\, W_\mu 
+\text{h.c.} =  \nonumber \\
=& \frac{1}{2} \frac{g}{\sqrt{2}} \hat{V}_{ud}
\:\overline{u}\gamma^\mu \left(1-\gamma^5
k_A^{ud} \right) d \: W_\mu 
+ \frac{1}{2} \frac{g}{\sqrt{2}} \hat{V}_{us}
\:\overline{u}\gamma^\mu \left(1-\gamma^5
k_A^{us} \right) s \:  W_\mu 
\label{ccudus}
\end{align}
where, from eq. (\ref{vckmR}):
\begin{align}
&  V_{R ud}=V_{R\, 4u}^*V_{R\, 4d}  \: ,  \quad \qquad V_{R us}=V_{R\, 4u}^*V_{R\, 4s}   
\label{vrudus}
\end{align}
and similarly for the mixing with the bottom quark.
The vector and axial-vector couplings are respectively: 
$\hat{V}_{u\alpha}= V_{L\,u\alpha}+V_{R\, u\alpha}$
and $k_A^{u\alpha} \hat{V}_{u\alpha} = V_{L\,u\alpha} - V_{R\, u\alpha}$,
with $\alpha =d,s,b$.
The most precise determination of SM $V_{ud}$ 
comes from superallowed beta decays.
Superallowed $0^+$ - $ 0^+$ beta decays
are Fermi transitions, that is they uniquely depend on the vector part of the hadronic weak interaction
$G_V=G_F V_{ud}$.
This means that the determination of the weak coupling in superallowed beta decays gives 
$|\hat{V}_{ud}|$ appearing in (\ref{ccudus}).
Regarding $|V_{us}|$, it is determined both from semileptonic kaon decays ($K\ell 3$)
and from the ratio of leptonic kaon decays ($K\mu 2$) and leptonic pion decays. 
It is assumed that only the vector current contributes to semileptonic kaon decays, that is
the coupling is given by $\hat{V}_{us}$ in (\ref{ccudus}). 
Leptonic decays instead depend on the axial-vector coupling, which corresponds to 
$k_A^{us}\hat{V}_{us}$ in (\ref{ccudus}).
Then, in this scenario, 
the determinations of $V_{us}$ obtained from eqs. (\ref{A}), (\ref{rapporto}), 
and (\ref{vudmedio}) correspond to the following observables:
\begin{align}
\label{AD}
& A\, : \qquad 
|V_{L\, us}+V_{R\, us}| = 0.22326(55)  
\\
\label{BD}
& B\, : \qquad 
\frac{|V_{L\, us}-V_{R\, us}|}{|V_{L\, ud}-V_{R\, ud}|}=0.23130(49)  \\
\label{CD}
&C\, : \qquad 
|V_{L\, ud}+V_{R\, ud}| = 0.97376(16) 
\end{align}
As regards $V_\text{CKM,L}$, 
as already stated (\ref{unleft}), it is a unitary matrix, so for the first row it holds that:
\begin{align}
\label{Lun}
& |V_{L\, ud}|^2+|V_{L\, us}|^2+|V_{L\, ub}|^2=1-|V_{L\,ub'}|^2 \approx 1
\end{align}
In fact,
$V_\text{CKM,L}$ can be parameterized as: 
\begin{align}
& V_\text{CKM,L} \approx  
 \left(\begin{array}{cccc}
c_{L 1}^u  & 0 &  0   & \tilde{s}^{u}_{L 1}   \\
0  &c^u_{L 2} & 0 & \tilde{s}^{u}_{L 2} \\
0  & 0 & c^u_{L 3} & \tilde{s}^{u}_{L 3}  \\
-\tilde{s}^{u*}_{L 1} & -\tilde{s}^{u*}_{L 2} & -\tilde{s}^{u*}_{L 3}  & 1
\end{array}\right) 
\left(\begin{array}{c@{\hspace{-1\tabcolsep}}  c@{\hspace{-1\tabcolsep}} cc}
 & &  & 0  \\
 & V_{\text{CKM}\, 3\times 3} &  & 0\\
 &  &  & 0 \\
0 & 0 & 0 & 1
\end{array}\right) 
\left(\begin{array}{cccc}
c_{L 1}^d  & 0 & 0    & -\tilde{s}^d_{L 1} \\
0 & c^d_{L 2} & 0    & -\tilde{s}^d_{L 2} \\
0 &0 &  c^d_{L 3} & 
-\tilde{s}^d_{L 3} \\
\tilde{s}^{d*}_{L 1}  & \tilde{s}^{d*}_{L 2} & \tilde{s}^{*d}_{L 3} & 
1
\end{array}\right) 
\end{align}
where
$V_{\text{CKM}\,3\times 3} $ is basically the same as the $3\times 3$ submatrix of
$V_\text{CKM,L}$ and contains $3$ angles and one phase. 
The complex sines $\tilde{s}^{u,d}_{Li}$ are defined in eq. (\ref{siL}) (and
also the phases in $\tilde{s}^{u,d}_{L1}$ and $\tilde{s}^{u,d}_{L2}$
can be absorbed in quark fields).
Then,
$V_{L\, ub'}\sim -\tilde{s}^d_{L 1}+  \tilde{s}^{u}_{L 1} \sim O(y^{u,d}_1y^{u,d}_{41}v_w^2/M_q^2)$ is totally negligible. 
$|V_{L\, ub}|$ has almost no influence 
and the value of $|V_{ub}|$ as exctracted in the SM can be used.
Hence, $|V_{L\, ud}|$ can be determined from $|V_{L\, us}|$ by using the unitarity relation (\ref{Lun}). 
Then, the system of the three different determinations A, B, C
is exactly solved by three real parameters $V_{Lus}$, $V_{Rud}$, $V_{Rus}$.
Using
the dataset (\ref{AD}), (\ref{BD}), (\ref{CD}),
the 
solution 
gives:
\begin{align}
& V_{R\, us}=V_{R4u}^{*}V_{R4s}= - 1.16(37) \cdot 10^{-3} \, , && 
V_{R\, ud}=V_{R4u}^{*}V_{R4d}= -0.73(18) \cdot 10^{-3}
\label{sol}
\end{align}
with $V_{L us}=0.22441(35) $ and $ V_{L ud} = 0.97449(8)$.
Figure \ref{VLRusplot} basically shows the interpretation of the
determinations of $V_{us}$ obtained from the three different processes
in the scenario with the extra isodoublet of quarks, as given in eqs. (\ref{AD}), (\ref{BD}), (\ref{CD}).
The element of the unitary matrix $V_\text{CKM,L}$
$V_{Lus}$ lies in between determinations A and B, and
the gap among the two is explained by the splitting due to $\pm V_{Rus}$ (the shift of determination
B is dominated by $V_{Rus}$).
On the other hand, $V_{Rud}$ explains the apparent lack of unitarity. In fact, 
the values of $V_{Lus}$ and $V_{Lud}$ are linked by the unitarity of $V_\text{CKM,L}$,
but the vector coupling giving determination C includes the additional coupling 
$V_{Rud}$.

By substituting determination C with the determinations 
$C_{1}$ or $C_{2}$ (see table \ref{TableC}),
the needed value of $V_{Rus}$ remains almost the same.
Using determination $C_{1}$: 
$|V_{ud}|=0.97370(14)$ with the dataset (\ref{AD}), (\ref{BD}), the 
result would be
$V_{Rus}=- 1.16(37)\cdot 10^{-3} $, $V_{Rud}=- 0.78(16)\cdot 10^{-3} $ with 
$V_{L us}=0.22442(35)$,
while using determination $C_{2}$: $|V_{ud}|=0.97389(18)$, the solution is
$V_{Rus}=- 1.14(37)\cdot 10^{-3} $, $V_{Rud}=- 0.60(19)\cdot 10^{-3} $ with 
$V_{L us}=0.22442(35)$.

However also in this scenario flavour changing neutral currents appear at tree level.
The neutral current interactions are described by the Lagrangian:
\begin{align}
\mathcal{L}_\text{nc}=& \frac{g}{\cos\theta_W}Z^\mu\left[
\left(\frac{1}{2}-\frac{2}{3}\sin^2\theta_W\right)\overline{\mathbf{u}_L}\gamma_\mu \mathbf{u}_L+
\left(-\frac{1}{2}+\frac{1}{3}\sin^2\theta_W\right)\overline{\mathbf{d}_L}\gamma_\mu \mathbf{d}_L+ \right. \nonumber \\
&\left. -\frac{2}{3}\sin^2\theta_W\overline{\mathbf{u}_R}\gamma_\mu \mathbf{u}_R
+\frac{1}{3}\sin^2\theta_W\overline{\mathbf{d}_R}\gamma_\mu \mathbf{d}_R
+\frac{1}{2}\overline{u_{R4}}\gamma_\mu u_{R4}-\frac{1}{2}\overline{d_{R4}}\gamma_\mu d_{R4}
\right]
\end{align}
where $\mathbf{u}$ and $\mathbf{d}$ are the vectors whose components are
$u_{L/R \, i}$, $d_{L/R \, i}$, with $i=1,...,4$, 
or, in terms of the mass eigenstates: 
\begin{align}
\label{ncdo}
\mathcal{L}_\text{nc}=& \frac{g}{\cos\theta_W}Z^\mu \left(
g^\alpha_L\overline{q_{\alpha L}}\gamma_\mu q_{\alpha L} + 
g^{\alpha \beta}_R\overline{q_{\alpha R}}\gamma_\mu q_{\beta R}  \right)
\end{align}
with 
$q_\alpha=u,c,t,t',d,s,b,b'$, and:
\begin{align}
\label{glr}
& g^\alpha_L=T_{3}^\alpha -Q_\alpha \sin^2\theta_W  \nonumber \\
& g^{\alpha\beta}_R=T_{3}^\alpha V_{R4\alpha}^*V_{R4\beta} -
Q_\alpha \sin^2\theta_W \delta_{\alpha\beta} 
\end{align}
where $T_{3}^\alpha$ is the weak isospin and $Q_\alpha$ is the charge in units of $e$.
Clearly, FCNC arise from:
\begin{align}
\mathcal{L}_\text{fcnc}=
& \frac{1}{2}\frac{g}{\cos\theta_W}Z^\mu(\overline{u_{R4}}\gamma_\mu u_{R4}-\overline{d_{R4}}\gamma_\mu d_{R4}) = \nonumber \\
= & \frac{g}{\cos\theta_W}Z^\mu 
 T_{3}^\alpha V_{R4\alpha}^*V_{R4\beta} \overline{q_{\alpha R}}\gamma_\mu q_{\beta R} 
 \end{align}
 that is from the mixing of the fourth weak isodoublet with the SM families. 
 Explicitly:
 \begin{align}
\label{fcncdo}
\mathcal{L}_\text{fcnc}= &\frac{1}{2}\frac{g}{\cos\theta_W}Z^\mu\left(\begin{array}{cccc}
\overline{u_R} & \overline{c_R} & \overline{t_R} & \overline{t'_R} 
\end{array}\right)\gamma^\mu V_R^{(u)\dag}\text{diag}(0,0,0,1)V_R^{(u)} \left(\begin{array}{c}
u_R \\ c_R \\ t_R \\ t'_R
\end{array} \right)+ \nonumber  \\
&-\frac{1}{2}\frac{g}{\cos\theta_W}Z^\mu\left(\begin{array}{cccc}
\overline{d_R} & \overline{s_R} & \overline{b_R} & \overline{b'_R} 
\end{array}\right)\gamma^\mu V_R^{(d)\dag}\text{diag}(0,0,0,1)V_R^{(d)} \left(\begin{array}{c}
d_R \\ s_R \\ b_R \\ b'_R
\end{array} \right) 
\end{align}
where
\begin{align}
&  
V_R^{(u)\dag}\text{diag}(0,0,0,1)V_R^{(u)}=
\left(\begin{array}{cccc}
|V_{R\, 4u}|^2 & V_{R\, 4u}^*V_{R\, 4c} & V_{R\, 4u}^*V_{R\, 4t} & V_{R\, 4u}^*V_{R\, 4t'} \\
V_{R\, 4c}^*V_{R\, 4u} & |V_{R\, 4c}|^2 & V_{R\, 4c}^*V_{R\, 4t} & V_{R\, 4c}^*V_{R\, 4t'} \\
V_{R\, 4t}^*V_{R\, 4u} & V_{R\, 4t}^*V_{R\, 4c} & |V_{R\, 4t}|^2 &  V_{R\, 4t}^*V_{R\, 4t'} \\
V_{R\, 4t'}^*V_{R\, 4u} & V_{R\, 4t'}^*V_{R\, 4c} &  V_{R\, 4t'}^*V_{R\, 4t} & |V_{R\, 4t'}|^2
\end{array}\right)
\nonumber \\
&  
V_R^{(d)\dag}\text{diag}(0,0,0,1)V_R^{(d)}=
\left(\begin{array}{cccc}
|V_{R\, 4d}|^2 & V_{R\, 4d}^*V_{R\, 4s} & V_{R\, 4d}^*V_{R\, 4b} & V_{R\, 4d}^*V_{R\, 4b'} \\
V_{R\, 4s}^*V_{R\, 4d} & |V_{R\, 4s}|^2 & V_{R\, 4s}^*V_{R\, 4b} & V_{R\, 4s}^*V_{R\, 4b'} \\
V_{R\, 4b}^*V_{R\, 4d} & V_{R\, 4b}^*V_{R\, 4s} & |V_{R\, 4b}|^2 &  V_{R\, 4b}^*V_{R\, 4b'} \\
V_{R\, 4b'}^*V_{R\, 4d} & V_{R\, 4b'}^*V_{R\, 4s} &  V_{R\, 4b'}^*V_{R\, 4b} & |V_{R\, 4b'}|^2
\end{array}\right)
\label{vncdou}
\end{align}
where, again, $V_{R4\alpha}$ are the elements of the fourth row of the 
matrices $V_{R}^{(d)}$, $V_{R}^{(u)}$ in eq. (\ref{vrd2}).

\subsection{Limits from flavour changing neutral currents}

\begin{figure}[t]
\centering
\includegraphics[width=0.7\textwidth]{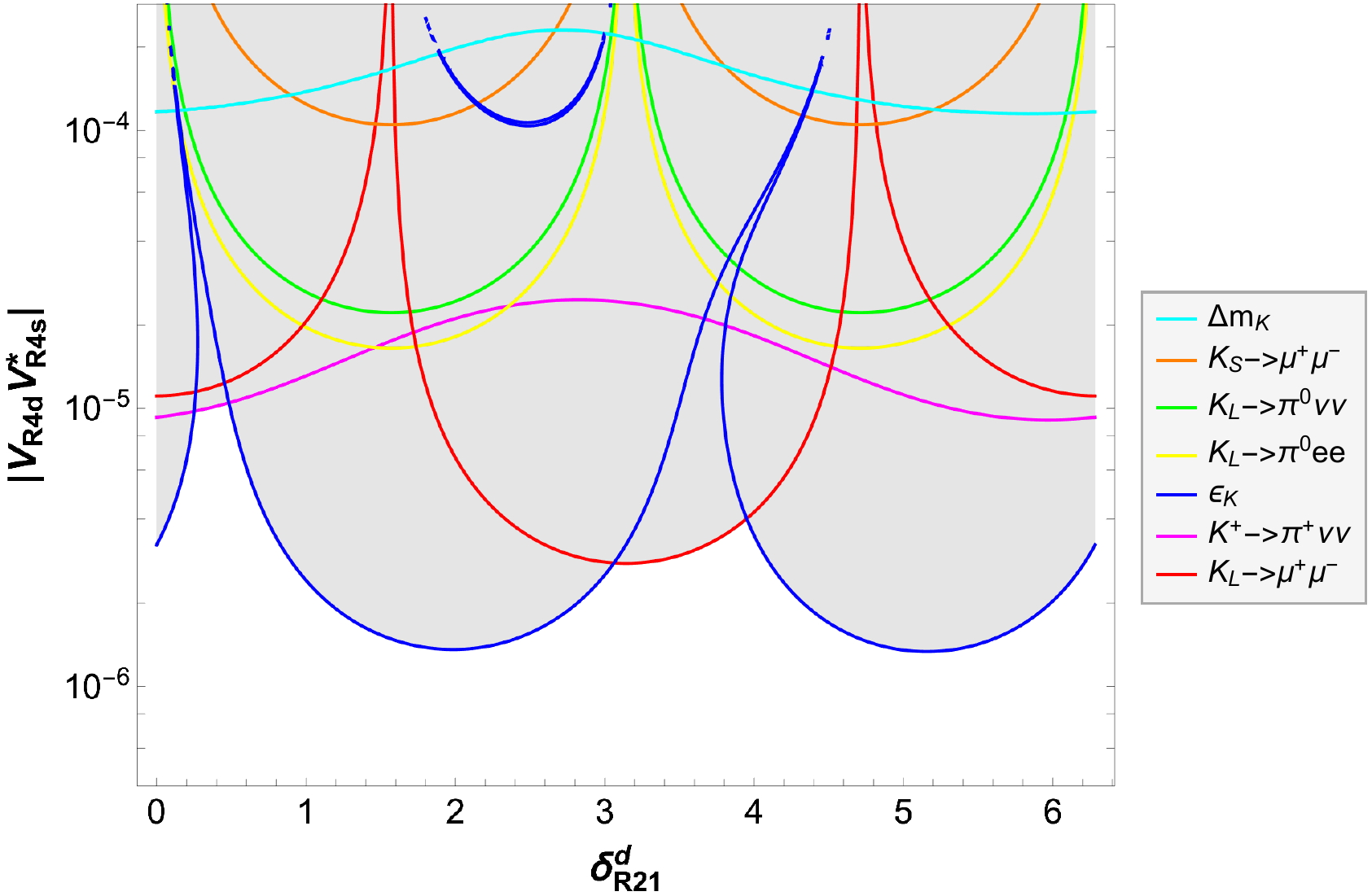}
\caption{\label{fcplotdoub} 
Upper limits on the product $|V_{R4s}^*V_{R4d}|$ from flavour changing
kaon decays and neutral kaon mixing, 
as a function of the relative phase $\delta^d_{R21}$,
for the mass of the extra quarks $M_{q}=1$ TeV.
}
\end{figure}

The new operators in eq. (\ref{fcncdo}), 
due to the mixing of the extra vector-like quark isodoublet with SM families,
give flavour changing 
processes at tree level. Therefore, they should be compared with experimental constraints.
In particular, the hadronic vector current contribute to semileptonic mesons decays, while 
the hadronic axial-vector current gives leptonic mesons decays.
By comparing the Lagrangian in eq. (\ref{fcncdo}) and the matrices in eq. 
(\ref{vncdou})
with Lagrangians in eqs. (\ref{nc}), (\ref{ncu}) and matrices in eqs. (\ref{vnc}), (\ref{vunc}),
it can be noticed that the same effective operators are generated, with the substitution in the
hadronic vector coupling:
\begin{align}
& -V_{L4\alpha}^{*}V_{L4\beta} \rightarrow V_{R4\alpha}^{*}V_{R4\beta}
\end{align}
and in the axial-vector coupling:
\begin{align}
& V_{L4\alpha}^{*}V_{L4\beta} \rightarrow V_{R4\alpha}^{*}V_{R4\beta}
\end{align}
It follows that there is a new contribution at tree level to the
same flavour changing neutral current processes 
examined in the previous sections and the same 
analysis can be applied to the mixing elements of the matrices $V_R^{(u)}$ and $V_R^{(d)}$.
In fact,
aside from the substitution $L\rightarrow R$,
only the sign of the interference in semileptonic decays is modified, 
that is only the phase dependence of the constraints from flavour changing semileptonic decays.
In particular, upper limits from flavour changing kaon decays are summarized in figure \ref{fcplotdoub}
(which is the analogous to figure \ref{fcplotd}). It results:
\begin{align}
& |V_{R4d}V_{R4s}^*|\lesssim (0.3 \div 1.7) \times 10^{-5}
\label{vdsdoub}
\end{align}
depending on the relative phase $\delta^d_{R21}=\text{Arg}(V_{R4d}V_{R4s}^*)$.

Concerning $K^{0}$ mixing, the new contribution from RH weak currents is:
\begin{align}
  \mathcal{L}^\text{right}_{\Delta S=2}\approx &-\frac{G_{F}}{\sqrt{2}}(V_{R4d}^{*}V_{R4s})^{2}
 \left(1+\frac{G_{F}M^{2}_{q}}{4\sqrt{2}\pi^{2}}\right)(\overline{d_{R}}\gamma^{\mu}s_{R})^{2}
 +\text{h.c.}=\nonumber \\
=&-\frac{G_{F}}{\sqrt{2}}(V_{R4d}^{*}V_{R4s})^{2}
 \left[ 1+\left(\frac{M_{q}}{2.2\,\text{TeV}}\right)^{2}\right](\overline{d_{R}}\gamma^{\mu}s_{R})^{2}
 +\text{h.c.}
\end{align}
However, in this case also mixed L-R contributions can in principle be relevant.
In fact, the new contribution to the mixing mass term includes:
\begin{align}
 M_{12}^\text{new} \approx &
\frac{1}{3}f_{K}^{2}m_{K^{0}}B_{K}\frac{G_{F}}{\sqrt{2}}(V_{R4d}^{*}V_{R4s})^{2}+\nonumber \\
&\; + \frac{G_{F}^{2}}{4\pi^{2}}\left[  \frac{1}{2}M^{2}_{q}(V_{R4d}^{*}V_{R4s})^{2}
-27.5\, m_{W}^{2}f(x_{q},x_{t})V_{R4d}^{*}V_{R4s}V_{td}^{*}V_{ts}
\right]
\end{align}
where $x_{q}=M_{q}^{2}/m_{W}^{2}$, $x_{t}=m_{t}^{2}/m_{W}^{2}$, and $f(x_{q},x_{t})\approx 5$
for $M_{q}=1$ TeV.
Then, applying the constraints 
$|M_{12}^\text{new}|<|M_{12}^\text{SM}|$,
$|\text{Im}M_{12}^\text{new}|<0.4\, |\text{Im}M_{12}^\text{SM}| $ 
(\ref{mkcon1}), (\ref{epskcon1}), for $M_{q}=1$ TeV, as shown in figure \ref{fcplotdoub},
it is obtained:
\begin{align}
& |V_{R4d}^*V_{R4s}| <   1.3 \times 10^{-6} \div 2.2 \times 10^{-4}
\end{align}
depending on the relative phase of the elements.
Together with the limits from flavour changing kaon decays, we obtain:
\begin{align}
& |V_{R4d}V_{R4s}^*|\lesssim (0.14 \div 1.0) \times 10^{-5}
\label{vdsdoub2}
\end{align}
We can consider that, if eq. \ref{sol} should be satisfied, the product $V_{R4d}^*V_{R4s}$ should be real and positive
(as can be seen by multiplying side by side the first equation with the complex conjugate of the second). Then the constraint from flavour changing processes of kaons would be:
\begin{align}
\label{v4ds0}
& |V_{R4d}^*V_{R4s}| <   3.2 \times 10^{-6}
\end{align}
for $M_{q}=1$ TeV.

The contribution to $D^{0}$ mesons system is:
\begin{align}
  \mathcal{L}^\text{new}_{\Delta C=2}\approx 
  &-\frac{G_{F}}{\sqrt{2}}(V_{R4u}^{*}V_{R4c})^{2}
 \left[ 1+\left(\frac{M_{q}}{2.2\,\text{TeV}}\right)^{2}\right](\overline{u_{R}}\gamma^{\mu}c_{R})^{2}
 +\text{h.c.}
\end{align}
giving the condition analogous to the results in eqs. (\ref{DDupcon}), (\ref{vcbpvubpdd}):
\begin{align}
&|V_{R4u}^*V_{R4c}|< 1.5 \times 10^{-4} \left[\frac{f_{q}(1\,\text{TeV})}{f_{q}(M_{q})} \right]^{1/2}
\left[ \frac{\Delta m_{D\text{max}}}{1.2\cdot 10^{-14}} \right] 
\end{align}
where 
\begin{align}
& f_{q}(M_{q})=1+\left(\frac{M_{q}}{2.2\,\text{TeV}}\right)^{2}
\end{align}



\subsection{Limits from low energy electroweak observables}
The insertion of the extra isodoublet also changes the diagonal couplings of weak neutral-current interactions,
as follows from eqs. (\ref{ncdo}) and (\ref{glr}). In terms of the axial-vector and vector couplings,
the diagonal Lagrangian can be written as:
\begin{align}
& \frac{g}{2\cos\theta_W}Z^\mu\left(
g^q_V\overline{q}\gamma_\mu q-
g^q_A\overline{q}\gamma_\mu\gamma_5 q  \right)
\end{align}
where $q=u,d,c,s,t,b$ and:
\begin{align}
\label{gva}
& g^q_V=T_{3}^q(1+|V_{R4q}|^2)-2Q_q \sin^2\theta_W  \\
& g^q_A=T_{3}^q(1-|V_{R4q}|^2) \nonumber
\end{align}
SM couplings at tree level are obtained for $|V_{R4q}|=0$.
As a consequence of
the variation of the diagonal couplings, predictions on low energy 
electroweak precision observables are modified.
At low momentum transfer ($Q^{2}\ll M^{2}_{Z}$), the parity violating part of
four-fermion Lagrangian corresponding to
$e$-hadron processes with Z-boson exchange can be written as \cite{PDG18}:
\begin{align}
\mathcal{L}& =
\frac{G_F}{\sqrt{2}}\sum_q \left[g^{eq}_{AV}\bar{e}\gamma_\mu\gamma^5 e\bar{q}\gamma^\mu q+g^{eq}_{VA}\bar{e}\gamma_\mu e\bar{q}\gamma^\mu\gamma^5 q\right]
\end{align}
and from the relations (\ref{gva}):
\begin{align}
\label{AVdo}
& g^{eu}_{AV}=-\frac{1}{2}(1+|V_{R4u}|^2)+\frac{4}{3}\sin^2\theta_W &&  
g^{ed}_{AV}=\frac{1}{2}(1+|V_{R4d}|^2)-\frac{2}{3}\sin^2\theta_W  \\
& g^{eu}_{VA}=(-\frac{1}{2}+2\sin^2\theta_W)(1-|V_{R4u}|^2) &&  
g^{ed}_{VA}=-(-\frac{1}{2}+2\sin^2\theta_W )(1-|V_{R4d}|^2)
\end{align}

The weak charge of the proton, $Q_{W}^{p}$,
is proportional to $g^{ep}_{AV}=2 g^{eu}_{AV}+g^{ed}_{AV}$, $Q_{W}^{p}=-2g^{ep}_{AV}$.
In the SM 
we have $g^{ep}_{AV,\text{SM}}=-0.0357$
(where, after including higher orders corrections, $g^{eu}_{AV,\text{SM}}=-0.1888$, $g^{ed}_{AV,\text{SM}}=0.3419$
 \cite{PDG18}).
Experimentally, the weak charge of the proton  \cite{PDG18}
can be extracted from the parity violating right-left asymmetry in $e^{-}p\rightarrow e^{-}p$ scattering,
from which it is obtained the constraint \cite{PDG18}: 
\begin{align}
&g^{ep}_{AV,\text{exp}}= 2g^{eu}_{AV}+g^{ed}_{AV} = 0.0356 \pm 0.0023
\end{align}
which is in agreement with the SM expectation.
%
%
%
After considering the existence of the extra isodoublet, the expected weak charge changes.
The additional contribution to $ g^{ep}_{AV,\text{SM}}$ is:
\begin{align}
& 
\Delta g^{ep}_{AV}=-|V_{R4u}|^2+\frac{1}{2}|V_{R4d}|^2
\end{align}
In order to stay in the $95\%$ C.L. of the experimental value, it should be:
\begin{align}
& | -|V_{R4u}|^2+\frac{1}{2}|V_{R4d}|^2 | < 0.0045 
\end{align}

%
Nuclear weak charges $Q^{Z,N}_W$ can be extracted from measurements of 
atomic parity violation. They
are defined as \cite{PDG18}:
\begin{align}
\label{QWZN}
& Q^{Z,N}_W=-2[Z(g^{ep}_{AV}+0.00005)+N(g^{en}_{AV}+0.00006)](1-\frac{\alpha}{2\pi})
\end{align}
where $Z$ and $N$ are the 
numbers of protons and neutrons in the nucleus,
$g^{ep}_{AV}=2 g^{eu}_{AV}+g^{ed}_{AV}$, $g^{en}_{AV}= g^{eu}_{AV}+2g^{ed}_{AV}$,
and $\alpha$ is the fine structure constant, $\alpha^{-1}\approx 137.036$. The most precise measurement of atomic parity violation is in Cesium \cite{PDG18}:
\begin{align}
& Q^{55,78}_W(Cs)_\text{exp}=-72.82\pm 0.42
\end{align}
corresponding to the constraint $55 g^{ep}_{AV}+78 g^{en}_{AV}=36.45\pm 0.21$,
while the SM prediction is 
$Q^{55,78}_W(Cs)_\text{SM}=-73.23\pm 0.01$ 
($55 g^{ep}_{AV}+78 g^{en}_{AV}=36.65$). 
The contribution of the extra isodoublet to the weak charge of cesium is:
\begin{align}
& \Delta Q_W^{55,78}(Cs)=Q_W^{55,78}(Cs)_\text{tot}-Q_W^{55,78}(Cs)_\text{SM}=
-2\, [-94|V_{R4u}|^2+105.5|V_{R4d}|^2](1-\frac{\alpha}{2\pi})
\end{align}
At $95\%$ C.L. we have the condition:
\begin{align}
& 
-0.0022\left[ \frac{|\Delta Q_W (Cs)|}{0.41} \right]
<|V_{R4u}|^2-1.12|V_{R4d}|^2<0.0066 \left[ \frac{\Delta Q_W (Cs)}{1.23} \right]
\end{align}

Other neutral current parameters include \cite{PDG18}:
\begin{align}
& (g^{eu}_{AV}+2g^{ed}_{AV})_\text{exp}=0.4927\pm 0.0031  \\
&  (2 g^{eu}_{AV}-g^{ed}_{AV})_\text{exp}=-0.7165\pm 0.0068
\end{align}
to be compared with the SM expectations $g^{eu}_{AV}+2g^{ed}_{AV}=0.4950 $ 
$ 2 g^{eu}_{AV}-g^{ed}_{AV}=-0.7195 $.
Regarding the quantity $g^{eu}_{AV}+2g^{ed}_{AV}$, the mixing with the extra doublet brings an extra contribution $-\frac{1}{2}|V_{R4u}|^2+|V_{R4d}|^2$. 
Then, at $95\%$ C.L. we obtain the constraint:
\begin{align}
-0.0084<-\frac{1}{2}|V_{R4u}|^2+|V_{R4d}|^2<0.0038
\end{align}
As regards the quantity $2 g^{eu}_{AV}-g^{ed}_{AV}$, the prediction is lowered with the extra doublet 
by $-|V_{R4u}|^2-\frac{1}{2}|V_{R4d}|^2$.
Then at $95\%$ C.L. it can be imposed that:
\begin{align}
& |V_{R4u}|^2+\frac{1}{2}|V_{R4d}|^2<0.010
\end{align}


Constraints from low energy electroweak observables are summarized in table \ref{tabdoZ}.

\subsection{Limits from $Z$-boson physics}
\label{Zdoub}

The presence of additional vector-like quarks affects 
the diagonal couplings of Z-boson with quarks, changing the prediction of many observables 
related to the $Z$ boson physics 
e.g. the $Z$ total width $\Gamma_Z$,
the partial decay width into hadrons $\Gamma(Z\rightarrow\text{had})$,
the partial decay widths 
$R_c=\Gamma(c\bar{c})/\Gamma(Z\rightarrow\text{had})$, $R_b=\Gamma(b\bar{b})/\Gamma(Z\rightarrow\text{had})$, 
$\Gamma(Z\rightarrow q\bar{q})$, $q=u,d,s,c,b$,  
Z-pole asymmetries.

Experimental values and SM predictions for Z pole quantities are taken from
Particle Data Group \cite{PDG18}, as reported in table \ref{Zvalues}. 

The SM predictions for the Z decay rate and partial decay rate into hadrons are
$\Gamma(Z)_\text{SM}=2.4942\pm0.0009$ GeV, 
$\Gamma(Z\rightarrow\text{hadr})_\text{SM}=1.7411\pm0.0008$ GeV, to be compared with the experimental results 
$\Gamma(Z)_\text{exp}=2.4952\pm0.0023 \,  \text{GeV}$, $\Gamma(Z\rightarrow\text{hadr})_\text{exp}=1.7444\pm0.0020  \,  \text{GeV}$
\cite{PDG18}.

In this model the deviation of the Z decay rate from the SM prediction is:
\begin{align}
&\Gamma(Z\rightarrow\text{had})-\Gamma(Z\rightarrow\text{had})_\text{SM}=
 \Gamma(Z)-\Gamma(Z)_\text{SM}
\approx  \nonumber \\
\approx&\frac{G_FM^3_Z}{\sqrt{2}\pi}   \left[
-\frac{2}{3}\sin^2\theta_W\left(|V_{R4u}|^2+|V_{R4c}|^2\right)
-\frac{1}{3}\sin^2\theta_W \left(|V_{R4d}|^2+|V_{R4s}|^2+|V_{R4b}|^2\right)\right]<0
\label{Zdo}
\end{align}
We can impose that this deviation is less than a chosen quantity $\Delta\Gamma_{Z}$:
\begin{align}
\frac{G_FM^3_Z}{\sqrt{2}\pi}  & \left[
\frac{2}{3}\sin^2\theta_W\left(|V_{R4u}|^2+|V_{R4c}|^2\right)
\frac{1}{3}\sin^2\theta_W \left(|V_{R4d}|^2+|V_{R4s}|^2+|V_{R4b}|^2\right) 
\right] < \Delta\Gamma_{Z}
\label{Zdocon}
\end{align}
where QCD corrections should also be included, that is a factor $1.050$ for $u,c$-quarks,
$1.041$ for $d,s$-quarks and $1.021$ for $b$-quark \cite{PDG18}.
As shown in eq. (\ref{Zdo}) the prediction for the decay rate is lower than the SM expectation
$\Gamma(Z\rightarrow\text{hadr})_\text{SM}$. Then, since the SM expectation (\ref{GZsm}) is below
the experimental result (\ref{GZexp}), we can choose to use the limit value of the SM prediction
$\Gamma(Z\rightarrow\text{had})_\text{SM}= 1.7419$ GeV in eq. (\ref{Zdocon}). Then,
at $95\%$ C.L. of the experimental result we can impose:
\begin{align}
|V_{R4u}|^2+|V_{R4c}|^2+0.5 \left(|V_{R4d}|^2+|V_{R4s}|^2+|V_{R4b}|^2\right) 
<4.4\times 10^{-3} \left[ \frac{\Delta\Gamma_{Z}}{1.4\times 10^{-3} \,\text{GeV}} \right]  
\label{Zdou}
\end{align}
This condition is compatible with the needed solution (\ref{sol}). However, let us notice
that, even setting $V_{R4c}=V_{R4b}=0$, this limit alone
would rule out the possibility to explain the CKM anomalies with the extra weak doublet
with a reduction of a factor less than $2$ of the experimental error if the central values do not change.
On the other hand, if the weak isodoublet is the solution to the CKM anomalies, 
anomolous $Z$-boson couplings with light fermions (in the ``right'' direction) should be detected if the
error-bars are reduced by a factor $\sim 4$.  

As regards the couplings of $b$ and $c$ quarks, we can confront the new prediction with
the experimental partial rates
for the decays $Z\rightarrow b\bar{b}$ and $Z\rightarrow c\bar{c}$.
by using data from PDG \cite{PDG18}, as reported in table \ref{Zvalues}.
At $95\%$ C.L. they give the limits: $|V_{R4b}|<0.074$, $|V_{R4c}|<0.18$.


Constraints are expected also from Z-pole asymmetry analyses of $e^+e^-\rightarrow f\bar{f}$.
processes. 
In particular, 
left-right asymmetries $A_{LR}$, forward-backward asymmetries $A_{FB}$ 
and left-right forward-backward asymmetries $A_{LRFB}$ 
\cite{ALEPH:2005ab} 
were measured at LEP.
Cross sections for $Z$-boson exchange are usually written 
in terms of the asymmetry parameters $A_f$, 
$f=e,\mu,\tau,b,c,s,q$, which contain final-state couplings. 
For example, they are related as 
$A_{FB}^{(0,f)}=\frac{3}{4}A_eA_f$, 
$A_{LRFB}^{(0,f)}=\frac{3}{4}A_f$ 
(where the superscript $0$ indicates the quantity corrected for radiative effects).
The presence of an additional isodoublet changes the couplings of quarks with $Z$ boson 
as in eq. (\ref{gva}),
and consequently the predictions for the asymmetries are also changed:
\begin{align}
\label{Abnew}
& A_b=\frac{1-\frac{4}{3}\bar{s}^2_b(1-|V_{R4b}|^2)-|V_{R4b}|^4}
{1-\frac{4}{3}\bar{s}^2_b(1+|V_{R4b}|^2)+\frac{8}{9}\bar{s}^4_b+|V_{R4b}|^4} \, , \quad
 A_c=\frac{1-\frac{8}{3}\bar{s}^2_c(1-|V_{R4c}|^2)-|V_{R4c}|^4}
{1-\frac{8}{3}\bar{s}^2_c(1+|V_{R4c}|^2)+\frac{32}{9}\bar{s}^4_c+|V_{R4c}|^4}
\end{align}
where $\bar{s}_f^2$ are the effective weak angles which take into account EW radiative corrections.
Taking the data from Particle Data Group \cite{PDG18}, also reported in table \ref{Zvalues},
regarding b-quark final state, 
the mixing of the extra-doublet increases the prediction of $A_{b}$, thus not going to
the ``right'' direction with respect to the experimental determinations
of $A_b$ and especially of $A_{FB}^{(0,b)}$. However, with
$V_{R4b}<0.19$ the expected value of $A_b$ remains in the $95\%$ C.L. interval of the
experimental result $A_b=0.923\pm 0.020$.

Similarly, as regards c-quark final state, the prediction for $A_c$ is increased by
the mixing of the extra-doublet. 
However, the expected value of $A_c$ remains in the $95\%$ C.L. interval of the 
experimental determination $A_c=0.670\pm 0.027$ with $V_{R4c}<0.18$.
and it also stands in the $95\%$ C.L. interval of 
the determination $A_c=0.628\pm 0.032$ (which can be obtained from 
$A_{FB}^{(0,c)}=0.0707\pm 0.0035$ using $A_e=0.1501\pm 0.0016$)
with $V_{R4c}<0.11$.

Constraints from Z-boson physics are summarized in table \ref{tabdoZ}.

\begin{table}
\centering
\begin{tabular}{| ll |}
\hline
Process & Constraint  \\
\hline
$Z\rightarrow \text{hadrons}$, $\Gamma_{Z}$ & 
$|V_{R4u}|^2+|V_{R4c}|^2+0.5 \left(|V_{R4d}|^2+|V_{R4s}|^2+|V_{R4b}|^2\right)<4.4 \times 10^{-3}$
 \\ 
$Q_W(Cs)$ & 
$-0.0022 
<|V_{R4u}|^2-1.12|V_{R4d}|^2< 0.0066 
$  \\ 
$Q_W(p)$ & $ \left| -|V_{R4u}|^2+\frac{1}{2}|V_{R4d}|^2 \right| <0.0045$    \\ 
$g^{eu}_{AV}+2g^{ed}_{AV}$ & 
$-0.0084<-\frac{1}{2}|V_{R4u}|^2+|V_{R4d}|^2<0.0038$  \\
$2g^{eu}_{AV}-g^{ed}_{AV}$ &  
$ |V_{R4u}|^2+\frac{1}{2} |V_{R4d}|^2<0.010$  \\
\hline
\end{tabular}
\caption{\label{tabdoZ} Limits on the mixing of the first three families with an extra
vector-like isodoublet from low energy electroweak observables and Z physics. 
For details see the text.
 }
\end{table}

\subsection{Summary of experimental limits }
\label{doubend}

\begin{figure}[t]
\centering
\includegraphics[width=0.6\textwidth]{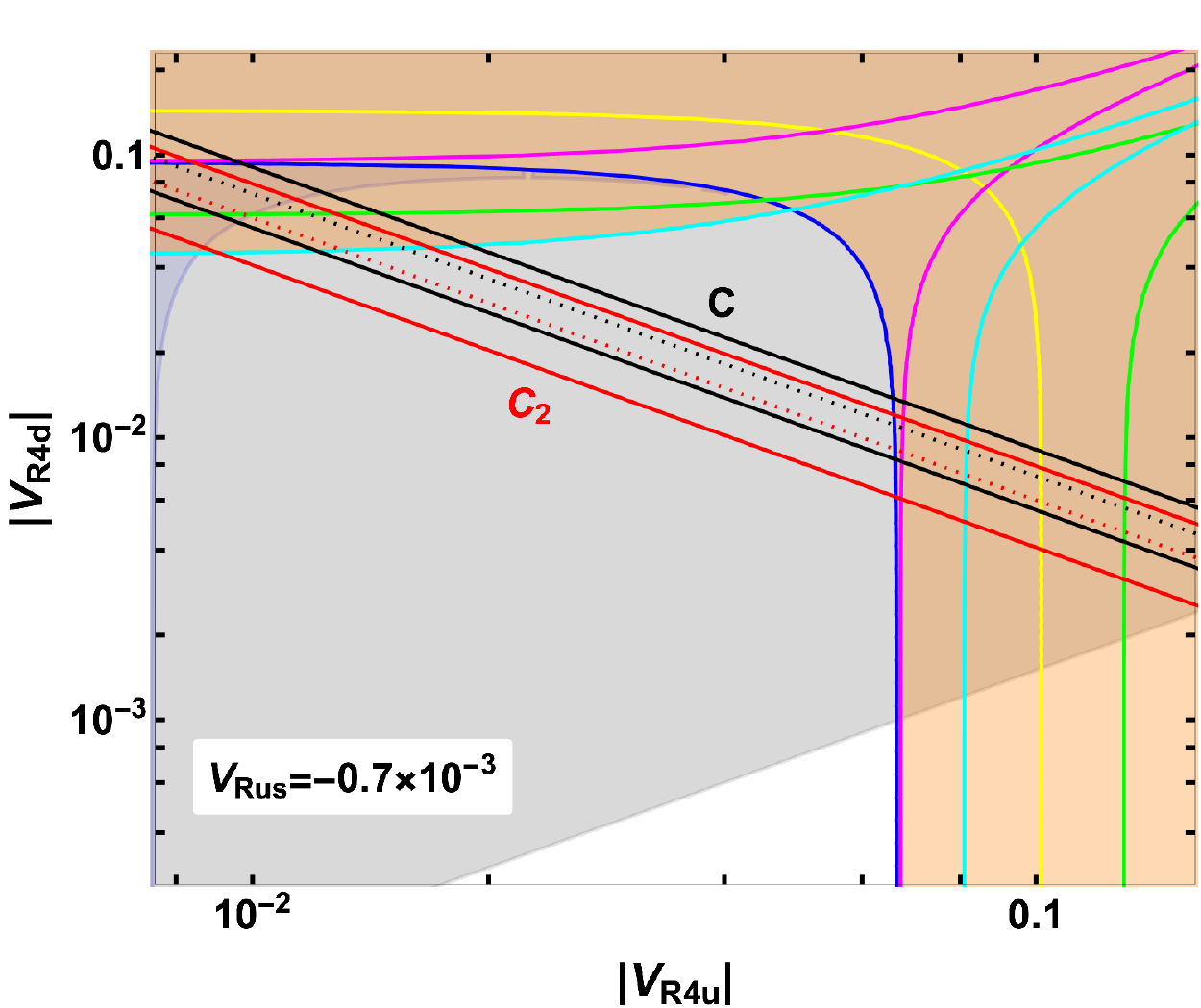} 
\caption{\label{v4dv4u2} Excluded area in the $|V_{R4d}|$ and $|V_{R4u}|$ parameter space.
The orange area is excluded by the following low energy electroweak quantities:
$Q_W(p)$ (magenta contour), $Q_W(Cs)$ (cyan),  
the couplings $g^{eu}_{AV}+2g^{ed}_{AV}$ (green) and
$2g^{eu}_{AV}-g^{ed}_{AV}$ (yellow), $Z$-boson decay rate (blue).
The black curve stands for the solution 
$|V_{Rud}|=|V_{R4u}^*V_{R4d}|=0.73(18)\times 10^{-3}$ in eq. (\ref{sol}).
The red curve stands for the conservative 
solution $|V_{Rud}|=|V_{R4u}^*V_{R4d}|=0.60(19)\times 10^{-3}$ obtained using determination $C_{2}$: $|V_{ud}|_{C_{2}}=0.97389(18)$, instead
of determination C: $|V_{ud}|_{C}=0.97376(16)$ in the dataset (\ref{CD}) (see table \ref{TableC}).
A conservative value $|V_{Rus}|=|V_{R4u}^*V_{R4s}|=0.7\cdot 10^{-3}$ is used
in order to exclude the blue area 
with the experimental determination of $Z$ decay rate, and the gray area which 
must be excluded to not contradict the constraint from
flavour changing kaon decays and mixing
$|V_{R4d}V_{R4s}^*|<1.0 \cdot 10^{-5}$  
in eq. (\ref{vdsdoub2}) obtained with
 the mass of the extra quark $M_{q}=1$ TeV.
}
\end{figure}

\begin{figure}[t]
\centering
\includegraphics[width=0.6\textwidth]{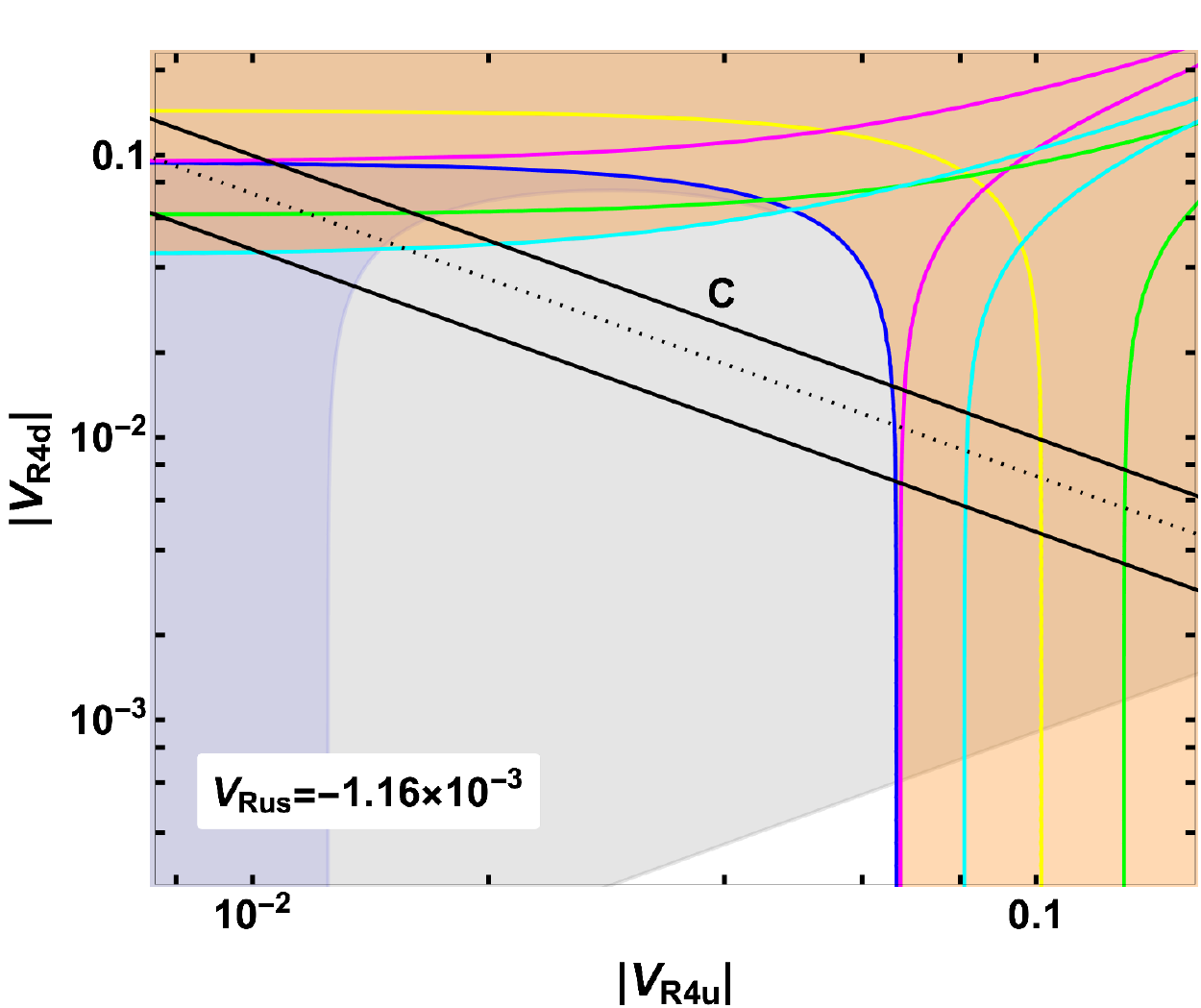} 
\caption{\label{v4dv4u} Excluded area in the $|V_{R4d}|$ and $|V_{R4u}|$ parameter space.
The orange area is excluded by the following low energy electroweak quantities:
$Q_W(p)$ (magenta contour), $Q_W(Cs)$ (cyan),  
the couplings $g^{eu}_{AV}+2g^{ed}_{AV}$ (green) and
$2g^{eu}_{AV}-g^{ed}_{AV}$ (yellow), $Z$-boson decay rate (blue).
In order to satisfy the second relation of the solution in eq. (\ref{sol})
$|V_{Rus}|=|V_{R4u}^*V_{R4s}|=1.16\cdot 10^{-3}$, the blue area should be 
excluded by the experimental determination of $Z$ decay rate, while
the gray area must be excluded to not contradict the constraint from
flavour changing kaon decays and mixing
for the mass of the extra quark $M_{q}=1$ TeV
$|V_{R4d}V_{R4s}^*|<1.0 \cdot 10^{-5}$  
in eq. (\ref{vdsdoub2}).
The black curve stands for the solution 
$|V_{Rud}|=|V_{R4u}^*V_{R4d}|=0.73(27)\times 10^{-3}$ in eq. (\ref{sol}), whose width correspond
to the $68\%$ C.L. allowed range of the parameters 
$V_{Rud}=-0.73(27)\times 10^{-3}$, $V_{Lus}=0.22441(51)$.
%
}
\end{figure}



In this section we analyzed the scenario in which an extra vector-like weak doublet
of quarks $(u_{4},d_{4})_{L,R}$ is mixing with the SM families, 
with mixing matrices $V_{R}^{(d,u)}$ given in eq. (\ref{vrd2}).
Although the presence of this extra species 
can in principle address the whole problem of CKM anomalies,
it seems not possible to obtain an acceptable solution satisfying both 
eq. (\ref{sol}) and the experimental
constraints from kaons flavour changing processes (figure \ref{fcplotdoub}) 
and electroweak observables
(table \ref{tabdoZ}) 
by introducing one extra isodoublet of quarks.

We can focus on 
the elements $V_{R4u}$, $V_{R4d}$, $V_{R4s}$ of the mixing matrices $V_{R}^{(d,u)}$,
which are the elements affecting 
the first row of CKM and consequently CKM anomalies 
(for our concern the other mixing elements can also be zero).
Summarizing all the results, 
the mixings of
the fourth doublet of quarks 
with SM families 
should be large enough
in order to justify the anomalies between the different determinations of $V_{us}$,
as found in eq. (\ref{sol}):
\begin{align}
& |V_{Rud}|=|V_{R4u}^*V_{R4d}|=0.73(18)\times 10^{-3} \, , &&
|V_{Rus}|=|V_{R4u}^{*}V_{R4s}|= 1.16(37)\times 10^{-3}
\end{align}
but at the same time 
they should not contradict
the constraints from FCNC in kaon decays and mixing in eq. (\ref{vdsdoub2}) 
and from low energy electroweak quantities and $Z$-boson physics in table \ref{tabdoZ}:
\begin{align}
& |V_{R4u}|^2+0.50 \left(|V_{R4d}|^2+|V_{R4s}|^2\right)<4.4 \times 10^{-3} \, , &&
|V_{R4d}V_{R4s}^*|\lesssim (0.13 \div 1.0) \times 10^{-5}
\end{align}
where the first constraint comes from the decay rate of $Z$-boson into hadrons.
Let us 
underline that flavour changing conditions are obtained 
for the extra quarks mass $M_{q}=1$ TeV and they become more stringent for larger masses.
Figure \ref{v4dv4u2} shows the excluded area in $V_{R4u}$, $V_{R4d}$ parameter space, 
after adopting some conservative choices.
In particular, as it will be described below, it is shown that even 
by choosing the very conservative condition
$|V_{R4u}^{*}V_{R4s}|= 0.7\times 10^{-3}$, there remains no allowed area in the parameter space.


The black curve shows the condition (\ref{sol}) 
$|V_{Rud}|=|V_{R4u}^*V_{R4d}|=0.73(18)\times 10^{-3} $,
obtained as solution of the dataset (\ref{AD}), (\ref{BD}), (\ref{CD}).
The red curve shows the most conservative solution 
$|V_{Rud}|=|V_{R4u}^*V_{R4d}|=0.60(19)\times 10^{-3}$
obtained 
using determination $C_{2}$: $|V_{ud}|_{C_{2}}=0.97389(18)$, instead
of determination C: $|V_{ud}|_{C}=0.97376(16)$ in eq. (\ref{CD}) (see table \ref{TableC}).

The orange area is excluded by low energy electroweak quantities, as reported in table \ref{tabdoZ}.

Then, 
we should consider the strong restrictions coming from flavor changing kaon decays.
As obtained in eq. (\ref{vdsdoub}),
the result is that the product $|V_{R4d}V_{R4s}^*|$ cannot exceed the limit
$ |V_{R4d}V_{R4s}^*|\lesssim 1.0 \times 10^{-5}$,
which is obtained for specific values of the relative phase 
$\delta^d_{R21}=\text{Arg}(V_{R4d}V_{R4s}^*)$.
On the other hand, the condition $V_{Rus}=V_{R4u}^{*}V_{R4s}= -1.16(37)\times 10^{-3}$
is needed in order to explain the discrepancy between the determinations of $V_{us}$
from leptonic and semileptonic kaon decays. 
In order to both explain the anomaly 
using the needed value of $V_{Rus}$ and satisfy
the flavor changing conditions, the gray area in the parameter space must be excluded.
In figure \ref{v4dv4u2} we show the excluded area adopting the conservative choices:
\begin{align}
\label{solcfc}
&  \left\lbrace \begin{array}{l}
V_{R4u}^{*}V_{R4s}= -0.7\times 10^{-3}   \\
|V_{R4d}V_{R4s}^*|\lesssim 1.0 \times 10^{-5}
\end{array} \right. 
\end{align}
%

With the same conservative choice for $V_{Rus}$, the blue area in figure \ref{v4dv4u2} is excluded
in order to have a solution not contradicting experimental measurements of hadronic $Z$
decay rate:
\begin{align}
& \left\lbrace \begin{array}{l}
V_{R4u}^{*}V_{R4s}=  -0.7\times 10^{-3}  \\
|V_{R4u}|^2+0.50 \left(|V_{R4d}|^2+|V_{R4s}|^2\right)<4.4 \times 10^{-3} \end{array} \right. 
\end{align}

It is clear that the values needed as solutions (black curve or also the most conservative red curve) 
are unachievable 
without contradiction with flavor changing experimental limits.

Namely, the best allowed point in figure \ref{v4dv4u2} gives:
 \begin{align}
 & |V_{R4u}| =6.6\times 10^{-2}  && |V_{R4d}|=1.6\times 10^{-3} && |V_{R4s}|=1.1\times 10^{-2}
 \end{align}
that is
 \begin{align}
 & |V_{Rus}|=|V_{R4u}^*V_{R4s}|=0.70\times 10^{-3} && 
    |V_{Rud}|=|V_{R4u}^*V_{R4d}|=0.07\times 10^{-3}
 \end{align}
Alternatively, 
by performing a fit of the values in eqs. (\ref{A}), (\ref{rapporto}), 
(\ref{vudmedio}), with real parameters $V_{Lus}$, $V_{R4d}$, $V_{R4s}$, $V_{R4u}$, 
but constraining them 
with experimental results,
 the best fit point ($\chi^2_\text{dof}=4.4$) is obtained 
in $V_{Lus}=0.22463 $ with:
 \begin{align}
 & V_{R4u}=- 6.60\times 10^{-2} && V_{R4d}=0.96\times 10^{-2} && V_{R4s}=0.11\times 10^{-2}
 \end{align}
and consequently
 \begin{align}
 & V_{Rus}=V_{R4u}^*V_{R4s}=-0.07\cdot 10^{-3} &&  V_{Rud}=V_{R4u}^*V_{R4d}=-0.63\cdot 10^{-3}
 \end{align}
In any case,
in order to not violate FC constraints only one of the two discrepancies,
between determination B (\ref{B}) and determination A (\ref{A}) or 
between determination B and determination C (\ref{c}), can be solved.

Moreover,
the LHC limit on extra vector-like quarks coupling to light quarks 
$M_{q} \gtrsim 700$~GeV \cite{CMS} 
implies that 
$\vert V_{R4u}\vert  \simeq  0.066$ 
can be obtained if $|y_{41}^u| \sim 0.27$, much larger than the 
Yukawa constant of the bottom quark. Further, with 
$| V_{R4u}| \sim  0.066$,  
assuming (for the perturbativity) $|y_{41}^{u}|< 1$, 
there is an upper limit on the extra doublet mass, 
$M_{q} \lesssim 2.6 $ TeV. 

\subsection{Possible extensions}
\label{other}

Looking at the results of previous sections, we can conclude that 
the presence of one single extra species cannot explain
all the CKM anomalies without contradicting 
other experimental constraints.
In order to improve the situation, it can be considered the case in which 
for example there exist two or more vector-like doublets mixing with SM families:
\begin{align}
& q_{R4}=\left(\begin{array}{c}
u_{R4} \\ d_{R4} 
\end{array}\right)  ,\,
q_{ L4}=\left(\begin{array}{c}
u_{L4} \\ d_{L4} 
\end{array}\right) ; \qquad \quad  q_{R5}=\left(\begin{array}{c}
u_{R5} \\ d_{R5} 
\end{array}\right)  ,\,
q_{ L5}=\left(\begin{array}{c}
u_{L5} \\ d_{L5} 
\end{array}\right) 
\end{align}
The Yukawa couplings and the mass terms can be written in the following basis:
\begin{align}
&\sum_{i=1}^5\sum_{j=1}^3\left[ y_{ij}^u\tilde{\varphi}\overline{q_{Li}}u_{Rj}+
y_{ij}^d\varphi\overline{q_{Li}}d_{Rj}\right]+
M_4\overline{q_{L4}}q_{R4}+M_5\overline{q_{L5}}q_{R5}+\text{h.c.}
\end{align}
Then, in order to avoid flavour changing effects, 
not all the couplings $y_{ij}^{u,d}$ should be non-zero.
Let us suppose that there can be a pattern of couplings, so that
the first doublet has couplings $y^d_{41}\neq 0$, $y^d_{42}=0$, $y^u_{41}\neq 0$,
while for the second doublet $y^d_{51}=0$, $y^d_{52}\neq 0$,  $y^u_{51}\neq 0$.
For simplicity, we also take $y^d_{43},y^d_{53}=0$
and $y^u_{4i},y^u_{5i}=0$ for $i\neq 1$.
The mass matrices for down and up quarks would result: 
\begin{align}
&\sum_{i=1}^4\sum_{j=1}^3 \overline{d_{Li}}\mathbf{m}^{(d)}_{ij}d_{Rj}
+\overline{q_{L4}}q_{R4}M_4+\overline{q_{L5}}q_{R5}M_5  
= \nonumber \\
&=\left(\begin{array}{ccccc} \overline{d_{L1}} & \overline{d_{L2}} & \overline{d_{L3}} & \overline{d_{L4}}
& \overline{d_{L5}} \end{array}\right)\left(\begin{array}{ccccc}
& & & 0 & 0 \\ & \mathbf{y}^{(d)}_{3\times 3}\, v_{w} & & 0 & 0 \\& & & 0 & 0 \\
 y^d_{41}v_{w} & 0 & 0 & M_4 & 0\\
 0 & y^d_{52}v_{w} & 0 & 0 & M_5
\end{array}\right)\left(\begin{array}{c} d_{R1} \\ d_{R2} \\ d_{R3} \\ d_{R4} \\ d_{R5} \end{array}\right)
\label{massdoub2} \\
&\sum_{i=1}^4\sum_{j=1}^3\overline{u_{Li}}\mathbf{m}^{(u)}_{ij}u_{Rj}
+\overline{q_{L4}}q_{R4}M_4+\overline{q_{L5}}q_{R5}M_5  
= \nonumber \\
&=\left(\begin{array}{ccccc} \overline{u_{L1}} & \overline{u_{L2}} & \overline{u_{L3}} & \overline{u_{L4}}
& \overline{u_{L5}}
\end{array}\right)\left(\begin{array}{ccccc}
& & & 0 & 0 \\ & \mathbf{y}^{(u)}_{3\times 3}\, v_{w} & & 0 & 0 \\& & & 0 & 0 \\
 y^u_{41}v_{w} & 0 & 0 & M_4 & 0 \\
 y^u_{51}v_{w} & 0 & 0 & 0 & M_5
\end{array}\right)\left(\begin{array}{c} u_{R1} \\ u_{R2} \\ u_{R3} \\ u_{R4} \\ u_{R5} \end{array}\right)
\label{massdoubu}
\end{align}
Then $V^{(d,u)}_{L,R}$ diagonalizing the mass matrices are unitary $5\times 5 $ matrices.
Weak eigenstates in terms of mass eigenstates are:
\begin{align}
& \left(\begin{array}{c} d_1 \\ d_2 \\ d_3 \\ d_4 \\ d_5 \end{array}\right)_{L,R}
= V^{(d)}_{L,R}\left(\begin{array}{c}
d \\ s \\ b \\ b'  \\ b'' \end{array}\right)_{L,R} \, , && 
\left(\begin{array}{c} u_1 \\ u_2 \\ u_3 \\ u_4 \\ u_5 \end{array}\right)_{L,R}
= V^{(u)}_{L,R}\left(\begin{array}{c}
u \\ c \\ t \\ t' \\ t'' \end{array}\right)_{L,R}
\end{align}
In this case $V^{(d)}_R$ can be parameterized as:
\begin{align}
&V^{(d)}_R=\left(\begin{array}{ccccc}
V_{R 1d} & V_{R 1s} & V_{R 1b} & V_{R 1b'} & V_{R 1b''} \\
V_{R 2d} & V_{R 2s} & V_{R 2b} & V_{R 2b'} & V_{R 2b''} \\
V_{R 3d} & V_{R 3s} & V_{R 3b} & V_{R 3b'} & V_{R 3b''} \\
V_{R 4d} & V_{R 4s} & V_{R 4b} & V_{R 4b'} & V_{R 4b''}\\
V_{R 5d} & V_{R 5s} & V_{R 5b} & V_{R 5b'} & V_{R 5b''} 
\end{array}\right)\approx
\left(\begin{array}{ccccc}
1 & 0 & 0 & -\tilde{s}^d_{R14} & 0 \\
 0   & 1 & 0 & 0 & -\tilde{s}^d_{R25} \\
 0   & 0 & 1 & 0 &0\\
\tilde{s}^{d*}_{R14} & 0 &0 & 1 & 0 \\
0 & \tilde{s}^{d*}_{R25} & 0 & 0 &  1  \\
\end{array}\right) 
\label{vrdd}
\end{align}
$c^d_{Ri}$ are cosines and
$\tilde{s}^d_{Ri}$ are complex sines of angles parameterizing the mixing of the first three families with the vector-like quarks, as in eq. (\ref{tildesdo}) and 
(\ref{sidou}), and similarly for up-type quarks.
The charged-current Lagrangian is changed in:
\begin{align}
& \mathcal{L}_{cc}= 
\frac{g}{\sqrt{2}}\sum_{i=1}^5(\overline{u_{Li}}\gamma^\mu d_{Li})W_\mu^+
+\frac{g}{\sqrt{2}}\overline{u_{4R}}\gamma^\mu d_{4R}W_\mu 
+\frac{g}{\sqrt{2}}\overline{u_{5R}}\gamma^\mu d_{5R}W_\mu +\text{h.c.}  = \nonumber \\
=& \frac{g}{\sqrt{2}}W_\mu^+ \! \left[\left(\begin{array}{ccccc} 
\overline{u_L} & \overline{c_L} & \overline{t_L} & \overline{t_L'} & \overline{t_L''}\end{array}\right)
 \gamma^\mu V_\text{CKM,L}\!
\left(\begin{array}{c} d_L \\ s_L \\ b_L \\ b_L'\\b''_L \end{array}\right)\!  
+\!  
\left(\begin{array}{ccccc} 
\overline{u_R} & \overline{c_R} & \overline{t_R} & \overline{t_R'}& \overline{t_R''}\end{array}\right)
 \gamma^\mu  
V_\text{CKM,R}\! \left(\begin{array}{c} d_R \\ s_R \\ b_R \\ b_R' \\b''_R\end{array}\right) 
\right]  
\nonumber \\
& + \text{h.c.} 
\label{doubletcc5}
\end{align}
where $V_\text{CKM,L}$ is unitary and the mixings of the first three famillies with the vector-like isodoublets are negligibly small, while 
$V_\text{CKM,R}$ is given by:
\begin{align}
V_\text{CKM,R} &= V_R^{(u)\dag}\text{diag}(0,0,0,1,0)V_R^{(d)}
+V_R^{(u)\dag}\text{diag}(0,0,0,0,1)V_R^{(d)}\approx 	\nonumber \\
&\approx 
\left(\begin{array}{c @{\hspace{1\tabcolsep}}  c @{\hspace{2\tabcolsep}}   c @{\hspace{2\tabcolsep}} c @{\hspace{1\tabcolsep}} c}
V_{R\, 4u}^*V_{R\, 4d} &V_{R\, 5u}^*V_{R\, 5s} & 0 & V_{R\, 4u}^*V_{R\, 4b'} & V_{R\, 5u}^*V_{R\, 5b''} \\
0 & 0 & 0  & 0 & 0 \\
0 & 0 & 0  &  0 & 0 \\
V_{R\, 4t'}^*V_{R\, 4d} & 0 &  0  & V_{R\, 4t'}^*V_{R\, 4b'}  & 0 \\
0 & V_{R\, 5t''}^*V_{R\, 5s} &  0 & 0  & V_{R\, 5t''}^*V_{R\, 5b''} 
\end{array}\right) 
\end{align}
Then, the condition (\ref{sol}) is requiring:
\begin{align}
\label{soldd}
& V_{R\, 5u}^*V_{R\, 5s}= - 1.16(37) \times 10^{-3} &&
V_{R\, 4u}^*V_{R\, 4d}= -0.73(18)\times 10^{-3}
\end{align}
The additional terms in the neutral current Lagrangian are:
\begin{align}
\label{fcncdo2}
& \mathcal{L}_\text{fcnc}= 
\frac{1}{2}\frac{g}{\cos\theta_W}Z^\mu(\overline{u_{R4}}\gamma_\mu u_{R4}-\overline{d_{R4}}\gamma_\mu d_{R4}+\overline{u_{R5}}\gamma_\mu u_{R5}-\overline{d_{R5}}\gamma_\mu d_{R5})= \nonumber\\
=&\frac{1}{2}\frac{g}{\cos\theta_W}Z^\mu \! \left[ 
\left(\begin{array}{ccccc}
\overline{u_R} & \overline{c_R} & \overline{t_R} & \overline{t'_R}  & \overline{t''_R} 
\end{array}\right)\gamma^\mu V_\text{nc}^{(u)}\left(\begin{array}{c}
u_R \\ c_R \\ t_R \\ t'_R \\t''_R
\end{array} \right)\! 
- \!  
\left(\begin{array}{ccccc}
\overline{d_R} & \overline{s_R} & \overline{b_R} & \overline{b'_R} & \overline{b''_R}
\end{array}\right)\gamma^\mu V_\text{nc}^{(d)}\left(\begin{array}{c}
d_R \\ s_R \\ b_R \\ b'_R \\b''_R
\end{array} \right) \right]
\end{align}
where in this case the matrix $V_\text{nc}^{(d)}$ is:
\begin{align}
V_\text{nc}^{(d)}= & V_R^{(d)\dag}\text{diag}(0,0,0,1,1)V_R^{(d)}\approx \nonumber \\ &\approx 
\left(\begin{array}{cc@{\hspace{1\tabcolsep}}c @{\hspace{1\tabcolsep}} cc}
|V_{R4d}|^2              & 0                             & 0       & V_{R4d}^*V_{R4b'}   & 0 \\
0                               & |V_{R5s}|^2            & 0       & 0       & V_{R 5s}^*V_{R 5b''}\\
0 & 0 &0 & 0 & 0 \\
V_{R 4b'}^*V_{R 4d} & 0                             &  0     & |V_{R 4b'}|^2 & 0 \\
0                              & V_{R 5b''}^*V_{R 5s} & 0   &  0                  & |V_{R 5b''}|^2  
\end{array}\right)
\end{align}
Therefore, at first order there are no FCNC between the 
first two SM families. 
Then, the solution explaining the anomalies in the first row of CKM matrix in eq. (\ref{sol}), 
and equivalently in eq. (\ref{soldd}), can be obtained 
without contradiction with experimental constraints on flavour changing phenomena. 
Regarding flavour conserving observables, the constraint from $Z$-boson decay into hadrons
gives (\ref{Zdou}):
\begin{align}
& |V_{R4u}|^{2}+|V_{R5u}|^{2}+0.50( |V_{R4d}|^{2}+|V_{R5s}|^{2})<4.4\times 10^{-3}
\end{align}
which can be satisfied together with the relations in eq. (\ref{soldd}).
However, let us notice that a reduction of the experimental error 
by a factor less than $2$, with the same central values,
would rule out the possibility to explain all the anomalies with this kind of solution.
On the other hand, if the weak isodoublets are the solution to the CKM anomalies, 
anomolous $Z$-boson couplings with light fermions (in the ``right'' direction) should be detected if the
error-bars are reduced by a factor of about $4$.  

Again,
the LHC limit on the mass extra vector-like quark mixing with light families 
$M\gtrsim 700$~GeV \cite{CMS} implies that 
$|V_{R4d}|,|V_{R5s}|,|V_{R4,5u} |  \sim  0.03$ 
can be obtained if $|y^{u,d}_{ij}| \sim 0.12$, much larger than the bottom Yukawa coupling.
Moreover,
in order to have (for the perturbativity) $|y^{u,d}_{ij} |< 1$, the mass of the extra quarks
should be no more than $M_{4,5} \sim 6$~TeV. 

Alternatively it can be imagined that there exist
a vector-like isodoublet together with a down-type or up-type isosinglet, 
or both of them, assembling a complete vector-like fourth
family with the doublet, mixing with SM families:
\begin{align}
& q_{R4}=\left(\begin{array}{c}
u_{R4} \\ d_{R4} 
\end{array}\right)  ,\,
q_{ L4}=\left(\begin{array}{c}
u_{L4} \\ d_{L4} 
\end{array}\right)\, ; \qquad d_{L5}\, , d_{R5} \,;\qquad u_{L5}\, , u_{R5}
\end{align}
with mass matrices:
 \begin{align}
&\overline{d_{Li}}\mathbf{m}^{(d)}_{ij}d_{Rj} 
= \nonumber \\
&=\left(\begin{array}{ccccc} \overline{d_{L1}} & \overline{d_{L2}} & \overline{d_{L3}} & \overline{d_{L4}} & \overline{d_{L5}}
 \end{array}\right)\left(\begin{array}{ccccc}
& & & 0 & y^d_{15}v_{w}  \\ 
& \mathbf{y}^{(d)}_{3\times 3}\, v_{w} & & 0 &  y^d_{25}v_{w}  \\
& & & 0 & y^d_{35}v_{w}  \\
 y^d_{41}v_{w} & y^d_{42}v_{w} & y^d_{43}v_{w}  & M_q & 0 \\
  0 & 0 & 0  & 0 & M_{b''} 
\end{array}\right)\left(\begin{array}{c} d_{R1} \\ d_{R2} \\ d_{R3} \\ d_{R4} \\d_{R5} \end{array}\right) 
\nonumber  \\
&\overline{u_{Li}}\mathbf{m}^{(u)}_{ij}u_{Rj}  
= \nonumber \\
&=\left(\begin{array}{ccccc} \overline{u_{L1}} & \overline{u_{L2}} & \overline{u_{L3}} & \overline{u_{L4}} & \overline{u_{L5}}
 \end{array}\right)\left(\begin{array}{ccccc}
& & & 0 & y^u_{15}v_{w}  \\ 
& \mathbf{y}^{(u)}_{3\times 3}\, v_{w} & & 0 & y^u_{25}v_{w} \\
& & & 0 &  y^u_{35}v_{w} \\
y^u_{41}v_{w} & y^u_{42}v_{w} & y^u_{43}v_{w}  & M_q & 0 \\
  0 & 0 & 0  & 0 &  M_{t''}  
\end{array}\right)\left(\begin{array}{c} u_{R1} \\ u_{R2} \\ u_{R3} \\ u_{R4} \\u_{R5} \end{array}\right)
\label{massdoub3u} 
\end{align}
In this way, 
the non-zero couplings of the doublet with SM families
$y^d_{42}, y^u_{41}\neq 0$ can cancel the discrepancy
among the determinations of $|V_{us}|$ obtained from leptonic and semileptonic kaon decays 
and 
the mixing with the first family 
of a down-type or up-type singlet (or both of them) $y^d_{15},y^u_{15} \neq 0$
would fix the lack of unitarity of the first row by removing 
the discrepancy between the determination obtained from beta decays with the
the determinations from kaon decays.
The anomalies are explained with:
\begin{align}
& 
V_{R4u}^{*}V_{R4s}=-0.99(36) \times 10^{-3} \, , && 
|V_{L5d}|^{2}+|V_{L5u}|^{2}=1.50(36)\times 10^{-3}
\label{soldosi}
\end{align}
with $V_{Lus}=0.22424(36) $ and using
determinations (\ref{A}), (\ref{rapporto}), (\ref{vudmedio}) in eqs. (\ref{AD}), (\ref{BD}), (\ref{CD})
respectively. Here, analogously to eqs. (\ref{v4h}), (\ref{v4hu}), (\ref{sidou}), we have:
\begin{align}
& V_{L5d}\approx -\frac{y_{15}^{d*}v_{w}}{M_{b''}} \, , \qquad 
V_{L5u}\approx -\frac{y_{15}^{u*}v_{w}}{M_{t''}}\, , 
\qquad
V_{R4u}\approx -\frac{y^{u}_{41} v_w}{M_q} \, , \qquad 
V_{R4s} \approx -\frac{y^{d}_{42} v_w}{M_q} 
\end{align}
and similarly for the other elements.
Then, also in this case
the masses of the extra vector-like family cannot exceed few TeV in order 
to mantain perturbativity.
In fact, for example in order to have $|V_{R4u}|\approx |V_{R4s}| \gtrsim 0.025 $ it should be that
$M_{q}\lesssim 7 $ TeV to have $|y^{d,u}_{15}|<1$.

%

Regarding weak charged currents, both LH and RH states interact with $W$-boson,
with mixing matrices $V_\text{CKM,L}$ and $V_\text{CKM,R}$, which are not unitary.
$V_\text{CKM,R}$ is generated by the mixing of the first three RH families with the
vector-like weak doublets, as in the previous case. As regards $V_\text{CKM,L}$,
analogously to eq. (\ref{4colu}), the fifth row of the enlarged CKM matrix has mixing elements:
\begin{align}
& V_{Lt''d}\approx -V_{L5u}  \, , &&
 V_{Lt''s}\approx -V_{L5u} V_{Lus}-V_{L5c}V_{Lcs} -V_{L5t}V_{Lts} \, , && V_{Lt''b}\approx -V_{L5t} 
\end{align}
generated by the mixing of the first three LH families with the vector-like up-type isosinglet.
As for the elements of the fifth column,
analogously to eq. (\ref{4col}), (\ref{Vcbp}), (\ref{4col3}),
the elements
generated by the mixing of the first three LH families with the vector-like down-type isosinglet
are:
\begin{align}
& V_{Lub''}\approx -V_{L5d}^{*} \, , &&
 V_{Lcb''}\approx -V_{L5d}^{*} V_{Lcd}-V_{L5s}^{*}V_{Lcs}-V_{L5b}^{*}V_{Lcb} \, , && 
 V_{tb''}\approx -V_{L5b}^{*}
\end{align}
while the mixings $V_{L\alpha b'}$, $V_{L t' \alpha}$ with the vector-like isodoublet
are negligibly small.


As regards experimental constraints, the upper bounds from
flavour changing 
decays of mesons are the same as in section
\ref{sec-down}, \ref{sec-up}, \ref{sec-doub}, 
applied to vector and axial couplings of semileptonic and leptonic decays respectively,
as shown in table \ref{tabdosi}. 
\input{tab/tabelladosi}

Regarding neutral mesons mixing,
tree level and loop level contributions from both LH and RH currents are present, including
mixed contributions. 
By taking $M_{b''}=M_{t''}=M_{q}=M$ for simplicity, the new additional terms in
mixing mass of $\bar{K}^{0}\rightarrow K^{0}$ transition would be:
\begin{align}
M_{12}^\text{new}=\frac{1}{3}f^{2}_{K}m_{K^{0}}&\bigg\{
\frac{G_{F}}{\sqrt{2}}\left[(V_{L5d}^{*}V_{L5s})^{2}+(V_{R4d}^{*}V_{R4s})^{2}
-27.5(V_{R4d}^{*}V_{R4s})(V_{L5d}^{*}V_{L5s})\right]+
\nonumber \\
& 
\; \frac{G_{F}^{2}}{16\pi^{2}}M^{2}\Big[(V_{L5d}^{*}V_{L5s})^{2}+
2 (V_{R4d}^{*}V_{R4s})^{2}
-54.9 (V_{R4d}^{*}V_{R4s})(V_{L5d}^{*}V_{L5s})+
\nonumber \\ &
\qquad \qquad\; + 
\left(|V_{L5u}|^{2}V_{Lus}+V_{L5u}^{*}V_{L5c}+V_{L5u}^{*}V_{L5t}V_{Lts}  \right)^{2}
\Big]\bigg\} 
\end{align}
(where we neglected terms growing logarithmically with the mass of the extra quarks)
and for $D^{0}$ mesons system:
\begin{align}
M_{12\,\text{new}}^{D}=\frac{1}{3}f^{2}_{D}m_{D^{0}}&\bigg\{
\frac{G_{F}}{\sqrt{2}}\left[(V_{L5u}^{*}V_{L5c})^{2}+(V_{R4u}^{*}V_{R4c})^{2}
-3.65 (V_{R4u}^{*}V_{R4c})(V_{L5u}^{*}V_{L5c})\right]+
\nonumber \\
& 
\; \frac{G_{F}^{2}}{16\pi^{2}}M^{2}\Big[(V_{L5u}^{*}V_{L5c})^{2}+
2 (V_{R4u}^{*}V_{R4c})^{2}
-7.3 (V_{R4u}^{*}V_{R4c})(V_{L5u}^{*}V_{L5c})+
\nonumber \\ &
\qquad \qquad\; + 
\left(|V_{L5d}|^{2}V_{Lcd}^{*}+V_{L5d}^{*}V_{L5s}+V_{L5d}^{*}V_{L5b}V_{Lcb}^{*}  \right)^{2}
\Big]\bigg\} 
\end{align}
and similarly for neutral B-mesons systems.
Therefore, 
in this case there can also be cancellations.
However, also constraints on flavour conserving processes should be taken into account.
The results on $Z$-boson decay rate into hadrons
give the bound (see sections \ref{Zdown}, \ref{Zup}, \ref{Zdoub}): 
\begin{align}
\label{Zdosi}
& |V_{R4u}|^{2}+|V_{R4c}|^{2}+0.50(|V_{R4s}|^{2}+|V_{R4d}|^{2}+|V_{R4b}|^{2})+\nonumber \\
&\qquad +2.72(|V_{L5d}|^{2}+|V_{L5s}|^{2}+|V_{L5b}|^{2})+2.24(|V_{L5u}|^{2}+|V_{L5c}|^{2})<4.4\times 10^{-3}
\end{align}

\begin{figure}[t]
\centering
\includegraphics[width=0.55\textwidth]{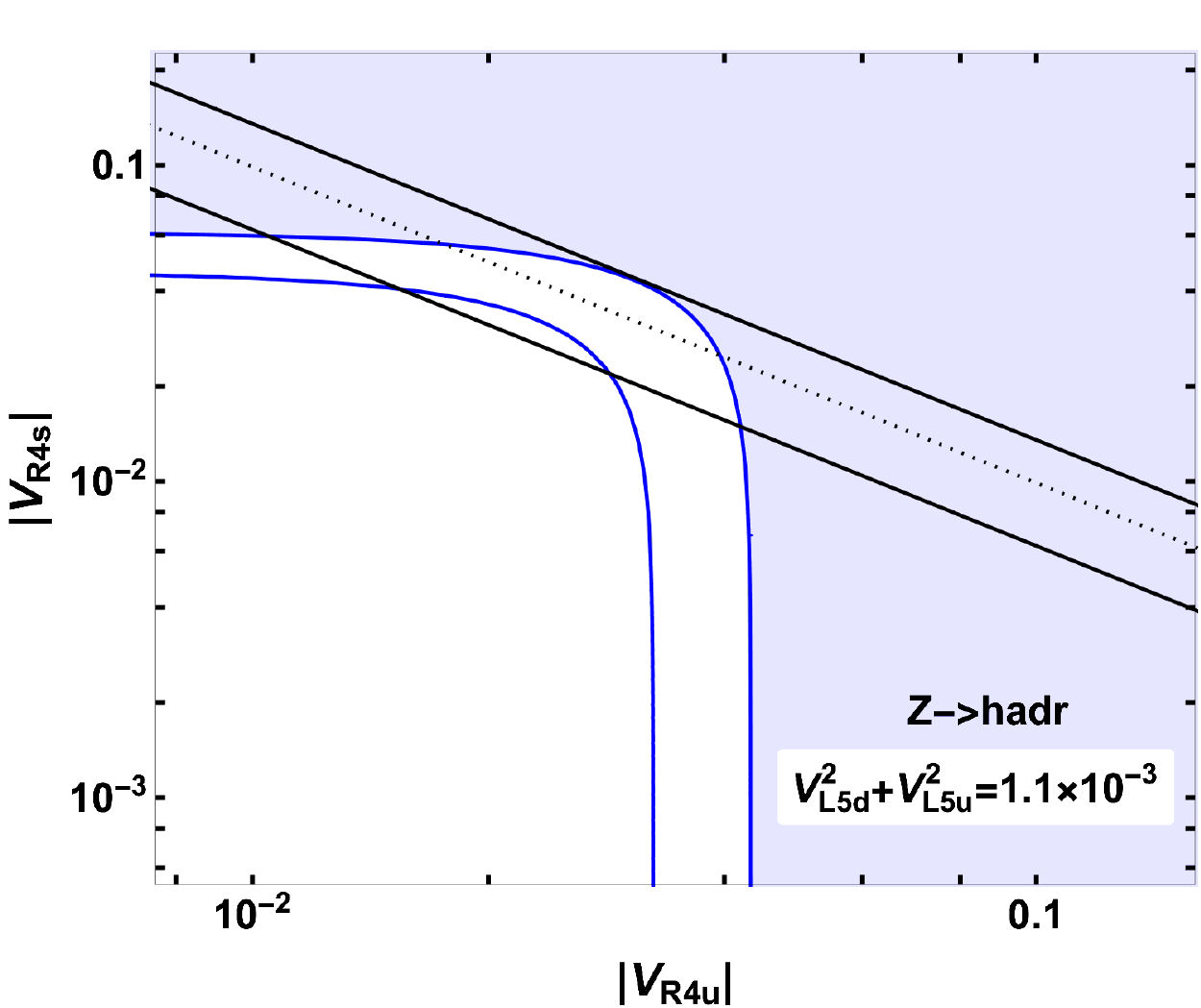}
\caption{\label{vr4svr4u} Excluded area in the $|V_{R4u}|$ and $|V_{R4s}|$ parameter space.
The blue area is excluded by experimental constraints from $Z$-boson decay rate into hadrons
(\ref{Zdosi}) using $|V_{L5d}|^{2}+|V_{L5u}|^{2}=1.1\times 10^{-3}$.
The blue curve indicates where the bound is shifted if the condition
$|V_{L5d}|^{2}+|V_{L5u}|^{2}=1.5\times 10^{-3}$  (\ref{soldosi}) is used.
The black curve stands for the solution 
$|V_{R4u}^{*}V_{R4s}|=0.99(36) \times 10^{-3}$ in eq. (\ref{soldosi}).
}
\end{figure}
Figure \ref{vr4svr4u} shows the parameter space in a simplified case
in which $y_{41}^{d}=y_{25}^{d}=y_{42}^{u}=y_{25}^{u}=0$
in eq. (\ref{massdoub3u}), corresponding to $V_{R4d}=V_{L5s}=V_{R4c}=V_{L5c}=0$.
The blue area is excluded by the condition (\ref{Zdosi}) using the conservative choice
$|V_{L5d}|^{2}+|V_{L5u}|^{2}=1.1\times 10^{-3}$ for the solution in eq. (\ref{soldosi}).
The solution 
$|V_{R4u}^{*}V_{R4s}|=0.99(36) \times 10^{-3}$ in eq. (\ref{soldosi}) is indicated by the 
black curve.
In this scenario with $V_{R4d}=V_{L5s}=V_{R4c}=V_{L5c}=0$,
the constraint from neutral $D$-mesons mixing would give (\ref{vcbpvubpdd}): 
\begin{align}
& |V_{Lub''}V_{Lcb''}^{*}|\approx |V_{L5d}|^{2}|V_{cd}|
<4.6\times 10^{-4}\left[ \frac{1\,\text{TeV}}{M_{b''}}\right] \nonumber \\
& |V_{L5d}|<0.046 \left[ \frac{1\,\text{TeV}}{M_{b''}}\right]^{1/2}
\end{align}
Kaon mixing would receive a real contribution at loop level giving the constraint (\ref{vtsvtdkk}):
\begin{align}
& |V_{Lt''d}V_{Lt''s}^{*}|\approx \left| |V_{L5u}|^{2}V_{Lus}+V_{L5u}V_{L5t}^{*}V_{Lts}\right|
<5.2\times 10^{-4}\left[ \frac{1\,\text{TeV}}{M_{b''}}\right] 
\end{align}
which together with the constraints from B-mesons decays and mixing
gives:
\begin{align}
& |V_{L5u}|\lesssim 0.049 \left[ \frac{1\,\text{TeV}}{M_{t''}}\right]^{1/2}
\end{align}
This means that, in order to have
$|V_{L5d}|^{2}+|V_{L5u}|^{2}>1.1\times 10^{-3}$ (\ref{soldosi}), it should be for example
$M_{b''}\approx M_{t''}< 4.1$ TeV.


\section{Conclusion}
\label{conclusion}

Present situation of the determination of the CKM matrix elements is very intriguing. 
From one side, there is a controversy in the determination of $\vert V_{us} \vert$.  
Namely, given the present experimental accuracy in kaon leptonic and semileptonic decays 
and present theoretical precision in the calculation of form factors (lattice QCD), 
there is about $3\sigma$ tension between the  $\vert V_{us} \vert$ values 
extracted from $K\ell3$ (determination A) and $K\mu2$ (determination B) decays, 
as discussed in section \ref{situation}. Although the recent high precision calculations  
in the kaon decays $K\ell3$ \cite{Seng:2021boy} and $K\mu2$ \cite{Martinelli} demonstrate 
that the radiative corrections should not be responsible for it, this tension  per s\`e 
cannot be considered that alarming since still there can be some theoretical loopholes 
in the interpretation of the lattice QCD results, or perhaps some unfixed systematics  
in the measurements of the kaon decay rates. In this case, the real value of $\vert V_{us} \vert$
could be near the average $A+B$ between the two above determinations. 

On the other side, the recent calculations of the short distance radiative corrections 
in the neutron $\beta$-decay with improved hadronic uncertainties \cite{Seng,Marciano2} 
leads to high accuracy in the determination of $\vert V_{ud} \vert$ from superallowed $0^+-0^+$ 
nuclear transitions (determination C). The obtained $\vert V_{ud} \vert$ value, 
in combination with the determinations of $\vert V_{us} \vert$, indicates towards a
violation of the CKM unitarity at about $4\sigma$ level. Namely, there appears a significant 
deficit $\delta_{\rm CKM}$ in the first row unitarity, as given by Eq. (\ref{newundelta}). 

These anomalies, if confirmed with future high precision data,  
would indicate towards some new physics beyond the Standard Model. 
In Ref. \cite{Belfatto:2019swo} we pointed out that the anomalies could be originated by mixing 
of ordinary light quarks to some extra vector-like quarks with masses at the TeV scale.  
In this paper we gave more detailed study of these scenarios, analyzing one by one 
the implications of vector-like quarks in the weak isosinglet or isodoublet representations. 
By introducing a weak isosinglet quark of the up ($U$-type)  or down ($D$-type) quarks, 
 one can explain the CKM unitarity anomaly, i.e. deficit of the first row unitarity (\ref{newundelta}), 
 but the tension between the two determinations of  $\vert V_{us} \vert$ cannot be explained 
 in this case. However, if  the latter discrepancy is taken seriously one has to look for a
a solution addressing the whole situation. By introducing a weak isodoublet  ($Q$-type) of
quarks one could potentially explain both  anomalies.  

However, there are strong phenomenological limits on the hypothetical vector-like quarks:  
participation of vector-like quark species in the quark mixing induces
flavor changing effects at the tree level as well as via box diagrams involving heavy species. 
In fact, we show that the latter give a bigger contribution than the tree level one
if the masses of heavy quarks are larger than 3 TeV or so.  
However, tree level effects are independently testable in $Z$-boson physics  
since $Z$-boson acquires small flavor non-diagonal couplings with quarks while also its 
flavor-diagonal couplings get isospin violating contributions which can be 
confronted with the 
limits on many observables related to $Z$-boson physics.  

Two approaches to the above anomalies can be considered.
The incompatibility between two determinations A and B from kaon physics 
may be attributed to some uncertainties which perhaps will disappear 
with more precise calculations. Neglecting this problem, one can focus instead 
on the unitarity violation problem, i.e. on the unitarity deficit which emerges by  
confronting the average $A+B$ value of $\vert V_{us} \vert$  obtained from kaons, 
with the value of $\vert V_{ud} \vert$ extracted from superallowed nuclear transitions. 

The just mentioned scenarios involving extra isosinglets quarks are on this line.
Namely, the CKM problem can be solved provided that the first family of quarks has 
a mixing $\vert V_{ub'} \vert \simeq 0.04$ with extra $D$-type quark $b'$, or alternatively 
the same size mixing $V_{t'd}$ with $U$-type quark $t'$. 
For this scenario the most severe limits come from the flavor changing 
phenomena, but comparable limits emerge from the precision data on $Z$-boson decays.
There still remains some available parameter space, although not very large.
In fact, we show that constraints
from $K^0-\overline{K}^0$ and $D^0-\overline{D}^0$ systems become 
more stringent for larger masses of $b'$ or $t'$. In particular,
e.g. for $M_{b'} > 1.5$~TeV they would 
exclude the possibility of having extra mixing as large as $\vert V_{ub'} \vert \sim 0.04$
if the unitarity deficit is due to a down-type species or, in the case of up-type extra quark, 
$M_{t'} > 2.5$~TeV would 
exclude the possibility of having the extra mixing $\vert V_{t'd} \vert \sim 0.04$
(see sections \ref{downend}, \ref{upend}). 
Therefore, we claim that if the CKM unitarity anomaly is due to the mixing with extra isosinglet 
($U$ or $D$ type) quarks, then ``4th family" states $b'$ or $t'$ should be discovered with a mass 
below a couple of TeV or so. 

As we noted above, both $V_{us}$ and CKM unitarity anomalies can be solved by 
introducing $Q$-type extra quarks having substantial mixings with both 1st and 2nd families 
of the normal quarks. In fact, kaon semileptonic  decays $K\ell3$ measure the quark vector current 
coupling to $W$-boson while leptonic decays $K\mu 2$ measure the axial current coupling. 
While both vector and axial couplings should be identical in the SM frames where $W$-boson 
couples solely to LH quarks, their difference can be induced via the mixing with weak 
isodoublet $Q$-type quarks.  The latter in fact induces some small $W$-boson couplings 
with the RH quarks which can be at the origin of the discrepancy between the two determinations 
A and B of $\vert V_{us} \vert$ element.  
However, we show that this solution with only one $Q$-type species is 
fully excluded by the flavor changing limits together with $Z$ boson physics and electroweak low energy observables (section \ref{doubend}). 

This brings us to conclude that the full solution cannot be achieved by 
introducing the only one species of extra $Q$-type quarks: it should be complemented 
by isosinglet quarks of $U$ or/and $D$ type, or by another isodoublet species. 
In this case somewhat larger available parameter space can be found with
the present limits on flavor-changing 
phenomena and on anomalous  $Z$-couplings. 
However, these scenarios are testable with the future experimental limits (section \ref{other}). 
In particular, 
in the scenario with two isodoublets, flavour changing limits can be softened, and then
the masses of the extra quarks would not be strictly constrained, apart from a perturbativity bound
(Yukawa couplings less than $1$), which implies 
$M_{q}<7$ TeV or so. However, in this case the main issue
would come from limits in $Z$-boson physics.
In the case of one extra isodoublet with isosinglets (up or/and down type), also
constraints from flavour changing phenomena should be considered.
We conclude that if the CKM anomalies are due to extra vector-like 
quarks, then ``4th family" quarks should be discovered with masses of few TeV, and 
anomalous $Z$-couplings should be detected by improving the experimental 
precision. 

For the minimality reasons, we did not consider extra vector-like species of leptons. 
However, if they also exist at the TeV scale and have the same size mixings with the electron and 
muon, then the experimental limits from the flavor changing phenomena involving leptons 
or limits on $Z$-boson flavor-nondiagonal couplings as $Z \overline{e} \mu$ 
would be much more stringent.

\acknowledgments

The work of B.B. was supported by the ERC research grant NEO-NAT no. 669668.
The work of Z.B. was supported in part by Ministero dell'Istruzione, Universit\`a e della Ricerca (MIUR)
under the program PRIN 2017, grant no. 2017X7X85K ``The Dark Universe: A Synergic Multimessenger Approach",  
and in part by Shota Rustaveli National Science Foundation 
(SRNSF) of Georgia, grant DI-18-335 ``New Theoretical Models for Dark Matter Exploration".

\end{document}

%% file: tab/tabelladown2.tex
\begin{table}
{\small 
\begin{tabular}{| l   @{\hspace{1\tabcolsep}} 
l  
@{\hspace{2.\tabcolsep}} l  
|}
\hline
Process & Constraint & 
\\
\hline
$K^+\rightarrow \pi^+\nu\bar{\nu}$ & 
$|V_{L4s}^{*}V_{L4d}| < 2.5\times 10^{-5} $ 
  & $ |V_{L4s}| < 8.2 \times 10^{-4} \left[ \frac{0.03}{|V_{ub'}|} \right]$ 
  \\
$K_\text{L}\rightarrow \pi^0\nu\bar{\nu}$ & 
$  |V_{L4s}^{*}V_{L4d}| |\sin\delta^d_{L21}|< 2.2\times 10^{-5}  $
& $|V_{L4s}| 
< \frac{7.4\times 10^{-4}}{|\sin\delta^d_{L21}|} \left[  \frac{0.03}{|V_{ub'}|} \right]$ 
\\
  $K_\text{L}\rightarrow \pi^0e^+e^-$ & 
  $|V_{L4s}^{*}V_{L4d}||\sin\delta^d_{L21}|<1.7\times 10^{-5} $
 & $|V_{L4s}| < \frac{5.5\times 10^{-4}}{|\sin\delta^d_{L21}|} \left[  \frac{0.03}{|V_{ub'}|} \right] $ 
 \\
  $K_\text{L}\rightarrow \mu^+\mu^-$ &  
 $-0.3\times 10^{-5}< |V_{L4s}^{*}V_{L4d}| \cos\delta^d_{L21}<1.1\times 10^{-5} $
 & $|V_{L4s}| < \frac{3.7\times 10^{-4}}{|\cos\delta^d_{L21}|} 
\left[  \frac{0.03}{|V_{ub'}|} \right]$ 
\\
$K_S\rightarrow\mu^+\mu^-$ &   
 $|V_{L4s}^{*}V_{L4d}||\sin\delta^d_{L21}|<1.0\times 10^{-4}  $
& $|V_{L4s}|< \frac{3.5 \times 10^{-3}}{|\sin\delta^d_{L21}|}
 \left[  \frac{0.03}{|V_{ub'}|} \right]  $ 
 \\
$K^0$-$\bar{K}^0$ & $ 
 |V_{L4s}^{*}V_{L4d}|  
 < 1.7 \times 10^{-4} 
 $
 & $|V_{L4s}| < 
 5.8 \times 10^{-3}
 \left[ \frac{0.03}{|V_{ub'}|}\right] $
  \\
 &  $ |V_{L4s}^{*}V_{L4d}|\sqrt{|\sin(2\delta^d_{L21})|}<1.3\times 10^{-5}  
 $
 & $|V_{L4s}|< \frac{4.4\times 10^{-4}}{\sqrt{|\sin (2\delta^d_{L21})|}} \left[ \frac{0.03}{|V_{ub'}|} \right] $ 
\\
\hline
$B^0$-$\bar{B}^0$ &
$|V_{L4b}V_{L4d}^*| <3.3 \times 10^{-4} $
& $ |V_{L4b}| < 1.1\times 10^{-2} \left[\frac{0.03}{|V_{ub'}|}\right]$ 
\\
$B^0\rightarrow \pi^0\ell^+\ell^-$ & 
$|V_{L4b}V_{L4d}^*| < 4.3 \times 10^{-4} $
& $|V_{L4b}| <   1.4 \times 10^{-2}  \left[ \frac{0.03}{|V_{ub'}|} \right]$
 \\
$B^{\pm}\rightarrow \pi^{\pm}\ell^+\ell^-$ & 
$|V_{L4b}V_{L4d}^*| < 2.6 \times 10^{-4} $
& $ |V_{L4b}| < 8.7\times 10^{-3} \left[ \frac{0.03}{|V_{ub'}|} \right]$
 \\
$B^0\rightarrow\mu^+\mu^-$ & 
$ |V_{L4b}V_{L4 d}^*| <2.2 \times 10^{-4} $
& $ |V_{L4b}| <  7.4\times 10^{-3} \left[ \frac{0.03}{|V_{ub'}|} \right] $
\\
\hline
$B^0_{s}$-$\bar{B}_{s}^0$ 
&  $|V_{L4b}V_{L4s}^*| < 1.6 \times 10^{-3}  $ &
\\
$B\rightarrow X_s\ell^+\ell^-$ &
 $|V_{L4b}V_{L4s}^*| <  1.9\times 10^{-3} $ &
  \\
$B^0_s\rightarrow\mu^+\mu^-$ &  
$|V_{L4b}V_{L4 s}^*|  < 8.1 \times 10^{-4}  $ &
\\
\hline
$D^0$-$\bar{D}^0$ & 
$ |V_{ub'}V_{cb'}^*|< 4.6 \times 10^{-4} \left[ \frac{1\,\text{TeV}}{M_{b'}}  \right] $
& $ |V_{cb'}| < 1.5 \times 10^{-2} \left[ \frac{0.03}{|V_{ub'}|} \right]$
 \\
\hline
$Z \rightarrow b\bar{b}$ & 
$|V_{L4b}|< 3.2\cdot 10^{-2}$ &
 \\
 \hline
$\Gamma_Z$, $Z \rightarrow \text{hadr}$ \: & $|V_{L4d}|< 0.041$ &
 \\
\hline
\end{tabular}
}
\caption{\label{limiti} Limits on the mixing of the SM three families with a fourth
down-type vector-like isosinglet. As in eq. (\ref{tildes}), $(\delta^d_{Li}-\delta^d_{Lj})=\delta^d_{Lij}$.
Regarding the elements of CKM matrix, $V_{ub'}\approx -V_{L4d}^{*}$,
$V_{cb'}\approx  -V_{L4d}^{*}V_{cd} -V_{L4s}^{*}V_{cs} -V_{L4b}^{*}V_{cb}$,
$V_{tb'}\approx -V_{L4b}^{*}$ as in eqs (\ref{4col}), (\ref{Vcbp}), (\ref{4col3}).
For details see the text.
}
\end{table}

%% file: tab/tabellaup2.tex
\begin{table}
\centering
{\small 
\begin{tabular}{| l   @{\hspace{2\tabcolsep}} 
l  
@{\hspace{4.\tabcolsep}} l  
|}
\hline
Process & Constraint &
\\
\hline
$D^+\rightarrow \pi^+ \mu^+ \mu^-$ & 
$ |V_{L4c}V_{L4u}^*|< 2.0 \times 10^{-3}
$
&  $|V_{L4c}|< 6.7 \times 10^{-2} \left[ \frac{0.03}{|V_{L4u}|} \right]$
\\
$D^0\rightarrow \pi^0 e^+ e^-$ & 
$|V_{L4c}V_{L4u}^*|<3.3\times 10^{-2}
$ & \\
  $D^0\rightarrow \mu^+\mu^-$ & 
  $ |V_{L4c}V_{L4u}^*|  < 2.0 \times 10^{-3}
$  & $|V_{L4c}| < 6.8\times 10^{-2} \left[ \frac{0.03}{|V_{L4u}|} \right]$ 
 \\
$D^0$-$\bar{D}^0$ & $ 
|V_{L4c}V_{L4u}^*|< 1.5 \times 10^{-4} 
$ & $ |V_{L4c}|< 5.0 \times 10^{-3} \left[ \frac{0.03}{|V_{L4u}|} \right] $
  \\
\hline
$K^0$-$\bar{K}^0$ & 
$ |V_{t's}^*V_{t'd}|<( 0.1  \div 5.8) \times 10^{-4} $
& $ |V_{t's}|< 1.9 
 \times 10^{-2} \left[ \frac{0.03}{|V_{L4u}|}\right]  $ 
\\
\hline
$B^0$-$\bar{B}^0$ 
&  $|V_{t'b}^*V_{t'd}| < ( 0.4  \div 2.6) \times 10^{-3}
$ & $  |V_{t'b}| <  8.5 \times 10^{-2} 
\left[\frac{0.03}{|V_{L4u}|}\right]$
\\
\hline
$B^0_{s}$-$\bar{B}_{s}^0$ 
&  $|V_{t'b}^*V_{t's}| < ( 0.2  \div 1.2) \times 10^{-2} $ &
\\
 \hline
$\Gamma_{Z}$, $\Gamma\rightarrow\text{hadr}\,$ & $|V_{L4u}|^2+|V_{L4c}|^2<  2.0\cdot 10^{-3} $, 
$|V_{L4u}|< 0.044$ &
 \\
\hline
\end{tabular}
}
\caption{\label{limiti2} Limits on the mixing of the SM three families with a 
vector-like up-type isosinglet $t'$. 
The upper bounds obtained from neutral mesons mixing depend
on the relative phases of the mixing elements and on the extra quark mass.
Here the limits are computed for $M_{t'}=1$ TeV (see the text for details). 
Regarding the elements of CKM matrix, $V_{t'd}\approx -V_{L4u}$,
$V_{t's}\approx  -V_{L4u}V_{us} -V_{L4c}V_{cs} -V_{L4t}V_{ts}$ and
$V_{t'b}\approx -V_{L4t}$, as in eq. (\ref{4col}).
}
\end{table}

%% file: tab/tabelladosi.tex
\begin{table}
\centering
{\small 
\begin{tabular}{| l   @{\hspace{1\tabcolsep}} 
l  |}
\hline
Process & Constraint  
\\
\hline
$K^+\rightarrow \pi^+\nu\bar{\nu}$ & 
$|V_{R4s}^{*}V_{R4d}+V_{L5s}^{*}V_{L5d}| < 2.5\times 10^{-5} $ 
  \\
$K_\text{L}\rightarrow \pi^0\nu\bar{\nu}$ & 
$| \text{Im}( V_{R4s}^{*}V_{R4d}+V_{L5s}^{*}V_{L5d})|< 2.2\times 10^{-5}  $
\\
  $K_\text{L}\rightarrow \pi^0e^+e^-$ & 
  $| \text{Im}( V_{R4s}^{*}V_{R4d}+V_{L5s}^{*}V_{L5d})|<1.7\times 10^{-5} $
 \\
  $K_\text{L}\rightarrow \mu^+\mu^-$ &  
 $-0.3\times 10^{-5}< | \text{Re}( V_{L5s}^{*}V_{L5d}-V_{R4s}^{*}V_{R4d})|<1.1\times 10^{-5} $
\\
$K_S\rightarrow\mu^+\mu^-$ &   
 $|\text{Im}( V_{L5s}^{*}V_{L5d})-V_{R4s}^{*}V_{R4d})|<1.0\times 10^{-4}  $
 \\
\hline
$B^{\pm}\rightarrow \pi^{\pm}\ell^+\ell^-$ & 
$|V_{L5b}V_{L5d}^*+V_{R4b}V_{R4d}^*| < 2.6 \times 10^{-4} $
 \\
$B^0\rightarrow\mu^+\mu^-$ & 
$ |V_{L5b}V_{L5d}^*-V_{R4b}V_{R4d}^*| <2.2 \times 10^{-4} $
\\
\hline
$Z \rightarrow b\bar{b}$ & 
$|V_{L5b}|^{2}+0.18|V_{R4b}|^{2}< 1.0 \times 10^{-3}$ 
 \\
 \hline
\end{tabular}
}
\caption{\label{tabdosi} Limits on the mixing of the SM three families with the extra vector-like 
weak doublet $q_{L4},q{R4}$ and the weak singlets $u_{L5},u_{R5}$, $d_{L5},d_{R 5}$.
See the text for constraints obtained from neutral mesons mixing and $Z$ decay into hadrons.
}
\end{table}